\documentclass[10pt,oneside,letterpaper]{book}
\usepackage[utf8]{inputenc}

\usepackage{tocloft,calc}

\makeatletter
\g@addto@macro\appendix{%
	\addtocontents{toc}{%
		\protect%
	}%
}
\makeatother

\usepackage{amsmath}
\usepackage{amsfonts}
\usepackage{amssymb}
\usepackage{graphicx}
\usepackage{slashed}

\usepackage{color}
\usepackage{soul}

\usepackage{anyfontsize}
\usepackage{setspace}
\usepackage{calrsfs}
\usepackage{tikz}
\usetikzlibrary{arrows,decorations.pathmorphing,backgrounds,positioning,fit,matrix}
\usepackage[left=1.50in, right=1.25in, top=1.25in, bottom=1.25in]{geometry}
\author{Shaoyang Jia}
\begin{document}
	\doublespacing
	\pagestyle{plain}
	\frontmatter
	\pagenumbering{gobble}
	\begin{titlepage}
		\centering
		{\LARGE\bfseries Formulating Schwinger--Dyson equations for QED propagators in Minkowski space\par}
		\vspace{2cm}
		{\Large Shaoyang Jia\par}
		\vspace{2cm}
		{\large Dissertation presented to the Graduate Faculty
		of The College of William \& Mary in Candidacy for the Degree of
		Doctor of Philosophy\par}
		\vspace{1.5cm}
		\vfill
		supervised by\par
		Michael \textsc{Pennington}
		\vfill	
		{\large \today\par}
	\end{titlepage}
	
	\section*{Abstract} 
		The Schwinger--Dyson equations (SDEs) are coupled integral equations for the Green's functions of a quantum field theory (QFT). The SDE approach is the analytic nonperturbative method for solving strongly coupled QFTs. When applied to QCD, this approach, also based on the first principle, is the analytic alternative to lattice QCD. However, the SDEs for the n-point Green's functions involves (n+1)-point Green's functions (sometimes (n+2)-point functions as well). Therefore any practical method for solving this infinitely coupled system of equations requires a truncation scheme. When considering strongly coupled QED as a modeling of QCD, naive truncation schemes violate various principles of the gauge theory. These principles include gauge invariance, gauge covariance, and multiplicative renormalizability. The combination of dimensional regularization with the spectral representation of propagators results in a tractable formulation of a truncation scheme for the SDEs of QED propagators, which has the potential to preserve the aforementioned principles and renders solutions obtainable in the Minkowski space.
		\newpage
	\pagenumbering{roman}	
			\renewcommand*{\contentsname}{\fontsize{14}{14}\selectfont \normalfont \hspace{4.5cm} TABLE OF CONTENTS\vspace{-1.35cm}}
			\fontsize{12}{12}\selectfont
			\tableofcontents
\chapter*{\fontsize{14}{14}\selectfont \normalfont\centering ACKNOWLEDGMENTS}
\addcontentsline{toc}{chapter}{Acknowledgments}
\vspace{-1.0cm} 
\begin{spacing}{1.0}
	\flushleft{\noindent\fontsize{12}{12}\selectfont
		Research contained in this dissertation could not be made possible without the wisdom and patience of my advisor Michael Pennington. I am grateful for his guidance throughout these years and feel privileged to have him as my supervisor. I would also like to than Prof Kostas Orginos, Prof Seth Aubin, Dr Ian Clo\"{e}t, and Prof Anatoly Radyushkin for their reviews and valuable suggestions to this work. 
		\newline
		\newline
		Additionally, I would like to thank Professor Keith Ellis and other members of the Institute for Particle Physics Phenomenology (IPPP) of Durham University for kind hospitality during my visit. I would also like to thank Dr Craig Roberts and Dr Ian Clo\"{e}t for many discussion we had during my visit to Argonne National Laboratory. This material is based upon work supported by the U.S. Department of Energy, Office of Science, Office of Nuclear Physics under contract DE-AC05-06OR23177 that funds Jefferson Lab research. 
	}
	\vspace{1ex}
\end{spacing}

\newpage

\addcontentsline{toc}{chapter}{Dedications}
\vspace*{\fill}
%
%
\begin{center}
	This dissertation is dedicated to the memory of my grandfather Rong Li. \\For the countless questions he answered inspired the science inside a young boy.
	\vskip .8in
\end{center}
\vspace*{\fill}
	\mainmatter
	\fontsize{10}{10}\selectfont
	\doublespacing
	\chapter{Introduction}
	The formalism of quantum field theories (QFTs) is intended to describe the mechanics of relativistic quantum particles. One perfect, yet unsolved, example of such a system is the confinement of quarks inside hadrons. Compared with the rest mass of nucleons and pions, the sums of their valence quark current masses are tiny. Therefore the majority mass of a light hadron is present in the form of binding energy, with their constituents moving at relativistic speeds. Because the strong nuclear force dominates the dynamics at the hadron scale, quantum chromodynamics (QCD) is the fundamental theory to describe the stable structures of hadrons. While decay phenomena are explained by the theory of electro-weak integrations in conjunction with QCD. While the electro-weak interaction remains weak at the hadron scale, the strong coupling constant is large, invalidating perturbative QCD expansions. Therefore nonperturbative approaches to QCD are required to understand the structure of hadrons. Among these approaches there are effective field theories \cite{Ebert:1994mf}, QCD sum rules \cite{Reinders:1984sr}, lattice QCD \cite{Beane:2008dv,Beane:2010em}, light-front quantization \cite{Vary:2009gt,Bakker:2013cea}, and Schwinger--Dyson equations \cite{Schwinger:1951ex,Schwinger:1951hq,Dyson:1949ha}.

Although the lattice QCD approach requires a significant amount of computing power, algorithm improvements combined with the Moore's law have allowed results to be computed more easily and with controlled systematic error. Alternatively the Schwinger--Dyson/Bethe--Salpeter/Faddeev approach \cite{Bashir:2012fs,Cloet:2013jya} to hadronic physics is far less numerically demanding than lattice QCD. Another known advantage of the SDE approach is that physics can be understood in the spacetime continuum free from effects of discretization. Better intuition about physics behind the theory can also be obtained when results are presented in analytic forms. However, the requirement of truncations introduces unknown uncertainties propagating to the final results. 

\begin{figure}
	\centering
	\includegraphics[width=0.65\linewidth]{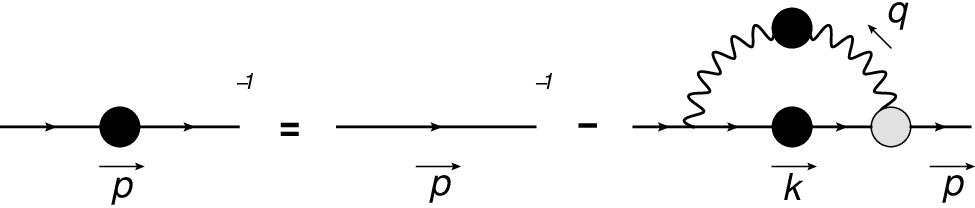}
	\caption{The diagrammatic representation of the SDE for the fermion propagator in momentum space. The fermion-photon vertex is unknown and an Ansatz is required to solve this equation.}
	\label{fig:DSE_fermion_ori}
\end{figure}

The SDE for the fermion propagator in momentum space is represented by Fig.~\ref{fig:DSE_fermion_ori}. 
It has been solved extensively using specific Ans\"{a}tze for the fermion-photon vertex \cite{Curtis:1993py,Kizilersu:2014ela} (also see Ref.~\cite{Fukuda:1976zb,Atkinson:1993mz,Curtis:1992jg,Kizilersu:2000qd,Kizilersu:2001pd,Bloch:1995dd,Bashir:1994az,Bashir:2011dp,Kondo:1990ky,Kondo:1990st,Bashir:2011ij,Kizilersu:2013hea}). Solutions in the Minkowski space have been obtained by \cite{Maris:1994ux} (also see Ref.~\cite{Maris:1991cb,Atkinson:1989fp,Kusaka:1997xd,Sauli:2001we,Gutierrez:2016ixt}). Alternatively, these equations can be solved using complex conjugate poles to represent the propagator functions \cite{Bhagwat:2002tx,Bhagwat:2003vw,Raya:2015gva,Chang:2013pq}.

Previous calculations of loops in the SDEs for fermion propagator and photon propagator were often regularized by a cut-off. For different regularization schemes, comparisons have been made in Refs.~\cite{Kizilersu:2000qd,Kizilersu:2001pd}. Since the cut-off regulator violates translational invariance, there is an ambiguity in the choices of loop momenta. One criterion on the correctness of the loop momentum is maintaining the transversality of the vacuum polarization tensor. Another improvement in the SDE approach is the introduction of the spectral representations of Green's functions, allowing SDEs to be solved directly in Minkowski space without the Wick rotation. In contrast, lattice QCD is formulated in the discretized Euclidean spacetime. The spectral representation also allows the direct application of Feynman parameterization and dimensional regularization. This set of procedures ensures the removal of the ambiguities in choosing the loop momenta.

To offset the most significant drawback of the SDE approach, the proper truncation of SDEs should respect various principles of the theory under study. For QCD in particular, the $SU(3)$ gauge group with its non-Abelian nature renders gauge symmetries complicated to preserve. While the $U(1)$ symmetry is much easier to handle, at the same time QED has gauge invariance, gauge covariance, and renormalizability built in. Therefore it is instructive to start with QED as a modeling of QCD to develop truncation schemes that preserve common principles of both gauge theories. The gauge invariance of QED manifests itself as Ward--Green--Takahashi identities (WGTIs). The original WGTI relates the longitudinal fermion-photon three-point function to the fermion propagator, which has been utilized to construct the Ball--Chiu vertex \cite{Ball:1980ay}. The transverse WGTIs \cite{He:1999hb,He:2000we,He:2006my} should, in principle, determine the transverse pieces of the fermion-photon vertex. However, these involve nonlocal terms and coupling to other QED vertices, which renders them impractical to solve. 

Because Green's functions are not physical observables, they can depend on the gauge fixing of the theory. The correct behavior of the Green's functions with changes in gauge fixing conditions is called gauge covariance. In covariant gauges, Green's functions in one gauge are related to themselves in another gauge by the Landau--Khalatnikov--Fradkin transform (LKFT) \cite{Landau:1955zz,Fradkin:1955jr}. Being a renormalizable theory, QED in four-dimensions only contains three primitive divergent diagrams. The ability to remove these divergences by multiplying the corresponding renormalization constants is named multiplicative renormalizability. The transverse supplement to the Ball--Chiu vertex can be constructed to satisfy multiplicative renormalizability \cite{Curtis:1993py}. Respecting these principles of QED helps control the systematics of any truncation scheme of SDEs. Achieving so paves the path towards the nonperturbative truncation of QCD SDEs.

Although Green's functions are not directly observable through experiments, they are part of the theory to understand the physical world, therefore cannot be arbitrary objects. The mathematical structures of the Green's functions should reflect our theory's understanding of the physical phenomena. Specifically for the propagators, their behavior near the mass shell corresponds to the asymptotic states observed by particle detectors. This can be seen through the well known Lehmann--Symanzik--Zimmermann reduction formula \cite{Lehmann:1954rq}. For confined particles, including quarks and gluons inside hadrons, their propagators should not have any on-shell behavior. Therefore the momentum space propagators of a QFT should not be more singular than the free-particle propagator. This observation lays the foundation of an integral representation of the momentum space propagators, namely the spectral representation. Specifically, the propagator function can be written as a linear combination of free-particle propagators weighted by the spectral function, which contains delta-functions and theta-functions. The delta-function terms of the spectral function correspond to the on-shell terms of the propagator, while the theta-function terms generate branch cuts along the positive real axis of the complex momentum plane corresponding to the timelike region. The spectral representation of the propagators is shown to be an elegant way to condense the analytic structures of the propagators into real functions. It also allows SDEs to be solved in Minkowski space directly.

Chapter \ref{cp:path_itg_QFT} discusses the path integral formulation of QED, where one derivation of the SDEs for the QED generating functional is provided. The (longitudinal) Ward--Green--Takahashi identity for the generating functional is also deduced from the path integral. The corresponding SDEs for the propagators and the WGTI for the vertex are obtained as the leading expansions of these functional identities. Additionally, the equivalence of SDEs to all-order perturbation theory in QED has been proved. Within this chapter, the real scalar $\phi^4$ theory is used as an example to explain several general properties of the path integral formalism.

Chapter \ref{cp:spec_repr} introduces the spectral representation for the propagators based on their analytic structures. The spectral representation allows the complete description of the propagator on the complex momentum plane by its imaginary part on the positive real axis. This spectral representation is the foundation of chapters that follow. 

Chapter \ref{cp:WGTIs} is a review of various Ward--Green--Takahashi identities for QED vertices. Here representations similar to the Gauge Technique are shown to include the fermion propagator contributions in the WGTIs. Because knowledge aside from the fermion propagator is required to solve WGTIs exactly, the WGTIs do not form a closed system. Additionally, the tensor WGTI is re-derived using the functional approach, and is shown to be anomaly free.

Chapter \ref{ch:SDE_oGT} starts with requirement on the fermion-photon vertex to insure loop-divergences from the fermion propagator SDE are removable by renormalization conditions. Although the Gauge Technique Ansatz does not meet such a requirement, with a minimal modification, the spectral functions for the fermion propagator have been solved from their SDEs in the quenched approximation in the Landau gauge. The solution of the photon SDE in the Minkowski space with the Gauge Technique is also discussed.

Chapter \ref{cp:div_QED} explores how the divergences in QED in different orders are closely related based on the dimensional regularization and a mass-independent renormalization scheme. Recurrence relations for the expansions of QED primitive divergences are derived. Results in this chapter aim to provide an insight into the multiplicative renormalizability of QED.

Chapter \ref{cp:LKFT} establishes the group nature of the Landau--Khalatnikov--Fradkin transform for the fermion propagator. The LKFT is then reformulated such that the exact dependences of the fermion propagator spectral functions on the covariance gauge-fixing parameter are obtained. Results in this chapter are also presented in Ref.~\cite{Jia:2016wyu}.

Chapter \ref{cp:GC_SDE_QED} employs both the spectral representation and the Ward identity to rewrite the  momentum space SDEs for the fermion propagator into linear equations for the spectral functions. Combined with results in Chapter \ref{cp:LKFT}, the necessary and sufficient condition for any truncation scheme of the SDEs to respect the gauge covariance of the fermion propagator is derived. A similar requirement to maintain the gauge independence of the vacuum polarization is also obtained. Results in this chapter are also presented in Refs.~\cite{Jia:2016udu,Pennington:2016vxv}.
\paragraph{Notations and conventions}
Throughout this article, the Bjorken--Drell metric for 4-vectors $x^\mu=(x_0,~x_1,~x_2,~x_3)$ is adopted. Therefore the metric tensor is given by
\begin{equation}
g^{\mu\nu}=\mathrm{diag}\{1,\,-1,\,-1,\,-1\}.
\end{equation}
As a consequence, the momentum $p^2$ is timelike if $p^2>0$. While the Euclidean momentum is defined as $p_E^2=-p^2$. Dirac gamma matrices satisfy anti-commutation relations $\lbrace\gamma^\mu,~\gamma^\nu\rbrace=2g^{\mu\nu}$. The fermion-photon vertex $\Gamma^{\mu}(k,p)$ is defined as the QED vector vertex with $k$ being the momentum of fermion flowing in and $p$ being the momentum of fermion flowing out. Meanwhile, define $q=k-p$ as the photon momentum. For notational convenience, we also define $t=k+p$. The tensor gamma matrices are defined as $\sigma^{\mu\nu}=\frac{i}{2}(\gamma^\mu\gamma^\nu-\gamma^\nu\gamma^\mu)$. 

The uncial epsilon ($\epsilon$) usually stands for how far the number of spacetime dimensions is away from 4 through $d=4-2\epsilon$. While the lowercase epsilon ($\varepsilon$) is specifically reserved for the Feynman prescription.

We adopt the set of transverse bases in Ref.~\cite{Ball:1980ay}. They are transverse with respect to photon momentum $q^\mu$. Explicitly, these basis are defined as
\begin{align}
& T^\mu_{1}(k,p)=p^\mu(k\cdot q)-k^\mu(p\cdot q) 
&& T^\mu_{2}(k,p)=T^\mu_{1}(k,p)\slashed{t} \nonumber\\
& T^\mu_{3}(k,p)=q^2\gamma^\mu-q^\mu\slashed{q} 
&& T^\mu_{4}(k,p)=T^\mu_{1}(k,p)\dfrac{k_\lambda p_\tau}{i}\sigma^{\lambda\tau} \nonumber\\
& T^\mu_{5}(k,p)=\dfrac{1}{i}\sigma^{\mu\nu}q_\nu 
&& T^\mu_{6}(k,p)=\gamma^\mu(p^2-k^2)+t^\mu\slashed{q} \nonumber\\
& T^\mu_{7}(k,p)=\dfrac{1}{2}(p^2-k^2)(\gamma^\mu\slashed{t}-t^\mu)+t^\mu k_\lambda p_\tau\frac{1}{i}\sigma^{\lambda\tau} \nonumber\\
& T^\mu_{8}(k,p)=-\gamma^\mu k_\lambda p_\tau\frac{1}{i}\sigma^{\lambda\tau}+k^\mu\slashed{p}-p^\mu\slashed{k} .
\end{align}
	\chapter{Path integral formulation of QFTs\label{cp:path_itg_QFT}}
	\section{The generating functionals}
\subsection{QED gauge fixing with the path integral}
Let's start with the QED Lagrangian
\begin{equation}
\mathcal{L}_{\mathrm{QED}}(x)=\dfrac{1}{2}A_\mu (x)\left[g^{\mu\nu}\partial^2-\partial^\mu \partial^\nu \right]A_\nu(x)+\overline{\psi}(x)(i\slashed{\partial}-m)\psi(x)+e\overline{\psi}(x)\gamma^\mu \psi(x)A_\mu (x).\label{eq:L_QED}
\end{equation}
This Lagrangian is constructed to be invariant under the following transform of the $U(1)$ gauge symmetry group:
\begin{equation}
\begin{cases}
\psi(x)\rightarrow e^{i\theta(x)}\psi(x)\\[0.5mm]
\overline{\psi}(x)\rightarrow \overline{\psi}(x) e^{-i\theta(x)}\\[0.5mm]
A^\mu(x)\rightarrow A^\mu(x)+\dfrac{1}{e}\partial^\mu\theta(x).\label{eq:local_U1_gauge}
\end{cases}
\end{equation}
The predictive power of a QFT is embedded within its correlation functions. To represent all correlation functions, the generating functional is then defined with the introduction of external sources. Explicitly, we define
\begin{equation}
Z[\eta,\overline{\eta},J^\mu]=\int\mathcal{D}\overline{\psi}\mathcal{D}\psi\mathcal{D}A\,\exp\bigg\{i \int d^4x \left[\mathcal{L}(x)+J_\mu (x)A^\mu (x)+\overline{\psi}(x)\eta (x)+\overline{\eta}(x)\psi(x)\right]\bigg\}.\label{eq:def_Z}
\end{equation}
Effectively, the action that defines $Z[\eta,\overline{\eta},J^\mu]$ consists of two parts, $S$ of the Lagrangian and $S_E$ of the external sources. They are defined as
\begin{equation}
S=\int d^4x \mathcal{L}(x),\quad S_E=\int d^4x\left[J_\mu (x)A^\mu (x)+\overline{\psi}(x)\eta (x)+\overline{\eta}(x)\psi(x)\right].\label{eq:S_E}
\end{equation}
The functional measure $\mathcal{D}A$ covers all four components of the photon field. However, since a shift does not modify the path-integral measure, we have
\begin{equation}
\mathcal{D}A^\mu(x)=\mathcal{D}\left(A^\mu(x)+\dfrac{1}{e}\partial^\mu\theta(x)\right).
\end{equation}
Combined with Eq.~\eqref{eq:local_U1_gauge}, this means the functional integral over the photon field counts physically identical configurations of the photon field. As a direct result of this overcounting, the free-photon propagator cannot be solved from the generating functional defined through Eq.~\eqref{eq:def_Z}, with the Lagrangian given in Eq.~\eqref{eq:L_QED}.

One way to confine the functional integration into one gauge orbit (one physically unique configuration of the photon field) is to apply the Faddeev--Popov procedure \cite{peskin1995introduction}. For covariant gauges, this procedure starts by imposing a covariant gauge fixing condition, which is given by
\begin{equation}
G(A)=\partial_\mu A^\mu-w(x)=0.
\end{equation}
Then, the following functional identity is inserted into the definition of $Z[\eta,\overline{\eta},J^\mu]$ in Eq.~\eqref{eq:def_Z},
\begin{equation}
1=\int \mathcal{D}\theta\,\det \left(\dfrac{\delta G(A^\theta)}{\delta\theta} \right)\delta[G(A^\theta)],
\end{equation}
where $A^\theta_\mu=A_\mu+\dfrac{1}{e}\partial_\mu\theta$, and $\delta[~]$ is the functional version of the Dirac delta-function. While in the case of QED, one immediately has $\delta G(A^\theta)/\delta\theta=\frac{1}{e}\partial^2$. The resulting determinant of $\partial^2$ can be relocated onto the exponential by the Faddeev--Popov ghost;
\begin{equation}
\det (\partial^2)=\int \mathcal{D}\overline{c}\mathcal{D}c\,\exp\left[-i\int d^4x\,\overline{c}(x)\partial^2 c(x)\right].
\end{equation}
After choosing a Gaussian weight function, we arrive at 
\begin{align}
&\quad \int \mathcal{D}\theta\mathcal{D}w\,\det(\partial^2/e)\delta[\partial_\mu A^\mu-w(x)]\exp\left[-i\int d^4x\dfrac{w^2(x)}{2\xi} \right]\nonumber\\
& =\int \mathcal{D}\overline{c}\mathcal{D}c\,\exp\bigg\{i\int d^4x\left[-\dfrac{1}{2\xi}(\partial_\mu A^\mu)^2-\overline{c}\partial^2 c \right]\bigg\},\label{eq:Faddeev_Popov_QED}
\end{align}
as the gauge fixing modification to $Z$ in Eq.~\eqref{eq:def_Z}. Since in QED, the ghost fields do not couple to any other field, they are readily integrated out. However, this is no longer true in non-Abelian gauge theories. The ghost field is then an extra degree of freedom introduced by this particular gauge fixing procedure \cite{peskin1995introduction}. The problem of Gribov ambiguities for gauge fixing non-Abelian fields is not covered in this article.

The net effect of Eq.~\eqref{eq:Faddeev_Popov_QED} is that the QED generating functional is still defined by Eq.~\eqref{eq:def_Z} with the Lagrangian changed from Eq.~\eqref{eq:L_QED} to
\begin{equation}
\mathcal{L}_{\mathrm{QED\,GF}}(x)=\dfrac{1}{2}A_\mu (x)\left[g^{\mu\nu}\partial^2+\left(\dfrac{1}{\xi}-1\right)\partial^\mu \partial^\nu \right]A_\nu(x)+\overline{\psi}(x)(i\slashed{D}-m)\psi(x),\label{eq:L_QED_GF}
\end{equation}
where the covariant derivative is defined as $D_\mu(x)=\partial_\mu -ieA_\mu(x)$.
\subsection{Taylor series of the generating functional}
\subsubsection{The generating functional as a collection of n-point functions}
To explain the claim that the generating functional contains all the knowledge of the field theory in terms of correlation functions, first consider the theory with one real scalar field $\phi$. One of the simplest interacting Lagrangian for this type of fields is the $\phi^4$ theory:
\begin{equation}
\mathcal{L}_{\phi 4}=\dfrac{1}{2}\phi(-\partial^2-m^2)\phi -\dfrac{\lambda}{4!}\phi^4.\label{eq:lagrangian_phi4}
\end{equation}
The generating functional is then a functional of only one external source $J$ defined by
\begin{equation}
Z[J]=\int \mathcal{D}\phi\,\exp\bigg\{i\left[\int d^4x\, \mathcal{L}_{\phi 4}(x)+J(x)\phi(x) \right]\bigg\}.\label{eq:def_Z_phi4}
\end{equation}
Meanwhile, the n-point correlation function is defined as
\begin{equation}
\langle\Omega|T\phi(x_1)\phi(x_2)\dots\phi(x_n)|\Omega\rangle=\dfrac{\int \mathcal{D}\phi\,\phi(x_1)\phi(x_2)\dots\phi(x_n)\exp\bigg\{i\int d^4x\, \mathcal{L}_{\phi 4}(x)\bigg\}}{\int \mathcal{D}\phi\exp\bigg\{i\int d^4x\, \mathcal{L}_{\phi 4}(x)\bigg\}},
\end{equation}
where $|\Omega\rangle$ stands for the vacuum state of the theory. Taking the convention that the generating functional $Z$ is always normalized by itself with vanishing external sources, the N-point correlation function is apparently obtainable by 
\begin{equation}
\langle\Omega|T\prod_{n=1}^{N}\phi(x_n)|\Omega\rangle=\lim\limits_{J\rightarrow 0}\left[\prod_{n=1}^{N}\left(\dfrac{-i\delta}{\delta J(x_n)} \right)\right]Z[J].\label{eq:n_point_Z_phi}
\end{equation}
Therefore, once the generating functional $Z[J]$ is known explicitly, the theory is solved completely. Equivalently, we have the following functional Taylor series expansion of the generating functional,
\begin{align}
Z[J]& =1+\int d^4x \langle\Omega|T\phi(x)|\Omega\rangle J(x)+\dfrac{1}{2}\int d^4x\int d^4y \langle\Omega|T\phi(x)\phi(y)|\Omega\rangle J(x)J(y)\nonumber\\
& \quad +\dfrac{1}{3!}\int d^4x\int d^4y \int d^4z\langle\Omega|T\phi(x)\phi(y)\phi(z)|\Omega\rangle J(x)J(y)J(z)+\dots\nonumber\\
& =1+\sum_{N=1}^{+\infty}\dfrac{1}{N!}\left[\prod_{n=1}^{N} \int d^4x_m\,J(x_m) \right]\langle\Omega|T\prod_{m=1}^{N}\phi(x_m)|\Omega\rangle.\label{eq:Z_Taylor_phi4}
\end{align}

For QED, the generating functional written as $Z[\eta,\overline{\eta},J^\mu]$ apparently depends on two spinor and one vector external sources. The multivariable version of Eq.~\eqref{eq:Z_Taylor_phi4} exists as Eq.~\eqref{eq:def_Z}. Unlike the real scalar field $\phi$, the fermion fields $\psi,~\overline{\psi}$ and their external sources $\overline{\eta},~\eta$ are anticommuting Grassmann fields. The explicit correspondences between fields and functional derivatives are given by
\begin{equation}
\psi(x)\leftrightarrow \dfrac{-i\delta}{\delta \overline{\eta}(x)},\quad \overline{\psi}(x)\leftrightarrow \dfrac{+i\delta}{\delta\eta(x)},\quad A^\mu(x)\leftrightarrow \dfrac{-i\delta}{\delta J_\mu(x)}\label{eq:field_deltaJ_QED}
\end{equation}
while using Eq.~\eqref{eq:def_Z} to define the QED generating functional. Derivatives in Eq.~\eqref{eq:field_deltaJ_QED} are understood as partial derivatives. However, unlike the differentials of functions, distinctions of partial and full differentials of functionals are not made explicit. The reference of $\delta$ to either partial or full differentials is understood by the context.
\subsubsection{The perturbation expansions of QED}
The perturbation expansions of a QFT manifest themselves as series expansions on the interaction term of the Lagrangian. In the case of QED, this term is 
\begin{equation}
\mathcal{L}_{\mathrm{QED\,int}}(x)=e\overline{\psi}(x)\gamma^\mu \psi(x)A_\mu(x).\label{eq:L_QED_int}
\end{equation}
In practical calculations, the strength of the interaction is more conveniently described by $\alpha=e^2/(4\pi)$. With the absence of this interaction, the generating functional of QED is given exactly by 
\begin{equation}
\lim\limits_{\alpha\rightarrow 0}Z[\eta,\overline{\eta},J^\mu]=\exp\bigg\{\int d^4x\int d^4y\left[ \overline{\eta}(x)S_F^0(x-y)\eta(y)-\dfrac{1}{2}J^\mu(x)D^{0}_{\mu\nu}(x-y)J^\nu(y)\right]\bigg\},\label{eq:Z_free_QED}
\end{equation}
which can be derived from Eq.~\eqref{eq:def_Z} with $\mathcal{L}$ defined by Eq.~\eqref{eq:L_QED_GF} when $\alpha=e^2/(4\pi)=0$. The functional variable transforms required to derive Eq.~\eqref{eq:Z_free_QED} are
\begin{equation}
\begin{cases}
\psi'(x)=\psi(x)+i\int d^4y\,S_F^0(x-y)\eta(y),\\
\overline{\psi}'(x)=\overline{\psi}(x)-i\int d^4y\,\overline{\eta}(y)\overline{S}^0_F(x-y),\\
A_\mu'(x)=A_\mu(x)-i\int d^4dy\,D^0_{\mu\nu}(x-y)J^\nu(y),
\end{cases}\label{eq:QED_functional_variable_trans}
\end{equation}
where $S_F^{0}(x-y)$ and $D_{\mu\nu}^0(x-y)$ are the free-particle propagators solved from 
\begin{align}
&(i\slashed{\partial}_x-m)S_F(x-y)=-i\delta(x-y),\label{eq:def_SF0}\\
& \left[g^{\mu\nu}\partial^2_x+\left(\dfrac{1}{\xi}-1\right)\partial_x^\mu\partial_x^\nu \right]D^{0}_{\nu\rho}(x-y)=i\delta^\mu_\rho\delta(x-y).\label{eq:def_Dmunu0}
\end{align}

When interactions are present, applying Eq.~\eqref{eq:field_deltaJ_QED} to the Lagrangian of the interaction, given by Eq.~\eqref{eq:L_QED_int} before the functional variable transforms, produces the following closed form of $Z[\eta,\overline{\eta},J^\mu]$:
\begin{align}
Z[\eta,\overline{\eta},J^\mu]& =\exp\bigg\{i\int d^4z\,e\dfrac{i\delta}{\delta\eta(z)}\gamma^\mu\dfrac{-i\delta}{\delta\overline{\eta}(z)}\dfrac{-i\delta}{\delta J^\mu(z)}\bigg\}\nonumber\\
& \hspace{0.5cm}\times\exp\bigg\{\int d^4x\int d^4y\left[ \overline{\eta}(x)S_F^0(x-y)\eta(y)-\dfrac{1}{2}J^\mu(x)D^{0}_{\mu\nu}(x-y)J^\nu(y)\right]\bigg\}.\label{eq:Z_ptb_QED}
\end{align}
Then the Taylor series expansion of the first exponential in Eq.~\eqref{eq:Z_ptb_QED} produces well-known perturbative results.
\subsection{The generating functional for connected diagrams}
The generating functional for connected diagrams is given by the logarithm of $Z$;
\begin{equation}
W[\eta,\overline{\eta},J^\mu]=\ln Z[\eta,\overline{\eta},J^\mu].\label{eq:def_W}
\end{equation}
The generating functional $W$ can be viewed as the the collection of all Green's functions when the Taylor series expansions with respect to external sources are taken;
\begin{align}
W[\eta,\overline{\eta},J^\mu]& =\int d^4x\int d^4y\,\overline{\eta}(x)S_F(x-y)\eta(y)-\dfrac{1}{2}\int d^4x\int d^4y\,J^\mu(x)D_{\mu\nu}(x-y)J^\nu(y)\nonumber\\
& \quad +i\int d^4x\int d^4y\int d^4z\,\overline{\eta}(x)G^\mu(x,y,z)\eta(y)J_\mu(z)+\mathcal{O}(\eta,\overline{\eta},J^\mu)^4,\label{eq:Taylor_W}
\end{align}
where $S_F(x-y)$ is the coordinate space fermion propagator, $D_{\mu\nu}$ is the photon propagator, and $G^\mu(x,y,z)$ is the connected fermion-photon three-point function. The vanishing of the three-photon vertex is given by the Furry's theorem \cite{peskin1995introduction}. Such a Taylor series expansion of $W$ has, at lease, a finite radius of convergence. Because when external sources vanish altogether, $Z$ becomes a nonzero constant. As a consequence, there is a finite distance between the functional Taylor series expansion point and any possible singularity of $W[\eta,\overline{\eta},J^\mu]$. The main topic of this chapter is to derive the SDEs relating $S_F,~D_{\mu\nu}$ to $G^\mu$ in Eq.~\eqref{eq:Taylor_W}. 
\subsection{The generating functional for one-particle irreducible diagrams}
When perturbative calculations are performed, a Feynman diagram is 1PI if cutting any internal line of the diagram does not result in two disconnected diagrams. Straightforward one-loop perturbation calculation shows that connected fermion-photon three-point function $G^\mu(x,y,z)$ contains information about the propagators. This implies that one further step of reduction can be achieved through the definition of one-particle irreducible (1PI) diagrams. The relation between a connected diagram and its 1PI counterpart is that the 1PI diagram is obtained by truncating external lines of the connected diagram by factoring out propagators.

The formulation of 1PI diagrams in the language of path integration is accomplished by the Legendre transform on the generating functional $W$, depending on external sources $\eta,~\overline{\eta},$ and $J^\mu$. Then, define the first order derivatives of $W[\eta,\overline{\eta},J^\mu]$ as classical fields:
\begin{align}
& \psi_c(x)[\eta,\overline{\eta},J^\rho]=\dfrac{-i\delta}{\delta\overline{\eta}(x)}W[\eta,\overline{\eta},J^\rho],\label{eq:def_psi_c}\\
& \overline{\psi}_c[\eta,\overline{\eta},J^\rho]=\dfrac{i\delta}{\delta\eta(x)}W[\eta,\overline{\eta},J^\rho],\label{eq:def_psibar_c}\\
& A_c^\mu(x)[\eta,\overline{\eta},J^\rho]=\dfrac{-i\delta}{\delta J_\mu(x)}W[\eta,\overline{\eta},J^\rho].\label{eq:def_Amu_c}
\end{align}
Notice that these classical fields are obtained without setting external sources to zero. Therefore they are all functionals of the external sources, understood as the quantum field averaging with the presence of external sources. For QED, in the limit where external sources vanish simultaneously, the classical fields vanish as well. 

The generating functional $\Gamma$ is defined as a functional of the classical fields and obtained through the following Legendre transform:
\begin{equation}
\Gamma[\psi_c,\overline{\psi}_c,A^\rho_c]=W[\eta,\overline{\eta},J^\rho]-i\int d^4x\left[\overline{\eta}(x)\psi_c(x)+\overline{\psi}_c(x)\eta(x)+J_\mu(x)A^\mu_c(x) \right],\label{eq:def_Gamma_gen_fun}
\end{equation}
where after the Legendre transform, the dependence on external sources are written in terms of classical fields by the inverse of Eqs.~(\ref{eq:def_psi_c},~\ref{eq:def_psibar_c},~\ref{eq:def_Amu_c}). With the definition of $\Gamma[\psi_c,\overline{\psi}_c,A_c^\mu]$ by Eq.~\eqref{eq:def_Gamma_gen_fun} and the chain rule of functional derivatives, one can obtain the following identities for the first order derivatives of $\Gamma$:
\begin{align}
& \dfrac{\delta}{\delta\overline{\psi}_c(x)}\Gamma[\psi_c,\overline{\psi}_c,A_c^\rho]=-i\eta(x)[\psi_c,\overline{\psi}_c,A_c^\rho],\label{eq:D_Gamma_psibar_c}\\
& \dfrac{\delta}{\delta\psi_c(x)}\Gamma[\psi_c,\overline{\psi}_c,A_c^\rho]=+i\overline{\eta}(x)[\psi_c,\overline{\psi}_c,A_c^\rho],\label{eq:D_Gamma_psi_c}\\
& \dfrac{\delta}{\delta A^\mu_c(x)}\Gamma[\psi_c,\overline{\psi}_c,A_c^\rho]=-iJ_\mu(x)[\psi_c,\overline{\psi}_c,A_c^\rho].\label{eq:D_Gamma_Amu_c}
\end{align}
These three equations, with external sources written as functionals of classical fields, are the formal inverses of Eq.~(\ref{eq:def_psi_c},~\ref{eq:def_psibar_c},~\ref{eq:def_Amu_c}).

Setting all external sources to zero after taking another functional derivative with respect to external sources on Eqs.~(\ref{eq:D_Gamma_psibar_c},~\ref{eq:D_Gamma_psi_c},~\ref{eq:D_Gamma_Amu_c}) produces 
\begin{align}
& \lim\limits_{(\eta,\overline{\eta},J)\rightarrow 0}\int d^4y\,\dfrac{\delta^2W[\eta,\overline{\eta},J^\rho]}{\delta\eta(x)\,\delta\overline{\eta}(y)}\dfrac{\delta^2\Gamma[\psi_c,\overline{\psi}_c,A_c^\rho]}{\delta\psi_c(y)\,\delta\overline{\psi}_c(z)}=\delta(x-z)\label{eq:1PI_fermion_prop}\\
&
\lim\limits_{(\eta,\overline{\eta},J)\rightarrow 0}\int d^4y\,\dfrac{\delta^2W[\eta,\overline{\eta},J^\rho]}{\delta\overline{\eta}(x)\,\delta\eta(y)}\dfrac{\delta^2\Gamma[\psi_c,\overline{\psi}_c,A_c^\rho]}{\delta\overline{\psi}_c(y)\,\delta\psi_c(z)}=\delta(x-z)\label{eq:1IP_fermionbar_prop}\\
& \lim\limits_{(\eta,\overline{\eta},J)\rightarrow 0}\int d^4y\,\dfrac{\delta^2W[\eta,\overline{\eta},J^\rho]}{\delta J_\lambda(x)\,\delta J_\mu(y)}\dfrac{\delta^2\Gamma[\psi_c,\overline{\psi}_c,A_c^\rho]}{\delta A^\mu_c(y)\,\delta A^\nu_c(z)}=\delta^\lambda_\nu\delta(x-z).\label{eq:1PI_photon_prop}
\end{align}
Together with Eq.~\eqref{eq:Taylor_W}, Eqs.~(\ref{eq:1IP_fermionbar_prop},~\ref{eq:1IP_fermionbar_prop},~\ref{eq:1PI_photon_prop}) specify that the 1PI diagrams for propagators are just the inverses of the corresponding connected diagrams. 

\begin{figure}
	\centering
	\includegraphics[width=0.5\linewidth]{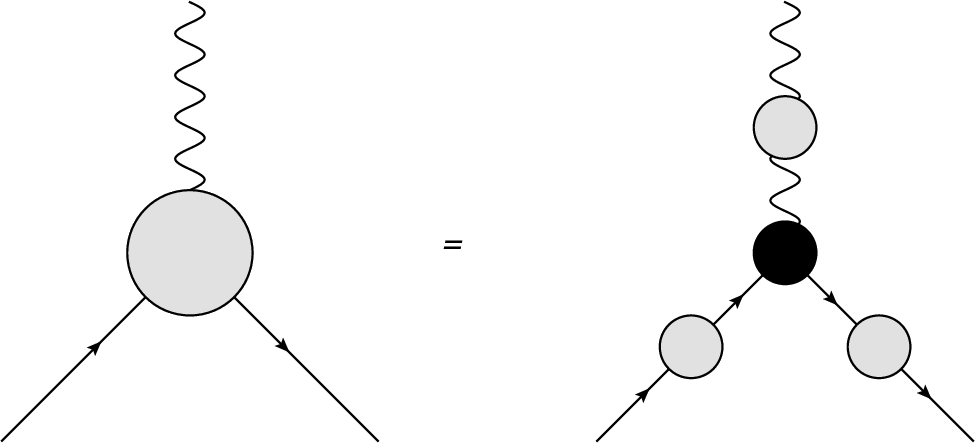}
	\caption{Relation between the connected fermion-photon three-point function and the 1PI three-point function.}
	\label{fig:ConnectexVertex}
\end{figure}

The fermion-photon three-point function generated by $\Gamma[\psi_c,\overline{\psi}_c,A_c^\mu]$ can be calculated by taking two other derivatives on any one of Eqs.~(\ref{eq:D_Gamma_psibar_c},~\ref{eq:D_Gamma_psi_c},~\ref{eq:D_Gamma_Amu_c}). Explicitly, taking $\delta/\delta\eta(y)$ on Eq.~\eqref{eq:D_Gamma_psibar_c} produces
\begin{align}
\dfrac{\delta^2 \Gamma[\psi_c,\overline{\psi}_c,A_c^\lambda]}{\delta\eta(y)\,\delta\overline{\psi}_c(x)}& =\int d^4w\,\left[\dfrac{\delta\psi_c(w)}{\delta\eta(y)}\dfrac{\delta}{\delta\psi_c(w)}+\dfrac{\delta\overline{\psi}_c(w)}{\delta\eta(y)}\dfrac{\delta}{\delta\psi_c(w)}+\dfrac{\delta A_c^\lambda(w)}{\delta\eta(y)}\dfrac{\delta}{\delta A_c^\lambda(w)} \right]\dfrac{\delta\Gamma}{\delta\overline{\psi}_c(x)}\nonumber\\
& =-i\delta(y-x),\label{eq:delta2_Gamma_eta_psic}
\end{align}
where the chain rule of derivatives has been applied. Utilizing $\delta AB=A\delta B+B\delta A$ and the chain rule again, subsequently taking another derivative with respect to $A_c^\mu(z)$ generates
\begin{align}
&\quad \lim\limits_{(\psi_c,\overline{\psi}_c,A^\lambda_c)\rightarrow 0}\dfrac{\delta^3\Gamma[\psi_c,\overline{\psi}_c,A_c^\lambda]}{\delta A_c^\mu(z)\,\delta\eta(y)\,\delta\overline{\psi}_c(x)}\nonumber\\
& =\lim\limits_{(\psi_c,\overline{\psi}_c,A^\lambda_c)\rightarrow 0}\int d^4w\,\left[\dfrac{\delta\psi_c(w)}{\delta\eta(y)}\dfrac{\delta}{\delta\psi_c(w)}+\dfrac{\delta\overline{\psi}_c(w)}{\delta\eta(y)}\dfrac{\delta}{\delta\overline{\psi}_c(w)}+\dfrac{\delta A_c^\lambda(w)}{\delta\eta(y)}\dfrac{\delta}{\delta A_c^\lambda(w)} \right]\dfrac{\delta^2\Gamma}{\delta A^\mu_c(z)\,\delta \overline{\psi}_c(x)}\nonumber\\
& \quad +\lim\limits_{(\psi_c,\overline{\psi}_c,A^\lambda_c)\rightarrow 0}\Bigg\{\int d^4v\left[\dfrac{\delta\eta(v)}{\delta A^\mu_c(z)}\dfrac{\delta}{\delta\eta(v)}+\dfrac{\delta\overline{\eta}(v)}{\delta A_c^\mu(z)}\dfrac{\delta}{\delta\overline{\eta}(v)}+\dfrac{\delta J^\rho(v)}{\delta A_c^\mu(z)}\dfrac{\delta}{\delta J^\rho(v)} \right]\nonumber\\
& \quad\quad \times\int d^4w\,\left[\dfrac{\delta\psi_c(w)}{\delta\eta(y)}\dfrac{\delta}{\delta\psi_c(w)}+\dfrac{\delta\overline{\psi}_c(w)}{\delta\eta(y)}\dfrac{\delta}{\delta\overline{\psi}_c(w)}+\dfrac{\delta A_c^\lambda(w)}{\delta\eta(y)}\dfrac{\delta}{\delta A_c^\lambda(w)} \right]\Bigg\}\dfrac{\delta\Gamma}{\delta\overline{\psi}_c(x)}\nonumber\\
& =\lim\limits_{(\psi_c,\overline{\psi}_c,A^\lambda_c)\rightarrow 0}\Bigg\{\int d^4w\,\dfrac{\delta\psi_c(w)}{\delta\eta(y)}\dfrac{\delta^3\Gamma}{\delta\psi_c(w)\,\delta A^\mu_c(z)\,\delta\overline{\psi}_c(x)}\nonumber\\
&\quad+\int d^4v\int d^4w\,\dfrac{\delta J^\rho(v)}{\delta A_c^\mu(z)}\dfrac{-i\delta^3W}{\delta J^\rho(v)\,\delta \eta(y)\,\delta\overline{\eta}(w)}\dfrac{\delta^2\Gamma}{\delta\psi_c(w)\,\delta\overline{\psi}_c(x)}\Bigg\}\nonumber\\
& =\lim\limits_{(\psi_c,\overline{\psi}_c,A^\lambda_c)\rightarrow 0}\int d^4w\Bigg\{\dfrac{-i\delta^2W}{\delta\eta(y)\,\delta\overline{\eta}(w)}\dfrac{\delta^3\Gamma}{\delta\psi_c(w)\,\delta A^\mu_c(z)\,\delta\overline{\psi}_c(x)}\nonumber\\
& \quad +\int d^4v\,\dfrac{i\delta^2\Gamma}{\delta A^\mu_c(z)\,\delta A_{c\,\rho}(v)}\dfrac{-i\delta^3W}{\delta J^\rho(v)\,\delta \eta(y)\,\delta\overline{\eta}(w)}\dfrac{\delta^2\Gamma}{\delta\psi_c(w)\,\delta\overline{\psi}_c(x)}\Bigg\}\nonumber\\
& =0\label{eq:Gamma_fermion_photon_3point}
\end{align}
Next, applying Eqs.~(\ref{eq:1IP_fermionbar_prop},~\ref{eq:1PI_photon_prop}) to Eq.~\eqref{eq:Gamma_fermion_photon_3point} produces
\begin{align}
\lim\limits_{(\eta,\overline{\eta},J^\lambda)\rightarrow 0}\dfrac{i\delta^3W[\eta,\overline{\eta},J^\lambda]}{\delta J^\rho(z)\,\delta\eta(y)\,\delta\overline{\eta}(x)}& = \lim\limits_{(\eta,\overline{\eta},J^\lambda)\rightarrow 0}\quad \int d^4x'\int d^4y'\int d^4z'\,\dfrac{\delta^2W}{\delta\eta(y)\delta\overline{\eta}(y')}\nonumber\\
& \quad\times\dfrac{\delta^3\Gamma}{\delta \psi_c(y')\,\delta A_c^\nu(z')\,\delta\overline{\psi}_c(x')}\dfrac{\delta^2W}{\delta\eta(x')\,\delta\overline{\eta}(x)}\dfrac{\delta^2W}{\delta J^\rho(z)\delta J_\nu(z')}.\label{eq:W_Gamma_fermion_photon_3point}
\end{align}
Diagrammatically, Eq.~\eqref{eq:Gamma_fermion_photon_3point} represents that the 1PI fermion-photon vertex is given by the connected three-point function truncated, as illustrated in Fig. \ref{fig:ConnectexVertex}.

Applying chain rules to deduce Eqs.~(\ref{eq:Gamma_fermion_photon_3point},~\ref{eq:delta2_Gamma_eta_psic}) is necessary because only after setting external sources or classical fields to zero are simple inverse relations in Eqs.~(\ref{eq:1IP_fermionbar_prop},~\ref{eq:1PI_fermion_prop},~\ref{eq:1PI_photon_prop}) valid. In general when the external sources or the classical fields are nonzero, the inverse relations for second order derivatives of $W$ and $\Gamma$ can be deduced from the functional version of the Jacobian matrix identity. Explicitly, consider a change of variable from external sources to classical fields. The functional integral measure changes according to 
\begin{equation}
\mathcal{D}\eta\mathcal{D}\overline{\eta}\mathcal{D}J^\nu=\mathcal{D}\psi_c\mathcal{D}\overline{\psi}_c\mathcal{D}A_c^\mu \det\bigg\{\dfrac{\delta[\eta,\overline{\eta},J^\nu]}{\delta[\psi_c,\overline{\psi}_c,A_c^\mu]}\bigg\}.
\end{equation}
where the functional Jacobian is defined as
\begin{equation}
\dfrac{\delta[\eta(y),\overline{\eta}(y),J^\nu(y)]}{\delta[\psi_c(x),\overline{\psi}_c(x),A_c^\mu(x)]}=\begin{pmatrix}
\dfrac{\delta\eta(y)}{\delta\psi_c(x)}& \dfrac{\delta\overline{\eta}(y)}{\delta\psi_c(x)} & \dfrac{\delta J^\nu(y)}{\delta \psi_c(x)} \\ 
\dfrac{\delta\eta(y)}{\delta\overline{\psi}_c(x)} & \dfrac{\overline{\eta}(y)}{\delta\overline{\psi}_c(x)} & \dfrac{\delta J^\nu(y)}{\delta \overline{\psi}_c(x)} \\ 
\dfrac{\delta \eta(y)}{\delta A_c^\mu(x)} & \dfrac{\delta \overline{\eta}(y)}{\delta A_c^\mu(x)} & \dfrac{\delta J^\nu(y)}{\delta A_c^\mu(x)}
\end{pmatrix} .
\end{equation}
Then we have the following relations among functional derivatives
\begin{equation}
\begin{pmatrix}
\dfrac{\delta}{\delta\psi_c(x)} \\ 
\dfrac{\delta}{\delta\overline{\psi}_c(x)} \\ 
\dfrac{\delta}{\delta A_c^\mu(x)}
\end{pmatrix} 
=\int d^4y\,
\dfrac{\delta[\eta(y),\overline{\eta}(y),J^\nu(y)]}{\delta[\psi_c(x),\overline{\psi}_c(x),A_c^\mu(x)]}
\begin{pmatrix}
\dfrac{\delta}{\delta\eta(y)} \\ 
\dfrac{\delta}{\delta\overline{\eta}(y)} \\ 
\dfrac{\delta}{\delta J^\nu(y)}
\end{pmatrix} .
\end{equation}
Similarly, we also have
\begin{equation}
\begin{pmatrix}
\dfrac{\delta}{\delta\eta(x)} \\ 
\dfrac{\delta}{\delta\overline{\eta}(x)} \\ 
\dfrac{\delta}{\delta J^\mu(x)}
\end{pmatrix} 
=\int d^4y\,
\dfrac{\delta[\psi_c(y),\overline{\psi}_c(y),A_c^\nu(y)]}{\delta[\eta(x),\overline{\eta}(x),J^\mu(x)]}
\begin{pmatrix}
\dfrac{\delta}{\delta\psi_c(y)} \\ 
\dfrac{\delta}{\delta\overline{\psi}_c(y)} \\ 
\dfrac{\delta}{\delta A_c^\nu(y)}
\end{pmatrix} .
\end{equation}
Since there is no net effect of compounding two functional variable transforms which are the inverse of each other, we obtain 
\begin{align}
& \quad \int d^4y\dfrac{\delta[\eta(y),\overline{\eta}(y),J^\mu(y)]}{\delta[\psi_c(x),\overline{\psi}_c(x),A_c^\lambda(x)]}\dfrac{\delta[\psi_c(z),\overline{\psi}_c(z),A_c^\nu(z)]}{\delta[\eta(y),\overline{\eta}(y),J^\mu(y)]}\nonumber\\
& =\int d^4y \dfrac{\delta[\psi_c(y),\overline{\psi}_c(y),A_c^\mu(y)]}{\delta[\eta(x),\overline{\eta}(x),J^\lambda(x)]}\dfrac{\delta[\eta(z),\overline{\eta}(z),J^\nu(z)]}{\delta[\psi_c(y),\overline{\psi}_c(y),A_c^\mu(y)]}\nonumber\\
& =\mathrm{diag}\{1,~1,~1\} \delta^\nu_\lambda \delta(x-z).\label{eq:Jacobian_identity_fun_var_trans}
\end{align}
Equation \eqref{eq:Jacobian_identity_fun_var_trans} relates all second order derivatives of $W[\eta,\overline{\eta},J^\mu]$ to those of $\Gamma[\psi_c,\overline{\psi}_c,A_c^\nu]$ without setting external sources or classical fields to zero. When there is only one active field, the single variable version of Eq.~\eqref{eq:Jacobian_identity_fun_var_trans} is expected.
\section{The longitudinal Ward--Green--Takahashi identity for the vector vertex\label{ss:derivation_WGTI_longi}}
Recall that the original QED Lagrangian given by Eq.~\eqref{eq:L_QED} is invariant under the local $U(1)$ gauge transformation given by Eq.~\eqref{eq:local_U1_gauge}. However, formulating QED in the language of path integration requires the gauge fixing procedure and external sources to properly define the generating functional $Z[\eta,\overline{\eta},J^\mu]$. This results in the action consisting of the part with the gauge fixed Lagrangian by Eq.~\eqref{eq:L_QED_GF} and another part from external sources as in Eq.~\eqref{eq:S_E}. 

While the invariance of $Z[\eta,\overline{\eta},J^\mu]$ under the $U(1)$ gauge transform indicates certain relations among the correlation functions. Such a gauge invariant requirement results in what is known as the Ward--Green--Takahashi identities. To derive such relations, consider the infinitesimal version of Eq.~\eqref{eq:local_U1_gauge}: 
\begin{equation}
\psi(x)\rightarrow (1+i\theta(x))\psi(x)+\mathcal{O}(\theta^2),\quad \overline{\psi}(x)=\overline{\psi}(x)(1-i\theta(x))+\mathcal{O}(\theta^2),\quad A^\mu(x)\rightarrow A^\mu(x)+\dfrac{1}{e}\partial^\mu\theta(x).\label{eq:gauge_U1_small_theta}
\end{equation}
The Jacobian of Eq.~\eqref{eq:gauge_U1_small_theta} is merely a constant, not affecting the action. Meanwhile, the total action for the generating functional changes according to 
\begin{align}
\dfrac{\delta (S+S_E)}{\delta\theta(x)}& =\dfrac{\delta}{\delta\theta(x)}\int d^4y\bigg\{\dfrac{1}{\xi}A_\mu(y)\partial_y^2\partial^\mu_y \theta(y)+\dfrac{1}{e}J_\mu(y)\partial_y^\mu\theta(y)+i\theta(y)[\overline{\eta}(y)\psi(y)-\overline{\psi}(y)\eta(y)] \bigg\}\nonumber\\
& =-\dfrac{1}{\xi}\partial^2\partial^\mu A_\mu(x)-\dfrac{1}{e}\partial^\mu J_\mu(x)+i[\overline{\eta}(x)\psi(x)-\overline{\psi}(x)\eta(x)].
\end{align}
The requirement of $Z[\eta,\overline{\eta},J^\mu]$ to be independent of $\theta(x)$ results in 
\begin{equation}
\dfrac{i}{\xi}\partial^2\partial^\mu \dfrac{\delta Z}{\delta J^\mu(x)}-\dfrac{Z}{e}\partial^\mu J_\mu +\overline{\eta}(x)\dfrac{\delta Z}{\delta\overline{\eta}(x)}+ \left[ \dfrac{\delta Z}{\delta\eta(x)}\right]\eta(x)=0,\label{eq:WGTI_Z}
\end{equation} 
where the definition of $Z$ through Eq.~\eqref{eq:def_Z} has been used. Eq.~\eqref{eq:WGTI_Z} is the Ward--Green--Takahashi identity for the QED generating functional $Z$. 

Next, with Eq.~\eqref{eq:def_W} and Eqs.~(\ref{eq:def_psi_c},~\ref{eq:def_psibar_c},~\ref{eq:def_Amu_c}) as the definitions of $W$ and the classical fields, Eq.~\eqref{eq:WGTI_Z} becomes
\begin{equation}
-\dfrac{1}{\xi}\partial^2\partial_\mu A^\mu_c(x)-\dfrac{1}{e}\partial^\mu J_\mu(x)+i\overline{\eta}(x)\psi_c(x)-i\overline{\psi}_c(x)\eta(x)=0.\label{eq:WGTI_Gamma}
\end{equation}
When external sources are viewed as functionals of classical fields as in Eqs.~(\ref{eq:D_Gamma_psibar_c},~\ref{eq:D_Gamma_psi_c},~\ref{eq:D_Gamma_Amu_c}), Eq.~\eqref{eq:WGTI_Gamma} is the Ward--Green--Takahashi identity for the generating functional $\Gamma$ in Eq.~\eqref{eq:def_Gamma_gen_fun}. The corresponding identity for the fermion-photon vector vertex can then be derived by taking two derivatives on Eq.~\eqref{eq:WGTI_Gamma} with respect to the classical fermion fields followed by setting classical fields to zero. Explicitly, we have
\begin{equation}
\lim\limits_{(\psi_c,\overline{\psi}_c,A_c^\rho)\rightarrow 0}\Bigg\{-\dfrac{\partial^\mu_x}{e}\dfrac{\delta^2J_\mu(z)}{\delta\psi_c(y)\,\delta\overline{\psi}_c(x)}+i\dfrac{\delta\overline{\eta}(z)}{\delta\overline{\psi}_c(x)}\delta(y-z)-i\delta(x-z)\dfrac{\delta\eta(z)}{\delta\psi_c(y)}\Bigg\}=0.
\end{equation}
This translates into an equation for derivatives of $\Gamma$ after substituting in Eqs.~(\ref{eq:D_Gamma_psibar_c},~\ref{eq:D_Gamma_psi_c},~\ref{eq:D_Gamma_Amu_c}). The resulting identity is
\begin{equation}
\lim\limits_{(\psi_c,\overline{\psi}_c,A_c^\rho)\rightarrow 0}\Bigg\{-\dfrac{\partial^\mu_x}{e}\dfrac{\delta^3\Gamma}{\delta\psi_c(y)\,\delta\overline{\psi}_c(x)\,\delta A^\mu_c(z)}+i\dfrac{\delta^2\Gamma}{\delta\psi_c(z)\,\delta\overline{\psi}_c(x)}\delta(y-z)-i\delta(x-z)\dfrac{\delta^2\Gamma}{\delta\psi_c(y)\delta\overline{\psi}_c(z)}\Bigg\}=0.
\end{equation}
After taking the Fourier transform, we obtain the (longitudinal) Ward--Green--Takahashi identity for the fermion-photon vector vertex as
\begin{equation}
q_\mu \Gamma^\mu(k,p)=S_F^{-1}(k)-S_F^{-1}(p).\label{eq:WGTI_longitudinal}
\end{equation}

While longitudinal Ward--Green--Takahashi identities for the photon propagator and other higher n-point functions can also be derived from Eq.~\eqref{eq:WGTI_Gamma}. They are beyond the scope of this chapter. Meanwhile, other types of Ward--Green--Takahashi identities can be deduced from local $U(1)$ transforms with Dirac structures other than the trivial identity matrix in Eq.~\eqref{eq:local_U1_gauge}. These identities will be discussed in Subsection \ref{sc:WGTI}.
\section{The Schwinger--Dyson equations for QED}
\subsection{The path integral derivation of SDEs\label{ss:derivation_SDEs}}
Aside from allowing the Lagrangian to differ up to 4-divergences, the asymptotically vanishing requirement on the fermion and photon fields also allows the SDEs for QFT Green's functions to be derived from the path integral formalism. Specifically, after the functional integrations, the divergences of a functional on the fermion and photon fields must vanish. As a result, we have 
\begin{equation}
\int \mathcal{D}\overline{\psi}\dfrac{\delta}{\delta\overline{\psi}(x)}=0,\quad \int \mathcal{D}\psi\dfrac{\delta}{\delta\psi(x)}=0,\quad \int \mathcal{D}A\dfrac{\delta}{\delta A^\mu(x)}=0.
\end{equation}
Taking the complete differential of the action plus external sources results in 
\begin{align}
\delta (S+S_E)& =\bigg\{\left[g^{\mu\nu}\partial^2+\left(\dfrac{1}{\xi}-1\right)\partial^\mu\partial^\nu \right]A_\nu(x)+e\overline{\psi}(x)\gamma^\mu\psi(x)+J^\mu(x)\bigg\}\delta A_\mu(x)\nonumber\\
&\quad +(\delta \overline{\psi}(x))\left[(i\slashed{\partial}-m)\psi(x)+e\gamma^\mu A_\mu(x)\psi(x)+\eta(x)\right]\nonumber\\
&\quad  +\left[\overline{\psi}(x)\left(-i\overleftarrow{\slashed{\partial}}-m\right)+e\overline{\psi}(x)\gamma^\mu A_\mu+\overline{\eta}(x) \right]\delta\psi(x).
\end{align}
Then, from $\int \mathcal{D}\overline{\psi}\mathcal{D}\psi\mathcal{D}A\,i\delta(S+S_E)\exp[i(S+S_E)]=0$
we have the following functional differential equations for the generating functional $Z[\eta,\overline{\eta},J^\mu]$,
\begin{align}
& \bigg\{\left[g^{\mu\nu}\partial^2+\left(\dfrac{1}{\xi}-1\right)\partial^\mu\partial^\nu \right]\dfrac{-i\delta}{\delta J^\nu(x)}+e\dfrac{i\delta}{\delta \eta(x)}\gamma^\mu\dfrac{-i\delta}{\delta\overline{\eta}(x)}+J^\mu(x)\bigg\}Z[\eta,\overline{\eta},J^\mu]=0,\label{eq:SDE_Z_J}\\
& \bigg\{\left[(i\slashed{\partial}-m)+e\gamma^\mu \dfrac{-i\delta}{\delta J^\mu(x)} \right]\dfrac{-i\delta}{\delta\overline{\eta}(x)}+\eta(x)\bigg\}Z[\eta,\overline{\eta},J^\mu]=0,\label{eq:SDE_Z_etabar}\\
& \bigg\{\dfrac{i\delta}{\delta\eta(x)}\left[\left(-i\overleftarrow{\slashed{\partial}}-m\right)+e\gamma^\mu\dfrac{-i\delta}{\delta J^\mu(x)} \right]+\overline{\eta}(x)\bigg\}Z[\eta,\overline{\eta},J^\mu]=0.\label{eq:SDE_Z_eta}
\end{align}
Here the highest order of derivative is $2$ because the Lagrangian contains no higher powers of fields. Next, as another result of the chain rule, we have 
\begin{equation}
\delta^2 Z=\delta^2 e^W=\delta e^W\delta W=e^W[(\delta W)^2+\delta^2W].
\end{equation}
Equations (\ref{eq:SDE_Z_J},~\ref{eq:SDE_Z_etabar},~\ref{eq:SDE_Z_eta}) then produce the functional differential equations for $W$, which can be written as
\begin{align}
& \bigg\{\left[g^{\mu\nu}\partial^2+\left(\dfrac{1}{\xi}-1\right)\partial^\mu\partial^\nu \right]\dfrac{-i\delta}{J^\nu(x)}+e\dfrac{i\delta}{\delta \eta(x)}\gamma^\mu\dfrac{-i\delta}{\delta\overline{\eta}(x)}\bigg\}W[\eta,\overline{\eta},J^\mu]+J^\mu(x)=-e\dfrac{i\delta\, W}{\delta \eta(x)}\gamma^\mu\dfrac{-i\delta\, W}{\delta\overline{\eta}(x)}\label{eq:diff_W_J}\\
& \left[(i\slashed{\partial}-m)+e\gamma^\mu \dfrac{-i\delta}{\delta J^\mu(x)} \right]\dfrac{-i\delta}{\delta\overline{\eta}(x)}W[\eta,\overline{\eta},J^\mu]+\eta(x)=-e\gamma^\mu \dfrac{-i\delta\, W}{\delta J^\mu(x)}\dfrac{-i\delta\, W}{\delta\overline{\eta}(x)},\label{eq:diff_W_etab}\\
& \dfrac{i\delta}{\delta\eta(x)}\left[\left(-i\overleftarrow{\slashed{\partial}}-m\right)+e\gamma^\mu\dfrac{-i\delta}{\delta J^\mu(x)} \right]W[\eta,\overline{\eta},J^\mu]+\overline{\eta}(x)=-e\dfrac{i\delta\,W}{\delta\eta(x)}\gamma^\mu\dfrac{-i\delta\,W}{\delta J^\mu(x)}.\label{eq:diff_W_eta}
\end{align}
Equations (\ref{eq:diff_W_J},~\ref{eq:diff_W_etab},~\ref{eq:diff_W_eta}) are coupled nonlinear functional differential equations for the generating functional $W$, once solved, specify the dynamics of QED. Their boundary conditions are given by the leading two terms of Eq.~\eqref{eq:Taylor_W}, namely the fully dressed propagators.
	
To derive the SDEs for the propagators, take another derivative with respect to the external sources on Eqs.~(\ref{eq:diff_W_J},~\ref{eq:diff_W_etab},~\ref{eq:diff_W_eta}) before setting external sources to zero. SDEs for higher n-point functions are obtained by taking extra derivatives with respect to external sources. Such a procedure effectively substitutes Eq.~\eqref{eq:Taylor_W} into Eqs.~(\ref{eq:diff_W_J},~\ref{eq:diff_W_etab},~\ref{eq:diff_W_eta}) to obtain the recurrence relations for the Green's functions. These recurrence relations form the infinite towers of equations for the Green's functions, collectively known as the Schwinger--Dyson equations.

\begin{figure}
	\centering
	\includegraphics[width=0.8\linewidth]{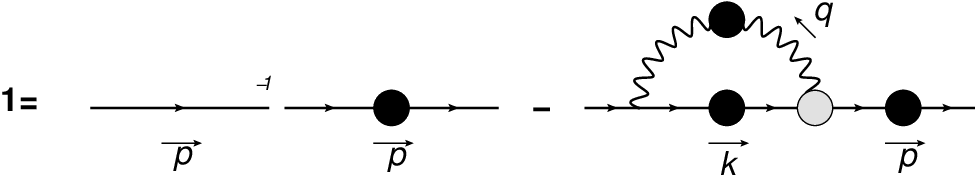}
	\caption{The diagrammatic representation of the SDE for the fermion propagator spectral functions.}
	\label{fig:DSE_fermion_rho}
\end{figure}

The leading SDEs are for the fermion and photon propagators. Setting external sources to zero after taking the functional derivative $i\delta/\delta\eta(y)$ on Eq.~\eqref{eq:diff_W_etab} results in the SDE for the fermion propagator as
\begin{equation}
(i\partial_x-m)S_F(x-y)+e\gamma^\mu G_\mu(x,y,x)+i\delta(x-y)=0,\label{eq:SDE_fermion_coord}
\end{equation}
where $G_\mu(x,y,x)$ is the connected fermion-photon three point function defined in Eq.~\eqref{eq:Taylor_W}. Subsequently, one substitute Eq.~\eqref{eq:W_Gamma_fermion_photon_3point} into Eq.~\eqref{eq:SDE_fermion_coord} and perform the Fourier transform to obtain the SDE for the fermion propagator in momentum space. Explicitly, these steps produce
\begin{equation}
1=(\slashed{p}-m)S_F(p)+ie^2\int d\underline{k}\gamma_\nu S_F(k)\Gamma_\mu(k,p)S_F(p)D^{\mu\nu}(q),\label{eq:SDE_fermion_momentum}
\end{equation}
where $S_F(p)$ is the fermion propagator in momentum space, $D_{\mu\nu}(q)$ is the photon propagator in momentum space, and $\Gamma_\mu(k,p)$ is the fermion-photon proper vertex, the Fourier transform of the 1PI vertex in Eq.~\eqref{eq:W_Gamma_fermion_photon_3point}. Recall for the proper vertex, $k$ is the momentum of the fermion flowing in, $p$ is the momentum of fermion flowing out, and $q=k-p$ is the photon momentum. The momentum measure of the loop integral is defined by $\int d\underline{k}=\mu^{4-d}\int d^dk/(2\pi)^d$, with $d$ being the number of spacetime dimensions and $\mu$ carrying the dimension of $e^2/(4\pi)$. The diagrammatic representation of Eq.~\eqref{eq:SDE_fermion_momentum} is given by Fig. \ref{fig:DSE_fermion_rho}.

\begin{figure}
	\centering
	\includegraphics[width=0.8\linewidth]{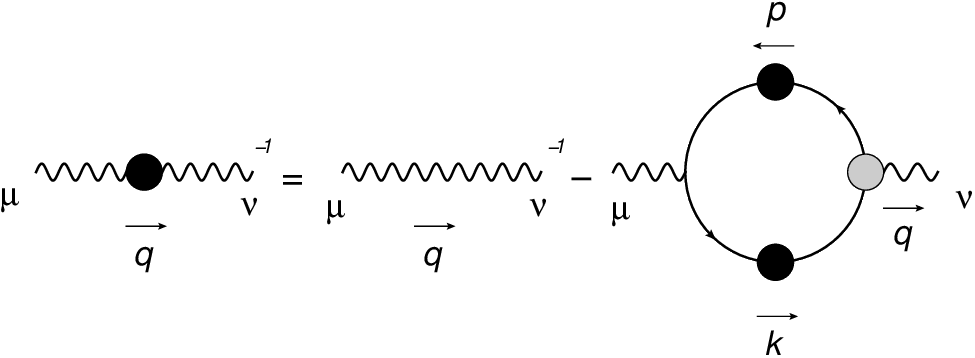}
	\caption{The diagrammatic representation of the SDE for the photon propagator.}
	\label{fig:SDE_photon}
\end{figure}

Similarly, taking $-i\delta/\delta J^\lambda(y)$ on Eq.~\eqref{eq:diff_W_J} and then setting external sources to zero gives
\begin{equation}
\left[g^{\mu\nu}+\left(\dfrac{1}{\xi}-1\right)\partial^\mu\partial^\nu \right]D_{\nu\lambda}(x-y)-e\mathrm{Tr}\,\{\gamma^\mu G^\nu(x,x,y) \}-i\delta ^\mu_\lambda\delta(x-y)=0,\label{eq:SDE_photon_coord}
\end{equation}
where we have used
\begin{equation}
\dfrac{i\delta}{\delta\eta(x)}\gamma^\mu\dfrac{-i\delta}{\delta\overline{\eta}(x)}W=\mathrm{Tr}\,\bigg\{\dfrac{i\delta}{\delta\eta(x)}\gamma^\mu\dfrac{-i\delta}{\delta\overline{\eta}(x)}W[\eta,\overline{\eta},J^\mu]\bigg\}=-\mathrm{Tr}\,\bigg\{\gamma^\mu\dfrac{i\delta}{\delta\eta(x)}\dfrac{-i\delta}{\delta\overline{\eta}(x)}W[\eta,\overline{\eta},J^\mu]\bigg\}.
\end{equation}
Again, substituting Eq.~\eqref{eq:W_Gamma_fermion_photon_3point} into Eq.~\eqref{eq:SDE_photon_coord} and subsequently taking the Fourier transform produces
\begin{equation}
D_{\mu\nu}^{-1}(q)=\left[g_{\mu\nu}q^2+\left(\dfrac{1}{\xi}-1\right)q_\mu q_\nu\right]-ie^2\mathrm{Tr}\,\int d\underline{k}\,\gamma_\nu S_F(k)\Gamma_\mu(k,p)S_F(p),\label{eq:SDE_photon_momentum}
\end{equation}
as the SDE for the photon propagator in momentum space. The diagrammatic representation of Eq.~\eqref{eq:SDE_photon_momentum} is given by Fig. \ref{fig:SDE_photon}. 

One Ward--Green--Takahashi identity specifies that the dressing of the photon propagator can only be transverse \cite{peskin1995introduction}. Therefore only one dressing function, $G(q^2)$, is required for $D_{\mu\nu}(q)$;
\begin{equation}
D_{\mu\nu}(q)=\dfrac{G(q^2)}{q^2+i\varepsilon}\left(g^{\mu\nu}-\dfrac{q^\mu q^\nu}{q^2}\right)+\xi\dfrac{q^\mu q^\mu}{q^4+i\varepsilon}. \label{eq:def_photon_propagator}
\end{equation}
Sometimes it is convenient to refer to the part of the photon propagator transverse to the momentum $q$ as 
\begin{equation}
\Delta_{\mu\nu}(q)=\frac{G(q^2)}{q^2+i\varepsilon}\left(g^{\mu\nu}-\frac{q^\mu q^\nu}{q^2}\right),\label{eq:def_Delta_munu}
\end{equation}
The inverse photon propagator in momentum space is given by
\begin{equation}
D_{\mu\nu}^{-1}(q)=\dfrac{1}{G(q^2)}(g_{\mu\nu}q^2-q_\mu q_\nu)+\dfrac{1}{\xi}q^\mu q^\nu.
\end{equation}
The $\xi$ dependent terms one on the left-hand side of Eq.~\eqref{eq:SDE_photon_momentum} cancels the one from the inverse of the bare propagator on the right-hand side. In fact, it can be shown exactly that the dressing function $G(q^2)$ is independent of $\xi$ to all orders in perturbation theory \cite{Breckenridge:1994gs}. 
\section{The equivalence of SDEs to all-order perturbation theory}
In previous sections, we derived two types of equations for the generating functional, the Ward--Green--Takahashi identity given by Eq.~\eqref{eq:WGTI_Gamma} and the Schwinger--Dyson equations given by Eqs.~(\ref{eq:SDE_Z_J},~\ref{eq:SDE_Z_etabar},~\ref{eq:SDE_Z_eta}). 
Two questions about the relations between these equations then arise. 
\begin{enumerate}
	\item Do these two types of equations contain information independently from each other?
	\item Do these functional differential equations encode all knowledge of QED? 
\end{enumerate}
Recall that WGTIs are results of the gauge symmetry of the theory, while SDEs are deduced from the more general presumption that the total divergences of a functional integral vanish. Intuitively we expect the SDEs to be more general than the WGTIs, indicating that all the knowledge of WGTIs is already contained in the SDEs. Under certain scenarios, a stronger statement can be made: the SDEs contain all the information of the theory. The proof of this statement relies on the following two assumptions.
\begin{enumerate}
	\item The perturbation calculation, when summed to all orders, contains all the knowledge of the theory.
	\item The generating functional can be adequately represented by a functional version of the Fourier transform.
\end{enumerate}
It will be shown, with Assumption \#2, the SDEs for the generating functional are equivalent to the perturbation theory to all orders. Then as a consequence of Assumption \#1, the SDEs know every aspect of the theory.
\subsection{Equivalence of SDE to all order perturbation for real scalar $\phi^4$ theory}
	\paragraph{The perturbative approach to $\phi^4$ theory}
	Let us start by proving the equivalence of SDEs for the real $\phi^4$ theory to all-order perturbation theory. The steps then generalize naturally to QED. 
	The Lagrangian of a real scalar theory with only $\phi^4$ interaction is given by Eq.~\eqref{eq:lagrangian_phi4}.
	The generating functional for this theory is defined by Eq.~\eqref{eq:def_Z_phi4}.
	Recall that The functional version of Taylor series expansion of $Z[J]$ gives every Green's function of the theory. Therefore $Z[J]$ contains all knowledge of the theory. 
	
	The formal solution to Eq.~\eqref{eq:def_Z_phi4} can be written as
	\begin{equation}
	Z[J]=\exp\bigg\{-\dfrac{i\lambda}{4!}\int d^4z\,\left[\dfrac{-i\delta}{\delta J(z)}\right]^4\bigg\}\exp\bigg\{-\dfrac{1}{2}\int d^4x\int d^4y\,J(x)D^0(x-y)J(y)\bigg\},\label{eq:Z_phi4_pert}
	\end{equation}
	where $D^0(x-y)$ is the free-particle propagator solved from
	\begin{equation}
	(-\partial_x^2-m^2)D^0(x-y)=i\delta(x-y).\label{eq:diff_D0}
	\end{equation}
	After shifting the $\phi$ fields in the interacting term of the Lagrangian into functional derivatives, Eq.~\eqref{eq:Z_phi4_pert} can be derived by applying the following functional variable transform
	\begin{equation}
	\phi(x)=\phi'(x)+i\int d^4y\,D^0(x-y)J(y)
	\end{equation}
	to Eq.~\eqref{eq:def_Z_phi4}. Perturbative results in the n-th order are obtained by Eq.~\eqref{eq:Z_phi4_pert} truncated to $\mathcal{O}(\lambda^n)$. 
	\paragraph{SDE for the $\phi^4$ theory generating functional}
	The integral of a total divergence is zero when there is no surface contribution. Since $\phi(x)$ is expected to vanish asymptotically, we have
	\begin{equation}
	\int\mathcal{D}\phi\dfrac{\delta}{\delta\phi(x)}\,\exp\bigg\{i\int d^4x\,[\mathcal{L}(x)+J(x)\phi(x)]\bigg\}=0.\label{eq:diff_Z_phi4}
	\end{equation}
	Substituting Eq.~\eqref{eq:lagrangian_phi4} into Eq.~\eqref{eq:diff_Z_phi4} produces the following functional differential equation for the generating functional $Z[J]$ defined by Eq.~\eqref{eq:def_Z_phi4}:
	\begin{equation}
	\left[ (-\partial_x^2-m^2)\dfrac{-i\delta}{\delta J(x)}-\dfrac{\lambda}{3!}\left(\dfrac{-i\delta}{\delta J(x)} \right)^3+J(x)\right]Z[J]=0.\label{eq:SDE_phi4_Z}
	\end{equation}
	Equation \eqref{eq:SDE_phi4_Z} is the SDE for the generating functional of the real scalar $\phi^4$ theory. The initial condition for Eq.~\eqref{eq:SDE_phi4_Z} is apparently $Z[0]=1$. The existence and uniqueness of solutions are not apparent by Eq.~\eqref{eq:SDE_phi4_Z} itself. 
	
	As for the existence of solutions, one naturally expects the perturbative solution given by Eq.~\eqref{eq:Z_phi4_pert} to satisfy Eq.~\eqref{eq:SDE_phi4_Z}. To show this explicitly, one functional operator relation needs to be proved first.	We start from the following simple operator identities:
	\begin{equation}
	\dfrac{\delta}{\delta J(y)}J(x)=\delta(x-y)+J(x)\dfrac{\delta}{\delta J(y)},\quad \left(\dfrac{\delta}{\delta J(y)}\right)^2J(x)=2\delta(x-y)\dfrac{\delta}{\delta J(y)}+J(x)\left(\dfrac{\delta}{\delta J(y)}\right)^2,\dots.
	\end{equation}
	Then by induction, this commutating relation holds:
	\begin{equation}
	\left[\left(\dfrac{\delta}{\delta J(y)}\right)^n,~J(x)\right]=n\delta(x-y)\left(\dfrac{\delta}{\delta J(y)}\right)^{n-1}.\label{eq:commu_dJn_J}
	\end{equation}
	Furthermore, since
	\begin{equation}
	\exp\bigg\{-\dfrac{i\lambda}{4!}\int d^4z\,\left[\dfrac{-i\delta}{\delta J(z)}\right]^4\bigg\}=1+\sum_{N=1}^{+\infty}\dfrac{1}{N!}\left(\dfrac{-i\lambda}{4!}\right)^n\prod_{n=1}^{N}\int d^4z_n\,\left[\dfrac{-i\delta}{\delta J(z_n)} \right]^4,
	\end{equation}
	we have
	\begin{align}
	&\quad \bigg\{\prod_{n=1}^{N}\int d^4z_n\left[\dfrac{-i\delta}{\delta J(z_n)}\right]^4\bigg\}J(x)\nonumber\\
	& =\bigg\{\prod_{n=1}^{N-1}\int d^4z_n\left[\dfrac{-i\delta}{\delta J(z_n)}\right]^4\bigg\}\bigg\{-4i\left[\dfrac{-i\delta}{\delta J(x)} \right]^3+J(x)\int d^4z_N\left[\dfrac{-i\delta}{\delta J(z_N)} \right]^4\bigg\}\nonumber\\
	& =\bigg\{\prod_{n=1}^{N-1}\int d^4z_n\left[\dfrac{-i\delta}{\delta J(z_n)}\right]^4\bigg\}(-4i)\left[\dfrac{-i\delta}{\delta J(x)} \right]^3+\bigg\{\prod_{n=1}^{N-2}\int d^4z_n\left[\dfrac{-i\delta}{\delta J(z_n)}\right]^4\bigg\}\nonumber\\
	&\quad \times\bigg\{-4i\left[\dfrac{-i\delta}{\delta J(x)} \right]^3+J(x)\int d^4z_{N-1}\left[\dfrac{-i\delta}{\delta J(z_{N-1})}\right]^4\bigg\}\int d^4z_N\left[\dfrac{-i\delta}{\delta J(z_N)}\right]^4\nonumber\\
	& =\dots\,\dots\nonumber\\
	& =-4iN\bigg\{\prod_{n=1}^{N-1}\int d^4z_n\left[\dfrac{-i\delta}{\delta J(z_n)} \right]^4\bigg\}\left[\dfrac{-i\delta}{\delta J(x)} \right]^3+J(x)\prod_{n=1}^{N}\int d^4z_n\left[\dfrac{-i\delta}{\delta J(z_n)} \right]^4.\label{eq:commu_DJ4_J}
	\end{align}
	Therefore the following commutation relation can be obtained:
	\begin{equation}
	\left[\exp\bigg\{-\dfrac{i\lambda}{4!}\int d^4z\,\left[\dfrac{-i\delta}{\delta J(z)}\right]^4\bigg\},\,J(x) \right]=-\dfrac{\lambda}{3!}\left(\dfrac{-i\delta}{\delta J(x)} \right)^3\exp\bigg\{-\dfrac{i\lambda}{4!}\int d^4z\,\left[\dfrac{-i\delta}{\delta J(z)}\right]^4\bigg\}.\label{eq:commu_expint_Jx}
	\end{equation}
	Here the commutation relations in Eqs.~(\ref{eq:commu_dJn_J},~\ref{eq:commu_expint_Jx}) can be understood by analogy with the commutation relations for coordinate operators and their conjugate momentum operator in quantum mechanics. Effectively, when commutators are calculated, one acts as the derivative to another.
	
	Substituting Eqs.~(\ref{eq:Z_phi4_pert},~\ref{eq:commu_expint_Jx}) into Eq.~\eqref{eq:SDE_phi4_Z} produces 
	\begin{equation}
	\left[ (-\partial^2_x-m^2)\dfrac{-i\delta}{\delta J(x)}+J(x)\right]\exp\bigg\{-\dfrac{1}{2}\int d^4x\int d^4y\,J(x)D^0(x-y)J(y)\bigg\}=0,
	\end{equation}
	which is apparently true considering Eq.~\eqref{eq:diff_D0}. Therefore we have shown that the solution to the SDE for the generating functional of the $\phi^4$ theory exists. One solution is Eq.~\eqref{eq:Z_phi4_pert}, the perturbation result.
	\paragraph{Solving the SDE for the generating functional using functional Fourier transform}
	The definition of generating functional $Z[J]$ by Eq.~\eqref{eq:def_Z_phi4} can be viewed as a functional version of the Fourier transform on the functional $\zeta[\phi]=\exp[i\int d^4x\,\mathcal{L}(x)]$. The functional Fourier transform relies on the following functional representation of the identity element in the linear functional operator space:
	\begin{align}
	& \quad \int \mathcal{D}\phi \exp\left[i\int d^4x\,J(x)\phi(x)\right]\nonumber\\
	& =\lim\limits_{\Delta x\rightarrow 0}\int \left[\prod_{n}d\phi(x_n) \right]\exp\left[i\sum_mJ(x_m)\phi(x_m)\Delta x \right]\nonumber\\
	& =\lim\limits_{\Delta x\rightarrow 0}\int_{-\infty}^{+\infty} \prod_{n}d\phi(x_n) \exp\left[iJ(x_n)\phi(x_n)\Delta x \right]\nonumber\\
	& =\lim\limits_{\Delta x\rightarrow 0}\prod_n(2\pi)\delta(J(x_n))\equiv \delta[J].
	\end{align}
	The linear functional operator (functional distribution) $\delta[J]$ is recognized as a generalized functional. As the Dirac delta-function is a distribution, a generalized function. The functional operator $\delta[J]$ is zero when there exist a measurable subset of $\{x_n\}$ such that $J(x_n)\neq 0$. The functional operator $\delta[J]$ diverges when $\forall x\in \{x_n\},~J(x)=0$. Meanwhile, $\delta[J]$ is normalized such that $\int\mathcal{D}J\,\delta[J]=1$.
	
	With $\delta[J]$ defined, we have the following Fourier transform and its inverse
	\begin{align}
	& Z[J]=\int\mathcal{D}\phi\,\zeta[\phi]\exp\left[i\int d^4x\,J(x)\phi(x) \right],\label{eq:Fourier_Z_zeta}\\
	& \zeta[\phi]=\int\mathcal{D}J\,Z[J]\exp\left[-i\int d^4x\,J(x)\phi(x) \right].\label{eq:Fourier_zeta_Z}
	\end{align}
	One can easily verify that Eqs.~(\ref{eq:Fourier_Z_zeta},~\ref{eq:Fourier_zeta_Z}) are inverse of each other using the definition of $\delta[J]$.
	
	Next, assuming that the solution to Eq.~\eqref{eq:SDE_phi4_Z} can be written as the Fourier transform of $J$, we then have
	\begin{equation}
	\left[ (-\partial_x^2-m^2)\dfrac{-i\delta}{\delta J(x)}-\dfrac{\lambda}{3!}\left(\dfrac{-i\delta}{\delta J(x)} \right)^3+J(x)\right]\int \mathcal{D}\phi\,\zeta[\phi]e^{iJ\phi}=0,
	\end{equation}
	Consequently, we obtain
	\begin{equation}
	\left[ (-\partial_x^2-m^2)\phi(x)-\dfrac{\lambda}{3!}\phi^3(x)+\dfrac{i\delta}{\delta\phi(x)}\right]\zeta[\phi]=0,\label{eq:diff_zeta}
	\end{equation}
	where we have used
	\begin{equation}
	\int \mathcal{D}\phi\left[e^{iJ\phi}\dfrac{\delta\zeta[\phi]}{\delta\phi(x)}+iJ(x)\zeta[\phi]e^{iJ\phi} \right]=\int \mathcal{D}\phi\dfrac{\delta}{\delta\phi(x)}\zeta[\phi]e^{iJ\phi}=0.
	\end{equation}
	While the solution to Eq.~\eqref{eq:diff_zeta} is apparently
	\begin{equation}
	\zeta[\phi]=\exp\bigg\{i\int d^4x\left[\dfrac{1}{2}\phi(x)(-\partial^2_x-m^2)\phi(x)-\dfrac{\lambda}{4!}\phi^4(x) \right]\bigg\}.\label{eq:solution_zeta}
	\end{equation}
	Substituting Eq.~\eqref{eq:solution_zeta} into Eq.~\eqref{eq:Fourier_Z_zeta} recovers Eq.~\eqref{eq:def_Z_phi4}. 
	
	Therefore under the assumption that $Z[J]$ is given by its functional Fourier transform, the solution to the SDE for the generating functional is unique. Furthermore, it is identical to the one obtained by the perturbation theory.
\subsection{Equivalence of QED SDEs to all-order perturbation theory}
	\paragraph{The perturbation theory solves SDEs}
	Similar to the proof in the $\phi^4$ theory, let us start by showing that the perturbation theory generating functional given by Eq.~\eqref{eq:Z_ptb_QED} solves Eqs.~(\ref{eq:SDE_Z_J},~\ref{eq:SDE_Z_etabar},~\ref{eq:SDE_Z_eta}) simultaneously. For Eq.~\eqref{eq:SDE_Z_J}, one starts with the commutation relation 
	\begin{align}
	&\quad \left[\exp\bigg\{i\int d^4z\,e\dfrac{i\delta}{\delta\eta(z)}\gamma^\nu\dfrac{-i\delta}{\delta\overline{\eta}(z)}\dfrac{-i\delta}{\delta J^\nu(z)}\bigg\},\,J^\mu(x) \right]\nonumber\\
	& =\exp\bigg\{i\int d^4z\,e\dfrac{i\delta}{\delta\eta(z)}\gamma^\nu\dfrac{-i\delta}{\delta\overline{\eta}(z)}\dfrac{-i\delta}{\delta J^\nu(z)}\bigg\}\,e\dfrac{i\delta}{\delta\eta(x)}\gamma^\mu\dfrac{-i\delta}{\delta\overline{\eta}(x)},\label{eq:commu_exp_Jmu}
	\end{align}
	which can be proved utilizing $[\delta/\delta J^\nu(z),\,J^\mu(x)]=\delta^\mu_\nu\delta(x-z)$ and the series expansion of the exponential. Here the derivation of Eq.~\eqref{eq:commu_exp_Jmu} is simpler compared to that for Eq.~\eqref{eq:commu_expint_Jx}, because in the argument of the exponential, $\delta/\delta J^\mu(z)$ is only raised to the first power. With the assistance of Eq.~\eqref{eq:commu_exp_Jmu}, substituting Eq.~\eqref{eq:Z_ptb_QED} into Eq.~\eqref{eq:SDE_Z_J} produces 
	\begin{align}
	& \bigg\{\left[g^{\mu\nu}\partial^2+\left(\dfrac{1}{\xi}-1\right)\partial^\mu\partial^\nu \right]\dfrac{-i\delta}{J^\nu(x)}+J^\mu(x)\bigg\}\nonumber\\
	& \hspace{1cm}\times \exp\bigg\{\int d^4x\int d^4y\left[ \overline{\eta}(x')S_F^0(x'-y)\eta(y)-\dfrac{1}{2}J^\mu(x')D^{0}_{\mu\nu}(x-y)J^\nu(y)\right]\bigg\}=0,\label{eq:SDE_Z_J_pert}
	\end{align}
	which is apparently true considering Eq.~\eqref{eq:def_Dmunu0}.
	
	The following two commutation relations,
	\begin{align}
	&\quad \left[\exp\bigg\{i\int d^4z\,e\dfrac{i\delta}{\delta\eta(z)}\gamma^\nu\dfrac{-i\delta}{\delta\overline{\eta}(z)}\dfrac{-i\delta}{\delta J^\nu(z)}\bigg\},\,\eta(x) \right]\nonumber\\
	& =\exp\bigg\{i\int d^4z\,e\dfrac{i\delta}{\delta\eta(z)}\gamma^\nu\dfrac{-i\delta}{\delta\overline{\eta}(z)}\dfrac{-i\delta}{\delta J^\nu(z)}\bigg\}\,e\,\gamma^\mu\dfrac{-i\delta}{\delta\overline{\eta}(x)}\dfrac{-i\delta}{\delta J^\mu(x)},\label{eq:commu_exp_eta}\\[2mm]
	&\quad \left[\exp\bigg\{i\int d^4z\,e\dfrac{i\delta}{\delta\eta(z)}\gamma^\nu\dfrac{-i\delta}{\delta\overline{\eta}(z)}\dfrac{-i\delta}{\delta J^\nu(z)}\bigg\},\,\overline{\eta}(x) \right]\nonumber\\
	& =\exp\bigg\{i\int d^4z\,e\dfrac{i\delta}{\delta\eta(z)}\gamma^\nu\dfrac{-i\delta}{\delta\overline{\eta}(z)}\dfrac{-i\delta}{\delta J^\nu(z)}\bigg\}\,e\dfrac{i\delta}{\delta\eta(x)}\gamma^\mu\dfrac{-i\delta}{\delta J^\mu(x)},\label{eq:commu_exp_etabar}
	\end{align}
	can be derived similar to Eq.~\eqref{eq:commu_expint_Jx}. However, after expanding the exponentials, the anticommutation relations 
	\begin{equation}
	\bigg\{\dfrac{\delta}{\delta\eta(z)},\,\eta(x)\bigg\}=\delta(x-z),\quad \bigg\{\dfrac{\delta}{\delta\overline{\eta}(z)},\,\overline{\eta}(x)\bigg\}=\delta(x-z),
	\end{equation}
	apply. Therefore in the steps similar to Eq.~\eqref{eq:commu_dJn_J}, the anticommutative properties of Grassmann numbers are used.
	
	Next, with Eqs.~(\ref{eq:commu_exp_eta},~\ref{eq:commu_exp_etabar}), substituting Eq.~\eqref{eq:Z_ptb_QED} into Eqs.~(\ref{eq:SDE_Z_etabar},~\ref{eq:SDE_Z_eta}) produces 
	\begin{align}
	& \left[(i\slashed{\partial}-m)\dfrac{-i\delta}{\delta\overline{\eta}(x)}+\eta(x) \right]\nonumber\\
	& \hspace{1cm}\times\exp\bigg\{\int d^4x\int d^4y\left[ \overline{\eta}(x')S_F^0(x'-y)\eta(y)-\dfrac{1}{2}J^\mu(x')D^{0}_{\mu\nu}(x-y)J^\nu(y)\right]\bigg\}=0,\label{eq:SDE_Z_etabar_pert}\\[2mm]
	& \left[\dfrac{i\delta}{\delta\eta(x)}(-i\overleftarrow{\slashed{\partial}}-m)+\overline{\eta}(x) \right]\nonumber\\
	& \hspace{1cm}\times\exp\bigg\{\int d^4x\int d^4y\left[ \overline{\eta}(x')S_F^0(x'-y)\eta(y)-\dfrac{1}{2}J^\mu(x')D^{0}_{\mu\nu}(x-y)J^\nu(y)\right]\bigg\}=0,\label{eq:SDE_Z_eta_pert}
	\end{align}
	respectively. Eqs.~(\ref{eq:SDE_Z_etabar_pert},~\ref{eq:SDE_Z_eta_pert}) are true because of Eq.~\eqref{eq:def_SF0}. Combining Eqs.~(\ref{eq:SDE_Z_J_pert},~\ref{eq:SDE_Z_etabar_pert},~\ref{eq:SDE_Z_eta_pert}), we have shown that the perturbation solution given by Eq.~\eqref{eq:Z_ptb_QED} solves the SDEs for the QED generating functional given by Eq.~(\ref{eq:SDE_Z_J},~\ref{eq:SDE_Z_etabar},~\ref{eq:SDE_Z_eta}) simultaneously.
	\paragraph{The uniqueness of the solution to SDEs}
	To see that the solution of the SDEs for the QED generating functional is uniquely equivalent to Eq.~\eqref{eq:Z_ptb_QED}, consider taking the functional Fourier transform pairs of the generating functional;
	\begin{align}
	& Z[\eta,\overline{\eta},J^\mu]=\int\mathcal{D}\overline{\psi}\mathcal{D}\psi\mathcal{D}A^\nu\,\zeta[\overline{\psi},\psi,A^\nu]\exp\bigg\{i\int d^4x\,\left[J^\mu(x)A_\mu(x)+\overline{\psi}(x)\eta(x)+\overline{\eta}(x)\psi(x) \right]\bigg\},\label{eq:Fourier_Z_zeta_QED}\\
	& \zeta[\overline{\psi},\psi,A^\nu]=\int\mathcal{D}\eta\mathcal{D}\overline{\eta}\mathcal{D}J^\nu\,Z[\eta,\overline{\eta},J^\nu]\exp\bigg\{-i\int d^4x\,\left[J^\mu(x)A_\mu(x)+\overline{\psi}(x)\eta(x)+\overline{\eta}(x)\psi(x) \right]\bigg\}.\label{eq:Fourier_zeta_Z_QED}
	\end{align}
	Substituting Eq.~\eqref{eq:Fourier_Z_zeta_QED} into Eq.~(\ref{eq:SDE_Z_J},~\ref{eq:SDE_Z_etabar},~\ref{eq:SDE_Z_eta}) results in
	\begin{align}
	& \bigg\{\left[g^{\mu\nu}\partial^2\left(\dfrac{1}{\xi}-1 \right)\partial^\mu \partial^\nu \right]A_\nu(x)+e\overline{\psi}(x)\gamma^\mu \psi(x)+\dfrac{i\delta}{\delta A_\mu(x)} \bigg\}\,\zeta[\overline{\psi},\psi,A^\lambda]=0,\label{eq:zeta_delta_A}\\
	& \bigg\{\left[(i\slashed{\partial}-m)+i\gamma^\mu A_\mu(x) \right]\psi(x)+\dfrac{i\delta}{\delta\overline{\psi}(x)}\bigg\}\,\zeta[\overline{\psi},\psi,A^\lambda]=0,\label{eq:zeta_delta_psibar}\\
	& \bigg\{\overline{\psi}(x)\left[(-i\overleftarrow{\slashed{\partial}}-m)+e\gamma^\mu A_\mu(x)\right]-\dfrac{i\delta}{\delta\psi(x)} \bigg\}\,\zeta[\overline{\psi}\psi,A^\lambda]=0,\label{eq:zeta_delta_psi}
	\end{align}
	where 
	\begin{align}
	& \quad \int \mathcal{D}A^\lambda\bigg\{e^{iJ^\nu A_\nu}\dfrac{\delta}{\delta A_\mu(x)}\zeta[\overline{\psi},\psi,A^\lambda]+iJ^\mu(x)\zeta[\overline{\psi},\psi,A^\lambda]e^{iJ^\nu A_\nu}\bigg\}\nonumber\\
	& =\int \mathcal{D}A^\lambda\dfrac{\delta}{\delta A_\mu(x)}\zeta[\overline{\psi},\psi,A^\lambda]e^{iJ^\nu A_\nu}=0,\label{eq:path_itg_delta_A}\\[2mm]
	& \quad \int \mathcal{D}\overline{\psi}\bigg\{e^{i\overline{\psi}\eta}\dfrac{\delta}{\delta\overline{\psi}(x)}\zeta[\overline{\psi},\psi,A^\lambda]+i\eta(x)\zeta[\overline{\psi},\psi,A^\lambda]e^{i\overline{\psi}\eta}\bigg\}\nonumber\\
	& =\int \mathcal{D}\overline{\psi}\dfrac{\delta}{\delta\overline{\psi}(x)}\zeta[\overline{\psi},\psi,A^\lambda]e^{i\overline{\psi}\eta}=0,\label{eq:path_itg_delta_psibar}\\[2mm]
	& \quad \int \mathcal{D}\psi\bigg\{e^{i\overline{\eta}\psi}\dfrac{\delta}{\delta\psi(x)}\zeta[\overline{\psi},\psi,A^\lambda]-i\overline{\eta}(x)\zeta[\overline{\psi},\psi,A^\lambda]e^{i\overline{\eta}\psi}\bigg\}\nonumber\\
	& =\int \mathcal{D}\psi\dfrac{\delta}{\delta\psi(x)}\zeta[\overline{\psi},\psi,A^\lambda]e^{i\overline{\eta}\psi}=0,\label{eq:path_itg_delta_psi}
	\end{align}
	have been used. Eqs.~(\ref{eq:path_itg_delta_A},~\ref{eq:path_itg_delta_psibar},~\ref{eq:path_itg_delta_psi}) are valid because fields $A^\mu,~\overline{\psi},~\psi$ vanish asymptotically.
	
	Equations~(\ref{eq:zeta_delta_A},~\ref{eq:zeta_delta_psibar},~\ref{eq:zeta_delta_psi}) are coupled partial functional differential equations for $\zeta[\overline{\psi},\psi,A^\lambda]$. The solution is apparently given by 
	\begin{equation}
	\zeta[\overline{\psi},\psi,A^\lambda]=\exp\left[i\int d^4x\,\mathcal{L}_{\mathrm{QED\,QF}}(x) \right],\label{eq:zeta_QED}
	\end{equation}
	with the gauge-fixed Lagrangian given by Eq.~\eqref{eq:L_QED_GF}. The generating functional $Z[\eta,\overline{\eta},J^\mu]$ is then calculated according to Eq.~\eqref{eq:Fourier_Z_zeta_QED}. Next, with the functional variable transform given by Eq.~\eqref{eq:QED_functional_variable_trans}, Eq.~\eqref{eq:zeta_QED} reproduces the generating functional in perturbation theory given by Eq.~\eqref{eq:Z_ptb_QED}. Therefore, the solution to the SDEs for the QED generating functional is given by Eq.~\eqref{eq:Z_ptb_QED}.
	
	In conclusion, we have proved that the QED SDEs given by Eqs.~(\ref{eq:SDE_Z_J},~\ref{eq:SDE_Z_etabar},~\ref{eq:SDE_Z_eta}) is equivalent to the perturbation theory to all orders. 
	\chapter{The spectral representation of propagators\label{cp:spec_repr}}
	\section{The K\"{a}ll\'{e}n--Lehmann spectral representation of real scalar propagators}
Consider a scalar QFT with Lagrangian
\[\mathcal{L}=\dfrac{1}{2}(\partial_\mu\phi)\partial^\mu \phi-\dfrac{1}{2}m^2\phi^2+\mathcal{L}_{int}.\]
One example of the interaction term is the $\phi^4$ theory in Eq.~\eqref{eq:lagrangian_phi4}. While in the absence of any interaction, the propagator in momentum space is given by
\[D^0(p^2)=\dfrac{1}{p^2-m^2+i\varepsilon}, \]
which is also known as the free-particle propagator.

\begin{figure}
	\centering
	\includegraphics[width=0.5\linewidth]{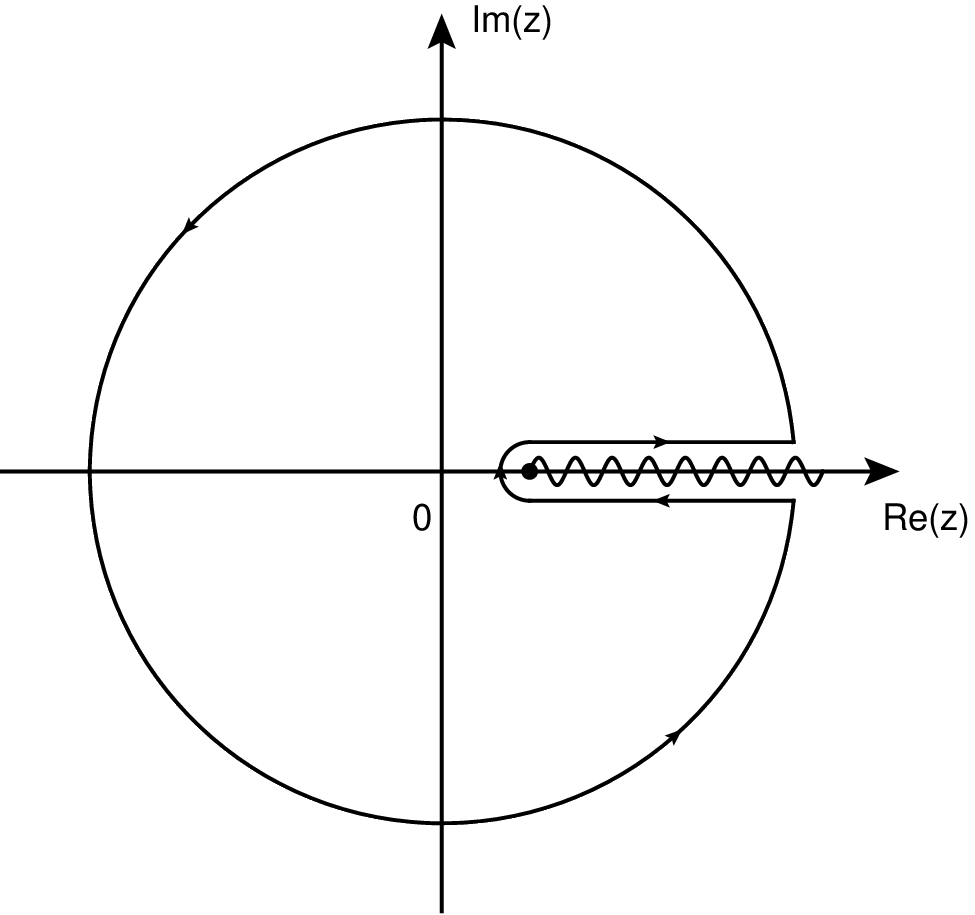}
	\caption{Illustration of the analytic structure of propagator function with dimensionless variables on the complex plane. The contour corresponds to evaluating Eq.~\eqref{eq:KLSR_scalar} using contour integral.}
	\label{fig:analytic_fz}
\end{figure}

The interaction $\mathcal{L}_{int}$ dresses up the scalar propagator. $D(p^2)$ is real for spacelike momentum $(p^2<0)$. When $p^2>0$, $D(p^2)$ becomes complex due to the production of real particles through loop corrections. The dressed propagator $D(p^2)$ can be written as a linear combination of free-particle propagators:
\begin{equation}
D(p^2)=\int_{m^2}^{+\infty}ds\dfrac{\rho(s)}{p^2-s+i\varepsilon},\label{eq:KLSR_scalar}
\end{equation}
where the weight function $\rho(s)$ is known as the spectral function of $D(p^2)$. For the bare spectral function, canonical quantization requires $\int_{m^2}^{+\infty}ds~\rho(s)=1$ \cite{Weinberg:1995mt}. The renormalization for this scalar propagator is given by $D_B(p^2)=ZD_R(p^2)$, so that $\rho_B(s)=Z\rho_R(s)$, where $\rho_R(s)$ is the renormalized spectral function. One can then derive that $\int_{m^2}^{+\infty}ds~\rho_R(s)=Z^{-1}$. 
The integral contour to prove Eq.~\eqref{eq:KLSR_scalar} is illustrated in Fig. \ref{fig:analytic_fz}.

The propagator function $D(p^2)$ and its spectral function $\rho(s)$ are interconnected. On one hand, the spectral function $\rho(s)$ is given by the imaginary part of the propagator function $D(p^2)$:
\begin{equation}
\rho(s)=-\dfrac{1}{\pi}\mathrm{Im}\{D(s+i\varepsilon) \}.\label{eq:rho_D}
\end{equation}
The $i\varepsilon$ modification to the argument of propagator function is essential because the propagator is expected to develop a branch cut when $p^2>m^2$. On the other hand, the spectral function determines the propagator function everywhere on the complex momentum plane. Although this is stated apparently in Eq.~\eqref{eq:KLSR_scalar}, with a specific $\rho(s)$ we are still faced with the difficulty of evaluating Eq.~\eqref{eq:KLSR_scalar} directly when $p^2$ is close to the branch cut. 

Alternatively, one can safely assume that the propagator function $D(p^2)$ is holomorphic everywhere on the complex $p^2$ plane except for the branch cut and perhaps a finite number of poles on the positive real axis. Meanwhile, the conjugation of $p^2$ is equivalent to the complex conjugation of $D(p^2)$, or $D(p^{2*})=\overline{D}(p^2)$ for any $p^2\in \mathbf{C}$ excluding where $D(p^2)$ is singular. The combination of these two properties of $D(p^2)$ allows us to determine $D(p^2)$ itself from only its imaginary part on the branch cut. Specifically, where $D(p^2)$ is holomorphic, the function satisfies Cauchy-Riemann equations. Therefore the imaginary part of $D(p^2)$ satisfies its Laplace equation. Once the imaginary part is known, the real part is given, up to a real constant, by integrating the Cauchy-Riemann equations. Since the imaginary part of $D(p^2)$ on its branch cut serves as the boundary condition for the Laplace equation, the imaginary part of $D(p^2)$ is completely determined by $\rho(s)$ almost everywhere (excluding singularities) on the complex plane. The arbitrary real integral constant is fixed by one specific evaluation of Eq.~\eqref{eq:KLSR_scalar} on a given $p^2$. Hence $D(p^2)$ in Minkowski space is determined by spectral function $\rho(s)$. Therefore solving for $D(p^2)$ in Minkowski space is equivalent to finding out its spectral function. 

Therefore, we have established that for any propagator function of a QFT, there is a unique spectral function, and vice versa. For more detailed discussion on the spectral representation of complex functions, see Appendix~\ref{ss:sr_fz}. An alternative way to understand the relation between a propagator function and its spectral function can be achieved through Mellin transforms discussed in Appendix~\ref{ss:Mellin}.
\section{The spectral representation of the fermion propagator}
In QED, due to the Dirac structure of the fermion field, the fermion propagator corresponds to two spectral functions $\rho_{1,2}(s)$. Specifically, we have
\begin{equation}
S_F(p)=\dfrac{\mathcal{F}(p^2)}{\slashed{p}-\mathcal{M}(p^2)}=\slashed{p}S_1(p^2)+S_2(p^2)=\slashed{p}\int_{m^2}^{+\infty}ds\dfrac{\rho_1(s)}{p^2-s+i\varepsilon}+\int_{m^2}^{+\infty}ds\dfrac{\rho_2(s)}{p^2-s+i\varepsilon}.\label{eq:SF_Dirac_struct}
\end{equation}
Following Ref.~\cite{Delbourgo:1977jc}, after effectively taking the square root of the integration variable $s$, the spectral functions $\rho_1$ and $\rho_2$ can be combined into one function;
\[\rho(W)=\mathrm{sign}(W)[W\rho_1(W^2)+\rho_2(W^2)].\]
Then the spectral representation of the fermion propagator can be written as
\begin{equation}
S_F(p)=\int_{|W|\geq m}^{+\infty}dW\dfrac{\mathrm{sign}(W)[W\rho_1(W^2)+\rho_2(W^2)]}{\slashed{p}-W+i\varepsilon}.
\end{equation}
Sometimes it is convenient to decompose the inverse of the fermion propagator into two Dirac components:
\begin{equation}
S_F^{-1}(p)=\slashed{p}A(p^2)+B(p^2).\label{eq:def_SF_AB}
\end{equation}
However, neither $A(p^2)$ nor $B(p^2)$ is linear in $\rho(W)$.

The introduction of the spectral representation for the fermion propagator allows the construction of a spectral representation for the fermion-photon vertex that respect the longitudinal Ward--Green--Takahashi identity. This construction is known as the Gauge Technique of Delbourgo, Salam, and Strathdee \cite{Delbourgo:1977jc,PhysRev.130.1287,PhysRev.135.B1398,PhysRev.135.B1428}. Explicitly, the Gauge Technique construction of fermion-photon vertex is given by
\begin{equation}
U[\gamma^\mu]=(S_F(k)\Gamma^\mu(k,p)S_F(p))_{\textrm{GT}}=\int dW \dfrac{1}{\slashed{k}-W}\gamma^\mu\dfrac{1}{\slashed{p}-W}\rho(W).\label{eq:original_GT}
\end{equation}
Here the subscript GT stands for the Gauge Technique. Writing a spectral representation for fermion propagator and relating it to the structure $S_F\Gamma^\mu S_F$ rather than to the one particle irreducible (1PI) vertex $\Gamma^\mu$ is shown to simplify manipulations of various WGTIs in Chapter \ref{cp:WGTIs}. 

The Gauge Technique construction of the $(S_F(k)\Gamma^\mu(k,p)S_F(p))_{\textrm{GT}}$ corresponds to a $\Gamma^\mu(k,p)$ that is free of kinematic singularity when $q^2\rightarrow 0$. Furthermore, one can easily verify that Eq.~\eqref{eq:original_GT} satisfies Eq.~\eqref{eq:WGTI_longitudinal}. Explicitly since
\begin{equation}
q_\mu\left(\Gamma^\mu(k,p)-U[\gamma^\mu] \right)=0,
\end{equation}
the Gauge Technique automatically satisfies the longitudinal Ward--Green--Takahashi identity. Therefore it contains the Ball--Chiu vertex as its longitudinal part. Meanwhile, we can calculate the additional transverse piece to Ball--Chiu vertex. Explicitly, one finds
\begin{equation}
U[\gamma^\mu]-\Gamma_{BC}^\mu (k,p)=\dfrac{A(k^2)-A(p^2)}{2(k^2-p^2)}T_3^\mu(k,p)-\dfrac{B(k^2)-B(p^2)}{k^2-p^2}T_5^\mu(k,p)+\dfrac{A(k^2)-A(p^2)}{k^2-p^2}T_8^\mu(k,p),\label{eq:Gamma_mu_transvese_GT}
\end{equation}
which agrees with the result obtained in Ref.~\cite{Qin:2013mta} after setting all scalar functions $X_i(k,p)$ to zero (up to convention and metric differences).

One can imagine there exist generalizations of K\"{a}ll\'{e}n--Lehmann spectral representation to any Green's functions (similar to the Nakanishi representation for scattering amplitudes \cite{nakanishi1971graph}). When solving SDEs for fermion and photon propagators, truncating SDEs is then equivalent to finding out the spectral representation of fermion-photon vertex $\Gamma^\mu(k,p)$ in terms of propagator spectral functions.
\section{The spectral representation of the photon propagator\label{sc:sr_photon}}
When the effects of vacuum polarization become significant, nonperturbative representations of the photon propagator in Minkowski space are required. For such representations, one naturally considers introducing a spectral function for the photon propagator. Similar to the fermion case, the existence of photon spectral function depends on the analytic structures of the photon propagator. 

The analytic structures of the photon propagator are completely determined by Eq.~\eqref{eq:SDE_photon_momentum}, the corresponding SDE. Notice that the vacuum polarization, defined by
\begin{equation}
\Pi(q^2)(g^{\mu\nu}q^2-q^\mu q^\nu)=\Pi^{\mu\nu}(q)=ie^2\int d\underline{k}\,\mathrm{Tr}\{\gamma^\mu S_F(k)\Gamma^\nu(k,p)S_F(p) \},\label{eq:def_vacuum_polarization_Pi}
\end{equation}
depends only on the fermion propagator and the vertex. Next, combining Eq.~\eqref{eq:SDE_photon_momentum} with Eq.~\eqref{eq:def_vacuum_polarization_Pi}, the photon propagator dressing function $G(q^2)$ can be written as
\begin{equation}
\dfrac{G(q^2)}{q^2+\varepsilon}=\dfrac{1}{(q^2+i\varepsilon)[1+\Pi(q^2)]}.\label{eq:photon_dressing}
\end{equation}
In the one-loop calculation, with dimensional regularization, the contribution to the vacuum polarization from the loop integral is given by
\begin{align}
\Pi_{1-\mathrm{loop}}(q^2)& =-\dfrac{\alpha}{4\pi}\bigg\{\dfrac{4}{3}\left[\dfrac{1}{\epsilon}-\gamma_E+\ln\left(\dfrac{4\pi\mu^2}{m^2} \right) \right]+ \dfrac{4}{3}\left(\dfrac{5}{3}+\dfrac{4}{z} \right)\nonumber\\
& \quad +\dfrac{8(z+2)}{3z^2}\sqrt{z(z-4)}\mathrm{arctanh}\left(\sqrt{\dfrac{z}{z-4}} \right)\bigg\}+\mathcal{O}(\epsilon^1),\label{eq:Pi_1loop_detlarho}
\end{align}
where $m$ is the fermion mass and $z=p^2/m^2$. One can easily verify that for the finite part 
\begin{equation}
\lim\limits_{z\rightarrow 0}\bigg\{\dfrac{4}{3}\left(\dfrac{5}{3}+\dfrac{4}{z} \right) + \dfrac{8(z+2)}{3z}\sqrt{\dfrac{z-4}{z}}\mathrm{arctanh}\left(\sqrt{\dfrac{z}{z-4}} \right)\bigg\}=0.
\end{equation}
Because the photon is strictly massless, the only singular structure of $\Pi(q^2)$ is the branch cut along $q^2\geq4m^2$. Admittedly, Eq.~\eqref{eq:Pi_1loop_detlarho} is only the result at one-loop. However, we expect it to elucidate the general analytic structures of the photon propagator.

Notice that vacuum polarization $\Pi(q^2)$ determines the photon dressing function $G(q^2)$ through Eq.~\eqref{eq:photon_dressing}, a nonlinear relationship. Consequently their analytic structures do not translate directly. However, the function $G(q^2)/q^2$ should behave similarly to a scalar propagator based on the following observations.
\begin{enumerate}
	\item It has a finite number of simple poles.
	\item It has a branch cut similar to that of $\Pi(q^2)$ for $q^2>4m^2$.
	\item Other than these singularities, the function is homomorphic everywhere else on the complex momentum plane.
\end{enumerate}
Observation \#1 is apparent since the denominator on the right-hand side of Eq.~\eqref{eq:photon_dressing} has one zero at $q^2=0$ and other possible zeros where $1+\Pi(q^2)=0$ is satisfied. To recover Observation \#2, one starts with the polar decomposition of complex function
\begin{equation}
1+\Pi(z)=1+u(x,y)+iv(x,y)=r(x,y)e^{i\phi(x,y)},
\end{equation}
where $z=x+iy$ and $u(x,y),~v(x,y),~r(x,y),~\phi(x,y)$ are real functions.
Then 
\begin{equation}
\dfrac{1}{1+\Pi(z)}=\dfrac{1+u(x,y)-iv(x,y)}{[1+u(x,y)]^2+[v(x,y)]^2}=\dfrac{e^{-i\phi(x,y)}}{r(x,y)}.
\end{equation}
For the $\Pi(z)$ having a regular branch cut on the real axis with $r(x,y)$ being finite and nonzero along the branch cut, $1/[1+\Pi(z)]$ also has a regular branch cut at the same position as that of the $\Pi(z)$. However, in the case that modulus function $r(x,y)$ is divergent along the branch cut, the corresponding branch cut for $1/(1+\Pi(z))$ vanishes.

Observation \#3 relies on the fact that the composition of holomorphic functions is also holomorphic giving the derivative of the composition exists. This can be shown by considering the general properties of a complex function. Explicitly, the function $f(z)=u(x,y)+iv(x,y)$ is holomorphic if and only if the following Cauchy-Riemann equations are satisfied,
\begin{equation}
\partial_x u=\partial_y v,\quad \partial_y u=-\partial_x v.
\end{equation}
Therefore 
\begin{equation}
\partial_x f=\partial_x u+i\partial_x v=\partial_y v-i\partial_y u=-i \partial_y f,
\end{equation}
which can also be shown equivalent to the Cauchy-Riemann equations.\footnote{Employing Wirtinger derivative $\partial/\partial\overline{z}=(\partial_x+i\partial_y)/2$ gives $\partial f/\partial\overline{z}=0$}
For a composite function, the following identity holds when $f(z)$ is holomorphic and $g(\zeta)$ is holomorphic at $\zeta\rightarrow f(z)$:
\begin{equation}
\partial_x g(f(z))=g'(f(z))\partial_xf(z)=g'(f(z))(-i\partial_y f(z))=-i\partial_yg(f(z)).
\end{equation}
Therefore the composite of holomorphic functions is holomorphic. In addition, the multiplication of two holomorphic functions is also holomorphic because 
\begin{align}
& \quad \partial_x f(z)g(z)=(\partial_xf(z))g(z)+f(z)\partial_x g(z)\nonumber\\
& =-i(\partial_y f(z))g(z)+f(z)(-i)\partial_yg(z)=-i\partial_yf(z)g(z).
\end{align}

Since the gauge-fixing term of the photon propagator in Eq.~\eqref{eq:def_photon_propagator} is not affected by the interaction of QED, we can postulate that only one scalar spectral function $\rho_\gamma(t)$ is required to represent $D^{\mu\nu}(q)$. Explicitly, we have 
\begin{equation}
\dfrac{G(q^2)}{q^2+i\varepsilon}=\int dt \dfrac{\rho_\gamma(t)}{q^2-t+i\varepsilon},\label{eq:def_spectral_rep_photon_propagator}
\end{equation}
which indicates
\begin{equation}
\rho_\gamma(t)=\dfrac{-1}{\pi}\mathrm{Im}\bigg\{\dfrac{G(t)}{t+i\varepsilon}\bigg\}=\dfrac{-1}{\pi}\mathrm{Im}\bigg\{\dfrac{1}{(t+i\varepsilon)[1+\Pi(t)]}\bigg\}.\label{eq:rho_gamma}
\end{equation}
Under the assumption that the analytic structure of $\Pi(q^2)$ is adequately represented by the one-loop calculation, $\Pi(q^2)$ is real and finite at $q^2=0$. 
Therefore Eq.~\eqref{eq:rho_gamma} becomes
\begin{equation}
\rho_\gamma(t)=\dfrac{1}{1+\Pi(0)}\delta(t)-\dfrac{1}{\pi t}\mathrm{Im}\bigg\{\dfrac{1}{1+\Pi(t)}\bigg\}.
\end{equation}
In addition, depending on the strength of the interaction, there may exists $t_k>0$ such that ${1+\Pi(t_k)=0}$, which corresponds to the point where $1+\Pi(t)$ crosses zero. Considering this possibility, the spectral function for the photon propagator is then calculated from the $\Pi(q^2)$ through
\begin{align}
\rho_\gamma(t)& =\dfrac{1}{1+\Pi(0)}\delta(t)+\sum_{k}^{0<t_k<t_{th}}\dfrac{\delta(t-t_k)}{t_k}\mathrm{Res}_{t'\rightarrow t_k}\bigg\{\dfrac{1}{1+\Pi(t')}\bigg\}\nonumber\\
& \quad -\dfrac{\theta(t-t_{th})}{\pi t}\mathrm{Im}\bigg\{\dfrac{1}{1+\Pi(t)}\bigg\},\label{eq:rho_gamma_Pi}
\end{align}
where $t_{th}$ is the branch point of $\Pi(q^2)$.
	\chapter{Ward--Green--Takahashi identities for transverse and non-vector vertices\label{cp:WGTIs}}
	\section{Relation between Ward--Green--Takahashi identities and the Gauge Technique}
\subsection{Gauge Technique transformations\label{ss:GaugeTechniqueTransf}}
Although the Gauge Technique satisfies the longitudinal WGTI, it does not contain the correct transverse vertex to respect renormalizability and gauge covariance, as will be shown in later chapters. We therefore propose the following generalizations of the original Gauge Technique:
\begin{align}
& u[\mathcal{K}]=\int dW\dfrac{1}{\slashed{k}-W}\mathcal{K}(k,p,W^2)\dfrac{1}{\slashed{p}-W}\rho(W)\label{eq:GTT_u}\\
& z[\mathcal{K}]=\int dW\dfrac{1}{\slashed{k}-W}W\mathcal{K}(k,p,W^2)\dfrac{1}{\slashed{p}-W}\rho(W),\label{eq:GTT_z}
\end{align}
which create functions of momenta $k$ and $p$. One immediately realizes that both $u[\mathcal{K}]$ and $z[\mathcal{K}]$ are linear in the kernel function $\mathcal{K}$ and the fermion spectral functions $\rho(W)$. Meanwhile, when $\mathcal{K}\rightarrow \gamma^\mu$, Eq.~\eqref{eq:GTT_u} produces the original Gauge Technique.

To make a connection with the 1PI vertex, inverses of fermion propagator are multiplied both to the left and to the right of Eqs.~(\ref{eq:GTT_u},~\ref{eq:GTT_z}), resulting in 
\begin{align}
& U[\mathcal{K}]=S_F^{-1}(k)\left[\int dW\dfrac{1}{\slashed{k}-W}\mathcal{K}(k,p,W^2)\dfrac{1}{\slashed{p}-W}\rho(W)\right]S_F^{-1}(p),\label{eq:GGT_U}\\
& Z[\mathcal{K}]=S_F^{-1}(k)\left[\int dW\dfrac{1}{\slashed{k}-W}W\mathcal{K}(k,p,W^2)\dfrac{1}{\slashed{p}-W}\rho(W)\right]S_F^{-1}(p).\label{eq:GGT_Z}
\end{align}
Straightforward calculation shows that
\begin{align}
& U[\mathcal{K}]=C_1\mathcal{K}+C_2\slashed{k}\mathcal{K}+C_3\mathcal{K}\slashed{p}+C_4\slashed{k}\mathcal{K}\slashed{p}\\
& Z[\mathcal{K}]=D_1\mathcal{K}+D_2\slashed{k}\mathcal{K}+D_3\mathcal{K}\slashed{p}+D_4\slashed{k}\mathcal{K}\slashed{p}.
\end{align}
Specifically, when the kernel function $\mathcal{K}$ does not depend on $W$, the coefficient functions are readily given by 
\begin{align*}
& C_1=\dfrac{k^2A(k^2)-p^2A(p^2)}{k^2-p^2} & C_2=\dfrac{B(k^2)-B(p^2)}{k^2-p^2} \\
& C_3=C_2 & C_4=\dfrac{A(k^2)-A(p^2)}{k^2-p^2}; \\
& D_1=\dfrac{p^2B(k^2)-k^2B(p^2)}{k^2-p^2} & D_2=\dfrac{p^2[A(k^2)-A(p^2)]}{k^2-p^2}\\
& D_3=\dfrac{k^2[A(k^2)-A(p^2)]}{k^2-p^2} & D_4=\dfrac{B(k^2)-B(p^2)}{k^2-p^2},
\end{align*}
where the functions $A$ and $B$ are defined by the inverse of fermion propagator in Eq.~\eqref{eq:def_SF_AB}.

Based on these results, one can easily derive that the original Gauge Technique construction of $\Gamma^\mu(k,p)$ contains the Ball--Chiu vertex as its longitudinal part, plus three transverse pieces in terms of transverse basis $T^\mu_i(k,p)$ as in Eq.~\eqref{eq:Gamma_mu_transvese_GT}.
\subsection{Ward--Green--Takahashi identities\label{sc:WGTI}}
We will demonstrate that Gauge Technique transformations, \textit{i.e.} Eqs.~(\ref{eq:GTT_u},~\ref{eq:GTT_z}), naturally include the propagator parts of Ward--Green--Takahashi identities.
\paragraph{Longitudinal WGTI} The longitudinal Ward identity ${q_\mu\Gamma^\mu(k,p)=S_F^{-1}(k)-S_F^{-1}(p)}$ is well known 
\cite{Ball:1980ay}. One derivation of Eq.~\eqref{eq:WGTI_longitudinal} has been presented in Section \ref{ss:derivation_WGTI_longi}. Since the right-hand side of Eq.~\eqref{eq:WGTI_longitudinal} only contains the propagator, we are prompted to introduce the spectral function $\rho(W)$. Explicitly, we have
\begin{align}
S_F^{-1}(k)-S_F^{-1}(p)& =S_F^{-1}(k)\left[S_F(p)-S_F(p)\right]S_F^{-1}(p)\nonumber\\
& =S_F^{-1}(k)\int dW\rho(W)\dfrac{1}{\slashed{k}-W}\left(\slashed{k}-W-\slashed{p}+W \right)\dfrac{1}{\slashed{p}-W}S_F^{-1}(p)\nonumber\\
& =q_\mu U[\gamma^\mu].\label{eq:SF_U_gamma}
\end{align}
Therefore Eq.~\eqref{eq:WGTI_longitudinal} is rewritten as
\begin{equation}
q_\mu\left(\Gamma^\mu(k,p)-U[\gamma^\mu] \right)=0,\label{eq:longi}
\end{equation}
a consequence of combining the longitudinal WGTI with the Gauge Technique. Equation~\eqref{eq:longi} implies that $U[\gamma^\mu]$ already contains the longitudinal vector vertex as described in Ref.~\cite{Ball:1980ay}. In fact, Eq.~\eqref{eq:SF_U_gamma} is the original motivation for the Gauge Technique \cite{Delbourgo:1977jc}. However, the longitudinal WGTI is insufficient to specify the fermion-photon vector vertex because it leaves any vector transverse to the photon momentum $q$ undetermined.
\paragraph{Axial WGTI}
From Ref.~\cite{He:2002jg}, the axial WGTI reads
\begin{equation}
q_\mu\Gamma_A^\mu(k,p)=S_F^{-1}(k)\gamma_5+\gamma_5S_F^{-1}(p)+2m\Gamma_5(k,p)+i\dfrac{g^2}{16\pi^2}F(k,p),\label{eq:AWI}
\end{equation}
where $\Gamma_5(k,p)$ is the pseudoscalar vertex, and $F(k,p)$ is the contribution from the axial anomaly.

Similar to Eq.~\eqref{eq:SF_U_gamma}, the propagator terms of the axial WGTI can be rewritten using our generalizations of the Gauge Technique given by Eqs.~(\ref{eq:GGT_U},~\ref{eq:GGT_U}). Let's start with
\begin{align}
S_F^{-1}(k)\gamma_5+\gamma_5S_F^{-1}(p)& =S_F^{-1}(k)\left(\gamma_5 S_F(p)+S_F(k)\gamma_5  \right)S_F^{-1}(p)\nonumber\\
& =S_F^{-1}(k)\int dW\rho(W)\dfrac{1}{\slashed{k}-W}\left[\gamma_5\left(\slashed{p}-W\right)+\left(\slashed{k}-W\right)\gamma_5 \right]\dfrac{1}{\slashed{p}-W}S_F^{-1}(p)\nonumber\\
& =q_\mu U[\gamma^\mu\gamma_5]-2Z[\gamma_5].
\end{align}
Equation~\eqref{eq:AWI} then becomes
\begin{equation}
q_\mu\left(\Gamma_A^\mu(k,p)-U[\gamma^\mu\gamma_5] \right)=2\left(m\Gamma_5(k,p)-Z[\gamma_5]\right)+i\dfrac{g^2}{16\pi^2}F(k,p).
\end{equation}
Here the longitudinal part of axial-vector vertex couples to the pseudoscalar vertex, unless the pseudoscalar vertex is identical to $Z[\gamma_5]/m$. Meanwhile, the anomalous term $F(k,p)$ is an unknown element of the axial Ward identity. Therefore the longitudinal part of the axial-vector vertex remains unknown except for the straightforward contribution from the fermion propagator. 
\paragraph{Transverse WGTI}
From Ref.~\cite{He:2002jg}, the transverse WGTI for the vector vertex is given by
\begin{align}
iq^\mu\Gamma^\nu(k,p)-iq^\nu\Gamma^\mu(k,p)=& S_F^{-1}(k)\sigma^{\mu\nu}+\sigma^{\mu\nu}S_F^{-1}(p)+2m\Gamma_T^{\mu\nu}(k,p)\nonumber\\
& +t_\lambda\epsilon^{\lambda\mu\nu\rho}\Gamma_{A\rho}(k,p)-\int\dfrac{d^4\kappa}{(2\pi)^4}2\kappa_\lambda\epsilon^{\lambda\mu\nu\rho}\Gamma_{A\rho}(k,p;\kappa).
\end{align}
If we rewrite the propagator part of this equation using Eqs.~(\ref{eq:GGT_U},~\ref{eq:GGT_Z}), we obtain
\begin{align}
S_F^{-1}(k)\sigma^{\mu\nu}+\sigma^{\mu\nu}S_F^{-1}(p)& =S_F^{-1}(k)\left[\sigma^{\mu\nu}S_F(p)+S_F(k)\sigma^{\mu\nu}\right]S_F^{-1}(p)\nonumber\\
& =S_F^{-1}(k)\left[\int dW\rho(W)\dfrac{1}{\slashed{k}-W}\left(\sigma^{\mu\nu}\slashed{p}+\slashed{k}\sigma^{\mu\nu}-2W\sigma^{\mu\nu}\right)\dfrac{1}{\slashed{p}-W}   \right]S_F^{-1}(p)\nonumber\\
& = U\left[\frac{1}{2}\lbrace\slashed{t},\sigma^{\mu\nu}\rbrace+\frac{1}{2}\left[\slashed{q},\sigma^{\mu\nu}\right] \right]-2Z[\sigma^{\mu\nu}]\nonumber\\
& =-\epsilon^{\lambda\mu\nu\rho}t_\lambda U[\gamma_\rho\gamma_5]+U[iq^\mu\gamma^\nu-iq^\nu\gamma^\mu]-2Z[\sigma^{\mu\nu}].
\end{align}
The transverse WGTI then becomes
\begin{align}
iq^\mu\Gamma^\nu(k,p)-iq^\nu\Gamma^\mu(k,p)=& iq^\mu U[\gamma^\mu]-iq^\nu U[\gamma^\nu]+2\left(m\Gamma_T^{\mu\nu}(k,p)-Z[\sigma^{\mu\nu}]\right)\nonumber\\
& +t_\lambda\epsilon^{\lambda\mu\nu\rho}\left(\Gamma_{A\rho}(k,p)-U[\gamma_\rho\gamma_5] \right)-\int\dfrac{d^4\kappa}{(2\pi)^4}2\kappa_\lambda\epsilon^{\lambda\mu\nu\rho}\Gamma_{A\rho}(k,p;\kappa)\label{eq:transvers_WGTI_GT}
\end{align}
Notice that except for the propagator terms, the tensor vertex, the axial-vector vertex, and a nonlocal term are also known to contribute to this identity. The transverse part of the fermion-photon vertex is, in principle, determined from Eq.~\eqref{eq:transvers_WGTI_GT}. However in practice, other than those already embedded in $U[\gamma^\mu]$, there is no more apparent transverse contribution from the fermion propagator. Equivalently speaking, the complete knowledge of $\Gamma^{\mu\nu}_T(k,p)$, $\Gamma_{A\rho}(k,p)$ and the nonlocal term are also required to solve the transverse WGTI.

Similar to steps in Ref.~\cite{Qin:2014vya}, one way to eliminate the vector vertex terms of the transverse WGTI is by contracting it with $\epsilon_{\alpha\mu\nu\beta}q^\beta/2$. The result is
\begin{align}
& \quad\left(t\cdot q~\delta_\alpha^\rho-t_\alpha q^\rho \right)\left(\Gamma_{A\rho}(k,p)-U[\gamma_\rho\gamma_5]\right)+2\epsilon_{\alpha\mu\nu\beta}q^\beta\left(m\Gamma_T^{\mu\nu}(k,p)-Z[\sigma^{\mu\nu}]\right)\nonumber\\
& =\int\dfrac{d^4\kappa}{(2\pi)^4}2\left(\kappa\cdot q\delta_\alpha^\rho-\kappa_\alpha q^\rho\right)\Gamma_{A\rho}(k,p;\kappa).\label{eq:trans_WGTI_projected}
\end{align}
Unlike Ref.~\cite{Qin:2013mta}, we do not parameterize the right-hand side of Eq.~\eqref{eq:trans_WGTI_projected} by scalar functions. Notice that Eq.~\eqref{eq:trans_WGTI_projected} is an unexpected identity for the axial vertex.
\paragraph{Transverse Axial WGTI}
From Ref.~\cite{He:2002jg}, the transverse axial WGTI is given by 
\begin{align}
iq^\mu\Gamma_A^\nu(k,p)-iq^\nu\Gamma_A^\mu(k,p)=& S_F^{-1}(k)\sigma^{\mu\nu}\gamma_5-\sigma^{\mu\nu}\gamma_5S_F^{-1}(p)+t_\lambda\epsilon^{\lambda\mu\nu\rho}\Gamma_\rho(k,p)\nonumber\\
& -\int\dfrac{d^4\kappa}{(2\pi)^2}2\kappa_\lambda\epsilon^{\lambda\mu\nu\rho}\Gamma_\rho(k,p;\kappa)+\dfrac{g^2}{16\pi^2}F_{(T)}^{\mu\nu}(k,p),\label{eq:TA_WGTI}
\end{align}
where $F_{(T)}^{\mu\nu}(k,p)$ comes from the transverse axial anomaly. We can rewrite the propagator part of the equation by our generalization of the Gauge Technique. Doing so produces 
\begin{align}
S_F^{-1}(k)\sigma^{\mu\nu}\gamma_5-\sigma^{\mu\nu}\gamma_5S_F^{-1}(p)& =S_F^{-1}(k)\left(\sigma^{\mu\nu}\gamma_5S_F(p)-S_F(k)\sigma^{\mu\nu}\gamma_5\right)S_F^{-1}(p)\nonumber\\
& =U\left[\slashed{k}\sigma^{\mu\nu}\gamma_5-\sigma^{\mu\nu}\gamma_5\slashed{p} \right]\nonumber\\
& =U\left[\dfrac{1}{2}\lbrace\slashed{t},\sigma^{\mu\nu}\rbrace\gamma_5+\dfrac{1}{2}\left[\slashed{q},\sigma^{\mu\nu}\gamma_5 \right] \right]\nonumber\\
& =-\epsilon^{\lambda\mu\nu\rho}t_\lambda U[\gamma_\rho]+iq^\mu U[\gamma^\nu\gamma_5]-iq^\nu U[\gamma^\mu\gamma_5].
\end{align}
Then the transverse axial WGTI becomes
\begin{align}
iq^\mu\Gamma_A^\nu(k,p)-iq^\nu\Gamma_A^\mu(k,p)=& iq^\mu U[\gamma^\nu\gamma_5]-iq^\nu U[\gamma^\mu\gamma_5]+t_\lambda\epsilon^{\lambda\mu\nu\rho}\left(\Gamma_\rho(k,p)-U[\gamma_\rho]\right)\nonumber\\
& -\int\dfrac{d^4\kappa}{(2\pi)^2}2\kappa_\lambda\epsilon^{\lambda\mu\nu\rho}\Gamma_\rho(k,p;\kappa)+\dfrac{g^2}{16\pi^2}F_{(T)}^{\mu\nu}(k,p).\label{eq:transverse_axial_WGTI_GT}
\end{align}
Similar to the transverse WGTI in Eq.~\eqref{eq:transvers_WGTI_GT}, the transverse axial WGTI formally determines the transverse components of the axial vector vertex. However, it couples to the vector vertex, a nonlocal term, and an anomalous term.

Following Ref.~\cite{Qin:2013mta}, contracting Eq.~\eqref{eq:transverse_axial_WGTI_GT} with $\frac{1}{2}\epsilon_{\alpha\mu\nu\beta}q^\beta$ produces
\begin{align}
\left(t\cdot q\delta_\alpha^\rho-t_\alpha q^\rho\right)\left(\Gamma_\rho(k,p)-U[\gamma_\rho]\right)=& \int\dfrac{d^4\kappa}{(2\pi)^4}2\left(\kappa\cdot q\delta_\alpha^\rho-\kappa_\alpha q^\rho\right)\Gamma_\rho(k,p;\kappa)\nonumber\\
& -\dfrac{g^2}{32\pi^2}\epsilon_{\alpha\mu\nu\beta}q^\beta F_{(T)}^{\mu\nu}(k,p).
\end{align}
Further substituting in Eq.~\eqref{eq:longi}, the projected transverse axial WGTI simplifies into
\begin{equation}
(k^2-p^2)\left(\Gamma_\alpha(k,p)-U[\gamma_\alpha]\right)=\int\dfrac{d^4\kappa}{(2\pi)^4}2\left(\kappa\cdot q\delta_\alpha^\rho-\kappa_\alpha q^\rho\right)\Gamma_\rho(k,p;\kappa)-\dfrac{g^2}{32\pi^2}\epsilon_{\alpha\mu\nu\beta}q^\beta F_{(T)}^{\mu\nu}(k,p).\label{eq:TWA}
\end{equation}
If one follows Ref.~\cite{Qin:2013mta}, Eq.~\eqref{eq:TWA} is used to solve for the transverse part of the vertex with a parametric representation of the nonlocal term together with the anomalous term. However, unlike in Ref.~\cite{Qin:2013mta}, we do not parameterize the right-hand side of Eq.~\eqref{eq:TWA} by scalar functions. Our analysis on this equation is further discussed in Section \ref{ss:proj_trans_axi_Ward}.
\paragraph{Tensor WGTI}
According to Ref.~\cite{PhysRevC.63.025207}, The tensor WGTI is given by 
\begin{align}
& \quad q^\mu\Gamma^{\nu\alpha}_T(k,p)+q^\nu\Gamma^{\alpha\mu}_T(k,p)+q^\alpha\Gamma_T^{\mu\nu}(k,p)\nonumber\\
&=-S_F^{-1}(k)\epsilon^{\mu\nu\alpha\rho}\gamma_\rho\gamma_5+\epsilon^{\mu\nu\alpha\rho}\gamma_\rho\gamma_5S_F^{-1}(p)+t_\lambda\epsilon^{\lambda\mu\nu\alpha}\Gamma_5(k,p).\label{eq:TensorWA}
\end{align}
Notice that unlike Eqs.~(\ref{eq:AWI},~\ref{eq:transverse_axial_WGTI_GT}), there is no anomalous term in Eq.~\eqref{eq:TensorWA}. Nor is there any nonlocal term. Ref.~\cite{Sun:2003ia} confirms that there is no anomaly for this tensor identity. However, Ref.~\cite{He:2001cu} produced the transverse axial WGTI with a missing term using the same approach applied in Ref.~\cite{Sun:2003ia}. Therefore Eq.~\eqref{eq:TensorWA} needs to be reexamined. The separation of the fermion propagator from the tensor WGTI is then postponed to Subsection \ref{ss:tensor_WGTI_momentum} when the ambiguities about Eq.~\eqref{eq:TensorWA} are clarified.
\section{A projection of the transverse axial WGTI\label{ss:proj_trans_axi_Ward}}
\subsection{The transverse axial WGTI after projection\label{ss:proj_trans_axi_Ward_terms}}
We know from Eq.~\eqref{eq:TWA}, a result following Ref.~\cite{Qin:2013mta}, that the transverse axial WGTI appears to determine the vector vertex. This is counterintuitive because one expects an identity for the axial vector vertex should specify the axial vector vertex itself. Our further investigation reveals that the result of the projection given by Eq.~\eqref{eq:TWA} is in fact an identity for another vertex. Let's start by deriving the momentum factor from the contracted transverse axial WGTI in Eq.~\eqref{eq:TWA}. This gives
\begin{equation}
\Gamma_\alpha(k,p)-U[\gamma_\alpha]=\dfrac{1}{k^2-p^2}\left[\int\dfrac{d^4\kappa}{(2\pi)^4}2\left(\kappa\cdot q\delta_\alpha^\rho-\kappa_\alpha q^\rho\right)\Gamma_\rho(k,p;\kappa)-\dfrac{g^2}{32\pi^2}\epsilon_{\alpha\mu\nu\beta}q^\beta F_{(T)}^{\mu\nu}(k,p)\right].\label{eq:vector}
\end{equation}
Formally when $k\neq \pm p$, Eq.~\eqref{eq:vector} specifies the difference between $\Gamma^\mu(k,p)$ and $U[\gamma^\mu]$ in terms of transverse vectors. The transverse vector vertex can be parameterized by eight scalar functions for eight transverse vectors. Therefore one could, in principle, specify all eight scalar functions from Eq.~\eqref{eq:vector}, as in Ref.~\cite{Qin:2013mta}. While in this section, we analyze the meaning of 
Eq.~\eqref{eq:vector} without a total parameterization of its right-hand side.

A curious observation of Eq.~\eqref{eq:vector} is that both terms on the right-hand side involve the Fourier transform of photon field $A^\mu$. Equation~\eqref{eq:vector} then indicates that $\Gamma_\mu(k,p)$ depends on photon dressing function $G(q^2)$ as well. However, in the following definition\footnote{For QED, it has been argued that the definition of $\Gamma^\mu(k,p)$ in Eq.~\eqref{eq:W_Gamma_fermion_photon_3point} is identical to that in Eq.~\eqref{eq:def_Gamma_mu_jmu} \cite{He:introQCD}.} of $\Gamma_\mu(k,p)$
\begin{equation}
(2\pi)^4\delta^4(k-p-q)iS_F(k)\Gamma_\mu(k,p)iS_F(p)=\int d^4xd^4yd^4ze^{i(k\cdot y-p\cdot z-q\cdot x)}\langle \Omega|Tj_\mu(x)\psi(y)\overline{\psi}(z)|\Omega\rangle,\label{eq:def_Gamma_mu_jmu}
\end{equation}
there is no apparent $G(q^2)$ dependence. Such a discrepancy can be understood by the nonperturbative equations of motion for photons field that relates photon field to fermion field: $\partial_\nu F^{\nu\mu}(x)=g\overline{\psi}(x)\gamma^\mu\psi(x)=gj^\mu(x)$.
\paragraph{Contribution from the tensor anomaly}
From Ref.~\cite{He:2002jg}, the anomalous term $F_{(T)}^{\mu\nu}(k,p)$ is defined by the following Fourier transform:
\begin{align}
& \quad\int d^4xd^4yd^4ze^{i(k\cdot y-p\cdot z-q\cdot x)}\langle \Omega|T\psi(y)\overline{\psi}(z)\left[\epsilon^{\alpha\beta\mu\rho}F^{\nu}_{~ \rho}(x)-\epsilon^{\alpha\beta\nu\rho}F^\mu_{~\rho}(x) \right]F_{\alpha\beta}(x)|\Omega\rangle\nonumber\\
& =(2\pi)^4\delta^4(k-p-q)iS_F(k)F^{\mu\nu}_{(T)}(k,p)iS_F(p).
\end{align}
Since taking a Fourier transform is linear, we suggest evaluating the anomalous term in coordinate space. Doing so bypasses the difficulty of not knowing ${\langle \Omega|T\psi(y)\overline{\psi}(z)F^{\nu}_{~\rho}(x)F_{\alpha\beta}(x)|\Omega\rangle}$ directly. Explicitly, we have
\begin{equation}
\epsilon_{\lambda\mu\nu\sigma}F^{\mu\nu}_{(T)}(k,p)=\mathcal{F}\lbrace\epsilon_{\lambda\mu\nu\sigma}\langle \Omega|T\psi(y)\overline{\psi}(z)\left[\epsilon^{\alpha\beta\mu\rho}F^{\nu}_{~\rho}(x)-\epsilon^{\alpha\beta\nu\rho}F^\mu_{~\rho}(x) \right]F_{\alpha\beta}(x)|\Omega\rangle\rbrace,
\end{equation}
where $\mathcal{F}$ stands for taking the Fourier transform in this subsection. Additionally, straightforward tensor algebra shows that
\begin{equation}
\epsilon_{\lambda\mu\nu\sigma}\left[\epsilon^{\alpha\beta\mu\rho}F^{\nu}_{~\rho}(x)-\epsilon^{\alpha\beta\nu\rho}F^\mu_{~\rho}(x) \right]F_{\alpha\beta}(x)=2\epsilon_{\lambda\mu\nu\sigma}\epsilon^{\alpha\beta\mu\rho}F^\nu_{~\rho}(x) F_{\alpha\beta}(x).
\end{equation}
Notice both $\epsilon_{\lambda\nu\nu\sigma}$ and the tensor inside the square bracket are antisymmetric upon exchanging $\mu\,\nu$.
Next, since ${F_{\alpha\beta}=\partial_\alpha A_\beta-\partial_\beta A_\alpha}$ is also antisymmetric, we obtain
\begin{equation}
2\epsilon_{\lambda\mu\nu\sigma}\epsilon^{\alpha\beta\mu\rho}F^\nu_{~\rho} F_{\alpha\beta}=4\epsilon_{\lambda\mu\nu\sigma}\epsilon^{\alpha\beta\mu\rho}F^\nu_{~\rho} \partial_\alpha A_\beta=4\epsilon_{\mu\nu\lambda\sigma}\epsilon^{\mu\alpha\beta\rho}F^\nu_{~\rho}\partial_\alpha A_\beta.
\end{equation}
From this contraction identity of $\epsilon_{\mu\nu\lambda\sigma}$
\begin{equation}
\epsilon_{\mu\nu\lambda\sigma}\epsilon^{\mu\alpha\beta\rho}=-\delta_{\nu\lambda\sigma}^{\alpha\beta\rho}+\delta_{\nu\lambda\sigma}^{\alpha\rho\beta}-\delta_{\nu\lambda\sigma}^{\beta\rho\alpha}+\delta_{\nu\lambda\sigma}^{\beta\alpha\rho}-\delta_{\nu\lambda\sigma}^{\rho\alpha\beta}+\delta_{\nu\lambda\sigma}^{\rho\beta\alpha},
\end{equation}
where $\delta_{\nu\lambda\sigma}^{\alpha\beta\rho}=\delta^\alpha_\nu\delta^\beta_\lambda\delta^\beta_\sigma$,
we can show
\begin{align}
2\epsilon_{\lambda\mu\nu\sigma}\epsilon^{\alpha\beta\mu\rho}F^\nu_{~\rho} F_{\alpha\beta}& =4(-\delta_{\nu\lambda\sigma}^{\alpha\beta\rho}+\delta_{\nu\lambda\sigma}^{\alpha\rho\beta}-\delta_{\nu\lambda\sigma}^{\beta\rho\alpha}+\delta_{\nu\lambda\sigma}^{\beta\alpha\rho}-\delta_{\nu\lambda\sigma}^{\rho\alpha\beta}+\delta_{\nu\lambda\sigma}^{\rho\beta\alpha})F^\nu_{~\rho}\partial_\alpha A_\beta\nonumber\\
&=4(-\delta_{\nu\lambda\sigma}^{\alpha\beta\rho}+\delta_{\nu\lambda\sigma}^{\alpha\rho\beta}-\delta_{\nu\lambda\sigma}^{\beta\rho\alpha}+\delta_{\nu\lambda\sigma}^{\beta\alpha\rho})F^\nu_{~\rho}\partial_\alpha A_\beta\nonumber\\
&= 4(-F^\alpha_{~\sigma}\partial_\alpha A_\lambda+F^\alpha_{~\lambda}\partial_\alpha A_\sigma-F^\beta_{~\lambda}\partial_\sigma A_\beta+F^\beta_{~\sigma}\partial_\lambda A_\beta)\nonumber\\
&= 4(-F^\alpha_{~\sigma}\partial_\alpha A_\lambda+F^\alpha_{~\lambda}\partial_\alpha A_\sigma-F^\alpha_{~\lambda}\partial_\sigma A_\alpha+F^\alpha_{~\sigma}\partial_\lambda A_\alpha)\nonumber\\
&= 4[F^\alpha_{~\sigma}(-\partial_\alpha A_\lambda+\partial_\lambda A_\alpha)+F^\alpha_{~\lambda}(\partial_\alpha A_\sigma-\partial_\sigma A_\alpha)]\nonumber\\
&= 4(F^\alpha_{~\sigma} F_{\lambda\alpha}+F^\alpha_{~\lambda} F_{\alpha\sigma})=4F^\alpha_{~\sigma}(F_{\lambda\alpha}+F_{\alpha\lambda})=0
\end{align}
Thus we obtain 
\begin{equation}
\epsilon_{\lambda\mu\nu\sigma}F^{\mu\nu}_{(T)}(k,p)=\mathcal{F}\lbrace 0\rbrace=0.\label{eq:F_T_munu_projected}
\end{equation}
Therefore combined with Eq.~\eqref{eq:vector}, we have shown that the tensor term does not contribute to the projected transverse axial WGTI.
\paragraph{Contribution from the Wilson line}
Known through its definition \cite{peskin1995introduction}, the Wilson line is given by 
\begin{equation}
U_P(x',x)=P\exp \left[-ig\int_x^{x'}dy^\lambda A_\lambda(y)\right].\label{eq:Wilson_line}
\end{equation}
Its Fourier transform defines $\Gamma_\rho(k,p;\kappa)$ through \cite{He:2002jg} 
\begin{align}
& \quad \int d^4xd^4x'd^4yd^4ze^{i(k\cdot y-p\cdot z+(p-\kappa)\cdot x-(k-\kappa)\cdot x')}\langle \Omega|T\overline{\psi}(x')\gamma_\rho U_P(x',x)\psi(x)\psi(y)\overline{\psi}(z)|\Omega\rangle\nonumber\\
& =(2\pi)^4\delta^4(k-p-q)iS_F(k)\Gamma_\rho(k,p;\kappa)iS_F(p).\label{eq:nonlocal_transverse_axial_WGTI}
\end{align}
At the same time we rewrite the kernel of the $\kappa$ integral in Eq.~\eqref{eq:vector} as
\begin{equation}
(\kappa\cdot q\delta^\rho_\alpha-\kappa_\alpha q^\rho)\Gamma_\rho(k,p;\kappa)=-q^\beta\left[\kappa_\alpha\Gamma_\beta(k,p;\kappa)-\kappa_\beta\Gamma_\alpha(k,p;\kappa)\right].
\end{equation}
We are only interested in the $\kappa$ dependence of $\Gamma_\rho(k,p;\kappa)$ in order to get rid of the $\kappa$ integral. Therefore, let's consider the most general decomposition of vectors generated by two vectors $\kappa_\rho,~u_\rho(k,p)$ and two scalars $1,~\slashed{\kappa}$, where $u_\rho(k,p)$ involves some combination of vectors $k_\rho$ and $p_\rho$. We then have, in general, 
\begin{equation}
\Gamma_\rho(k,p;\kappa)=\kappa_\rho f(\kappa^2;k,p)+u_\rho(k,p)g(\kappa^2)+\slashed{\kappa}u_\rho(k,p)\tilde{g}(\kappa^2)+\slashed{\kappa}\kappa_\rho\tilde{f}(\kappa^2;k,p).
\end{equation}
We can show that first two terms do not contribute to the $\kappa$ integral due to symmetry.
Because it is an odd integral over an even domain, apparently the second term of $\Gamma_\rho(k,p;\kappa)$ does not contribute to the integral. While the first term of $\Gamma_\rho(k,p;\kappa)$ vanishes due to the following symmetry:
\begin{equation}
\int\dfrac{d^4 \kappa}{(2\pi)^4}\left(\kappa_\alpha\kappa_\beta-\kappa_\beta\kappa_\alpha\right)f(\kappa^2;k,p)\propto g_{\alpha\beta}-g_{\beta\alpha}=0.
\end{equation}
The last term does not contribute also because we are integrating an odd function over an even domain:
\begin{equation}
\int \dfrac{d^4\kappa}{(2\pi)^4}\kappa_\alpha\slashed{\kappa}\kappa_\beta\tilde{f}(\kappa^2;k,p)=\int \dfrac{d^4\kappa}{(2\pi)^4}\kappa_\alpha\kappa_\lambda\kappa_\beta\gamma^\lambda\tilde{f}(\kappa^2;k,p)=0.
\end{equation}
However, the third term does not have to vanish after the integral:
\begin{equation}
\int \dfrac{d^4\kappa}{(2\pi)^4}(\kappa_\alpha\slashed{\kappa}u_\beta(k,p)-\kappa_\beta\slashed{\kappa}u_\alpha(k,p))\tilde{g}(\kappa^2)\propto\gamma_\alpha u_\beta-\gamma_\beta u_\alpha.\\
\end{equation}
So if the coefficient functions $\tilde{g}(\kappa^2)$ did not vanish, they would have contributed to the transverse part of $\mathcal{K}^\mu$ to represent the 1PI vertex through Eq.~\eqref{eq:GGT_U}, adding more terms to the Gauge Technique construction of the fermion-photon vertex.

In this subsection, we have shown by Eq.~\eqref{eq:F_T_munu_projected} that the tensor anomaly does not contribute to the projected axial WGTI in Eq.~\eqref{eq:vector}. We have also demonstrated that the nonlocal term in Eq.~\eqref{eq:vector} is nonzero. These two results supplement discussions made in Ref.~\cite{Qin:2013mta}.
\subsection{Variable transform analysis}
We have demonstrated in the Subsection \ref{ss:proj_trans_axi_Ward_terms} that one contraction the transverse axial WGTI leads to a closed equation that formally specifies the transverse part of the vector vertex, which confirms Eqs.~(12,~13) in Ref.~\cite{Qin:2013mta}. It was also demonstrated that the nonlocal term $\Gamma_\alpha(k,p;\kappa)$ contributes essentially to the contracted identity. Realizing that explicitly knowing the nonlocal term with the Wilson line is not required; only the integral over the momentum $\kappa$ matters, we are motivated to investigate further into this term.
\paragraph{The contracted axial WGTI}
considering we have established that the anomalous term does not contribute, Eq.~\eqref{eq:vector} is reduced to
\begin{equation}
\Gamma_\alpha(k,p)-U[\gamma_\alpha]=\dfrac{1}{k^2-p^2}\int\dfrac{d^4\kappa}{(2\pi)^4}2q^\beta\left(\kappa_\beta\Gamma_\alpha(k,p;\kappa)-\kappa_\beta\Gamma_\beta(k,p;\kappa)\right),\label{eq:PAWI}
\end{equation}
where the nonlocal term $\Gamma_\alpha(k,p;\kappa)$ is defined by Eq.~\eqref{eq:nonlocal_transverse_axial_WGTI}.
The gauge link in Eq.~\eqref{eq:nonlocal_transverse_axial_WGTI} is defined through Eq.~\eqref{eq:Wilson_line}. Let's assume that we know nothing about this nonlocal term. For the interest of solving for the transverse part of the vector vertex, according to Subsection \ref{ss:proj_trans_axi_Ward_terms}, only one component of the nonlocal term, $\slashed{\kappa}v_\rho(k,p)\tilde{g}(\kappa^2)$, contributes to the overall integral. However, without knowing this component exactly, our analysis ends here in this direction.
\paragraph{Localization} Because for QED the gauge field is Abelian, we can ignore the path ordering in Eq.~\eqref{eq:Wilson_line}. Since the vertex function $\Gamma_\mu$ is local, so is its transverse part. There must exist localization procedures to finalize the $\kappa$ integral. Doing so will generate a delta-function whose argument is integrated.

First, let's look at the phase factor $e^{i\phi}$ of Eq.~\eqref{eq:nonlocal_transverse_axial_WGTI}. This factor is given by 
\begin{equation}
\phi=k\cdot y-p\cdot z+(p-\kappa)\cdot x-(k-\kappa)\cdot x'.
\end{equation}
The following substitution of variables,
\begin{equation}
u=x'-x,\quad v=\dfrac{x'+x}{2},\quad \eta=y-v-\dfrac{u}{2},\quad\zeta=z-v+\dfrac{u}{2},\label{eq:localization_variable_trans}
\end{equation}
has a unit Jacobian:
\begin{equation}
\mathrm{abs}\left( \mathrm{det} \left(\dfrac{\partial(x',x,y,z)}{\partial(u,v,\eta,\zeta)}\right)\right)=\begin{Vmatrix}
\mathbf{1/2} & \mathbf{-1/2} & \mathbf{1/2} & \mathbf{-1/2} \\ 
\mathbf{1} & \mathbf{1} & \mathbf{1} & \mathbf{1} \\
\mathbf{0} & \mathbf{0} & \mathbf{1} & \mathbf{0} \\ 
\mathbf{0} & \mathbf{0} & \mathbf{0} & \mathbf{1}
\end{Vmatrix}=1.
\end{equation}
With this change of variables, we obtain
\begin{align}
& \quad (2\pi)^4\delta^4(k-p-q)iS_F(k)\Gamma_\rho(k,p;\kappa)iS_F(p)\nonumber\\
& =\int d^4ud^4vd^4\eta d^4\zeta e^{i(k\cdot \eta-p\cdot \zeta+\kappa\cdot u)}\langle \Omega|T\overline{\psi}(v+u/2)\gamma_\rho U_P(v+u/2,v-u/2)\nonumber\\
& \quad \hspace{5cm}\times \psi(v-u/2)\psi(\eta+v+u/2)\overline{\psi}(\zeta+v-u/2)|\Omega\rangle.
\end{align}
Subsequently, $\kappa_\mu$ in momentum space can be converted into $i\dfrac{\partial}{\partial u^\mu}$ in coordinate space, after which there will be no $\kappa$ dependence except for a pure phase in the $\kappa$ part of the Fourier transform. We then use the following identity for the delta-function:
\begin{equation}
\int\dfrac{d^4\kappa}{(2\pi)^4}e^{i\kappa\cdot u}=\delta^4(u).
\end{equation}
Meanwhile, taking the derivative with respect to the variable $u$ in coordinate space produces six terms. Explicitly, we have 
\begin{align}
i\dfrac{\partial}{\partial u^\mu}U_P(v+u/2,v-u/2)& =i\dfrac{\partial}{\partial u^\mu}\exp\left[-ig\int_{v-u/2}^{v+u/2}dy^\lambda A_\lambda(y)\right]\nonumber\\
& =U_P~i\dfrac{\partial}{\partial u^\mu}\left[-ig\int_{v-u/2}^{v+u/2}dy^\lambda A_\lambda(y)\right]\nonumber\\
&= U_P~\dfrac{g}{2}\left[A_\mu(v+u/2)+A_\mu(v-u/2) \right].
\end{align}
For notational convenience, define the following rule of partial derivative to fermion field as the derivative acting on the natural argument of the field operator only\footnote{This rule only applies in this subsection.}:
\begin{equation}
\partial_\mu \psi(x)=\dfrac{\partial \psi(x)}{\partial x^\mu},
\end{equation}
such that
\begin{equation}
\dfrac{\partial}{\partial u^\mu}\psi(v\pm u/2)=\pm\dfrac{1}{2}\partial_\mu\psi(v\pm u/2).
\end{equation}
Within this notation, we obtain
\begin{align}
&\quad \int d^4 u\delta^4(u)i\dfrac{\partial}{\partial u^\beta}\langle \Omega|T\overline{\psi}(v+u/2)\gamma_\alpha U_P(v+u/2,v-u/2)\nonumber\\
& \quad \hspace{5cm}\times\psi(v-u/2)\psi(\eta+v+u/2)\overline{\psi}(\zeta+v-u/2)|\Omega\rangle\nonumber\\
& =\dfrac{-1}{2}\langle \Omega|T\left(\overline{\psi}(v)\gamma_\alpha\left(-i\overleftarrow{\partial}_\beta+i\overrightarrow{\partial}_\beta-2gA_\beta(v)\right)\psi(v) \right)\psi(\eta+v)\overline{\psi}(\zeta+v)|\Omega\rangle\nonumber\\
&\quad +\dfrac{i}{2}\langle \Omega|T\overline{\psi}(v)\gamma_\alpha\psi(v)\left(\partial_\beta\psi(\eta+v)\right)\overline{\psi}(\zeta+v)|\Omega\rangle\nonumber\\
& \quad -\dfrac{i}{2}\langle \Omega|T\overline{\psi}(v)\gamma_\alpha\psi(v)\psi(\eta+v)\left(\partial_\beta\overline{\psi}(\zeta+v)\right)|\Omega\rangle.
\end{align}
Next, we perform the change of integral variables going back into the original coordinate variables:
\begin{equation}
v=x,\quad \eta=y-x,\quad \zeta=z-x,
\end{equation}
again with a unit Jacobian. Next, after recognizing
\begin{equation}
-i\overleftarrow{\partial}_\beta+i\overrightarrow{\partial}_\beta-2gA_\beta=i\overrightarrow{D}_\beta-i\overleftarrow{D}_\beta=i\overleftrightarrow{D}_\beta,
\end{equation}
we get
\begin{align}
&\quad  (2\pi)^4\delta^4(k-p-q)iS_F(k)\int\dfrac{d^4\kappa}{(2\pi)^4}\kappa_\beta\Gamma_\alpha(k,p;\kappa)iS_F(p)\nonumber\\
&=\int d^4xd^4yd^4z e^{i(k\cdot y-p\cdot z-q\cdot x)} \Big\lbrace \dfrac{-1}{2}\langle \Omega|T\left(\overline{\psi}(x)\gamma_\alpha i\overleftrightarrow{D}_\beta\psi(x) \right)\psi(y)\overline{\psi}(z)|\Omega\rangle\nonumber\\
&\quad +\dfrac{i}{2}\langle \Omega|T\overline{\psi}(x)\gamma_\alpha\psi(x)\left(\partial_\beta\psi(y)\right)\overline{\psi}(z)|\Omega\rangle-\dfrac{i}{2}\langle \Omega|T\overline{\psi}(x)\gamma_\alpha\psi(x)\psi(y)\left(\partial_\beta\overline{\psi}(z)\right)|\Omega\rangle \Big\rbrace.
\end{align}
For notational convenience, define the vertex function $\Xi_{\alpha\beta}(k,p)$ through
\begin{align}
& \quad(2\pi)^4\delta^4(k-p-q)iS_F(k)\Xi_{\alpha\beta}(k,p)iS_F(p)\nonumber\\
& =\int d^4xd^4yd^4z e^{i(k\cdot y-p\cdot z-q\cdot x)}\langle \Omega|T\left(\overline{\psi}(x)\gamma_\alpha i\overleftrightarrow{D}_\beta\psi(x) \right)\psi(y)\overline{\psi}(z)|\Omega\rangle.\label{eq:def_Xi_alphabeta}
\end{align}
After converting $i\dfrac{\partial}{\partial y^\beta}-i\dfrac{\partial}{\partial z^\beta}$ into $k_\beta+p_\beta$ in momentum space, we obtain
\begin{align}
& \quad\int\dfrac{d^4\kappa}{(2\pi)^4}2q^\beta\left(\kappa_\beta\Gamma_\alpha(k,p;\kappa)-(\alpha\leftrightarrow\beta)\right)\nonumber\\
& =q^\beta\left(-\Xi_{\alpha\beta}+t_\beta\Gamma_\alpha(k,p)-(\alpha\leftrightarrow\beta)\right)=q\cdot t \Gamma_\alpha(k,p)-t_\alpha q\cdot\Gamma(k,p)-q^\beta(\Xi_{\alpha\beta}-\Xi_{\beta\alpha})\nonumber\\
& =(k^2-p^2)\Gamma_\alpha(k,p)-t_\alpha U[\slashed{q}]-q^\beta(\Xi_{\alpha\beta}-\Xi_{\beta\alpha}).
\end{align}
Then Eq.~\eqref{eq:PAWI}, the projected transverse axial WGTI, becomes
\begin{equation}
\Gamma_\alpha(k,p)-U[\gamma_\alpha]=\Gamma_\alpha(k,p)-\dfrac{t_\alpha}{k^2-p^2}U[\slashed{q}]-\dfrac{q^\beta}{k^2-p^2}(\Xi_{\alpha\beta}-\Xi_{\beta\alpha}),\label{eq:cancel}
\end{equation}
which is recognized as an identity for $\Xi_{\alpha\beta}(k,p)$ only, because $\Gamma_\alpha(k,p)$ on both sides apparently cancel out. After the cancellation, we have
\begin{equation}
q^\beta\left(\Xi_{\alpha\beta}(k,p)-\Xi_{\beta\alpha}(k,p) \right)=U[q\cdot t\gamma_\alpha-t_\alpha\slashed{q}]=q^\beta U[\gamma_\alpha t_\beta-\gamma_\beta t_\alpha],\label{eq:WGTI_Xi_alpha_beta}
\end{equation}
as a Ward--Green--Takahashi type of identity for $\Xi_{\alpha\beta}$. Eq.~\eqref{eq:WGTI_Xi_alpha_beta} may indicate
\begin{equation}
\Xi_{\alpha\beta}-\Xi_{\beta\alpha}=U[\gamma_\alpha t_\beta-\gamma_\beta t_\alpha].
\end{equation}
However, there may also be other transverse components in $\Xi_{\alpha\beta}$. Based on the analysis in this subsection, we conclude that knowing the nonlocal term $\Gamma_\rho(k,p,;\kappa)$ exactly, the $\slashed{\kappa}u_\rho(k,p)\tilde{g}(\kappa^2)$ component in particular, is required to determine the vector vertex directly from Eq.~\eqref{eq:vector}. Otherwise, our localization procedures indicate Eq.~\eqref{eq:vector} only contains information about another vertex that is $\Xi_{\alpha\beta}$ in Eq.~\eqref{eq:def_Xi_alphabeta}.
	\section{Functional integral derivation of tensor Ward--Green--Takahashi identity\label{ss:FunctionalDerivation}}
\subsection{Does the tensor Ward--Green--Takahashi identity contain anomalies?\label{ss:anomaly_tensor}}
Following the scheme described in Ref.~\cite{He:introQCD},
Ward--Green--Takahashi identities are deduced by requiring the generating functional to be invariant under gauge transforms. The usual (longitudinal) WGTI is the result of the following transform: 
\begin{align}
\psi'(x)=e^{i\theta(x)}\phi(x)=[1+i\theta(x)]\psi(x)+\mathcal{O}(\theta^2),\label{eq:vector_gauge_trans}
\end{align}
which gives rise to Eq.~\eqref{eq:gauge_U1_small_theta}, as in the beginning of Section \ref{ss:derivation_WGTI_longi}. Meanwhile Eq.~\eqref{eq:AWI}, the axial WGTI, is derived based on another transform, which is explicitly given by
\begin{equation}
\begin{cases}
\psi'(x)=[1+i\theta(x)\gamma_5]\psi(x)\\
\overline{\psi}'(x)=\overline{\psi}(x)[1+i\theta(x)\gamma_5].\label{eq:chiral_gauge_trans}
\end{cases}
\end{equation}
Equations~(\ref{eq:vector_gauge_trans},~\ref{eq:chiral_gauge_trans}) are known to Ref.~\cite{He:introQCD}. In the case of Eq.~\eqref{eq:chiral_gauge_trans}, the functional integral measure is modified by Eq.~\eqref{eq:chiral_gauge_trans}, resulting in a non-trivial (not just a number) Jacobian \cite{He:introQCD}. This Jacobian leads to the Adler--Bell--Jackiw anomalous term in the axial WGTI \cite{PhysRevD.21.2848}. This nontrivial Jacobian can be calculated according to the Fujikawa approach \cite{PhysRevLett.42.1195}.

Based on Ref.~\cite{He:introQCD}, the gauge transform leading to the transverse WGTI is 
\begin{equation}
\begin{cases}
\psi'(x)=\left[1+\dfrac{g}{4}\theta(x)\epsilon^{\mu\nu}\sigma_{\mu\nu}\right]\psi(x)\\
\overline{\psi}'(x)=\overline{\psi}(x)\left[1+\dfrac{g}{4}\theta(x)\epsilon^{\mu\nu}\sigma_{\mu\nu}\right].\label{eq:trans_gauge_trans}
\end{cases}
\end{equation}
where $\epsilon^{\mu\nu}$ is an arbitrary constant tensor without any Dirac structures. Here in Eq.~\eqref{eq:trans_gauge_trans}, the symmetric part of epsilon, $(\epsilon^{\mu\nu}+\epsilon^{\nu\mu})/2$, does not contribute. Meanwhile, Eq.~\eqref{eq:trans_gauge_trans} produces no anomalous term to the transverse Ward--Green--Takahashi identity because it keeps the functional measure invariant.

The transform leading to the transverse axial WGTI is also given by Ref.~\cite{He:introQCD} as
\begin{equation}
\begin{cases}
\psi'(x)=\left[1+\dfrac{g}{4}\theta(x)\epsilon^{\mu\nu}\sigma_{\mu\nu}\gamma_5\right]\psi(x)\\
\overline{\psi}'(x)=\overline{\psi}'(x)\left[1-\dfrac{g}{4}\theta(x)\epsilon^{\mu\nu}\sigma_{\mu\nu}\gamma_5\right],
\end{cases}
\end{equation}
which gives rise to an anomalous term as the last term on the right-hand side of Eq.~\eqref{eq:TA_WGTI}.

The tensor WGTI is derived by in Ref.~\cite{PhysRevC.63.025207} using the canonical operator approach, through which the existence of anomalous terms was unclear. Quantum anomalies of WGTI can be investigated by carefully taking derivatives with respect to locally gauge invariant operators \cite{He:2001cu}. It has been shown, through perturbative calculation, that the tensor WGTI identity for $\mathrm{U}(1)$ gauge theory is free of anomalies \cite{Sun:2003ia}. We will verify this statement about the anomalous term in the tensor WGTI through the functional approach.

We propose that the gauge transform on fermion fields to derive tensor WGTI is
\begin{equation}
\psi'(x)=[1+i\Theta(x)]\psi(x), \quad \overline{\psi}'(x)=\overline{\psi}(x)[1-i\Theta(x)],
\label{eq:tensor_gauge_trans}
\end{equation}
with
\begin{equation}
\Theta(x)=\theta(x)\tau_{\mu\nu\alpha}\epsilon^{\mu\nu\alpha\rho}\gamma_\rho\gamma_5.
\label{eq:def_tensor}
\end{equation}
Here $\tau_{\mu\nu\alpha}$ is an arbitrary antisymmetric constant tensor without any Dirac structures. The effectiveness of Eq.~\eqref{eq:tensor_gauge_trans} will be demonstrated in Subsection \ref{ss:Tensor_WGTI}.

Based on the discussion in Ref.~\cite{PhysRevD.21.2848}, anomalous terms in various Ward--Green--Takahashi identities come from the Jacobian of functional measure as a result of the gauge transforms on the fermion fields. Explicitly, when the gauge transform on the fermion fields is given by Eq.~\eqref{eq:tensor_gauge_trans}, we have the following change in the functional measure: 
\begin{equation}
\mathcal{D}\overline{\psi}'\mathcal{D}\psi'=\mathcal{D}\overline{\psi}\mathcal{D}\psi\exp\left[-2i\int d^4x\lim\limits_{M\rightarrow+\infty}\langle x|\mathrm{Tr}~\Theta(x)e^{i\slashed{D}^2/M^2}|x\rangle\right].\label{eq:WTI_starting}
\end{equation} 
meanwhile, we have $(i\slashed{D})^2=-D^2-\dfrac{e}{2}\sigma_{\mu\nu}F^{\mu\nu}$. The background field $A^\mu$ can be ignored. Substituting Eq.~\eqref{eq:def_tensor} into Eq.~\eqref{eq:WTI_starting} results in
\begin{align}
& \quad \lim\limits_{M\rightarrow+\infty}\langle x|\mathrm{Tr}~\Theta(x)e^{i\slashed{D}^2/M^2}|x\rangle\nonumber\\
& =\theta(x)\tau_{\mu\nu\alpha}\epsilon^{\mu\nu\alpha\rho}\lim\limits_{M\rightarrow+\infty}\mathrm{Tr}~\bigg\{\gamma_\rho\gamma_5 \exp\left(-\dfrac{e}{2M^2}\sigma:F\right)\bigg\}\langle x|e^{-\partial^2/M^2}|x\rangle\nonumber\\
& =\theta(x)\tau_{\mu\nu\alpha}\epsilon^{\mu\nu\alpha\rho}\lim\limits_{M\rightarrow+\infty}\dfrac{iM^4}{(4\pi)^2}\mathrm{Tr}~\bigg\{\gamma_\rho\gamma_5 \exp\left(-\dfrac{e}{2M^2}\sigma:F\right)\bigg\},
\end{align}
where $\sigma:F=\sigma_{\mu\nu}F^{\mu\nu}$. The trace operated term vanishes because there are only terms with odd number of gamma matrices in the arguments of the trace. Alternatively since $\gamma_5$ commutes with $\sigma_{\mu\nu}$, \[\gamma_5\exp\left(-\dfrac{e}{2M^2}\sigma:F\right)=\exp\left(-\dfrac{e}{2M^2}\sigma:F\right)\gamma_5,\]
we arrive at
\begin{align}
& \quad \mathrm{Tr}~\bigg\{\gamma_\rho\gamma_5 \exp\left(-\dfrac{e}{2M^2}\sigma:F\right)\bigg\}=\mathrm{Tr}~\bigg\{\gamma_\rho \exp\left(-\dfrac{e}{2M^2}\sigma:F\right)\gamma_5\bigg\}\nonumber\\
& =\mathrm{Tr}~\bigg\{\gamma_5\gamma_\rho \exp\left(-\dfrac{e}{2M^2}\sigma:F\right)\bigg\}=-\mathrm{Tr}~\bigg\{\gamma_\rho\gamma_5 \exp\left(-\dfrac{e}{2M^2}\sigma:F\right)\bigg\}.
\end{align}
Therefore $\mathrm{Tr}~\bigg\{\gamma_\rho\gamma_5 \exp\left(-\dfrac{e}{2M^2}\sigma:F\right)\bigg\}=0$. So the gauge transform given by Eq.~\eqref{eq:def_tensor} does not modify the functional integral measure. Since only the Jacobian of functional measure contributes to the anomalous terms in the WGTIs, there is no anomaly in the tensor identity.
\subsection{Tensor Ward--Green--Takahashi identity from the functional integral approach\label{ss:Tensor_WGTI}}
The Ward--Green--Takahashi identity for the generating functional can be derived following procedures discussed in Section \ref{ss:derivation_WGTI_longi}. When deriving the identity for one specific n-point function, an alternative method also based on the functional integral is well documented in Refs.~\cite{He:introQCD,PhysRevD.21.2848,PhysRevLett.42.1195}. Specifically, to derive the WGTIs for the three-point functions, let's start from the fermion two-point function being invariant under the local gauge transform given by Eq.~\eqref{eq:tensor_gauge_trans}, with an appropriate local phase function $\Theta$. Therefore considering an infinitesimal transform, the following equation must hold: 
\begin{equation}
\int \mathcal{D}\overline{\psi}'\mathcal{D}\psi'\mathcal{D}A^\mu \,\psi'(y)\overline{\psi}'(z)e^{iS'}=\int \mathcal{D}\overline{\psi}\mathcal{D}\psi\mathcal{D}A^\mu \,\psi(y)\overline{\psi}(z)e^{iS},\label{eq:WTI_functional}
\end{equation}
where $S=\int d^4x \mathcal{L}(x)$. Here one immediately notices that external sources are not introduced. For QED, prior to gauge fixing, we have
\begin{align}
& \mathcal{L}=\overline{\psi}(i\slashed{D}-m)\psi-\dfrac{1}{4}F_{\mu\nu}F^{\mu\nu},\\
& \mathcal{L'}=\overline{\psi}'(i\slashed{D}-m)\psi'-\dfrac{1}{4}F_{\mu\nu}F^{\mu\nu},
\end{align}
with $D^\mu=\partial^\mu+ieA^\mu, \quad F^{\mu\nu}=\partial^\mu A^\nu-\partial ^\nu A^\mu$. For the purpose of deducing the identity for the fermion-photon vertex, only fermion terms of the Lagrangian matter.

Since we have established in Subsection \ref{ss:anomaly_tensor} that the Jacobian of a gauge transform given by Eq.~\eqref{eq:def_tensor} is a constant, we only need to work out the implication of the gauge invariance requirement from Eq.~\eqref{eq:WTI_functional} implied by the local gauge transform on the action. First, the fermion part of the Lagrangian transforms in the following manner:
\begin{equation}
\overline{\psi}'(i\slashed{D}-m)\psi'=\overline{\psi}(1-i\Theta)(i\slashed{D}-m)(1+i\Theta)\psi=\overline{\psi}(i\slashed{D}-m)\psi+\overline{\psi}[\Theta,\slashed{D}]\psi+\mathcal{O}(\Theta^2).
\end{equation}
Next, because $D^\mu(x)$ has a differential piece and $\Theta(x)$ is a local function, we have
\begin{align}
-\slashed{D}\Theta & =-\tau_{\mu\nu\alpha}\epsilon^{\mu\nu\alpha\rho}(\slashed{\partial}+ie\slashed{A})\theta(x)\gamma_\rho\gamma_5 =-\tau_{\mu\nu\alpha}\epsilon^{\mu\nu\alpha\rho}[(\slashed{\partial}\theta)+\theta(\slashed{\partial}+ie\slashed{A})]\gamma_\rho\gamma_5\nonumber\\
& =-\tau_{\mu\nu\alpha}\epsilon^{\mu\nu\alpha\rho}[(\slashed{\partial}\theta)+\theta \slashed{D}]\gamma_\rho\gamma_5.
\end{align}
Then we obtain the following identity for the commutator $[\Theta,\slashed{D}]$: 
\begin{equation}
[\Theta,\slashed{D}] =\tau_{\mu\nu\alpha}\epsilon^{\mu\nu\alpha\rho}[\theta\gamma_\rho\gamma_5\slashed{D}-(\slashed{\partial}\theta)\gamma_\rho\gamma_5-\theta \slashed{D}\gamma_\rho\gamma_5]=\tau_{\mu\nu\alpha}\epsilon^{\mu\nu\alpha\rho}[-\theta(\gamma_\rho\slashed{D}+\slashed{D}\gamma_\rho)\gamma_5-(\partial^\lambda\theta)\gamma_\lambda\gamma_\rho\gamma_5].\label{eq:commutation_Theta_D}
\end{equation}
Combining
\begin{equation}
\gamma_\lambda\gamma_\rho=\dfrac{1}{2}\{\gamma_\lambda,\gamma_\rho \}+\dfrac{1}{2}[\gamma_\lambda,\gamma_\rho]=g_{\lambda\rho}-i\sigma_{\lambda\rho}
\end{equation}
with Eq.~\eqref{eq:commutation_Theta_D} gives
\begin{equation}
\overline{\psi}[\Theta,\slashed{D}]\psi=\tau_{\mu\nu\alpha}\epsilon^{\mu\nu\alpha\rho}\overline{\psi}[-2\theta D_\rho\gamma_5-(\partial_\rho\theta)\gamma_5+i(\partial^\lambda\theta)\sigma_{\lambda\rho}\gamma_5]\psi.
\end{equation}
Since everything is under the integral over $x$, applying integration by parts produces
\begin{align}
& \quad \overline{\psi}[-2\theta D_\rho\gamma_5-(\partial_\rho\theta)\gamma_5]\psi \nonumber\\
&  \rightarrow \theta(-2\overline{\psi}D_\rho\gamma_5 \psi+\overline{\psi}\partial_\rho\gamma_5 \psi+\overline{\psi}\overleftarrow{\partial}_\rho\gamma_5 \psi)\nonumber\\
& =\theta(\overline{\psi}\overleftarrow{\partial}_\rho\gamma_5 \psi-\overline{\psi}\overrightarrow{\partial}_\rho\gamma_5 \psi-2ieA_\rho\overline{\psi}\gamma_5 \psi)\nonumber\\
& =-\theta(\overline{\psi}\overleftrightarrow{D}_\rho\gamma_5 \psi)\nonumber\\
& =\theta(x)\lim\limits_{x'\rightarrow x}(\partial_{x'}-\partial_{x})_\rho\overline{\psi}(x')U_P(x',x)\gamma_5\psi(x),
\end{align}
where the Wilson line $U_P(x',x)=P\exp\left[-ie\int_{x}^{x'}dy\cdot A(y)\right]$ is introduced to ensure local gauge invariance.

For the remaining term of Eq.~\eqref{eq:commutation_Theta_D}, recall the following commutation relations for triple gamma matrices:
\begin{equation}
[\gamma^\mu,\sigma^{\nu\rho}]=2i(g^{\mu\nu}\gamma^\rho-g^{\mu\rho}\gamma^\nu), \quad \{\gamma^\lambda, \sigma^{\mu\nu}\}=-2\epsilon^{\lambda\mu\nu\rho}\gamma_\rho\gamma_5.
\end{equation}
We then have 
\begin{align}
& \quad\epsilon^{\mu\nu\alpha\rho}i(\partial^\lambda\theta)\sigma_{\lambda\rho}\gamma_5\nonumber\\
& =-\dfrac{(\partial^\lambda\theta)}{2}(\gamma_\lambda\epsilon^{\mu\nu\alpha\rho}\gamma_\rho\gamma_5+\epsilon^{\mu\nu\alpha\rho}\gamma_\rho\gamma_5\gamma_\lambda)=\dfrac{(\partial^\lambda\theta)}{4}\{\gamma_\lambda\,\{\gamma^\mu,\sigma^{\nu\alpha} \}\}\nonumber\\
& =\dfrac{(\partial_\lambda\theta)}{4}(\gamma^\lambda\gamma^\mu\sigma^{\nu\alpha}+\gamma^\lambda\sigma^{\nu\alpha}\gamma^\mu+\gamma^\mu\sigma^{\nu\alpha}\gamma^\lambda+\sigma^{\nu\alpha}\gamma^\mu\gamma^\lambda)\nonumber\\
& =\dfrac{(\partial_\lambda\theta)}{4}\big\{\gamma^\lambda\gamma^\mu\sigma^{\nu\alpha}+\sigma^{\nu\alpha}\gamma^\lambda\gamma^\mu+2i(g^{\lambda\nu}\gamma^\alpha-g^{\lambda\alpha}\gamma^\nu)\gamma^\mu +\gamma^\mu\gamma^\lambda\sigma^{\nu\alpha}\nonumber\\
& \quad\hspace{5cm} -2i\gamma^\mu(g^{\lambda\nu}\gamma^\alpha-g^{\lambda\alpha}\gamma^\nu)+\sigma^{\nu\alpha}\gamma^\mu\gamma^\lambda\big\}\nonumber\\
& =\dfrac{(\partial_\lambda\theta)}{4}[4g^{\lambda\mu}\sigma^{\nu\alpha}+4g^{\lambda\nu}\sigma^{\alpha\mu}+4g^{\lambda\alpha}\sigma^{\mu\nu}]\nonumber\\
& =(\partial^\mu\theta)\sigma^{\nu\alpha}+(\partial^\nu\theta)\sigma^{\alpha\mu}+(\partial^{\alpha}\theta)\sigma^{\mu\nu}.
\end{align}
Therefore, the action part of the gauge transform becomes, up to higher orders in $\theta$,
\begin{align}
e^{iS'}=& e^{iS}\Bigg\{1+i\int d^4x\Big\{\tau_{\mu\nu\alpha}\overline{\psi}(x)\left[(\partial^\mu\theta)\sigma^{\nu\alpha}+(\partial^\nu\theta)\sigma^{\alpha\mu}+(\partial^{\alpha}\theta)\sigma^{\mu\nu}\right]\psi(x)\nonumber\\
& +\theta(x)\tau_{\mu\nu\alpha}\epsilon^{\mu\nu\alpha\rho}\lim\limits_{x'\rightarrow x}(\partial_{x'}-\partial_{x})_\rho\overline{\psi}(x')U_P(x',x)\gamma_5\psi(x)\Big\}+\mathcal{O}(\theta^2)\Bigg\}.\label{eq:tensor_gauge_trans_action}
\end{align}
The remaining part is trivial and can be calculated according to
\begin{align}
\psi'(y)\overline{\psi}'(z) &=[1+i\Theta(y)]\psi(y)\overline{\psi}(z)[1-i\Theta(z)]\nonumber\\
& =\psi(y)\overline{\psi}(z)+i\Theta(y)\psi(y)\overline{\psi}(z)-i\psi(y)\overline{\psi}(z)\Theta(z)\nonumber\\
& =\psi(y)\overline{\psi}(z)+i\theta(y)\tau_{\mu\nu\alpha}\epsilon^{\mu\nu\alpha\rho}\gamma_\rho\gamma_5\psi(y)\overline{\psi}(z)-i\theta(z)\tau_{\mu\nu\alpha}\epsilon^{\mu\nu\alpha\rho}\psi(y)\overline{\psi}(z)\gamma_\rho\gamma_5.\label{eq:tensor_trans_prop}
\end{align}
Substituting Eqs.~(\ref{eq:tensor_gauge_trans_action},~\ref{eq:tensor_trans_prop}) into Eq.~\eqref{eq:WTI_functional}, expanding up-to linear orders in $\theta$, and then taking the functional derivative with respect to $\theta(x)$ yields the tensor WGTI in coordinate space. Explicitly, the tensor identity is
\begin{align}
& \quad \partial^\mu\langle \Omega|Tj^{\nu\alpha}(x)\psi(y)\overline{\psi}(z)|\Omega\rangle+ \partial^\nu\langle \Omega|Tj^{\alpha\mu}(x)\psi(y)\overline{\psi}(z)|\Omega\rangle+ \partial^\alpha\langle \Omega|Tj^{\mu\nu}(x)\psi(y)\overline{\psi}(z)|\Omega\rangle\nonumber\\
& =\delta^{(4)}(y-x)\epsilon^{\mu\nu\alpha\rho}\gamma_\rho\gamma_5\langle\Omega|T\psi(y)\overline{\psi}(z)|\Omega\rangle -\delta^{(4)}(z-x)\epsilon^{\mu\nu\alpha\rho}\langle\Omega|T\psi(y)\overline{\psi}(z)|\Omega\rangle\gamma_\rho\gamma_5\nonumber\\
& \quad +\epsilon^{\mu\nu\alpha\rho}\lim\limits_{x'\rightarrow x}(\partial_{x'}-\partial_{x})_\rho\langle\Omega|T\overline{\psi}(x')U_P(x',x)\gamma_5\psi(x)\psi(y)\overline{\psi}(z)|\Omega\rangle,\label{eq:Tensor_WTI_cord}
\end{align}
where the tensor current $j^{\mu\nu}(x)$ is defined by ${j^{\mu\nu}(x)=\overline{\psi}(x)\sigma^{\mu\nu}\psi(x)}$. Notice there is a sign difference between Wilson line terms in Eq.~\eqref{eq:Tensor_WTI_cord} and the corresponding term in Eq.~(6) of Ref.~\cite{PhysRevC.63.025207}.

The tensor WGTI was previously derived in Ref.~\cite{PhysRevC.63.025207} by the canonical approach. So far in this section we have adapted the method in Refs.~\cite{He:introQCD,PhysRevD.21.2848,PhysRevLett.42.1195} to derive the tensor WGTI based on the functional approach. Our analysis of the functional Jacobian confirms that the tensor WGTI given by Eq.~\eqref{eq:Tensor_WTI_cord} is free from any anomaly.
\subsection{Tensor Ward--Green--Takahashi identity in momentum space\label{ss:tensor_WGTI_momentum}}
In order to obtain the tensor WGTI in momentum space, we need to carry out the Fourier transform of Eq.~\eqref{eq:Tensor_WTI_cord}. The only nontrivial term is the one with the Wilson line. Following Ref.~\cite{PhysRevC.63.025207}, let us first define $\Gamma_5(k,p;\kappa)$ through
\begin{align}
&\quad 
\int d^4xd^4x'd^4yd^4ze^{i[k\cdot y-p\cdot z+(p-\kappa)\cdot x-(k-\kappa)\cdot x']}\langle\Omega|T\overline{\psi}(x')U_P(x',x)\gamma_5\psi(x)\psi(y)\overline{\psi}(z)|\Omega\rangle\nonumber\\
& =(2\pi)^4\delta^{(4)}(k-p-q)iS_F(k)\Gamma_5(k,p;\kappa)iS_F(p).
\end{align}
Considering $\lim\limits_{x'\rightarrow x}=\int dx'\delta^{(4)}(x-x')=\int dx'\int d\underline{\kappa}e^{-i\kappa\cdot(x-x')}$, we obtain
\begin{align}
& \quad\int dxdydz\,e^{i(k\cdot y-p\cdot z-q\cdot x)}\lim\limits_{x'\rightarrow x}(\partial_{x'}-\partial_{x})_\rho\langle\Omega|T\overline{\psi}(x')U_P(x',x)\gamma_5\psi(x)\psi(y)\overline{\psi}(z)|\Omega\rangle\nonumber\\
& =\int d\underline{\kappa}\int d^4xd^4x'd^4yd^4ze^{i[k\cdot y-p\cdot z+(p-\kappa)\cdot x-(k-\kappa)\cdot x']}\langle\Omega|T\overline{\psi}(x')U_P(x',x)\gamma_5\psi(x)\psi(y)\overline{\psi}(x)|\Omega\rangle\nonumber\\
& =(2\pi)^4\delta^{(4)}(k-p-q)iS_F(k)\left[\int d\underline{\kappa}~i(t-2\kappa)_\rho\Gamma_5(k,p;\kappa)\right]iS_F(p).
\end{align}
Further applying the obvious relation, $\int d\underline{\kappa}\Gamma_5(k,p;\kappa)=\Gamma_5(k,p)$, results in
\begin{align}
&\quad q^\mu\Gamma^{\nu\alpha}(k,p)+q^\nu\Gamma^{\alpha\mu}(k,p)+q^\alpha\Gamma^{\mu\nu}(k,p)\nonumber\\
& =\epsilon^{\mu\nu\alpha\rho}\left[-S_F^{-1}(k)\gamma_\rho\gamma_5+\gamma_\rho\gamma_5S_F^{-1}(p)+t_\rho\Gamma_5(k,p)-\int d\underline{\kappa}2\kappa_\rho\Gamma_5(k,p;\kappa)\right].\label{eq:tensor_WTI_momentum}
\end{align}
If we ignore the possible $\kappa$ odd terms in $\Gamma_5(k,p;\kappa)$, as is in Ref.~\cite{PhysRevC.63.025207}, we obtain Eq.~(9) of Ref.~\cite{PhysRevC.63.025207} with a minus sign difference in the pseudoscalar term. The term $\int d\underline{\kappa}\kappa_\rho\Gamma_5(k,p;\kappa)$ only vanishes if $\Gamma_5(k,p;\kappa)$ has no $\kappa^\mu$ dependence, or equivalently, ${\Gamma_5(k,p;\kappa)=\Gamma_5(k,p;\kappa^2)}$. In general, this is not true.

Next, Eq.~\eqref{eq:GGT_U} rewrites Eq.~\eqref{eq:tensor_WTI_momentum} into
\begin{equation}
q^\mu(\Gamma^{\nu\alpha}(k,p)-U[\sigma^{\nu\alpha}])+(\nu\alpha\mu)+(\alpha\mu\nu)=\epsilon^{\lambda\mu\nu\alpha}\left[t_\lambda\left(U[\gamma_5]-\Gamma_5(k,p)\right)+\int d\underline{\kappa}2\kappa_\lambda\Gamma_5(k,p;\kappa)\right].\label{eq:WGTI_tensor_coupled}
\end{equation}
Eq.~\eqref{eq:WGTI_tensor_coupled} is the Ward--Green--Takahashi identity for the tensor vertex deduced from the gauge transform specified by Eqs.~(\ref{eq:tensor_gauge_trans},~\ref{eq:def_tensor}).
We derive, for the first time, the Ward--Green--Takahashi identity for the pseudoscalar vertex:
\begin{equation}
(k^2-p^2)(\Gamma_5(k,p)-U[\gamma_5])=\int d\underline{\kappa}2q\cdot \kappa\Gamma_5(k,p;\kappa).\label{eq:WGTI_pseudoscalar}
\end{equation}
Equation~\eqref{eq:WGTI_pseudoscalar} is obtained by projecting Eq.~\eqref{eq:WGTI_tensor_coupled} with $\epsilon_{\mu\nu\alpha\beta}q^\beta$.
	\chapter{Solving the SDEs for the Minkowski-space propagators\label{ch:SDE_oGT}}
	\section{General discussions}
Having derived the SDEs for QED propagators in Chapter \ref{cp:path_itg_QFT} as Eqs.~(\ref{eq:SDE_fermion_momentum},~\ref{eq:SDE_photon_momentum}), we immediately encounter the question of how to truncate them. Specifically, the construction of an Ansatz for the fermion-photon three-point function is required. The gauge symmetry of QED relates this vertex to the fermion propagator in the form of the Ward--Green--Takahashi identity as in Eq.~\eqref{eq:WGTI_longitudinal}, which specifies the longitudinal part of the vertex. Unfortunately, as shown in Chapter \ref{cp:WGTIs}, the transverse WGTI couples to other identities and contains unknown terms, rendering it difficult to solve for the transverse vertex exactly from the transverse WGTI. Meanwhile, combining the spectral representation for the fermion propagator with the WGTI produces the original Gauge Technique discussed in Chapter \ref{cp:spec_repr}. Although the Gauge Technique is missing crucial transverse terms, it is still worthwhile to solve for the spectral functions from the SDEs using this Ansatz. 

Since for QED, electrons are particles visible to detectors, there has to be an on-shell component within the QED fermion propagator, based on the Lehmann--Symanzik--Zimmermann reduction theorem. In this chapter, the original Gauge Technique is modified such that on-shell renormalization of the fermion propagator SDE can be accomplished. After the on-shell renormalization, the spectral functions for the fermion propagator are solved analytically in the quenched approximation in the Landau gauge. Next, the unquenching of photons by introducing a spectral representation for the photon propagator is also briefly discussed.

The analytic properties of propagators have been discussed in Chapter \ref{cp:spec_repr}. They are required to solve for the propagators in Minkowski space from their SDEs. Just as the fermion propagator itself, the spectral function $\rho(W)$ is also renormalization scheme dependent. From Ref.~\cite{Delbourgo:1977jc}, one renormalization scheme is based on
\[1=Z_2\int dW\rho(W),\quad mZ_m=Z_2\int dW~W\rho(W).\]
Within this scheme, the spectral functions for the fermion propagator, when written as $\rho(W)=\delta(W-m)+r(W)$, can be solved with the original Gauge Technique. As given by Eq.~(20) of Ref.~\cite{Delbourgo:1977jc}, the result is
\begin{align}
r(W)& =-\mathrm{sign}(W)\theta(W^2-m^2)\dfrac{2a}{W}\left(\dfrac{W^2-m^2}{\mu^2}\right)^{-2a}\dfrac{m^2}{W^2-m^2}\nonumber\\
&\quad \hspace{1cm}\times\Bigg\{~_2F_1\left(-a,-a;-2a;1-\dfrac{W^2}{m^2}\right)+\dfrac{W}{m}~_2F_1\left(-a, 1-a, -2a, 1-\dfrac{W^2}{m^2}\right)\Bigg\}\label{eq:rW_Delbourgo}
\end{align}	
where $a=3\alpha/(4\pi)$. The singular behavior of hypergeometric functions in Eq.~\eqref{eq:rW_Delbourgo} near the mass shell results in functions more singular than the free-particle propagators. This can be verified by applying Eq.~(15.3.6)
of Ref.~\cite{abramowitz1964handbook} to Eq.~(21) in Ref.~\cite{Delbourgo:1977jc} in the $p^2\rightarrow m^2$ limit. Since physical propagators for electrons should not behave more singular than the free-particle propagator, we are motivated to solve the SDE for the fermion propagator within the on-shell renormalization scheme. Additional modifications to the divergent parts of the fermion equations are required by renormalizability, after which a different set of solutions to $r_{j}(s)$ can be obtained.
\section{Loop-renormalizability of the the fermion equation\label{ss:loop_ren_SF}}
The SDE for the fermion propagator is given by Eq.~\eqref{eq:SDE_fermion_momentum}. With the Gauge Technique, Eq.~\eqref{eq:SDE_fermion_momentum} becomes
\begin{equation}
1=(\slashed{p}-m)S_{F}(p)+ie^2\int d\underline{k}\int dW\gamma^\nu\dfrac{1}{\slashed{k}-W}\gamma^\mu\dfrac{1}{\slashed{p}-W}D_{\mu\nu}(q)\rho(W).\label{eq:SDE_fermion_bare}
\end{equation}
Next, for notational convenience, define the following functions of $p^2$ also linear in the spectral function $\rho(W)$:
\begin{equation}
\sigma(p)=\sigma_1(p^2)+\slashed{p}\sigma_2(p^2)=ie^2\int d\underline{k}\int dW\gamma^\nu\dfrac{1}{\slashed{k}-W}\gamma^\mu\dfrac{1}{\slashed{p}-W}D_{\mu\nu}(q)\rho(W).\label{eq:def_sigma_j_SDE}
\end{equation}
They are related to the fermion self-energy by merely a factor of fermion propagator. 

On one hand, the divergent part of Eq.~\eqref{eq:def_sigma_j_SDE} can be calculated. Specifically in the quenched approximation, the dressing function $G(q^2)$ of Eq.~\eqref{eq:def_photon_propagator} is set to unity as for a free photon. 
Then after adopting dimensional regularization, explicit calculation shows that the divergent part of Eq.~\eqref{eq:def_sigma_j_SDE} is given by 
\begin{equation}
\sigma(p)=-\dfrac{\alpha}{4\pi\epsilon}\left[3p^2S_1(p^2)-(\xi+3)+3S_2(p^2)\slashed{p} \right]+\mathcal{O}(\epsilon^0).\label{eq:div_sigma_j_SDE}
\end{equation}
On the other hand, based on the multiplicative renormalization of QED \cite{peskin1995introduction}, Eq.~\eqref{eq:SDE_fermion_bare} becomes
\begin{align}
& Z_2^{-1}+mZ_mS_2(p^2)=p^2S_1(p^2)+\sigma_1(p^2) \label{eq:SDE_R_Z2}\\ 
& mZ_mS_1(p^2)=S_2(p^2)+\sigma_2(p^2).\label{eq:SDE_R_Zm}
\end{align}
Equations~(\ref{eq:SDE_R_Z2},~\ref{eq:SDE_R_Zm}) are coupled identities for fermion propagator functions $S_{j}(p^2)$ with $j=1,~2$. In order to derive the corresponding equations for spectral functions $\rho_{j}(s)$, naturally we should proceed in finding out how to generate these $p^2$ dependences from the free-particle propagator. However, we are faced with a more immediate problem that the Gauge Technique Ansatz for the fermion-photon vertex produces divergences in $\sigma_{j}(p^2)$ not completely removable by the renormalization conditions\footnote{Compare Eq.~\eqref{eq:SDE_R_Z2} and Eq.~\eqref{eq:SDE_R_Zm} with themselves at any renormalization point.}, as will be demonstrated later in this section.

It is well known that the ability to remove the loop divergences in the SDEs constrains the fermion-photon vertex \cite{Curtis:1990zs}. To distinguish this from multiplicative renormalizability discussed in Chapter \ref{cp:div_QED}, we name it \textit{loop-renormalizability}. The principle of loop-renormalization is best illustrated going back to the SDE for the propagator functions. With the conventional definitions of dressing functions $\mathcal{F}(p^2)$ and $\mathcal{M}(p^2)$ given by Eq.~\eqref{eq:SF_Dirac_struct}, the relations among $S_{j}(p^2)$ and $\mathcal{F}(p^2)$, $\mathcal{M}(p^2)$ can be immediately worked out. Meanwhile, the fermion self-energy is defined as
\begin{equation}
\Sigma_1(p^2)\slashed{p}+\Sigma_2(p^2)=(\sigma_1+\slashed{p}\sigma_2)S_F^{-1}=\dfrac{\sigma_1S_1-\sigma_2S_2}{p^2S_1^2-S_2^2}\slashed{p}+\dfrac{p^2S_1\sigma_2-\sigma_1S_2}{p^2S_1^2-S_2^2}.
\end{equation}
Notice that ${p^2S_1^2(p^2)-S_2^2(p^2)=\mathcal{F}(p^2)S_1(p^2)}$. Within the Gauge Technique in the quenched approximation, divergent parts of fermion self-energy can then be calculated according to Eq.~\eqref{eq:div_sigma_j_SDE}. Explicitly, we have
\begin{align}
\Sigma_1(p^2)& =-\dfrac{3\alpha}{4\pi\epsilon}\left[p^2S_1^2-S_2^2-\left(\dfrac{\xi}{3}+1\right)S_1\right]\dfrac{1}{\mathcal{F}S_1}\nonumber+\mathcal{O}(\epsilon^0)\label{eq:Sigma_1_ori}\\
& =-\dfrac{3\alpha}{4\pi\epsilon}\left[1-\left(\dfrac{\xi}{3}+1\right)\dfrac{1}{\mathcal{F}(p^2)}\right]+\mathcal{O}(\epsilon^0)\\
\Sigma_2(p^2)&=-\dfrac{3\alpha}{4\pi\epsilon}\left[p^2S_1S_2-\left(p^2S_1-\left(\dfrac{\xi}{3}+1\right)\right)S_2\right]\dfrac{1}{\mathcal{F}S_1}+\mathcal{O}(\epsilon^0)\nonumber\\
& =-\dfrac{3\alpha}{4\pi\epsilon}\left(\dfrac{\xi}{3}+1\right)\dfrac{\mathcal{M}(p^2)}{\mathcal{F}(p^2)}+\mathcal{O}(\epsilon^0).\label{eq:Sigma_2_ori}
\end{align}
The renormalized SDEs for the fermion propagator dressing functions are written as
\begin{align}
& \dfrac{Z_2^{-1}}{\mathcal{F}(p^2)}=1+\Sigma_1(p^2),\label{eq:SDE_F}\\
& \dfrac{Z_2^{-1}\mathcal{M}(p^2)}{\mathcal{F}(p^2)}=m_RZ_m-\Sigma_2(p^2).\label{eq:SDE_MF}
\end{align}
Since with the $\Sigma_1(p^2)$ given by Eq.~\eqref{eq:Sigma_1_ori}, Eq.~\eqref{eq:SDE_F} cannot be renormalized when compared to itself at $p^2=\mu^2$, the Gauge Technique fails the loop-renormalizability requirement. Explicitly, this ratio of equations,
\begin{equation}
\dfrac{\mathcal{F}(\mu^2)}{\mathcal{F}(p^2)}=\dfrac{1+\Sigma_1(p^2)}{1-\Sigma_1(\mu^2)},
\end{equation}
contains divergences not taken care of.

Consider instead of Eqs.~(\ref{eq:Sigma_1_ori},~\ref{eq:Sigma_2_ori}), the divergences in the fermion self-energy are given by
\begin{align}
\Sigma_1(p^2)& =\dfrac{\alpha\xi}{4\pi\epsilon}\dfrac{Z_2^{-1}}{\mathcal{F}(p^2)}+\mathcal{O}(\epsilon^0),\label{eq:Sigma_1_alt}\\
\Sigma_2(p^2)& =-\dfrac{\alpha\xi}{4\pi\epsilon}\dfrac{Z_2^{-1}M(p^2)}{\mathcal{F}(p^2)}+\mathcal{O}(\epsilon^0).\label{eq:Sigma_2_alt}
\end{align}
Here in Eq.~\eqref{eq:Sigma_1_alt}, the divergent part of $\Sigma_1$ is homogeneous with respect to ${Z_2^{-1}/\mathcal{F}(p^2)}$, \textit{i.e.} the propagator term in the renormalized SDE. Satisfying Eq.~\eqref{eq:Sigma_1_alt} ensures Eq.~\eqref{eq:SDE_F} to be loop-renormalizable. Because Eq.~\eqref{eq:Sigma_1_alt} allows us to rewrite Eq.~\eqref{eq:SDE_F} as
\begin{equation}
\left(1-\dfrac{\alpha\xi}{4\pi\epsilon}\right)\dfrac{Z_2^{-1}}{\mathcal{F}(p^2)}=1+\overline{\Sigma}_1(p^2),
\end{equation}
where $\overline{\Sigma}_1(p^2)$ is $\Sigma_1(p^2)$ with its divergent terms subtracted. Now we can apply the renormalization condition to eliminate the $[1-\alpha\xi/(4\pi\epsilon)]Z_2^{-1}$ term and obtain
\begin{equation}
\dfrac{\mathcal{F}(\mu^2)}{\mathcal{F}(p^2)}=\dfrac{1+\overline{\Sigma}_1(p^2)}{1+\overline{\Sigma}_1(\mu^2)},
\end{equation}
which is free from divergences.

As for the other component of the renormalized SDE, combining Eq.~\eqref{eq:SDE_F} with Eq.~\eqref{eq:SDE_MF} produces,
\begin{equation}
m_RZ_m=\mathcal{M}(p^2)+\Sigma_1(p^2)\mathcal{M}(p^2)+\Sigma_2(p^2),\label{eq:SDE_M}
\end{equation} 
Next, with the help of Eqs.~(\ref{eq:Sigma_1_alt},~\ref{eq:Sigma_2_alt}), the loop divergence is given by
\begin{equation}
\Sigma_1(p^2)\mathcal{M}(p^2)+\Sigma_2(p^2)=\dfrac{\alpha\xi}{4\pi\epsilon}\left[\dfrac{ Z_2^{-1}\mathcal{M}}{\mathcal{F}}-\dfrac{Z_2^{-1}\mathcal{M}}{\mathcal{F}}\right]+\mathcal{O}(\epsilon^0)=0+\mathcal{O}(\epsilon^0).
\end{equation}
Therefore, comparing Eq.~\eqref{eq:SDE_M} with itself at $\mu^2$ produces
\begin{equation}
\mathcal{M}(p^2)=\dfrac{1+\overline{\Sigma}_1(\mu^2)}{1+\overline{\Sigma}_1(p^2)}\mathcal{M}(\mu^2)+\dfrac{\overline{\Sigma}_2(\mu^2)-\overline{\Sigma}_2(p^2)}{1+\overline{\Sigma}_1(p^2)},
\end{equation}
which means that the divergence in Eq.~\eqref{eq:SDE_MF} can also be removed.

Based on the previous discussion, in order to ensure Eqs.~(\ref{eq:SDE_F},~\ref{eq:SDE_MF}), namely the renormalized SDE for fermion propagator, being renormalizable by eliminating renormalization constants $Z_2$ and $Z_m$ at $\mu^2$, divergent parts of the fermion self-energy $\Sigma_1$ and $\Sigma_2$ are required to be homogeneous with respect to the propagator contribution in the SDE. This also ensures that divergences cancel after decoupling $\mathcal{M}(p^2)$ from the equation for the Dirac scalar component. These loop-renormalizability conditions are then formulated as 
\begin{equation}
\Sigma_1(p^2)=\dfrac{\lambda\alpha}{4\pi\epsilon}\dfrac{Z_2^{-2}}{\mathcal{F}(p^2)}+\overline{\Sigma}_1(p^2),\quad \Sigma_2(p^2)=\dfrac{-\lambda\alpha}{4\pi\epsilon}\dfrac{Z_2^{-1}\mathcal{M}(p^2)}{\mathcal{F}(p^2)}+\overline{\Sigma}_2(p^2),
\label{eq:loop_renormalizability}
\end{equation}
where $\overline{\Sigma}_{j}(p^2)$ are finite, therefore at $\mathcal{O}(\epsilon^0)$. In order to agree with one-loop perturbation calculation, the homogeneous coefficient is given by $\lambda=\xi$. Notice again that the fermion self-energy differs from the Gauge Technique loop counterpart simply by the fermion propagator:
\begin{equation}
[\slashed{p}\Sigma_1(p^2)+\Sigma_2(p^2)][\slashed{p}S_1(p^2)+S_2(p^2)]=\slashed{p}\sigma_2(p^2)+\sigma_1(p^2).
\end{equation}
Because the fermion propagator can be viewed as a linear transform with finite matrix elements, renormalizability conditions written as Eq.~\eqref{eq:loop_renormalizability} indicate for $\sigma_j(p^2)$ we have 
\begin{align}
\sigma_1(p^2)& =p^2\Sigma_1(p^2)S_1(p^2)+\Sigma_2(p^2)S_2(p^2)
=\dfrac{\lambda\alpha}{4\pi\epsilon}Z_2^{-1}+\overline{\sigma}_1(p^2),\label{eq:loop_ren_sigma_1}\\
\sigma_2(p^2)& =\Sigma_1(p^2)S_2(p^2)+\Sigma_2(p^2)S_1(p^2)=\overline{\sigma}_2(p^2),\label{eq:loop_ren_sigma_2}
\end{align}
where $\overline{\sigma}_{j}(p^2)$ are the finite parts of $\sigma_{j}(p^2)$. 

In this section, we have proposed the loop-renormalizability requirement in Eqs.~(\ref{eq:Sigma_1_ori},~\ref{eq:Sigma_2_ori}) for the fermion propagator SDE. This condition has been translated into Eqs.~(\ref{eq:loop_ren_sigma_1},~\ref{eq:loop_ren_sigma_2}) as the loop-renormalizability requirements for $\sigma_{j}(p^2)$.
\section{Solving for the fermion spectral functions}
\subsection{A renormalizable modification to the Gauge Technique}
Consider the Gauge Technique in the quenched approximation. The loop integral of the fermion propagator SDE can be easily calculated applying the well established perturbative procedures. After Feynman parameterization and dimensional regularization, we obtain
\begin{align}
&\sigma_1(p^2)=-\dfrac{3\alpha}{4\pi}\int ds\dfrac{sK(p^2,s)}{p^2-s}\rho_1(s)+\dfrac{\alpha\xi}{4\pi}\int ds\left(C_{div}+1+\ln\dfrac{\nu^2}{s-p^2}\right)\rho_1(s)\label{eq:Gauge_Technique_sigma_1}\\
&\sigma_2(p^2)=-\dfrac{3\alpha}{4\pi}\int ds\dfrac{K(p^2,s)}{p^2-s}\rho_2(s)+\dfrac{\alpha\xi}{4\pi}\int ds \dfrac{1}{p^2}\left(-1+\dfrac{s}{p^2}\ln\dfrac{s}{s-p^2}\right)\rho_2(s),\label{eq:Gauge_Technique_sigma_2}
\end{align}
where
\begin{equation}
K(p^2,s) =C_{div}+\dfrac{4}{3}+\ln\dfrac{\nu^2}{s-p^2}-\dfrac{s}{p^2}\ln\dfrac{s}{s-p^2},
\end{equation}
and ${C_{div}=1/\epsilon-\gamma_E+\ln 4\pi}$, with ${d=4-2\epsilon}$. 

Apparently, this result does not satisfy the loop-renormalizability requirements specified by Eqs.~(\ref{eq:loop_ren_sigma_1},~\ref{eq:loop_ren_sigma_2}). We therefore propose a minimal modification to meet this criterion, which replaces ${K(p^2,s)}$ by
\begin{equation}
\overline{K}(p^2,s)=\dfrac{4}{3}+\left(1-\dfrac{s}{p^2}\right)\ln\dfrac{s}{s-p^2}.\label{eq:minimum_LR_GaugeTechnique}
\end{equation}
This represents a scenario where transverse pieces additional to the Guage Technique only accomplish Eq.~\eqref{eq:minimum_LR_GaugeTechnique} by canceling the divergence in $K(p^2,s)$. Within the modified minimum subtraction scheme $(\overline{\mathrm{MS}})$, ${1/\epsilon-\gamma_E+\ln 4\pi}$ terms in $\sigma_{j}$ are removed altogether. Based on the definition of $\overline{\sigma}_{j}$ by Eqs.~(\ref{eq:loop_ren_sigma_1},~\ref{eq:loop_ren_sigma_2}), Eqs.~(\ref{eq:SDE_R_Z2},~\ref{eq:SDE_R_Zm}) then become
\begin{align}
& \left(1-\dfrac{\lambda\alpha}{4\pi\epsilon}\right)Z_2^{-1}+mZ_mS_2=p^2S_1+\overline{\sigma}_1\label{eq:SDE_oriR1}\\
& mZ_mS_1=S_2+\overline{\sigma}_2.\label{eq:SDE_oriR2}
\end{align} 
Next, we introduce the renormalization conditions to specify values of the propagator functions at the renormalization point. With these conditions $mZ_m$, $Z_2^{-1}$ and ${\lambda\alpha/(4\pi\epsilon)}$ are eliminated altogether;
\begin{align}
& \dfrac{S_2(p^2)+\overline{\sigma}_2(p^2)}{S_1(p^2)}=\dfrac{S_2(\mu^2)+\overline{\sigma}_2(\mu^2)}{S_1(\mu^2)},\label{eq:SDE_fqR_ori_1}\\
& p^2S_1(p^2)+\overline{\sigma}_1(p^2)-S_2(p^2)\dfrac{S_2(\mu^2)+\overline{\sigma}_2(\mu^2)}{S_1(\mu^2)}=\mu^2S_1(\mu^2)+\overline{\sigma}_1(\mu^2)-S_2(\mu^2)\dfrac{S_2(\mu^2)+\overline{\sigma}_2(\mu^2)}{S_1(\mu^2)}.\label{eq:SDE_fqR_ori_2}
\end{align}
The resulting Eqs.~(\ref{eq:SDE_fqR_ori_1},~\ref{eq:SDE_fqR_ori_2}) appear nonlinear in the spectral functions $\rho_{j}(s)$. We will see in Subsection \ref{ss:on_shell_SDE_SF} that the on-shell renormalization will linearize them.
\subsection{On-shell renormalization conditions\label{ss:on_shell_SDE_SF}}
The on-shell renormalization stipulates that propagator functions evaluated near the mass shell are dominated by their free-particle counterparts \cite{peskin1995introduction}. Mathematically, we translate this statement into
\begin{equation}
\lim\limits_{\mu^2\rightarrow m^2}(\mu^2-m^2)S_1(\mu^2)=1,\quad \lim\limits_{\mu^2\rightarrow m^2}(\mu^2-m^2)S_1(\mu^2)=m,\label{eq:lim_Sj_onshell}
\end{equation}
where the parameter $\mu$ is the renormalization scale, which is eventually set to $m$ by the limits. Equivalently for the propagator functions, when the free-particle contributions are subtracted according to
\begin{align}
& S_1(p^2)=\dfrac{1}{p^2-m^2}+P_1(p^2)\label{eq:OnShellCond1}\\
& S_2(p^2)=\dfrac{m}{p^2-m^2}+P_2(p^2).\label{eq:OnShellCond2}
\end{align}
The remaining functions must be less singular than the free-particle propagator in the vicinity of $m^2$. Therefore we have the following identities for $P_j(p^2)$:
\begin{equation}
\begin{cases}
\lim\limits_{\mu^2\rightarrow m^2}(\mu^2-m^2)P_1(\mu^2)=0,\\
\lim\limits_{\mu^2\rightarrow m^2}(\mu^2-m^2)P_2(\mu^2)=0.
\end{cases}
\end{equation}
Consequently, the spectral functions $\rho_j(s)$ cannot be more singular than the $\delta$-function in the limit ${s\rightarrow m^2}$. For the multipliers to the $\theta$-function terms in $\rho_j(s)$, they should not result in behaviors of the propagator functions more singular than the free-particle either. Therefore too strong singularities in the multipliers of the $\theta$-function in the $s\rightarrow m^2$ limit are not allowed. 

With these requirements in mind, Eqs.~(\ref{eq:OnShellCond1},~\ref{eq:OnShellCond2}) indicate
\begin{align}
& \rho_1(s)=\delta(s-m^2)+r_1(s)\theta(s-m^2),\\
& \rho_2(s)=m\delta(s-m^2)+r_2(s)\theta(s-m^2),
\end{align}
where $\theta(x)$ is the Heaviside step function, and $r_{j}(s)$ are supposed to be regular functions instead of distributions with exotic features. The $\delta$-functions contribute to the ${(\mu^2-m^2)^{-1}}$ singular parts of $S_{j}(\mu^2)$ while $r_{j}(s)$ give rise to $P_{j}(\mu^2)$, which are expected to be at most $\ln(m^2-\mu^2)$ divergent when $\mu^2\rightarrow m^2$. This requirement can be understood by the observation that for regularly behaving functions, the interchange of limits and integrations is allowed. Therefore we have
\begin{equation}
\lim\limits_{\mu^2\rightarrow m^2}(\mu^2-m^2)\int_{m^2}^{+\infty} ds \dfrac{r(s)}{\mu^2-s}=\int_{m^2}^{+\infty}ds~r(s)\lim\limits_{\mu^2\rightarrow m^2}\dfrac{\mu^2-m^2}{s-m^2}=0.
\end{equation}
Meanwhile, $\delta$-functions in the spectral functions $\rho_{j}$ contribute to free-propagator terms within $\overline{\sigma}_{j}(p^2)$ as well:
\begin{align}
& \overline{\sigma}_1(p^2)=\dfrac{-\lambda_1\alpha}{4\pi}\dfrac{m^2}{p^2-m^2}+q_1(p^2),\label{eq:def_SDE_Sigam_1}\\
& \overline{\sigma}_2(p^2)=\dfrac{-\lambda_2\alpha}{4\pi}\dfrac{m}{p^2-m^2}+q_2(p^2),\label{eq:def_SDE_Sigma_2}
\end{align}
with
\[\lim\limits_{\nu^2\rightarrow m^2}(\mu^2-m^2)q_1(\mu^2)=0,\quad \lim\limits_{\nu^2\rightarrow m^2}(\mu^2-m^2)q_2(\mu^2)=0.\]
Parameters $\lambda_1$ and $\lambda_2$ characterize the contribution to the free-particle terms from the loop diagram in Fig. \ref{fig:DSE_fermion_rho}.

Having separated the on-shell behavior of propagator functions $S_j$ and loop functions $\overline{\sigma}_{j}$ at $p^2=m^2$, the renormalized Eqs.~(\ref{eq:SDE_fqR_ori_1},~\ref{eq:SDE_fqR_ori_2}) simplify. Explicitly, the on-shell conditions simplify renormalization constants in the following way: 
\begin{equation}
\lim\limits_{\mu^2\rightarrow m^2}\dfrac{S_2(\mu^2)+\overline{\sigma}_2(\mu^2)}{S_2(\mu^2)}=\left(1-\dfrac{\lambda_2\alpha}{4\pi}\right)m,
\end{equation}
and
\begin{align}
& \quad \lim\limits_{\mu^2\rightarrow m^2}\Big\{\mu^2S_1(\mu^2)+\overline{\sigma}_1(\mu^2)-\dfrac{S_2(\mu^2)}{S_1(\mu^2)}[S_2(\mu^2)+\overline{\sigma}_2(\mu^2)] \Big\}\nonumber\\
& =\lim\limits_{\mu^2\rightarrow m^2}\Big\{(\lambda_2-\lambda_1)\dfrac{\alpha}{4\pi}\dfrac{m^2}{\mu^2-m^2}\Big\}+1+\left(2-\dfrac{\lambda_2\alpha}{4\pi}\right)[m^2P_1(m^2)-mP_2(m^2)]\nonumber\\
& \quad+q_1(m^2)-mq_2(m^2).
\end{align}
As a consequence, Eqs.~(\ref{eq:SDE_fqR_ori_1},~\ref{eq:SDE_fqR_ori_2}) both become linear in $\rho_j$:
\begin{align}
p^2S_1(p^2)+\overline{\sigma}_1(p^2)& =\left(1-\dfrac{\lambda_2\alpha}{4\pi}\right)mS_2(p^2)+\lim\limits_{\mu^2\rightarrow m^2}\Big\{(\lambda_2-\lambda_1)\dfrac{\alpha}{4\pi}\dfrac{m^2}{\mu^2-m^2}\Big\} +1\nonumber\\
& \quad +\left(2-\dfrac{\lambda_2\alpha}{4\pi}\right)[m^2P_1(m^2)-mP_2(m^2)]+q_1(m^2)-mq_2(m^2)\label{eq:SDE_RZ1_OnShell}\\[2mm]
S_2(p^2)+\overline{\sigma}_2(p^2)& =\left(1-\dfrac{\lambda_2\alpha}{4\pi}\right)mS_1(p^2).\label{eq:SDE_RZm_OnShell}
\end{align}
The next step is to separate the free-propagator terms, which can be accomplished through
\begin{align}
&\quad p^2P_1(p^2)+q_1(p^2)+(\lambda_2-\lambda_1)\dfrac{\alpha}{4\pi}\dfrac{m^2}{p^2-m^2}\nonumber\\
&  =\left(1-\dfrac{\lambda_2\alpha}{4\pi}\right)mP_2(p^2)+\lim\limits_{\mu^2\rightarrow m^2}\Big\{(\lambda_2-\lambda_1)\dfrac{\alpha}{4\pi}\dfrac{m^2}{\mu^2-m^2}\Big\}\nonumber\\
& \quad +\left(2-\dfrac{\lambda_2\alpha}{4\pi}\right)[m^2P_1(m^2)-mP_2(m^2)]+q_1(m^2)-mq_2(m^2)\label{eq:SDE_r1_OnShell}
\end{align}
and
\begin{equation}
P_2(p^2)+q_2(p^2) =\left(1-\dfrac{\lambda_2\alpha}{4\pi}\right)mP_1(p^2).\label{eq:SDE_r2_OnShell}
\end{equation}
Since the spectral functions $r_{j}(s)$ are effectively the spectral functions of $P_{j}$, we have
\begin{equation}
r_{j}(s)=\dfrac{-1}{\pi}\mathrm{Im}\{P_{j}(s+i\epsilon)\}.
\end{equation}
Although at $s=m^2$, $\rho_{j}(s)$ are dominated by the $\delta$-functions, this does not forbid $r_{j}(m^2)$ being nonzero but finite.
\subsection{SDEs for the fermion propagator spectral functions\label{ss:sr_SDE_SF}}
To derive the equations from which $\rho_j$ are solved, consider Eq.~\eqref{eq:minimum_LR_GaugeTechnique}, the minimum loop-renormalizable modification to the original Gauge Technique in the quenched approximation. In the case of $\overline{\mathrm{MS}}$, the loop integral in the fermion propagator SDE becomes
\begin{align}
& \overline{\sigma}_1(p^2)=\dfrac{-3\alpha}{4\pi}\int_{m^2}^{+\infty}ds\dfrac{s\overline{K}(p^2,s)}{p^2-s}\rho_1(s)+\dfrac{\alpha\xi}{4\pi}\int_{m^2}^{+\infty}ds\left(1+\ln\dfrac{\mu^2}{s-p^2}\right)\rho_1(s)\\
& \overline{\sigma}_2(p^2)=\dfrac{-3\alpha}{4\pi}\int_{m^2}^{+\infty}ds\dfrac{\overline{K}(p^2,s)}{p^2-s}\rho_2(s)+\dfrac{\alpha\xi}{4\pi}\int_{m^2}^{+\infty}ds\dfrac{1}{p^2}\left(-1+\dfrac{s}{p^2}\ln\dfrac{s}{s-p^2}\right)\rho_2(s).
\end{align}
From Eq.~\eqref{eq:minimum_LR_GaugeTechnique} one immediately sees $\lambda_1=\lambda_2=4$ . Then straightforwardly from Eqs.~(\ref{eq:def_SDE_Sigam_1},~\ref{eq:def_SDE_Sigma_2}), we obtain
\begin{align}
q_1(p^2)& =\overline{\sigma}_1(p^2)+\dfrac{\lambda_1\alpha}{4\pi}\dfrac{m^2}{p^2-m^2}\nonumber\\
& =-\dfrac{3\alpha}{4\pi}\dfrac{m^2}{p^2}\ln\dfrac{m^2}{m^2-p^2}+\dfrac{\alpha\xi}{4\pi}\left(1+\ln\dfrac{m^2}{m^2-p^2}\right)\nonumber\\
&\quad -\dfrac{3\alpha}{4\pi}\int_{m^2}^{+\infty}ds\dfrac{s\overline{K}(p^2,s)}{p^2-s}r_1(s)+\dfrac{\alpha\xi}{4\pi}\int_{m^2}^{+\infty}ds\left(1+\ln\dfrac{m^2}{s-p^2}\right)r_1(s),\\
q_2(p^2) & =\overline{\sigma}_2(p^2)+\dfrac{\lambda_2\alpha}{4\pi}\dfrac{m}{p^2-m^2}\nonumber\\
& =-\dfrac{3\alpha}{4\pi}\dfrac{m}{p^2}\ln\dfrac{m^2}{m^2-p^2}+\dfrac{\alpha\xi}{4\pi}\dfrac{m}{p^2}\left(-1+\dfrac{m^2}{p^2}\ln\dfrac{m^2}{m^2-p^2}\right)\nonumber\\
& \quad -\dfrac{3\alpha}{4\pi}\int_{m^2}^{+\infty}ds\dfrac{\overline{K}(p^2,s)}{p^2-s}r_2(s)+\dfrac{\alpha\xi}{4\pi}\int_{m^2}^{+\infty}ds\dfrac{1}{p^2}\left(-1+\dfrac{s}{p^2}\ln\dfrac{s}{s-p^2}\right)r_2(s).
\end{align}
The following relations are useful in order to obtain the imaginary part of $q_1$:
\begin{align}
& \dfrac{s'K(s,s')}{s-s'}=\dfrac{s'}{s-s'}\left(C_{div}+\dfrac{4}{3}+\ln\dfrac{\nu^2}{s'}\right)+\dfrac{s'}{s}\ln\dfrac{s'}{s'-s}\\
& -\dfrac{1}{\pi}\mathrm{Im}\Big\{\dfrac{1}{s-s'+i\epsilon}\Big\}=\delta(s-s')\\
& -\dfrac{1}{\pi}\mathrm{Im}\Bigg\{\int ds'\left(C_{div}+1+\ln\dfrac{\nu^2}{s'}\right)\rho_1(s')\Bigg\}=0.
\end{align}
After defining $z=s/s'$, the logarithmic terms of $\overline{\sigma}_1$ can be reparameterized by introducing an intermediate spectral function $\kappa(\zeta)$ for the nontrivial imaginary parts of kernel functions with the help of 
\begin{align}
& \dfrac{s'}{s}\ln\dfrac{s'}{s'-s}=\dfrac{1}{z}\ln\dfrac{1}{1-z}=\int d\zeta \dfrac{-\dfrac{1}{\zeta}\theta(\zeta-1)}{z-\zeta+i\epsilon}\\
& \ln\dfrac{s'}{s'-s}=\ln\dfrac{1}{1-z}=\int d\zeta\dfrac{-\theta(\zeta-1)}{z-\zeta+i\epsilon}.
\end{align}
Then $f(s/s')=\int d\zeta\, \kappa(\zeta)/(s/s'-\zeta+i\varepsilon)$. Therefore $\overline{R}_1(s)$, the spectral function of $q_1(p^2)$, is given by\footnote{Notice $\dfrac{-1}{\pi}\mathrm{Im}\Big\{\int_{m^2}^{+\infty}ds'\ln\dfrac{m^2}{s'}r_1(s')\Big\}=0 $.}, 
\begin{align}
\overline{R}_1(s)& =-\dfrac{1}{\pi}\mathrm{Im}\{q_1(s+i\epsilon) \}\nonumber\\
& =\dfrac{3\alpha}{4\pi}\dfrac{m^2}{s}\theta(s-m^2)-\dfrac{\alpha\xi}{4\pi}\theta(s-m^2)\nonumber\\
& \quad -\dfrac{3\alpha}{4\pi}\left[\dfrac{4}{3}sr_1(s)-\int_{m^2}^{s}ds'\dfrac{s'}{s}r_1(s) \right]-\dfrac{\alpha\xi}{4\pi}\int_{m^2}^{s}ds'r_1(s').
\end{align}
Similarly with the help of \[\dfrac{1}{s}\left(-1+\dfrac{s'}{s}\ln\dfrac{s'}{s'-s}\right)=\dfrac{1}{s'z}\left(-1+\dfrac{1}{z}\ln\dfrac{1}{1-z}\right)=\dfrac{1}{s'}\int d\zeta\dfrac{\dfrac{-1}{\zeta^2}\theta(\zeta-1)}{z-\zeta+i\epsilon},\]
the spectral function of $q_2(p^2)$ is given by
\begin{align}
\overline{R}_2(s)& =-\dfrac{1}{\pi}\mathrm{Im}\{q_2(s+i\epsilon) \}\nonumber\\
& =\dfrac{3\alpha}{4\pi}\dfrac{m}{s}\theta(s-m^2)-\dfrac{\alpha\xi}{4\pi}\dfrac{m^3}{s^2}\theta(s-m^2)\nonumber\\
& \quad -\dfrac{3\alpha}{4\pi}\left[\dfrac{4}{3}r_2(s)-\int_{m^2}^{s}ds'\dfrac{1}{s}r_2(s') \right]-\dfrac{\alpha\xi}{4\pi}\int_{m^2}^{s}ds'\dfrac{s'}{s^2}r_2(s').
\end{align}
Before deriving the SDE for functions $r_{j}(s)$, we need to calculate the imaginary parts of inhomogeneous terms in Eq.~\eqref{eq:SDE_r1_OnShell}. Although the value of $\theta$-functions at the threshold for the $\theta(s-m^2)$ remains undetermined, the linear combination $\overline{R}_1(m^2)-m\overline{R}_2(m^2)$ is free from such ambiguity;
\begin{align}
& \overline{R}_1(m^2)=\lim\limits_{s\rightarrow m^2}\left[\dfrac{3\alpha}{4\pi}\theta(s-m^2)-\dfrac{\alpha\xi}{4\pi}\theta(s-m^2) \right]-\dfrac{\alpha}{\pi}m^2r_1(m^2),\nonumber\\
& \overline{R}_2(m^2)=\lim\limits_{s\rightarrow m^2}\left[\dfrac{3\alpha}{4\pi}\dfrac{1}{m}\theta(s-m^2)-\dfrac{\alpha\xi}{4\pi}\dfrac{1}{m}\theta(s-m^2) \right]-\dfrac{\alpha}{\pi}r_2(m^2),\nonumber\\
& \overline{R}_1(m^2)-m\overline{R}_2(m^2)=-\dfrac{\alpha}{\pi}[m^2r_1(m^2)-mr_2(m^2)].
\end{align}
Therefore, taking the imaginary parts of Eqs.~(\ref{eq:SDE_r1_OnShell},~\ref{eq:SDE_r2_OnShell}) produces
\begin{align}
sr_1(s)+\overline{R}_1(s)& =\left(1-\dfrac{\alpha}{\pi}\right)mr_2(s)+2\left(1-\dfrac{\alpha}{\pi}\right)[m^2r_1(m^2)-mr_2(m^2)],\\
r_2(s)+\overline{R}_2(s)& =\left(1-\dfrac{\alpha}{\pi}\right)mr_1(s).
\end{align}
While at $s=m^2$, we have
\begin{align}
& [m^2r_1(m^2)-mr_2(m^2)]\left(1-\dfrac{\alpha}{\pi}\right)+\dfrac{(3-\xi)\alpha}{4\pi}\lim\limits_{s\rightarrow m^2}\theta(s-m^2)=2[m^2r_1(m^2)-mr_2(m^2)]\left(1-\dfrac{\alpha}{\pi}\right)\\
& [r_2(m^2)-mr_1(m^2)]\left(1-\dfrac{\alpha}{\pi}\right)+\dfrac{(3-\xi)\alpha}{4\pi}\dfrac{1}{m}\lim\limits_{s\rightarrow m^2}\theta(s-m^2)=0,
\end{align}
or equivalently, 
\begin{equation}
\left(1-\dfrac{\alpha}{\pi}\right)[m^2r_1(m^2)-mr_2(m^2)]=\dfrac{(3-\xi)\alpha}{4\pi}\lim\limits_{s\rightarrow m^2}\theta(s-m^2),\quad m\neq 0.\label{eq:SDE_ini_cond_1}
\end{equation}
Adopting ${\lim\limits_{s\rightarrow m^2}\theta(s-m^2)=1}$ produces non-trivial solutions. The choice of this particular limit of the $\theta$-functions will be explained later in Eq.~\eqref{eq:a0_limit_theta}. The equations for functions $r_{j}(s)$ then become
\begin{equation}
\begin{cases}
sr_1(s)+\overline{R}_1(s)=\left(1-\dfrac{\alpha}{\pi}\right)mr_2(s)+\dfrac{(3-\xi)\alpha}{2\pi} \\
r_2(s)+\overline{R}_2(s)=\left(1-\dfrac{\alpha}{\pi}\right)mr_1(s)
\end{cases}.\label{eq:SDE_LG_OnShell}
\end{equation}
The terms $(3-\xi)\alpha/(2\pi)$ on the right-hand side of Eq.~\eqref{eq:SDE_LG_OnShell} is then implicitly proportional to $\lim\limits_{s\rightarrow m^2}\theta(s-m^2)$, which needs to be recovered in Eq.~\eqref{eq:a0_limit_theta}.

Explicitly, we obtain the following integral equations for $r_j(s)$:
\begin{equation}
\begin{cases}
\quad  \left(1-\dfrac{\alpha}{\pi}\right)[s^2r_1(s)-msr_2(s)]+\dfrac{3\alpha}{4\pi}\left[m^2\theta(s-m^2)+\int_{m^2}^{s}ds's'r_1(s') \right]\\[1mm]
=\dfrac{(3-\xi)\alpha}{2\pi}s+\dfrac{\alpha\xi}{4\pi}\left[s\theta(s-m^2)+s\int_{m^2}^{s}ds'r_1(s') \right] \\[2mm]
\quad \left(1-\dfrac{\alpha}{\pi}\right)[-msr_1(s)+sr_2(s)]+\dfrac{3\alpha}{4\pi}\left[m\theta(s-m^2)+\int_{m^2}^{s}ds'r_2(s') \right]\\[1mm]
=\dfrac{\alpha\xi}{4\pi}\left[\dfrac{m^3}{s}\theta(s-m^2)+\dfrac{1}{s}\int_{m^2}^{s}ds's'r_2(s') \right].
\end{cases}
\label{eq:SDE_R_OnShell}
\end{equation}
We have derived Eq.~\eqref{eq:SDE_R_OnShell} as the coupled SDE for fermion spectral functions $\rho_{j}(s)$ with a loop-renormalizable modification to the Gauge Technique in the quenched approximation. Because spectral variables $s$ and $s'$ are separable, these integral equations can be converted into differential equations by taking derivatives with respect to $s$. 
\subsection{Solutions in the Landau gauge}
For simplicity, consider Eq.~\eqref{eq:SDE_R_OnShell} in the Landau gauge. For notational convenience, define coupling parameter $a$ as
\begin{equation}
a=\dfrac{3\alpha/(4\pi)}{1-\alpha/\pi}.\label{eq:def_a_alpha}
\end{equation}
After setting $\xi=0$ and taking one derivative with respect to $s$, Eq.~\eqref{eq:SDE_R_OnShell} becomes
\begin{equation}
\left[
\begin{pmatrix}
s & -ms \\ 
-m & s
\end{pmatrix}
\dfrac{d}{ds}
+
\begin{pmatrix}
a+1 & -m \\ 
0 & a+1
\end{pmatrix}
\right]
\begin{pmatrix}
sr_1(s) \\ 
r_2(s)
\end{pmatrix} 
=
\begin{pmatrix}
2a \\ 
0
\end{pmatrix} .\label{eq:SDE_R_OnShell_LG_Matrix}
\end{equation}

Straightforwardly as a first step to solve Eq.~\eqref{eq:SDE_R_OnShell_LG_Matrix}, one can eliminate diagonal matrix elements for the coefficient matrix for the non-differentiating terms through the following decompositions ${sr_1(s)=f_1(s)g_1(s)}$, and ${r_2(s)=f_2(s)g_2(s)}$.

Utilizing the inverse relation,
\[
\begin{pmatrix}
s & -ms \\ 
-m & s
\end{pmatrix}^{-1}
\begin{pmatrix}
a+1 & -m \\ 
0 & a+1
\end{pmatrix}
=
\dfrac{1}{s-m^2}
\begin{pmatrix}
a+1 & am \\ 
(a+1)m/s & a+1-m^2/s
\end{pmatrix}   
,\]
coverts Eq.~\eqref{eq:SDE_R_OnShell_LG_Matrix} into
\begin{align}
& 
\begin{pmatrix}
g_1(s) &  \\ 
& g_2(s)
\end{pmatrix} 
\dfrac{d}{ds}\begin{pmatrix}
f_1(s) \\ 
f_2(s)
\end{pmatrix} 
+
\begin{pmatrix}
g_1'(s) &  \\ 
& g_2'(s)
\end{pmatrix} 
\begin{pmatrix}
f_1(s) \\ 
f_2(s)
\end{pmatrix} 
\nonumber\\
& +
\dfrac{1}{s-m^2}
\begin{pmatrix}
a+1 & am \\ 
(a+1)m/s & a+1-m^2/s
\end{pmatrix} 
\begin{pmatrix}
g_1(s) &  \\ 
& g_2(s)
\end{pmatrix} 
\begin{pmatrix}
f_1(s) \\ 
f_2(s)
\end{pmatrix}
=
\begin{pmatrix}
\dfrac{2a}{s-m^2} \\ 
\dfrac{2am}{s(s-m^2)}
\end{pmatrix} .
\end{align}
Next, after multiplying ${\mathrm{diag}\{g_1^{-1}(s),~g_2^{-1}(s) \}}$
we obtain the coupled differential equations for $f_{j}(s)$ as 
\begin{equation}
\dfrac{d}{ds}
\begin{pmatrix}
f_1(s) \\ 
f_2(s)
\end{pmatrix} +
\begin{pmatrix}
\dfrac{g_1'(s)}{g_1(s)}+\dfrac{a+1}{s-m^2} & \dfrac{am}{s-m^2}\dfrac{g_2(s)}{g_1(s)} \\ 
\dfrac{(a+1)mg_1(s)}{(s-m^2)sg_2(s)} & \dfrac{g_2'(s)}{g_2(s)}+\dfrac{a+1-m/s}{s-m^2}
\end{pmatrix} 
\begin{pmatrix}
f_1(s) \\ 
f_2(s)
\end{pmatrix}=
\begin{pmatrix}
\dfrac{2a}{(s-m^2)g_1(s)} \\ 
\dfrac{2am}{s(s-m^2)g_2(s)}
\end{pmatrix}.\label{eq:SDE_ROS_LG_Matrix_Dcp}
\end{equation}
When diagonal elements of the matrix in Eq.~\eqref{eq:SDE_ROS_LG_Matrix_Dcp} vanish, taking another derivative with respect to the spectral variable $s$ decouples $f_{j}(s)$. In this scenario, $g_{j}(s)$ are required to satisfy their differential equations;
\begin{equation}
\dfrac{g_1'(s)}{g_1(s)}+\dfrac{a+1}{s-m^2}=0,\quad \dfrac{g_2'(s)}{g_2(s)}+\dfrac{a+1-m^2/s}{s-m^2}=0.
\end{equation}
One set of solutions for $g_j(s)$ is
\begin{equation}
g_1(s)=\left(\dfrac{m^2}{s-m^2}\right)^{a+1},\quad g_2(s)=\left(\dfrac{m^2}{s-m^2}\right)^a\dfrac{m}{s},
\end{equation}
where the integration constants are chosen such that $f_{j}(s)$ are dimensionless. 
Eq.~\eqref{eq:SDE_ROS_LG_Matrix_Dcp} then becomes
\begin{align}
& \dfrac{d}{ds}f_1(s)+\dfrac{a}{s}f_2(s)=\dfrac{2a}{m^2}\left(\dfrac{s-m^2}{m^2}\right)^a\label{eq:df1_f2}\\
& \dfrac{d}{ds}f_2(s)+\dfrac{(a+1)m^2}{(s-m^2)^2}f_1(s)=\dfrac{2a}{s-m^2}\left(\dfrac{s-m^2}{m^2}\right)^a\label{eq:df2_f1}.
\end{align}
Next, taking another derivative with respect to $s$ yields
\begin{align}
& \dfrac{d}{ds}s\dfrac{d}{ds}f_1(s)-\dfrac{a(a+1)m^2}{(s-m^2)^2}f_1(s)=\dfrac{2a(a+1)}{m^2}\left(\dfrac{s-m^2}{m^2}\right)^a,\label{eq:ODE_f1}\\
& \dfrac{d}{ds}(s-m^2)^2\dfrac{d}{ds}f_2(s)-\dfrac{a(a+1)m^2}{s}f_2(s)=0.\label{eq:ODE_f2}
\end{align}
Components in Eq.~\eqref{eq:SDE_R_OnShell_LG_Matrix} are now decoupled. 

The homogeneous part of Eq.~\eqref{eq:ODE_f1} can be solved using the Frobenius method. Refer to Appendix \ref{SS:Froenius} for details. The homogeneous solution to Eq.~\eqref{eq:f_1_homogeneous} inspires the following decomposition
\begin{equation}
f_1(x)=x^{a+1}\phi_1(x),
\end{equation}
where $x=s/m^2-1$. Eq.~\eqref{eq:ODE_f1} then becomes
\begin{equation}
x(x+1)\dfrac{d^2}{dx^2}\phi(x)+[2a+2+(2a+3)x]\dfrac{d}{dx}\phi(x)+(a+1)^2\phi(x)=2a(a+1).\label{eq:ODE_phi_1}
\end{equation}
The homogeneous part of Eq.~\eqref{eq:ODE_phi_1} is the hypergeometric differential equation for $\phi$ on the variable $-x$. After finding out one particular solution to the inhomogeneous equation, the general solution to Eq.~\eqref{eq:ODE_phi_1} with a finite initial condition is
\begin{equation}
\phi(x)=a_0~_2F_1(a+1,a+1;2a+2;-x)+\dfrac{2a}{a+1}.
\end{equation}
Therefore, the solution for Eq.~\eqref{eq:ODE_f1} is
\begin{equation}
f_1(x)=x^{a+1}\left[a_0~_2F_1(a+1,a+1;2a+2,-x)+\dfrac{2a}{a+1} \right].\label{eq:f1_exact}
\end{equation}
There is no need to solve Eq.~\eqref{eq:ODE_f2} separately because substituting Eq.~\eqref{eq:f1_exact} into Eq.~\eqref{eq:df1_f2} generates the solution for Eq.~\eqref{eq:ODE_f2} directly. The result is 
\begin{equation}
f_2(x)=-a_0\dfrac{(a+1)}{a}x^a~_2F_1(a,a+1;2a+2;-x).\label{eq:f2_exact}
\end{equation}
Based on Eqs.~(\ref{eq:f1_exact},~\ref{eq:f2_exact}) for functions $f_{j}(x)$, the following solutions are obtained, 
\begin{align}
& r_1(s)=\dfrac{a_0}{s}~_2F_1\left(a+1,a+1;2a+2;-\dfrac{s-m^2}{m^2}\right)+\dfrac{2a}{(a+1)s},\label{eq:r1_exact}\\
& r_2(s)=-a_0\dfrac{(1+a)}{am}~_2F_1\left(a+1,a+2;2a+2;-\dfrac{s-m^2}{m^2}\right).\label{eq:r2_exact}
\end{align}
The parameter $a_0$ can be determined by the the initial condition specified in Eq.~\eqref{eq:SDE_ini_cond_1}. Given that ${_2F_1(a+1,a+1;2a+2,0)=1}$, and ${~_2F_1(a+1,a+2;2a+2;0)=1}$, we have
\begin{equation}
a_0+\dfrac{2a}{a+1}\lim\limits_{s\rightarrow m^2}\theta(s-m^2)+a_0\dfrac{a+1}{a}=2a\lim\limits_{s\rightarrow m^2}\theta(s-m^2),\label{eq:a0_limit_theta}
\end{equation}
where the implicit dependence of the inhomogeneous term on the limit of $\theta$-functions has been restored. 

When the boundary values of $\theta$-function are chosen as $\lim\limits_{s\rightarrow m^2}\theta(s-m^2)=0$, only homogeneous parts of Eqs.~(\ref{eq:r1_exact},~\ref{eq:r2_exact}) survive. As a result, the only viable choice of parameter $a_0$ is $0$, which produces trial solutions. 

Therefore in order to obtain nontrivial solutions, the consistent limit is ${\lim\limits_{s\rightarrow m^2}\theta(s-m^2)=1}$. In this case, $a_0$ can be solved from Eq.~\eqref{eq:a0_limit_theta} as
\begin{equation}
a_0=\dfrac{2a^3}{(2a+1)(a+1)}.
\end{equation}
With this expression for $a_0$, we obtain the following set of solutions for Eq.~\eqref{eq:SDE_R_OnShell} in the Landau gauge:
\begin{equation}
\begin{cases}
r_1(s)=\dfrac{2a}{(a+1)s}\left[1+\dfrac{a^2}{(2a+1)}~_2F_1\left(a+1,a+1;2a+2;-\dfrac{s-m^2}{m^2}\right)\right] \\ 
r_2(s)=-\dfrac{2a^2}{(2a+1)m}~_2F_1\left(a+1,a+2;2a+2;-\dfrac{s-m^2}{m^2}\right)
\end{cases}.\label{eq:r12_2F1}
\end{equation}
Equation~\eqref{eq:r12_2F1} specifies the theta-function parts of the fermion propagator spectral function $\rho_{j}(s)$, with our interpretation of the on-shell renormalization conditions given by Eq.~\eqref{eq:lim_Sj_onshell}. 
Because Eq.~\eqref{eq:ODE_f1} is a second order ordinary differential equation, there is another solution linearly independent of Eq.~\eqref{eq:f1_exact}. The corresponding $r_1(s)$ is
\begin{equation}
r_1(s)=\dfrac{2a}{(a+1)s}+\dfrac{c}{s}(s/m^2-1)^{-2a-1}~_2F_1(-a,-a;-2a;1-s/m^2).\label{eq:r1_alternative}
\end{equation}
An interesting observation is that apart from the difference in the definitions of coupling parameter $a$, the homogeneous part in Eq.~\eqref{eq:r1_alternative} is identical to the corresponding $r_1(s)$ from Eq.~\eqref{eq:rW_Delbourgo}, the result based on the renormalization scheme in Ref.~\cite{Delbourgo:1977jc}. However, Eq.~\eqref{eq:r1_alternative} does not agree with the on-shell renormalization scheme because it fails to satisfy the finite boundary condition given by Eq.~\eqref{eq:SDE_ini_cond_1}. Specifically, it corresponds to a propagator function having a different $p^2\rightarrow m^2$ behavior from the on-shell propagators. 
\subsection{The Pad\'{e} approximation of hypergeometric functions}
\begin{figure}
	\centering
	\includegraphics[width=1\linewidth]{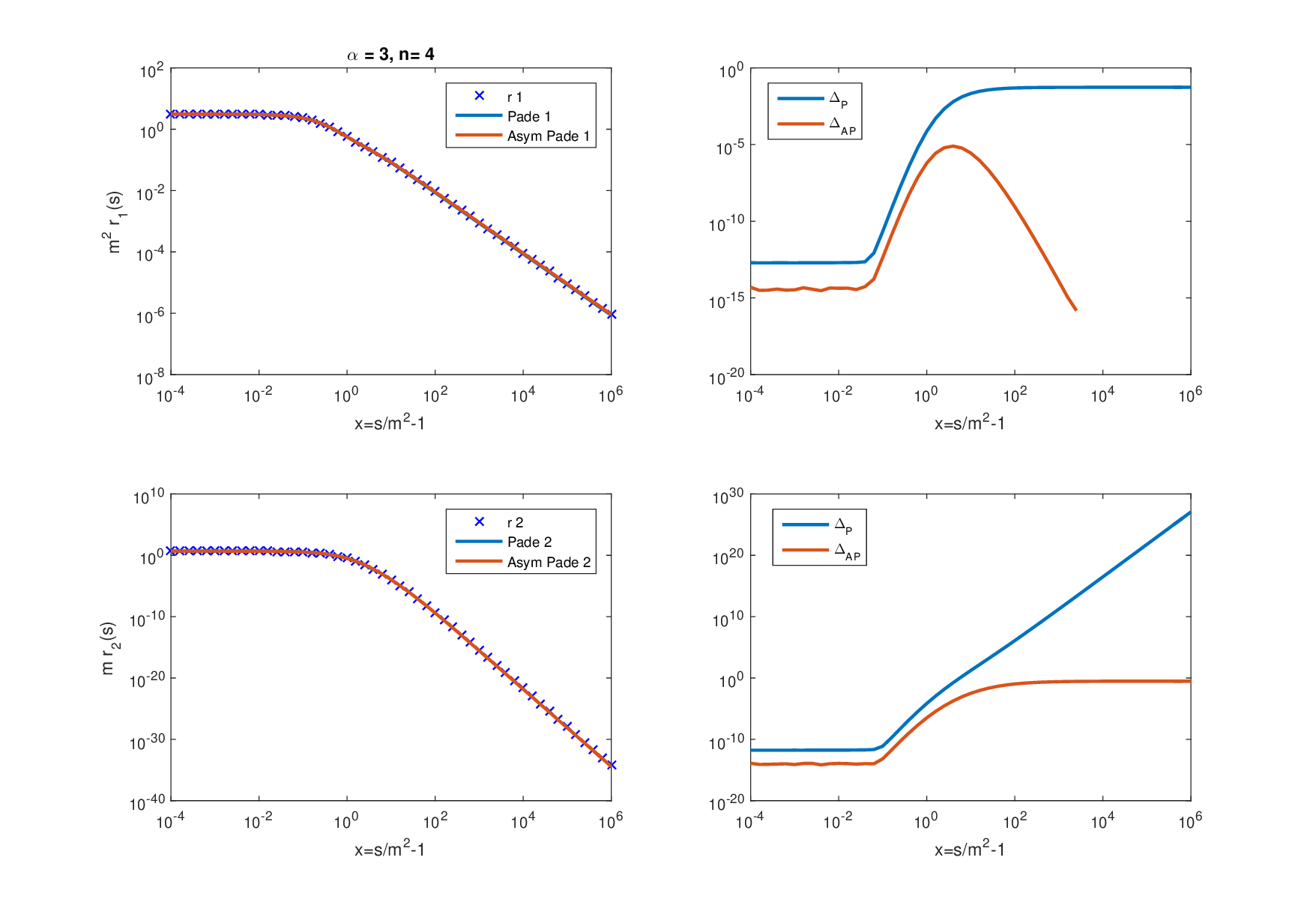}
	\caption{Pad\'{e} approximations to $r_j(s)$ in Eq.~\eqref{eq:r12_2F1}. The blue crosses represent the exact functions. The blue solid lines on the left figures are Pad\'{e} approximations directly applied to functions in Eq.~\eqref{eq:r12_2F1}. The red solid lines are approximations using Eqs.~(\ref{eq:asym_r1},~\ref{eq:asym_r2}). They are indistinguishable from the plots on the left. Figures on the right are relative errors defined by $\Delta=|[r(s)-p(s)]/r(s)|$, where $r(s)$ is the exact function while $p(s)$ is the approximation.}
	\label{fig:alpha3n4_Pade}
\end{figure}

The nontrivial parts of the fermion propagator spectral functions given by Eq.~\eqref{eq:r12_2F1} are the hypergeometric functions. Representing these nontrivial factors with elementary functions simplifies calculations where the fermion propagator spectral functions are used as input conditions. Since for positive $a$, the hypergeometric functions in Eq.~\eqref{eq:r12_2F1} are monotonous and asymptotically vanishing, one natural approximation to such functions is the Pad\'{e} approximation with the denominator polynomial at least one degree higher than the numerator polynomial. Since the Maclaurin series of a hypergeometric function is well defined, the natural variable for the Pad\'{e} polynomials is $x=s/m^2-1$. However, the asymptotic behaviors of the hypergeometric functions in Eq.~\eqref{eq:r12_2F1} cannot be well reproduced by the direct application of the Pad\'{e} approximation procedures with the variable $x$, as shown in Fig.~\ref{fig:alpha3n4_Pade}.

The asymptotic behavior of a hypergeometric functions in Eq.~\eqref{eq:r12_2F1} can be calculated using linear transformation formulae given by Eqs.~(15.3.3) through (15.3.9) of Ref.~\cite{abramowitz1964handbook}. However, for hypergeometric functions in $r_1(s)$ and $r_2(s)$ of Eq.~\eqref{eq:r12_2F1}, these integer differences in the first two parameters require Eq.~(15.3.13) and Eq.~(15.3.14) of Ref.~\cite{abramowitz1964handbook} to be apply respectively. Since asymptotically $s/m^2-1\simeq s/m^2$, we have the following limiting behaviors:
\begin{align}
& \lim\limits_{s\rightarrow +\infty}~_2F_1\left(a+1,a+1;2a+2;1-\dfrac{s}{m^2}\right)=\dfrac{\Gamma(2a+2)}{[\Gamma(a+1)^2]}\left(\dfrac{s}{m^2}\right)^{-a-1}\nonumber\\
& \hspace{7.5cm}\times\left[\ln\left(\dfrac{s}{m^2} \right)+2\gamma_E-2\psi(a+1) \right],\label{eq:asymp_2F1_r1}\\
& \lim\limits_{s\rightarrow+\infty}~_2F_1\left(a+1,a+2;2a+2;1-\dfrac{s}{m^2}\right)=\dfrac{\Gamma(2a+2)}{\Gamma(a+1)\Gamma(a+2)}\left(\dfrac{s}{m^2}\right)^{-a-1}.\label{eq:asymp_2F1_r2}
\end{align}
Then directly from Eq.~\eqref{eq:asymp_2F1_r2}, the asymptotic behavior of $r_2(s)$ in Eq.~\eqref{eq:r12_2F1} is given by $s^{-a-1}$. While due to the existence of the logarithm, the behavior described by Eq.~\eqref{eq:asymp_2F1_r1} is only weaker than $s^{-a}$. Therefore the asymptotic behavior of the second term of $r_1(s)$ in Eq.~\eqref{eq:r12_2F1} can be approximated by $s^{-a-1}$. 

Our analysis above for the asymptotic behaviors motivates the following approximations for $r_{j}(s)$:
\begin{align}
& r_1(s)\simeq \dfrac{2a}{(a+1)s}+\dfrac{1}{m^2}\left(\dfrac{s}{m^2}\right)^{-a}\dfrac{N_1(x)}{Q_1(x)},\label{eq:asym_r1}\\
& r_2(s)\simeq \dfrac{1}{m}\left(\dfrac{s}{m^2}\right)^{-a}\dfrac{N_2(x)}{Q_2(x)},\label{eq:asym_r2}
\end{align}
where $N_j(x)$ are polynomials of degree $n-1$ while $Q_j(x)$ are polynomials of degree $n$. Equations~(\ref{eq:asym_r1},~\ref{eq:asym_r2}) then allow $r_j(s)$ in Eq.~\eqref{eq:r12_2F1} to be well approximated using the following equations representing Pad\'{e} approximations, after factoring out the asymptotic behaviors. 
Base on Eqs.~(\ref{eq:r12_2F1},~\ref{eq:asym_r1},~\ref{eq:asym_r2}), we have 
\begin{align}
& \dfrac{N_1(x)}{Q_1(x)}\simeq \dfrac{2a^3}{(2a+1)(a+1)}(x+1)^{a-1}~_2F_1\left(a+1,a+1;2a+2;-x\right),\label{eq:Pade_r1_2F1} \\
& \dfrac{N_2(x)}{Q_2(x)}\simeq -\dfrac{2a^2}{2a+1}(x+1)^{a}~_2F_1\left(a+1,a+2;2a+2;-x\right),\label{eq:Pade_r2_2F1}
\end{align}
where the semiequal signs stand for taking the Pad\'{e} approximations. Since the Maclaurin series for terms on the right-hand sides of Eqs.~(\ref{eq:Pade_r1_2F1},~\ref{eq:Pade_r2_2F1}) are well defined, the standard procedure of Pad\'{e} approximations applies. To see how well this approximation shceme works, the example with $\alpha=3$ and quartic $Q_j(x)$ is given by Fig.~\ref{fig:alpha3n4_Pade}.
	\section{SDE for the photon spectral function\label{sc:SDE_rho_gamma}}
		\subsection{The Gauge Technique contribution to the vacuum polarization with arbitrary dimensions}
	Similar to the SDE for the fermion propagator, the SDE for the photon propagator also depends on the fermion-photon vertex through $S_F(k)\Gamma^\mu(k,p)S_F(p)$. The dependence of $S_F(p)\Gamma^\mu(k,p)S_F(p)$ on the spectral representations of the fermion propagator is only known up to the longitudinal components through the Gauge Technique. Through our limited knowledge of the vertex, we explore how the photon equation characterizes the nonlinear aspect of the SDEs for QED propagators.
	
	Within the original Gauge Technique, the vacuum polarization can be calculated by
	\begin{equation}
	\Pi(q^2)=\dfrac{-\alpha}{4\pi}\int ds\int dF^2\Gamma(\epsilon)\left(\dfrac{4\pi\mu^2}{s}\right)^\epsilon\dfrac{8xy}{(1-xyz)^\epsilon}\rho_1(s),\label{eq:Pi_OGT_preIdF}
	\end{equation}
	where $z=q^2/s$. The integral over the Feynman parameters is defined as $\int dF^2=\int dxdy~\delta(1-x-y)$. 
	In this subsection, $\epsilon=2-d/2$ is kept finite and explicit when required. Therefore Eq.~\eqref{eq:Pi_OGT_preIdF} needs to be evaluated directly. A straightforward transform to simplify the integration of the Feynman parameter in Eq.~\eqref{eq:Pi_OGT_preIdF} is difficult to find. Alternatively, knowing its analytic structure, specifically the branch cut along $z>4$, allows us to evaluate the imaginary part of $\Pi(q^2)$ first. 
	
	Let's start with a variable transform 
	\[\xi=x-1/2=1/2-y,\] 
	where $\xi$ in this subsection is not the gauge parameter but just a convenient parameter label. Together with the following identities,
	\begin{equation}
	\dfrac{-1}{\pi}\mathrm{Im}\bigg\{\dfrac{1}{[1-(1/4-\xi^2)z-i\varepsilon]^\epsilon}\bigg\}=\dfrac{-\sin(\pi\epsilon)}{\pi[(1/4-\xi^2)z-1]^\epsilon}\theta\left(z-\left(1/4-\xi^2\right)^{-1} \right)
	\end{equation}
	\begin{equation}
	\int_{-\frac{1}{2}\sqrt{\frac{z-4}{z}}}^{\frac{1}{2}\sqrt{\frac{z-4}{z}}}d\xi\dfrac{8(1/4-\xi^2)}{[(1/4-\xi^2)z-1]^\epsilon}=\dfrac{-2\sqrt{\pi}[-2+(\epsilon-1)z]\Gamma(1-\epsilon)}{\Gamma(5/2-\epsilon)z^{3/2}}\left(\dfrac{z-4}{4}\right)^{1/2-\epsilon},
	\end{equation}
	the imaginary part of Eq.~\eqref{eq:Pi_OGT_preIdF} can be calculated according to
	\begin{align}
	& \quad \dfrac{-1}{\pi}\mathrm{Im}\big\{\Pi(q^2+i\varepsilon)\big\}\nonumber\\
	& =\dfrac{-\alpha}{4\pi}\int ds\Gamma(\epsilon)\left( \dfrac{4\pi\mu^2}{s}\right)^\epsilon\rho_1(s)\dfrac{-1}{\pi}\mathrm{Im}\bigg\{\int_{-1/2}^{1/2}d\xi\dfrac{8(1/4-\xi^2)}{[1-(1/4-\xi^2)z-i\varepsilon]^\epsilon} \bigg\}\nonumber\\
	& =\dfrac{-\alpha}{4\pi}\int ds\Gamma(\epsilon)\left( \dfrac{4\pi\mu^2}{s}\right)^\epsilon\rho_1(s)\dfrac{-\sin(\pi\epsilon)}{\pi}\theta(z-4)\int_{-\frac{1}{2}\sqrt{\frac{z-4}{z}}}^{\frac{1}{2}\sqrt{\frac{z-4}{z}}}d\xi\dfrac{8(1/4-\xi^2)}{[(1/4-\xi^2)z-1]^\epsilon}\nonumber\\
	& =\dfrac{-\alpha}{4\pi}\int ds\left(\dfrac{4\pi\mu^2}{s}\right)^\epsilon\rho_1(s)\theta(z-4)\dfrac{2\sqrt{\pi}[-2+(\epsilon-1)z]}{\Gamma(5/2-\epsilon)z^{3/2}}\left(\dfrac{z-4}{4}\right)^{1/2-\epsilon}.
	\end{align}
	We presume that the analytic structure of $\Pi(q^2)$ is not altered discontinuously by changing the number of spacetime dimensions. Therefore for any $\epsilon$ where the integration over Feynman parameters in Eq.~\eqref{eq:Pi_OGT_preIdF} converges, $\Pi(q^2)$ is holomorphic in the complex $q^2$ plane except for a branch cut along the positive real axis. With the imaginary part of $\Pi(q^2)$ along this branch cut calculated, one can determine its real part, up to at most a real constant, from the spectral representation of $\Pi(q^2)$. 
	
	To calculate $\Pi(q^2)$ from its imaginary part using the spectral representation, the following integral identity is helpful,
	\begin{align}
	p_n(z)& \equiv \int_{4}^{+\infty}d\zeta\left(\dfrac{\zeta}{4}-1\right)^{1/2-\epsilon}\dfrac{\zeta^{n-3/2}}{z-\zeta+i\varepsilon} =4^{n-3/2}\int_{0}^{1}d\eta\dfrac{\eta^{\epsilon-n}(1-\eta)^{1/2-\epsilon}}{\eta z/4-1+i\varepsilon}\nonumber\\
	& =\dfrac{-4^{n-3/2}\Gamma(\epsilon-n+1)\Gamma(3/2-\epsilon)}{\Gamma(5/2-n)}~_2F_1(1,\epsilon-n+1;5/2-n;z/4),
	\end{align}
	where $\epsilon<3/2$ to ensure convergence. Next with the substitution of integral variable $\eta=1/\zeta$, one specifically obtains
	\begin{align}
	& p_0(z)=\dfrac{-\epsilon}{6\sqrt{\pi}}\Gamma(\epsilon)\Gamma(3/2-\epsilon)~_2F_1(1,\epsilon+1;5/2;z/4),\\
	& p_1(z)=\dfrac{-1}{\sqrt{\pi}}\Gamma(\epsilon)\Gamma(3/2-\epsilon)~_2F_1(1,\epsilon;3/2;z/4).
	\end{align}
	As a result, we have  
	\begin{align}
	\Pi(z)& =\int d\zeta \dfrac{\dfrac{-1}{\pi}\mathrm{Im}\big\{\Pi(\zeta+i\varepsilon)\big\}}{z-\zeta+i\varepsilon}\nonumber\\
	& =\dfrac{-\alpha}{4\pi}\int ds\left(\dfrac{4\pi\mu^2}{s}\right)^\epsilon\rho_1(s)\dfrac{2\sqrt{\pi}}{\Gamma(5/2-\epsilon)}[-2p_0(z)+(\epsilon-1)p_1(z)]\nonumber\\
	& =\dfrac{-\alpha}{4\pi}\int ds\rho_1(s)\left(\dfrac{4\pi\mu^2}{s}\right)^\epsilon\Gamma(\epsilon)\dfrac{2}{3/2-\epsilon}\times\nonumber\\
	& \quad\hspace{3cm} \left[\dfrac{\epsilon}{3}~_2F_1\left(1,\epsilon+1;\dfrac{5}{2};\dfrac{z}{4}\right)+(1-\epsilon)~_2F_1\left(1,\epsilon;\dfrac{3}{2};\dfrac{z}{4}\right) \right],
	\end{align}
	up to a real constant, which can be shown to be zero by matching $\Pi(0)$.
	
	To further simplify the expression for $\Pi(z)$, Eq.~(15.2.24) of Ref.~\cite{abramowitz1964handbook} with \newline ${b=\epsilon,~c=5/2,~a=1,~z\rightarrow z/4}$ is applied. Doing so produces
	\begin{equation}
	(3/2-\epsilon)~_2F_1(1,\epsilon;5/2;z/4)+\epsilon~_2F_1(1,\epsilon+1;5/2;z/4)=3/2~_2F_1(1,\epsilon;3/2;z/4).
	\end{equation}
	Therefore, we have
	\begin{align}
	& \quad \dfrac{\epsilon}{3}~_2F_1\left(1,\epsilon+1;\dfrac{5}{2};\dfrac{z}{4}\right)+(1-\epsilon)~_2F_1\left(1,\epsilon;\dfrac{3}{2};\dfrac{z}{4}\right)\nonumber\\
	& =\epsilon\left(1-\dfrac{2\epsilon}{3}\right)~_2F_1\left(1,\epsilon+1;\dfrac{5}{2};\dfrac{z}{4}\right)+(\epsilon-1)\left(1-\dfrac{2\epsilon}{3}\right)~_2F_1\left(1,\epsilon;\dfrac{5}{2};\dfrac{z}{4}\right).
	\end{align}
	Meanwhile, with $a=\epsilon,~b=1,~c=5/2,~z\rightarrow z/4$, Eq.~(15.2.14) of Ref.~\cite{abramowitz1964handbook} becomes
	\begin{equation}
	(1-\epsilon)~_2F_1(\epsilon,1;5/2;z/4)+\epsilon~_2F_1(\epsilon+1,1;5/2;z/4)=~_2F_1(\epsilon,2;5/2;z/4).
	\end{equation}
	Finally we obtain 
	\begin{equation}
	\Pi(q^2)=\dfrac{-\alpha}{4\pi}\int ds\rho_1(s)\left(\dfrac{4\pi\mu^2}{s}\right)^\epsilon \Gamma(\epsilon)\dfrac{4}{3}~_2F_1\left(\epsilon,2;\dfrac{5}{2};\dfrac{q^2}{4s}\right).\label{eq:Pi_OGT_epsilon}
	\end{equation}
	Equation \eqref{eq:Pi_OGT_epsilon} is the original Gauge Technique contribution to the QED vacuum polarization in $d=4-2\epsilon$ dimensions.
	
	Specifically in the case of small $\epsilon$, we have  
	\begin{equation}
	~_2F_1(\epsilon,2;5/2;z/4)=1+\epsilon\left[\dfrac{5}{3}+\dfrac{4}{z}+\dfrac{2(z+2)}{z^{3/2}}\sqrt{z-4}~\mathrm{arccsc}\left(\dfrac{2}{\sqrt{z}}\right)\right]+\mathrm{O}(\epsilon^2),
	\end{equation}
	Therefore in this limit, Eq.~\eqref{eq:Pi_OGT_epsilon} reproduces Eq.~\eqref{eq:Pi_1loop_detlarho} with $\rho_1(s)=\delta(s-m^2)$, as expected.
	\subsection{The $\theta$-function term of $\rho_\gamma(t)$}
	Equation~\eqref{eq:def_vacuum_polarization_Pi} states that with a specific Ansatz for the fermion-photon vertex, the vacuum polarization $\Pi(q^2)$ can be calculated. In the case of the Gauge Technique, $\Pi(q^2)$ is given by Eq.~\eqref{eq:Pi_OGT_epsilon}. The $\delta$-function terms of $\rho_\gamma(t)$ can be calculated by finding out the zeros of ${q^2[1+\Pi(q^2)]}$. The remaining part of $\rho_\gamma(t)$ should only consist of $\theta$-functions. When the $\delta$-functions are subtracted, denote the remaining term of $\rho_\gamma(t)$ to $r_\gamma(t)$. In general, $r_\gamma(t)$ is a functional of the fermion propagator spectral functions $\rho_j(s)$ with $j=1,~2$. It is desired to be able to write such dependences of $r_\gamma(t)$ on $\rho_j(s)$ explicitly. Because doing so solves the SDE for the photon propagator. 
	
	When $\Pi(q^2)$ is given by Eq.~\eqref{eq:Pi_OGT_epsilon}, $r_\gamma(t)$ is obtained by calculating the imaginary part of the Landau gauge photon propagator above the $4m^2$ threshold;
	\begin{equation}
	r_\gamma(t)=-\dfrac{1}{\pi}\mathrm{Im}\bigg\{\dfrac{\theta(t-4m^2)}{t(1+\Pi(t+i\varepsilon))} \bigg\}.
	\end{equation}
	Above this real-particle production threshold, $\Pi$ takes complex values, therefore we have the decomposition ${\Pi(t+i\varepsilon)=\Pi_R(t)+i\Pi_I(t)}$. Then, $\rho_\gamma$ is straightforwardly given by 
	\begin{equation}
	r_\gamma(t)=\dfrac{1}{\pi t}\dfrac{\Pi_I(t)}{\left[1+\Pi_R\left(t\right)\right]^2+[\Pi_I(t)]^2}.\label{eq:r_gamma}
	\end{equation}
	The next task to find out $\Pi_R$ and $\Pi_I$ as linear functionals of $\rho_j(s)$. 
	
	Specifically when $\Pi(q^2)$ is given by Eq.~\eqref{eq:Pi_OGT_epsilon}, according to Eq.~(15.3.7) of Ref.~\cite{abramowitz1964handbook} together with 
	${(-z-i\varepsilon)^{-\epsilon}\theta(z)=z^{-\epsilon}[\cos(\pi\epsilon)+i\,\sin(\pi\epsilon)]}$, when $z>1$ we have
	\begin{align}
	~_2F_1(\epsilon,2;5/2;z)& =\dfrac{3\sqrt{\pi}\Gamma(2-\epsilon)}{4\Gamma(5/2-\epsilon)}z^{-\epsilon}[\cos(\pi\epsilon)+i\,\sin(\pi\epsilon)]~_2F_1(\epsilon,\epsilon-3/2;\epsilon-1;1/z)\nonumber\\
	& \quad +\dfrac{3\Gamma(\epsilon-2)}{4\Gamma(\epsilon)}z^{-2}~_2F_1(2,1/2;3-\epsilon;1/z).
	\end{align}
	Consequently, with $z=t/(4s)$, 
	\begin{align}
	\Pi_R(t)& =-\dfrac{\alpha}{4\pi}\int ds\left(\dfrac{4\pi\mu^2}{s}\right)^\epsilon\Bigg\{ \dfrac{\sqrt{\pi}\Gamma(\epsilon)\Gamma(2-\epsilon)}{\Gamma(5/2-\epsilon)}\left(\dfrac{4s}{t}\right)^{\epsilon}\cos(\pi\epsilon)~_2F_1\left(\epsilon,\epsilon-\dfrac{3}{2};\epsilon-1;\dfrac{4s}{t}\right)\nonumber\\
	& \quad +\Gamma(\epsilon-2)\left(\dfrac{4s}{t}\right)^2~_2F_1\left(2,\dfrac{1}{2};3-\epsilon;\dfrac{4s}{t}\right)\Bigg\}\theta(t-4s)\rho_1(s),\label{eq:Pi_R_GT}\\
	\Pi_I(t)& =-\dfrac{\alpha}{4\pi}\int ds\left(\dfrac{16\pi\mu^2}{t}\right)^\epsilon\dfrac{\sqrt{\pi}\Gamma(\epsilon)\Gamma(2-\epsilon)}{\Gamma(5/2-\epsilon)}\sin(\pi\epsilon)~_2F_1\left(\epsilon,\epsilon-\dfrac{3}{2};\epsilon-1;\dfrac{4s}{t} \right)\theta(t-4s)\rho_1(s).\label{eq:Pi_I_GT}
	\end{align}
	In this section, we have derived Eqs.~(\ref{eq:Pi_R_GT},~\ref{eq:Pi_I_GT}) to specify two linear functionals of $\rho_1(s)$, which completely determine $r_\gamma(t)$ through Eq.~\eqref{eq:r_gamma}. Since $\rho_1(s)$ starts from $s=m^2$, the integrals in Eqs.~(\ref{eq:Pi_R_GT},~\ref{eq:Pi_I_GT}) only sample the region where $m^2\leq s\leq t/4$. When $t<m^2$, $\rho_\gamma(t)$ and $\Pi_I(t)$ vanish while $\Pi_R(t)$ is given directly by Eq.~\eqref{eq:Pi_OGT_epsilon}. Meanwhile, Eqs.~(\ref{eq:Pi_R_GT},~\ref{eq:Pi_I_GT}) are real functionals because hypergeometric functions inside the integrals are real when $t\geq 4s$. 
	
	Therefore, through Eqs.~(\ref{eq:rho_gamma_Pi},~\ref{eq:r_gamma}), the SDE for the photon propagator as an equation on the complex $q^2$ plane has been reduced to two real integrals and finding a finite number of roots. Here Eqs.~(\ref{eq:Pi_R_GT},~\ref{eq:Pi_I_GT}) are obtained with the Gauge Technique. In general they will be different with different Ans\"{a}tze.
	\chapter{Primitive divergences of QED\label{cp:div_QED}}
	\section{Minimum subtraction mass-independent renormalization scheme}
Recall that in order to preserve the longitudinal Ward--Green--Takahashi identity for the fermion-photon three-point function, the Ball--Chiu vertex was proposed \cite{Ball:1980ay}. It has the correct longitudinal (with respect to photon momentum) part of $\Gamma^\mu(k,p)$ but does not contain any transverse pieces \cite{Ball:1980ay}. One insufficiency of the Ball--Chiu vertex can be illustrated by the fact that solutions from the SDE for the fermion propagator with this vertex do not maintain the correct leading logarithmic divergence in the renormalizing the fermion field. Therefore the Ball--Chiu vertex violates the principle of multiplicative renormalizability of QED.

Multiplicative renormalizability of the fermion propagator can be maintained by the introduction of transverse pieces in addition to the Ball--Chiu vertex \cite{Ball:1980ay}. In the quenched approximation with leading logarithm divergences, this transverse part is given by the Curtis-Pennington vertex \cite{Curtis:1990zs}. In the massless unquenched case, the transverse part is given by the Kizilersu-Pennington vertex \cite{Kizilersu:2009kg}. There exist various other transverse vertices to maintain multiplicative renormalizability in different scenarios \cite{Curtis:1990zs,Kizilersu:2009kg}. Because we do not know the exact transverse vertex, guidance for building Ans\"{a}tze is provided by various principles of the gauge theory.

Recall the Gauge Technique Ansatz \cite{Delbourgo:1977jc} for the fermion-photon vertex translates into Eq.~\eqref{eq:Gamma_mu_transvese_GT} for $\Gamma^\mu(k,p)$. We already know from the discussion in Section \ref{ss:loop_ren_SF} that the Gauge Technique do not satisfy loop-renormalizability. 
Note that loop-renormalizability is a weaker condition than the multiplicative renormalizability, because only divergences from the loop integrals of SDEs are removed. Consequently these transverse pieces in Eq.~\eqref{eq:Gamma_mu_transvese_GT} do not to preserve the multiplicative renormalizability of the fermion propagator. A similar statement can be made based on the SDE for the photon propagator as well.

Both QED and QCD are renormalizable QFTs, meaning that there are only a finite number of primitive divergent diagrams by the means of counting superficial degrees of divergence \cite{peskin1995introduction}. For QED, these diagrams are the fermion self-energy, the vacuum polarization, and the fermion-photon 1PI vertex. Both the fermion self-energy and the fermion-photon 1PI vertex are logarithmic divergent by this power counting. Although the vacuum polarization has a superficial degree of divergence corresponding to the quadratic divergence, its Ward identity requires the vacuum polarization tensor to be transverse, reducing its divergence to logarithmic. 

These primitive divergences correspond to the divergent parts of renormalization constants $Z_1,~Z_2,~Z_3$ and $Z_m$. While the Ward identity specifies $Z_1=Z_2$ \cite{PhysRev.78.182}. All divergences in QED can then be absorbed into the three bare parameters: $Z_2,~\alpha_B$ and $m_B$. The mass function of fermion propagator has its divergence only coming from the bare mass. After treating the mass term in the QED Lagrangian as only perturbative, the divergence contributed by the mass term can only occur at most by one insertion of the bare mass operator. As a result, the divergence from the bare mass can always been handled separately from those of $Z_{2}$ and $\alpha_B$, allowing mass-independent renormalization schemes to be formulated.

When calculating loop diagrams, dimensional regularization treats the number of space-time dimensions as a continuous parameter to separate divergent pieces from finite parts. Define the number of space-time dimensions as $d=4-2\epsilon$. The minimum subtraction (MS) scheme removes divergences in loop diagrams by subtracting terms proportional to $\alpha^l/\epsilon^m$, with $l\ge m \ge 1$. Next, when all divergent loop integrals are regularized by dimensional regularization, the combination of this dimensional regulator, the minimum subtraction, and the mass-independent renormalization constitutes the minimum-subtraction mass-independent renormalization scheme (MS$m^0$). Within MS$m^0$, using the fact that bare Green's functions are independent of renormalization scale $\mu$, an analysis can be made through the Taylor expansion in the coupling constant $\alpha$ and the Laurent expansion in $\epsilon$, the regularization parameter. Recurrence relations for expansion coefficients can then be obtained. 

With dimensional regularization, divergences are represented as $1/\epsilon^n$ terms. At a certain level of divergence, $\alpha^{m+n}/\epsilon^n$ terms are summed up for all integer $n$ with a fixed integer $m$. The leading level of divergence corresponds to $m=0$. We will show that in the leading level of divergence,
\begin{align}
& Z_2=1~\mathrm{in~the~Landau~gauge}\\
& Z_m=m_0/m_R=\left(1-\dfrac{\alpha b_1}{2\epsilon}\right)^{\gamma_1/b_1}\\
& Z_3^{-1}=\left(1-\dfrac{\alpha b_1}{2\epsilon}\right)^{-1},\label{eq:MR_Z3}
\end{align}
with $\alpha$ being the renormalized coupling constant. While $b_i$, $\gamma_i$ are expansion coefficients given later in this chapter.

We follow the mass-independent renormalization scheme developed by Weinberg \cite{Weinberg:1951ss}. Instead of using a momentum space cut-off regulator, we use the method developed by Collins and Macfarlane \cite{Collins:1973yy} to apply dimensional regularization to Weinberg's mass-independent scheme. Elias and McKeon \cite{Elias:2002ry} showed that by requiring the bare coupling constant to be independent of renormalization scale, summing contributions from all orders of the renormalized coupling $\alpha$ to the bare coupling $\alpha_B$ is possible. 

However, instead of solving differential equations for contributions from different orders in $1/\epsilon$, we expand divergent parts of the bare quantities in a double series expansions of the renormalized couping $\alpha$ and the regularization parameter $1/\epsilon$. For simplicity, we limit ourselves to the Landau gauge ($\xi=0$) so that there is no need to renormalize the gauge parameter $\xi$. Moving towards other gauges only affects $Z_2$, which can be calculated through the Landau--Khalatnikov--Fradkin transform.
\section{A matrix representation of the double parameter expansion\label{ss:ori_expansion}}
\subparagraph{Equivalent definitions}
For an arbitrary function $f(\alpha,\epsilon)$ with a well defined Taylor expansion in $\alpha$ and a Laurent expansion in $\epsilon$, we can write
\begin{equation}
f(\alpha,\epsilon)=\sum_{i=0}^{\infty}\sum_{j=-\infty}^{\infty}f_{ij}\alpha^i\epsilon^j.
\end{equation}
A similar expansion for the bare charge has been provided by Eq.~(1) of Ref.~\cite{Elias:2002ry}. Since the part of $\epsilon^n$ with $n\ge 0$ for the bare quantities is of no interest for the discussion of renormalization, we truncate the Laurent expansion in $\epsilon$ to only keep the main branch. While the $\mathcal{O}(\epsilon^0)$ part of $f(\alpha,\epsilon)$ is usually specified by renormalization conditions, they are ignored here since we are interested in the divergent behavior only. Instead of using differential equations as in Ref.~\cite{Elias:2002ry}, we propose an algebraic approach based on matrix multiplications. The part of $f(\alpha,\epsilon)$ we are interested in is then given by
\begin{align}
f(\alpha,\epsilon)& =\sum_{i=1}^{\infty}\sum_{j=1}^{i}f_{ij}\alpha^i\dfrac{1}{\epsilon^j}\label{eq:f_Taylor_Laurent}\\
& =\begin{pmatrix}
\alpha^1 & \alpha^2 & \alpha^3 & \dots
\end{pmatrix}
\begin{pmatrix}
f_{11} & 0 & 0 & \cdots \\ 
f_{21} & f_{22} & 0 & \cdots \\ 
f_{31} & f_{32} & f_{33} & \cdots \\ 
\vdots & \vdots & \vdots & \ddots
\end{pmatrix} 
\begin{pmatrix}
1/\epsilon^1 \\ 
1/\epsilon^2 \\ 
1/\epsilon^3 \\ 
\vdots
\end{pmatrix}.\label{eq:matrix_expansion}
\end{align}
Denoting by $f$ the coefficient matrix in the middle of Eq.~\eqref{eq:f_Taylor_Laurent}. While $f(\alpha,\epsilon)$ with variables specified represents the function we are interested in. Imagine perturbation calculation to all orders has been performed. Therefore knowing the coefficient matrix $f$ is equivalent to having the complete knowledge of all divergences in $f(\alpha,\epsilon)$.
\subparagraph{Operations in the matrix space}
Matrix multiplications are well understood as row and column operations. While several relevant operations on $f(\alpha,\epsilon)$ include differentiation with respect to $\alpha$, multiplication by $\epsilon$, and multiplication by the Taylor series of $\alpha$. We will translate these operations into matrix multiplications in the coefficient space. Taking the derivative with respect to $\alpha$ becomes 
\begin{equation}
\alpha\dfrac{\partial}{\partial\alpha}f(\alpha,\epsilon)\Longleftrightarrow
\begin{pmatrix}
1 &  &  &  \\ 
& 2 &  &  \\ 
&  & 3 &  \\ 
&  &  & \ddots
\end{pmatrix} 
\begin{pmatrix}
f_{11} & 0 & 0 & \cdots \\ 
f_{21} & f_{22} & 0 & \cdots \\ 
f_{31} & f_{32} & f_{33} & \cdots \\ 
\vdots & \vdots & \vdots & \ddots
\end{pmatrix} \equiv Df.
\end{equation}
Multiplication by $\epsilon$ is given by
\begin{equation}
\epsilon f(\alpha,\epsilon)\Longleftrightarrow
\begin{pmatrix}
f_{11} & 0 & 0 & \cdots \\ 
f_{21} & f_{22} & 0 & \cdots \\ 
f_{31} & f_{32} & f_{33} & \cdots \\ 
\vdots & \vdots & \vdots & \ddots
\end{pmatrix}
\begin{pmatrix}
0 &  &  &  \\ 
1 & 0 &  &  \\ 
& 1 & 0 & \cdots \\ 
&  & \vdots & \ddots
\end{pmatrix} 
\equiv fE.
\end{equation}
Consider a function $b(\alpha)$ expanded as the Taylor series of $\alpha$:
\begin{equation}
b(\alpha)=b_0+b_1\alpha+b_2\alpha^2+\dots=\sum_{i=0}^{\infty}b_i\alpha^i.
\end{equation}
Then, one can easily show that
\begin{equation}
b(\alpha)f(\alpha,\epsilon)\Longleftrightarrow
\begin{pmatrix}
b_0 & 0 & 0 & \cdots \\ 
b_1 & b_0 & 0 & \cdots \\ 
b_2 & b_1 & b_0 & \cdots \\ 
\vdots & \vdots & \vdots & \ddots
\end{pmatrix} 
\begin{pmatrix}
f_{11} & 0 & 0 & \cdots \\ 
f_{21} & f_{22} & 0 & \cdots \\ 
f_{31} & f_{32} & f_{33} & \cdots \\ 
\vdots & \vdots & \vdots & \ddots
\end{pmatrix} \equiv Bf.
\end{equation}
Notice that operation matrices are represented by uppercase letters, while coefficient matrices remain lowercase letters in this representation.
\section{Bare quantities associated with primitive divergences}
Renormalization group equations can be derived based on the independence of bare propagators on the renormalization scale. Instead of studying the scale dependence of renormalized quantities, we investigate the divergent parts of primitive divergent quantities by the more tractable scale independence of bare quantities.
\subsection{The $\beta$-function in d=4-2$\epsilon$ dimensions}
In QED, the bare coupling constant carries a mass dimension depending on the number of spacetime dimensions. Explicitly, we have
\begin{equation}
\alpha_B=\mu^{2\epsilon}\left[\alpha+\mathcal{O}(\alpha^2)\right],
\end{equation}
where $\alpha_B$ is the bare coupling in $d=4-2\epsilon$ dimensions, and $\alpha$ is the renormalized dimensionless coupling. Meanwhile, we know that the Ward identity allows us to match divergences of the 1PI three-point function to those of the fermion propagator. Divergences in the photon propagator are then associated with the bare coupling. We are allowed to write
\begin{equation}
\alpha=Z_3\alpha_B\mu^{-2\epsilon}.
\end{equation}
Define $\hat{\beta}(\alpha)$ as the beta function in $d$ dimensions. Based on $\partial\,\alpha_B/\partial \ln\mu=0$, we obtain
\begin{align}
\hat{\beta}(\alpha)& =\dfrac{\partial}{\partial \ln \mu}Z_3\alpha_B\mu^{-2\epsilon}=\mu^{-2\epsilon}\left(-2\epsilon Z_3\alpha_B+\alpha_B\dfrac{\partial}{\partial \ln \mu}Z_3\right)\\\nonumber
& =-2\epsilon \alpha+\alpha\dfrac{\mu\partial Z_3}{Z_3\partial \mu}.
\end{align}
While in practical renormalization calculations, the renormalization constant $Z_3$ is expanded as the Taylor series of $\alpha$ and the Laurent series of $\epsilon$. As a result, $\dfrac{\mu\,\partial Z_3}{Z_3\,\partial\mu}=\dfrac{\beta(\alpha)}{\alpha}$, where $\beta(\alpha)$ is the usual $\beta$-function in the $\epsilon\rightarrow 0$ limit.

Therefore to maintain consistency, in the intermediate steps of renormalization where $\epsilon$ is nonzero, we should use
\begin{equation}
\mu\dfrac{\partial\alpha}{\partial \mu}=\hat{\beta}(\alpha)=-2\epsilon\alpha+\beta(\alpha)\label{eq:beta}
\end{equation}
as the derivative of the coupling constant with respect to the renormalization scale. Such a treatment is also consistent with the $\beta$-function used in Ref.~\cite{Elias:2002ry}.
\subsection{Divergences in the bare coupling constant $\alpha_B$\label{subss:alpha}}
Similar to discussions in Ref.~\cite{Elias:2002ry}, we can expand the bare coupling constant in its double series expansion. Define $a(\alpha,\epsilon)$ such that 
\begin{equation}
\alpha_B=\mu^{2\epsilon}\alpha[1+a(\alpha,\epsilon)],
\end{equation}
with the coefficient matrix $a$ defined by 
\begin{equation}
a(\alpha,\epsilon)=\sum_{i=1}^{\infty}\sum_{j=1}^{i}a_{ij}\alpha^i\dfrac{1}{\epsilon^j}\Longleftrightarrow a.
\end{equation}
Here ``$\Leftrightarrow$" means that the information in the double series expansion of function $a(\alpha,\epsilon)$ is equivalently contained in the matrix $a$. Since the bare coupling is independent of the renormalization scale, we have $\mu\,\partial\alpha_B/\partial \,\mu=0$. Consequently, we obtain
\begin{equation} 2\epsilon\alpha\left[1+a(\alpha,\epsilon)\right]+\hat{\beta}(\alpha)\left[1+a(\alpha,\epsilon)+\alpha\dfrac{\partial}{\partial\alpha}a(\alpha,\epsilon)\right]=0,
\end{equation}
or equivalently,
\begin{equation}
\left[1+a(\alpha,\epsilon) \right]\beta(\alpha)+\left[\beta(\alpha)-2\epsilon\alpha\right]\alpha\dfrac{\partial}{\partial\alpha}a(\alpha,\epsilon)=0.
\end{equation}
Next, define $b(\alpha)=b_1\alpha^1+b_2\alpha^2+\dots=\beta(\alpha)/\alpha$. Notice that for the notational convenience, coefficients $b_i$ are trivially different from their usual definitions for the QED $\beta$-function. We then write the renormalization group equations for $a(\alpha,\epsilon)$ as
\begin{equation}
b(\alpha)\left[1+a(\alpha,\epsilon)\right]+\left[b(\alpha)-2\epsilon\right]\alpha\dfrac{\partial}{\partial\alpha}a(\alpha,\epsilon)=0.\label{eq:RGE_a}
\end{equation}
Such an equation should be satisfied to all orders in the double expansion of $a(\alpha,\epsilon)$.

Notice that the only inhomogeneous term $b(\alpha)$ is at $\mathcal{O}(1/\epsilon^0)$. While the lowest order of $a(\alpha,\epsilon)$ is at $\mathcal{O}(1/\epsilon^1)$. Therefore in order to match coefficients of the $\alpha^i$ terms, at $\mathcal{O}(1/\epsilon^0)$, we must have
\begin{equation}
b(\alpha)=2\epsilon \alpha\dfrac{\partial}{\partial\alpha}a(\alpha,\epsilon)+\mathcal{O}(1/\epsilon^1),
\end{equation}
or equivalently
\begin{equation}
b_i=2ia_{i1},\qquad i\ge 1.\label{eq:RGE_a_i}
\end{equation}
With the inhomogeneous term cleaned out, we then have
\begin{equation}
b(\alpha)a(\alpha,\epsilon)+\left[b(\alpha)-2\epsilon\right]\alpha\dfrac{\partial}{\partial\alpha}a(\alpha,\epsilon)=0.\label{eq:RGE_a_h}
\end{equation}
Equation \eqref{eq:RGE_a_h} translated into matrix form becomes
\begin{equation}
B(\mathbf{1}+D)a=2DaE,\label{eq:RGE_a_M}
\end{equation}
or explicitly
\begin{align}
& \quad\begin{pmatrix}
0 &  &  &  \\ 
b_1 & 0 &  &  \\ 
b_2 & b_1 & 0 &  \\ 
\vdots & \vdots & \vdots & \ddots
\end{pmatrix} 
\begin{pmatrix}
2 &  &  &  \\ 
& 3 &  &  \\ 
&  & 4 &  \\ 
&  &  & \ddots
\end{pmatrix} 
\begin{pmatrix}
a_{11} &  &  &  \\ 
a_{21} & a_{22} &  &  \\ 
a_{31} & a_{32} & a_{33} &  \\ 
\vdots & \vdots & \vdots & \ddots
\end{pmatrix} \\\nonumber
& =
\begin{pmatrix}
2 &  &  &  \\ 
& 4 &  &  \\ 
&  & 6 &  \\ 
&  &  & \ddots
\end{pmatrix} 
\begin{pmatrix}
a_{11} &  &  &  \\ 
a_{21} & a_{22} &  &  \\ 
a_{31} & a_{32} & a_{33} &  \\ 
\vdots & \vdots & \vdots & \ddots
\end{pmatrix}
\begin{pmatrix}
0 &  &  &  \\ 
1 & 0 &  &  \\ 
& 1 & 0 & \cdots \\ 
&  & \vdots & \ddots
\end{pmatrix}. \label{eq:explicit_counting_alpha}
\end{align}
Because the existence of the B matrix on the left-hand side of Eq.~\eqref{eq:RGE_a_M} and the E matrix on the right-hand side, we obtain a matrix identity with only the triangle region below the diagonal. Therefore to the n-th order of perturbation calculation, we have $ U_a=n(n+1)/2 $ unknown parameters $a_{ij}$, $U_b=n$ unknown parameters $b_i$, with $K_{\alpha,h}=n(n-1)/2$ equations from Eq.~\eqref{eq:RGE_a_M} and $K_{\alpha,i}=n$ equations from Eq.~\eqref{eq:RGE_a_h}. As a result, the net number of unknown parameters is given by
\begin{equation}
u_\alpha=(U_a+U_b)-(K_{\alpha,h}+K_{\alpha,i})=\left[n(n+1)/2+n \right]-\left[n(n-1)/2+n\right]=n.\label{eq:counting_alpha}
\end{equation}

These parameters are related by the recurrence relations. Explicit calculation shows that for the leading level of divergences,
\begin{equation}
a_{nn}=\dfrac{b_1}{2}a_{n-1,n-1}.
\end{equation}
Given $b_1=2a_{11}$, from Eq.~\eqref{eq:RGE_a_i} we have 
\begin{equation}
\sum_{n=1}^{\infty}a_{nn}\left(\dfrac{\alpha}{\epsilon}\right)^n=\dfrac{1}{\dfrac{2\epsilon}{b_1\alpha}-1}.\label{eq:a_leading_level}
\end{equation}
Beyond the leading order, by induction, we obtain the following recurrence relations
\begin{align}
& 2na_{n,n-1}=na_{n-1,n-2}b_1+(n-1)a_{n-2,n-2}b_2\qquad (\mathrm{for}~n\geq 3),\\
& 2na_{n,n-2}=na_{n-1,n-3}b_1+(n-1)a_{n-2,n-3}b_2+(n-2)a_{n-3,n-3}b_3\qquad (\mathrm{for}~n\geq 4),\\
&\dots\, \dots\nonumber\\
& 2na_{n,n-m}=\sum_{j=0}^{m}(n-j)b_{j+1}a_{n-j-1,n-m-1}\qquad (\mathrm{for}~n\geq m+2,~m\geq 0).\label{eq:rec_alpha}
\end{align}
\subsection{Divergences in the fermion wavefunction renormalization\label{ss:div_F}}
The dimension of fermion propagator in coordinate space is given by
\begin{equation}
[S_F(y-x)]=[\overline{\psi}\psi]=d-1.
\end{equation}
Since the fermion propagator in momentum space is related to its coordinate space counterpart by the Fourier transform in d-dimensions, its dimension is calculated according to
\begin{equation}
[S_F(p)]=[S_F(y-x)]-d=-1,
\end{equation}
a result independent of $\epsilon$.

Therefore the fermion field renormalization $\mathcal{F}(p^2)$, given by Eq.~\eqref{eq:SF_Dirac_struct}, is strictly dimensionless. Meanwhile, in Weinberg's mass-independent scheme, the divergent part of $\mathcal{F}(p^2)$ does not depend on the fermion mass \cite{Weinberg:1951ss}. After denoting by $F_B(\alpha,\epsilon)$ the divergent part of $\mathcal{F}(p^2)$, we have
\begin{equation}
F_B(\alpha,\epsilon)=1+f(\alpha,\epsilon)=1+\sum_{i=1}^{\infty}\sum_{j=1}^{i}f_{i,j}\alpha^i\dfrac{1}{\epsilon^j}.
\end{equation}
Since the divergent part of this bare quantity is renormalization scale independent,
\begin{equation}
\mu\dfrac{d}{d\mu}F_B(\alpha,\epsilon)=\mu\dfrac{d}{d\mu}f(\alpha,\epsilon)=\hat{\beta}(\alpha)\dfrac{\partial}{\partial\alpha}f(\alpha,\epsilon)=0.\label{eq:diff_FB_mu}
\end{equation}
After substituting Eq.~\eqref{eq:beta} into, Eq.~\eqref{eq:diff_FB_mu}, we obtain
\begin{equation}
\left[b(\alpha)-2\epsilon\right]\alpha\dfrac{\partial}{\partial\alpha}f(\alpha,\epsilon)=0.\label{eq:REG_f}
\end{equation}
Notice that the resulting Eq.~\eqref{eq:REG_f} is homogeneous. By evaluating its $\mathcal{O}(1/\epsilon^0)$ part, we obtain
\begin{equation}
f_{i1}=0,\label{eq:f_epsilon^0}
\end{equation}
which agrees with the one-loop calculation in the Landau gauge when $i=1$.

Next, translating Eq.~\eqref{eq:REG_f} into matrix form produces
\begin{equation}
BDf=2DfE,
\end{equation}
or written explicitly as
\begin{align}
& \quad\begin{pmatrix}
0 &  &  &  \\ 
b_1 & 0 &  &  \\ 
b_2 & b_1 & 0 &  \\ 
\vdots & \vdots & \vdots & \ddots
\end{pmatrix} 
\begin{pmatrix}
1 &  &  &  \\ 
& 2 &  &  \\ 
&  & 3 &  \\ 
&  &  & \ddots
\end{pmatrix} 
\begin{pmatrix}
f_{11} &  &  &  \\ 
f_{21} & f_{22} &  &  \\ 
f_{31} & f_{32} & f_{33} &  \\ 
\vdots & \vdots & \vdots & \ddots
\end{pmatrix} \\\nonumber
& =
\begin{pmatrix}
2 &  &  &  \\ 
& 4 &  &  \\ 
&  & 6 &  \\ 
&  &  & \ddots
\end{pmatrix} 
\begin{pmatrix}
f_{11} &  &  &  \\ 
f_{21} & f_{22} &  &  \\ 
f_{31} & f_{32} & f_{33} &  \\ 
\vdots & \vdots & \vdots & \ddots
\end{pmatrix}
\begin{pmatrix}
0 &  &  &  \\ 
1 & 0 &  &  \\ 
& 1 & 0 & \cdots \\ 
&  & \vdots & \ddots
\end{pmatrix}. 
\end{align}
Analysis on the divergences of $F_B(\alpha,\epsilon)$ introduces $U_f=n(n+1)/2$ unknown $f_{ij}$ with \newline ${K_f=n(n-1)/2+n}$ equations. So the net increase in the number of unknown parameters is $\Delta u_f=0$.

\underline{Recurrence Relations}
Explicit calculation shows that in the leading level of divergence, 
\begin{equation}
2nf_{nn}=(n-1)b_1f_{n-1,n-1}\qquad (\mathrm{for}~n\geq 2),
\end{equation}
which indicates
\begin{equation}
\sum_{n=1}^{\infty}f_{nn}\left(\frac{\alpha}{\epsilon}\right)^n=\dfrac{2f_{11}}{b_1}\ln\left(\dfrac{1}{1-\dfrac{\alpha b_1}{2\epsilon}}\right).
\end{equation}
Since $f_{11}=0$, we know that $f_{nn}=0$ for $n\geq 2$.

Furthermore, induction reveals that
\begin{align}
& 2nf_{n,n-1}=(n-2)b_2f_{n-2,n-2}+(n-1)b_1f_{n-1,n-2}\qquad(n\geq 3)\\
& 2nf_{n,n-2}=(n-3)b_3f_{n-3,n-3}+(n-2)b_2f_{n-2,n-3}+(n-1)b_1f_{n-1,n-3}\qquad(n\geq 4)\\
&\dots\,\dots\nonumber\\
& 2nf_{n,n-m}=\sum_{j=0}^{m}(n-j-1)b_{j+1}f_{n-j-1,n-m-1}\qquad(n\geq m+2,~m\geq 0).\label{eq:rec_f}
\end{align}
We will see that such recurrence relations, together with Eq.~\eqref{eq:f_epsilon^0}, result in $f(\alpha,\epsilon)=0$. Therefore there is no divergence associated with the fermion field renormalization in the Landau gauge.
\subsection{Divergences in the bare mass}
The renormalization constant for the mass parameter is defined as $Z_m^{-1}=m_R/m_B$.
In Weinberg's scheme, $Z_m$ is independent of the renormalized mass \cite{Weinberg:1951ss}, therefore
\begin{equation}
Z_m=Z_m(\alpha,\epsilon)=1+\zeta(\alpha,\epsilon),
\end{equation}
where there exists a matrix representation $\zeta_{ij}$ for the function $\zeta(\alpha,\epsilon)$.

Meanwhile, the renormalized mass depends on the renormalization scale $\mu$. Such a dependence requires the introduction of the anomalous dimension $\gamma_m$ for the mass parameter. 
Since the bare mass is renormalization scale independent, we have
\begin{equation}
\gamma_m=-\dfrac{\mu}{m_R}\dfrac{dm_R}{d\mu}=-\dfrac{d\ln Z_m^{-1}}{d\ln \mu}=\dfrac{\mu}{Z_m}\dfrac{dZ_m}{d\mu}=\dfrac{1}{Z_m}\hat{\beta}\dfrac{\partial}{\partial \alpha}Z_m.\label{eq:gamma_m}
\end{equation}
Because $\gamma_m$ is dimensionless, it can only depend on $\alpha$. Based on this observation, we have the following Taylor series expansion of $\gamma_m$: 
\begin{equation}
\gamma_m(\alpha)=\gamma_1\alpha+\gamma_2\alpha^2+\dots=\sum_{i=1}^{\infty}\gamma_i\alpha^i.
\end{equation}
Equation \eqref{eq:gamma_m} then becomes
\begin{equation}
\gamma_m(\alpha)Z_m(\alpha,\epsilon)=\left[-2\epsilon\alpha+\beta(\alpha) \right]\dfrac{\partial}{\partial\alpha}Z_m(\alpha,\epsilon).\label{eq:RGE_m_ori}
\end{equation}
Writing in terms of $\zeta(\alpha,\epsilon)$, Eq.~\eqref{eq:RGE_m_ori} becomes 
\begin{equation}
\gamma_m(\alpha)[1+\zeta(\alpha,\epsilon)]=[-2\epsilon +b(\alpha)]\alpha\dfrac{\partial}{\partial\alpha}\zeta(\alpha,\epsilon).\label{eq:RGE_m}
\end{equation}
Since the inhomogeneous terms in Eq.~\eqref{eq:RGE_m} only contribute to $\mathcal{O}(1/\epsilon^0)$, we obtain
\begin{equation}
\gamma_m(\alpha)=-2\epsilon\alpha\dfrac{\partial}{\partial\alpha}\zeta(\alpha,\epsilon)+\mathcal{O}(1/\epsilon^1),\label{eq:RGT_m_i}
\end{equation}
or equivalently,
\begin{equation}
\gamma_i=-2i\zeta_{i1},\qquad i\ge 1.\label{eq:RGE_m_i}
\end{equation}
Next, when Eq.~\eqref{eq:RGT_m_i} is satisfied, the inhomogeneous terms can be removed, resulting in
\begin{equation}
\gamma_m(\alpha)\zeta(\alpha,\epsilon)=[-2\epsilon +b(\alpha)]\alpha\dfrac{\partial}{\partial\alpha}\zeta(\alpha,\epsilon).\label{eq:RGE_m_h}
\end{equation}
After translated into the matrix form, Eq.~\eqref{eq:RGE_m_h} becomes
\begin{equation}
\Gamma\zeta=BD\zeta-2D\zeta E,
\end{equation}
or written explicitly as 
\begin{align}
& \quad\begin{pmatrix}
0 &  &  &  \\ 
\gamma_1 & 0 &  &  \\ 
\gamma_2 & \gamma_1 & 0 &  \\ 
\vdots & \vdots & \vdots & \ddots
\end{pmatrix} 
\begin{pmatrix}
\zeta_{11} &  &  &  \\ 
\zeta_{21} & \zeta_{22} &  &  \\ 
\zeta_{31} & \zeta_{32} & \zeta_{33} &  \\ 
\vdots & \vdots & \vdots & \ddots
\end{pmatrix} \\\nonumber
& =\begin{pmatrix}
0 &  &  &  \\ 
b1 & 0 &  &  \\ 
b2 & b1 & 0 &  \\ 
\vdots & \vdots & \vdots & \ddots
\end{pmatrix} 
\begin{pmatrix}
1 &  &  &  \\ 
& 2 &  &  \\ 
&  & 3 &  \\ 
&  &  & \ddots
\end{pmatrix} 
\begin{pmatrix}
\zeta_{11} &  &  &  \\ 
\zeta_{21} & \zeta_{22} &  &  \\ 
\zeta_{31} & \zeta_{32} & \zeta_{33} &  \\ 
\vdots & \vdots & \vdots & \ddots
\end{pmatrix} \\\nonumber
& \quad -\begin{pmatrix}
2 &  &  &  \\ 
& 4 &  &  \\ 
&  & 6 &  \\ 
&  &  & \ddots
\end{pmatrix} 
\begin{pmatrix}
\zeta_{11} &  &  &  \\ 
\zeta_{21} & \zeta_{22} &  &  \\ 
\zeta_{31} & \zeta_{32} & \zeta_{33} &  \\ 
\vdots & \vdots & \vdots & \ddots
\end{pmatrix}
\begin{pmatrix}
0 &  &  &  \\ 
1 & 0 &  &  \\ 
& 1 & 0 & \cdots \\ 
&  & \vdots & \ddots
\end{pmatrix}. 
\end{align}
Similar to the analysis before, we have a net amount of $n$ unknown parameters for the divergent part  of the bare mass.

\underline{Recurrence Relations}
Explicit calculation shows that, by induction, parameters of $Z_m$ are related by 
\begin{align}
& 2n\zeta_{nn}=[(n-1)b_1-\gamma_1]\zeta_{n-1,n-1},\qquad (n\geq 2),\\
& 2n\zeta_{n,n-1}=[(n-2)b_2-\gamma_2]\zeta_{n-2,n-2}+[(n-1)b_1-\gamma_1]\zeta_{n-1,n-2},\qquad (n\geq 3)\\
&\dots\,\dots\nonumber\\
&2n\zeta_{n,n-m}=\sum_{j=0}^{m}[(n-j-1)b_{j+1}-\gamma_{j+1}]\zeta_{n-j-1,n-m-1}\qquad (n\geq m+2,~m\geq 0).\label{eq:rec_m}
\end{align}
\section{An alternative matrix representation\label{ss:alternative}}
\subsection{Definition of the alternative representation}
In the previous section, the matrix form defined by Eq.~\eqref{eq:matrix_expansion} is utilized to represent divergences of the bare quantities. Such a representation is convenient for the purpose of counting free parameters and consistently truncating to a certain order in perturbation theory. However, this representation is cumbersome in obtaining the coefficient vector at a specified level of divergence. Therefore the following expansion is proposed as an alternative. This rearrangement of expansions will be used in order to analyze recurrence relations obtained in the previous section. Keep in mind that truncation to a specific perturbation order will involve projection operations to eliminate higher order terms.

Alternative to Eq.~\eqref{eq:f_Taylor_Laurent}, we expand the divergent part of a bare quantity $f(\alpha,\epsilon)$ in the following way 
\begin{equation}
f(\alpha,\epsilon) =\sum_{i=1}^{\infty}\sum_{k=0}^{\infty}\left(\dfrac{\alpha}{\epsilon}\right)^jF_{jk}\alpha^k.\label{eq:dim_rel_alt_exp}
\end{equation}
While the previous version of expansion is given by
\begin{equation}
f(\alpha,\epsilon)=\sum_{i=1}^{\infty}\sum_{j=1}^{i}f_{ij}\alpha^{i-j}\left(\dfrac{\alpha}{\epsilon}\right)^j.
\end{equation}
Comparing these two equations produces
\begin{equation}
F_{jk}=f_{j+k-1,j}.
\end{equation}
Therefore, the alternative expansion by Eq.~\eqref{eq:dim_rel_alt_exp}, explicitly in the matrix form, is written as 
\begin{equation}
f(\alpha,\epsilon)=
\begin{pmatrix}
\left(\dfrac{\alpha}{\epsilon}\right)^1, & \left(\dfrac{\alpha}{\epsilon}\right)^2, & \left(\dfrac{\alpha}{\epsilon}\right)^3, & \cdots
\end{pmatrix} 
\begin{pmatrix}
f_{11} & f_{21} & f_{31} & \cdots \\ 
f_{22} & f_{32} & f_{42} & \cdots \\ 
f_{33} & f_{43} & f_{53} & \cdots \\ 
\vdots & \vdots & \vdots & \ddots
\end{pmatrix} 
\begin{pmatrix}
\alpha^0 \\ 
\alpha^1 \\ 
\alpha^2 \\ 
\vdots
\end{pmatrix} .
\end{equation}
We have, in the alternative representation, $f(\alpha,\epsilon)\Longleftrightarrow F$. Notice that the first column of $F$ is the coefficient for the leading $\epsilon$ divergence, the second column for the next-to-leading divergence, etc.

It is then straightforward to verify the following correspondences between operations on $f(\alpha,\epsilon)$ and those on $F$. Explicitly, we have
\begin{align}
\alpha\dfrac{\partial}{\partial\alpha}f(\alpha,\epsilon)& \Longleftrightarrow DF+F(D-\mathbf{1})\equiv F^\alpha,\\
\epsilon f(\alpha,\epsilon)& \Longleftrightarrow E^TFE^T,\\
b(\alpha)f(\alpha,\epsilon) &\Longleftrightarrow FB^T,
\end{align}
where 
``$T$" on the superscript stands for the matrix transpose. Operator matrices are otherwise defined identical to those in Section \ref{ss:ori_expansion}.

Furthermore, Eq.~\eqref{eq:RGE_a_h} can be rewritten as
\begin{equation}
(A+A^\alpha)B^T=2E^TA^\alpha E^T,
\label{eq:RGE_A}
\end{equation}
or explicitly
\begin{equation}
\begin{pmatrix}
0 & 2a_{11}b_1 & 2a_{11}b_2+3a_{21}b_1 & \cdots \\ 
0 & 3a_{22}b_1 & 3a_{22}b_2+4a_{32}b_1 & \cdots \\ 
0 & 4a_{33}b_1 & 4a_{33}b_2+5a_{43}b_1 & \cdots \\ 
\vdots & \vdots & \vdots & \ddots
\end{pmatrix} 
=
\begin{pmatrix}
0 & 4a_{22} & 6a_{32} & \cdots \\ 
0 & 6a_{33} & 8a_{43} & \cdots \\ 
0 & 8a_{44} & 10a_{54} & \cdots \\ 
\vdots & \vdots & \vdots & \ddots
\end{pmatrix} ,
\end{equation}
which agrees with the recurrence relation given by Eq.~\eqref{eq:rec_alpha}.

Similarly, we can rewrite equations for fermion field renormalization $f(\alpha,\epsilon)$ and mass renormalization $Z_m(\alpha,\epsilon)$ in the alternative representation. They are given by
\begin{align}
& F^\alpha B^T=2E^TF^\alpha E^T,\label{eq:RGE_F}\\
& Z^\alpha B^T-Z\Gamma^T=2E^T Z^\alpha E^T.\label{eq:RGE_Z}
\end{align}
\subsection{Solutions in the alternative representation}
Realize that Eqs.~(\ref{eq:rec_alpha},~\ref{eq:rec_f},~\ref{eq:rec_m}) are solutions to Eqs.~(\ref{eq:RGE_A},~\ref{eq:RGE_F},~\ref{eq:RGE_Z}), respectively. We then divide the coefficient matrix in the alternative representation into its column vectors:
\begin{equation}
F=(F_0,~F_1,~F_2,~\cdots),
\end{equation}
where $F_0$ is the coefficient vector for the leading divergent piece, $F_1$ for the next-to-leading part, etc.

Then based on Eq.~\eqref{eq:rec_alpha}, we have
\begin{equation}
2[D+(m+1)\mathbf{1}]E^TA_m=\sum_{j=0}^{m}b_{j+1}[D+(m+1-j)\mathbf{1}]A_{m-j}.
\end{equation}
Specifically we obtain, for $m=0$
\begin{equation}
2(D+\mathbf{1})E^TA_0=b_1(D+\mathbf{1})A_0;
\end{equation}
for $m=1$
\begin{equation}
2(D+\mathbf{2})E^TA_1=b_1(D+\mathbf{2})A_1+b_2(D+\mathbf{1})A_0;
\end{equation}
and for $m=2$
\begin{equation}
2(D+\mathbf{3})E^TA_2=b_1(D+\mathbf{3})A_2+b_2(D+\mathbf{2})A_1+b_3(D+\mathbf{1})A_0.
\end{equation}
By observation, the equation for $A_0$ is a homogeneous linear equation, while other algebraic equations for $A_m,~(m\geq 1)$ contain inhomogeneous terms depending on coefficient vectors for lower order divergences. While as long as the determinant of the linear operator
\begin{equation}
\mathcal{O}_A=2E^T-b_1\mathbf{1}
\end{equation}
is non-vanishing, we can always find its inverse and solve for $A_m^*$ from the inhomogeneous linear equation.

As for the homogeneous part of the solution, it is determined by
\begin{equation}
[D+(m+1)\mathbf{1}](2E^T-b_1\mathbf{1})\tilde{A}_m=\mathbf{0}.\label{eq:A_tilde_m}
\end{equation}
Because homogeneous equations for all $m\in \mathbf{N}$ are identical, we only need to solve one of them. In order to satisfy Eq.~\eqref{eq:RGE_a_i}, the following identity holds
\begin{equation}
\tilde{a}_{1+m,1}=\dfrac{b_{m+1}}{2(m+1)}-a^*_{m+1,1},
\end{equation}
where $a^*_{m+1,1}$ is the first coefficient of the solution vector $A_m^*$ to the inhomogeneous equation. Consequently, we have
\begin{align}
\tilde{a}_{n+m,n}&=\dfrac{b_1}{2}\tilde{a}_{n+m-1,n-1}\qquad (\mathrm{for}~n\geq 2)
\end{align}
from Eq.~\eqref{eq:A_tilde_m}. The solution is therefore given by
\begin{equation}
A_m=\tilde{A}_m+A_m^*.
\end{equation}
We have formally constructed the coefficient matrix A, with $n$ unknown coefficients if truncated to $\alpha^n$. At the same time we have showed that the equation counting given by Eq.~\eqref{eq:counting_alpha} is correct; there are $n(n-1)/2$ linearly independent equations for $a_{ij}$.

Next, based on the recurrence relation in Eq.~\eqref{eq:rec_f}, we obtain
\begin{equation}
2[D+(m+1)\mathbf{1}]E^TF_m=\sum_{j=0}^{m}b_{j+1}[D+(m-j)\mathbf{1}]F_{m-j}.
\end{equation}
The linear operator for $F_m$ is recognized as
\begin{equation}
\mathcal{O}_{F_m}=2[D+(m+1)\mathbf{1}]E^T-b_1(D+m\mathbf{1})=(D+m\mathbf{1})(2E^T-b_1\mathbf{1})+2E^T.
\end{equation}
We can show that
\begin{equation}
\mathrm{det}(\mathcal{O}_{F_m})\neq 0.
\end{equation}
Meanwhile, having established that $F_0=0$ and $f_{i1}=0$, by the same procedures used to construct $A_m$, we find out that $F_m=0$ for all $m\in N$. Therefore $f(\alpha,\epsilon)$ has been proved to be zero. 

Similar analysis based on Eq.~\eqref{eq:rec_m} shows that 
\begin{equation}
2[D+(m+1)\mathbf{1}]E^TZ_m=\sum_{j=0}^{m}\{b_{j+1}[D+(m-j)\mathbf{1}]-\gamma_{j+1}\mathbf{1}\}Z_{m-j},
\end{equation}
for the divergence of the mass renormalization.
\section{Summary of QED divergences}
By working in the Landau gauge, we have demonstrated that adapting the mass-independent renormalization scheme in Ref.~\cite{Weinberg:1951ss} to dimensional regularization strongly constrains primitive divergences in QED by expanding the bare parameters in the Taylor series expansion in the renormalized coupling constant and the Laurent expansion in the dimensional regularization parameter. Our result in the leading level divergence for the coupling constant in the form of Eq.~\eqref{eq:a_leading_level} agrees with Eq.~(13) of Ref.~\cite{Elias:2002ry}. We have derived additionally the recurrence relations for the primitive divergences of QED in the Landau Gauge at any level of divergence as Eqs.~(\ref{eq:rec_alpha},~\ref{eq:rec_f},~\ref{eq:rec_m}).

Naively counting the total number of independent coefficients of these three QED primitive divergences at n-th order in perturbation theory gives $3n(n+1)/2$ unknown parameters. While in the Landau gauge, combining MS$m^0$ with analytical requirements, the total number of independent coefficients for primitive divergences is reduced to $2n$. Specifically, because $f(\alpha,\epsilon)=0$, the fermion renormalization has no divergence.

Explicit calculation through the matrix representation shows that the leading divergent result agrees with the result in Ref.~\cite{Elias:2002ry} for $\alpha_B$, which works as both a cross-check and an indication that the double expansion approach is equivalent to the analytical approach. 

Once we know $\beta_i,~\gamma_i$ and $f_{i,1}$ for $i \geq 1$, the inhomogeneous relations in Eqs.~(\ref{eq:RGE_a_i},~\ref{eq:RGE_m_i}) and the recurrence relations in Eqs.~(\ref{eq:rec_alpha},~\ref{eq:rec_f},~\ref{eq:rec_m}) determine all coefficients associated with divergences using the MS$m^0$ scheme, as shown in Section \ref{ss:alternative}. Generalization of the MS$m^0$ to arbitrary covariant gauge can be accomplished by introducing another expansion parameter that is the renormalized gauge parameter. However, this is not required as going to other covariant gauges only influences the divergence of the $F_B$, which can be calculated through Landau--Khalatnikov--Fradkin transform in Chapter \ref{cp:LKFT} in a more elegant fashion.
	\chapter{The Landau--Khalatnikov--Fradkin transformation\label{cp:LKFT}}
	The relation between QED Green's functions evaluated in different covariant gauges is specified by the Landau--Khalatnikov--Fradkin transformation (LKFT) \cite{Landau:1955zz,Fradkin:1955jr,Zumino:1959wt,Sonoda:2000kn}. Differential forms of LKFT are also known as Nielsen identities \cite{Breckenridge:1994gs,Alavian:2000ud,PhysRevD.62.076002}. Incorporating LKFT into the construction of vertices in scalar QED has been studied in Refs.~\cite{Ahmadiniaz:2015kfq,Fernandez-Rangel:2016zac}. While in Ref.~\cite{Bashir:2002sp}, assuming the propagator is bare in one gauge, Fourier transforms have been used to show explicitly how the LKFT specifies the momentum space propagator in any other gauge. The gauge dependence for the momentum space fermion propagator has been recently shown calculable using diagrammatic cancellation identities \cite{Kissler:2016tne}.

QED in 4D being renormalizable, its divergences are best captured as long known by dimensional regularization \cite{tHooft:1972tcz}, which have been discussed in Chapter \ref{cp:div_QED}. Dimensional regularization also preserves gauge symmetry and translational invariance. Here we solve the LKFT for the gauge covariant behavior of fermion propagator independently from the form of the Landau gauge propagator using the spectral representation discussed in Chapter \ref{cp:spec_repr}. We demonstrate that continuing in the number of spacetime dimensions provides a convenient way to regularize behaviors more singular than free-particle propagators at the real particle production thresholds. Moreover keeping the number of spacetime dimensions explicit also allows simultaneous calculation of results in 3D and 4D, both of which are of current interest. As we will see the LKFT for fermion propagator in 3D is simpler than that in 4D. Explicit solutions to LKFT in 3D will be used to illustrate properties of the LKFT made in the general case. What is more the dependence of the solutions on $\epsilon=2-d/2$ provides insights into how gauge covariance of QED in different dimensions are connected explicitly.
\section{LKFT as group transforms\label{ss:LKFT_group}}
\subsection{LKFT in differential form}
To derive the LKFT for the QED fermion propagator in the momentum space, first consider that under two gauge fixing conditions the coordinate space photon propagator changes from $D_{\mu\nu}(z)$ to $D_{\mu\nu}'(z)$;
\begin{equation}
D_{\mu\nu}'(z)=D_{\mu\nu}(z)+\partial_\mu\partial_\nu \delta M(z).
\end{equation}
According to Zumino \cite{Zumino:1959wt}, coordinate space fermion propagators evaluated with the corresponding gauge fixing conditions are related by
\begin{equation}
S_F'(x-y)=\exp\big\{ie^2\left[\delta M(x-y)-\delta M(0)\right]\big\}S_F(x-y).\label{eq:LKFT_SF}
\end{equation}
Specifically for our interest, starting from the Landau gauge to any other covariant gauge, function $\delta M(z)$ becomes \cite{Bashir:2002sp}
\begin{equation}
\delta M(z)=\xi M(z)=-\xi \int d\underline{l}\;\frac{e^{-il\cdot z}}{l^4+i\epsilon},\label{eq:def_M_covariant}
\end{equation}
where $d\underline{l}$ denotes the $d$-dimensional momentum measure $d\underline{l}\equiv d^d l/(2\pi)^d$.
Substituting Eq.~\eqref{eq:def_M_covariant} into Eq.~\eqref{eq:LKFT_SF} produces the LKFT for the covariant gauge fermion propagator in coordinate space. In principle, taking the Fourier transform of Eq.~\eqref{eq:LKFT_SF} gives the LKFT for fermion propagators in momentum space. In practice this is difficult to accomplish because of the exponential factor in Eq.~\eqref{eq:LKFT_SF} defined by $M(z)$ in Eq.~\eqref{eq:def_M_covariant} remaining illusive. However, it has been shown that if the fermion propagator takes its free-particle form in the Landau gauge, the Fourier transform of Eq.~\eqref{eq:LKFT_SF} can be calculated~\cite{Bashir:2002sp}.

While we are interested in the scenario where the fermion propagator in the Landau gauge is more than the free-particle propagator, to circumvent the difficulty of performing Fourier transforms of implicit functions, consider taking a first order derivative with respect to $\xi$ of Eq.~\eqref{eq:LKFT_SF}. Noting that $S_F(x-y)$ on the right of Eq.~\eqref{eq:LKFT_SF} is in a specific gauge, we have the result,
\begin{equation}
\dfrac{\partial}{\partial\xi}S_F'(x-y)=ie^2[M(x-y)-M(0)]\,S_F'(x-y).\label{eq:Dxi_SF_cord}
\end{equation}
Notice that the exponential factor has been absorbed into the Landau gauge propagator using  Eq.~\eqref{eq:LKFT_SF}, giving rise to the $S'_F(x-y)$ factor on the right-hand side of  Eq.~\eqref{eq:Dxi_SF_cord}. Now that all propagators in  Eq.~\eqref{eq:Dxi_SF_cord} are primed, the prime notation can be dropped. We use $S_F(p;\xi)$ to denote the propagator in momentum space in any covariant gauge. Since there are no implicit functions left, taking the Fourier transform of  Eq.~\eqref{eq:Dxi_SF_cord} gives
\begin{equation}
\dfrac{\partial}{\partial\xi}S_F(p;\xi)=ie^2\int d\underline{l}\;\dfrac{1}{l^4+i\epsilon}\;[S_F(p;\xi)-S_F(p-l;\xi)].\label{eq:Dxi_SF}
\end{equation}
Eq.~\eqref{eq:Dxi_SF} is the LKFT for the momentum space fermion propagator, in differential form. Unlike  Eq.~\eqref{eq:LKFT_SF}, differentiating means there is no explicit dependence on the initial condition. Eq.~\eqref{eq:Dxi_SF} also agrees with the corresponding Nielsen identity given by Eq.~(11) of Ref.~\cite{Alavian:2000ud}.

However, when the propagator goes to a constant while $p^2\rightarrow \infty$, the following rewriting might be required
\begin{equation}
S_F(p;\xi)=R_F(\xi)+\tilde{S}_F(p;\xi).\label{eq:S_j_subtraction}
\end{equation}
Since the Fourier transform of a constant is $\delta$-function,  Eq.~\eqref{eq:S_j_subtraction} indicates, in the coordinate space,
\begin{equation}
S_F(x-y;\xi)=R_F(\xi)\delta(x-y)+\tilde{S}_F(x-y;\xi).
\end{equation}
Substituting it into  Eq.~\eqref{eq:LKFT_SF} gives $R_F(\xi)=R_F(0)$ and
\begin{equation}
\tilde{S}_F(x-y;\xi)=\exp\big\{ie^2[\delta M(x-y)-\delta M(0)] \big\}\tilde{S}_F(x-y;0).
\end{equation}
Therefore the LKFT for the subtracted propagator is identical to  Eq.~\eqref{eq:LKFT_SF}. Effectively we have confirmed that when the QED fermion propagator vanishes asymptotically, its Nielsen identity is equivalent to its LKFT.

As with any first order differential equations, solving for the fermion propagator from  Eq.~\eqref{eq:Dxi_SF} cannot be achieved without knowing initial conditions. However, the $\xi$ dependence of $S_F(p;\xi)$ can be deduced independently of the propagator itself at any specific gauge. 
In the next two subsections we expand on these properties.

	\subsection{Various representations of LKFT\label{ss:rep_LKFT}}
	\begin{figure}
		\centering
		\begin{tikzpicture}
		\node[draw,circle] (A) at (90:3) {$\rho(s;\xi)$};
		\node[draw,circle] (B) at (210:3) {$D(p^2;\xi)$};
		\node[draw,circle] (C) at (330:3) {$P(x^2;\xi)$};
		\draw[-open triangle 45] (A.225) -- node[rotate=60,above] {$\int ds\dfrac{1}{p^2-s+i\varepsilon}$} (B.75);
		\draw[open triangle 45-] (A.255) -- node[rotate=60,below] {$-\dfrac{1}{\pi}\mathrm{Im}\{\}$} (B.45);
		\draw[-open triangle 45] (A.285) -- node[rotate=300,below] {...} (C.135);
		\draw[open triangle 45-] (A.315) -- node[rotate=300,above] {...} (C.105);
		\draw[-open triangle 45] (B.345) -- node[rotate=0,below] {$\mathcal{F}^{-1}$} (C.195);
		\draw[open triangle 45-] (B.15) -- node[rotate=0,above] {$\mathcal{F}$} (C.165);
		\end{tikzpicture}
		\caption{Scalar particles propagators in coordinate space $P(x^2;\xi)$, momentum space $D(p^2;\xi)$ and its spectral function $\rho(s;\xi)$ with bijective relations among them illustrated. The Fourier transform is bijective. For momentum space propagators with branch cuts and poles as their singularities, the spectral representation is also bijective. Consequently there must be a bijective relation between the coordinate space propagator and its spectral function.}
		\label{fig:LKFT_group}
	\end{figure}
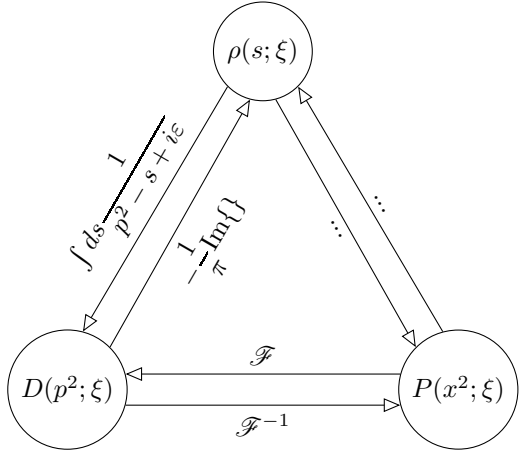
	Substituting the spectral representation of fermion propagator, Eq.~\eqref{eq:SF_Dirac_struct}, into Eq.~\eqref{eq:Dxi_SF} allows the effective one-loop integral to be evaluated explicitly. However, the spectral representation alone is not sufficient for us to solve for the dependence of fermion propagator on the gauge parameter $\xi$ from LKFT. 
	
	Observations about  Eq.~\eqref{eq:LKFT_SF} will provide insight into a more useful mathematical aspect of LKFT for a gauge covariant  fermion propagator. Formally, Eq.~\eqref{eq:LKFT_SF} states that the LKFT for the fermion propagator in coordinate space is simply a phase factor, which bears close resemblance to elements of a Lie group. 
	One can further verify that, when group multiplication is defined as function multiplication, this phase factor satisfies closure, associativity, and the existence of identity element and inverse elements. Therefore when considered as a linear transformation on coordinate space functions, the LKFT is indeed a group transform for coordinate space fermion propagators. 
	
	Fourier transforms are known to be {\it one-to-one} and {\it onto}. Since we have established in Chapter \ref{cp:spec_repr} that with certain assumptions about the analytic structure of the fermion propagator, the spectral representation is also {\it one-to-one} and {\it onto}. These correspondences, illustrated in Fig. \ref{fig:LKFT_group}, clearly indicate that, just as with the LKFT for coordinate space propagator, LKFT for momentum space propagators and for spectral functions should both be group transforms. In fact, the coordinate space representation, the momentum space representation and the spectral representation of LKFT are isomorphic representations of the same group. Additionally, since $\xi$ parameterizes the LKFT as a continuous group, the starting gauge of LKFT does not matter; only the difference in $\xi$ enters in calculation. Though the default initial gauge for LKFT can be conveniently chosen to be the Landau gauge,  for calculations with the  initial value of gauge parameter that is $\xi_0$, one simply replaces Landau gauge quantities by those at $\xi_0$ and replaces $\xi$ by $\xi-\xi_0$.
	
	Our observation that the LKFT in its spectral representation is a group transformation enables us to develop schemes for solving Eq.~\eqref{eq:Dxi_SF}. As illustrated in Fig.~\ref{fig:LKFT_group}, the correspondence between the fermion propagator in momentum space and its spectral function is linear, as a consequence of which LKFT in spectral form is also required to be linear. However, instead of a simple phase factor, we expect the LKFT in its spectral form to involve more complicated linear operations. Therefore, without loss of generality, we can write
	\begin{equation}
	\rho_j(s;\xi)=\int ds'\, \mathcal{K}_j(s,s';\xi)\,\rho_j(s';0),\label{eq:LKFT_linearity_spectral_rep}
	\end{equation}
	where distributions $\mathcal{K}_j(s,s';\xi)$ work as the Green's function for Eq.~\eqref{eq:Dxi_SF}. They represent linear operations that encode $\xi$ dependences of $\rho_j(s;\xi)$ to be determined by the LKFT, and so respect all group properties. Explicitly, denote $\mathbf{K}$ the set of distribution $\mathcal{K}(s,s';\xi)$, with group multiplication defined as integration over spectral variables. To verify that $\mathbf{K}$ is indeed a group, for any $\mathcal{K}(s,s';\xi)\in \mathbf{K}$ the following properties have to be satisfied:
	\begin{enumerate}
		\item \underline{Closure} $\int ds'\mathcal{K}(s,s';\xi)\mathcal{K}(s',s'';\xi')$ is also an element of $\mathbf{K}$;
		\item \underline{Associativity} 
		\begin{align*}&\quad \int ds'\mathcal{K}(s,s';\xi)\int ds''\mathcal{K}(s',s'';\xi')\mathcal{K}(s'',s''';\xi'')\\
		&=\int ds''\left[\int ds'\mathcal{K}(s,s';\xi)\mathcal{K}(s',s'';\xi')\right]\mathcal{K}(s'',s''';\xi'');
		\end{align*}
		\item \underline{Identity Element} $\exists~\mathcal{K}_I(s,s') \in \mathbf{K}$ such that \begin{equation*}
		 \int ds'\mathcal{K}_I(s,s')\mathcal{K}(s',s'';\xi)\ =\int ds'\mathcal{K}(s,s';\xi)\mathcal{K}_I(s',s'')=\mathcal{K}(s,s'';\xi);
		\end{equation*}
		\item \underline{Inverse Element} $\exists~\mathcal{K}_{inv}(s,s';\xi)$ such that
		\begin{equation*}
		\int ds'\,\mathcal{K}_{inv}(s,s';\xi)\,\mathcal{K}(s',s'';\xi)\;=\;\int ds'\,\mathcal{K}(s,s';\xi)\,\mathcal{K}_{inv}(s',s'';\xi)\;=\;\mathcal{K}_I(s,s'').
		\end{equation*}
	\end{enumerate}	
	Substituting Eqs.~(\ref{eq:SF_Dirac_struct}~,\ref{eq:LKFT_linearity_spectral_rep}) into Eq.~\eqref{eq:Dxi_SF} gives
	\begin{equation}
	\dfrac{\partial}{\partial\xi}\int ds\,\dfrac{\mathcal{K}_j(s,s';\xi)}{p^2-s+i\epsilon}=-\dfrac{\alpha}{4\pi}\int ds\,\dfrac{\Xi_j(p^2,s)}{p^2-s+i\epsilon}\,\mathcal{K}_j(s,s';\xi),\label{eq:LKFT_k12}
	\end{equation}
	where the $\Xi_j(p^2,s)$ are determined by the effective one-loop integral, which can be evaluated using Feynman parameterization for combining denominators, together with dimensional regularization. Apparently from Eq.~\eqref{eq:LKFT_linearity_spectral_rep}, the initial condition for distributions $\mathcal{K}_j$ is $\mathcal{K}_j(s,s';0)=\delta(s-s')$. In the remaining part of this subsection, two methods for solving Eq.~\eqref{eq:LKFT_k12} will be presented.
	\paragraph{Method 1: analogue to first-order ordinary differential equations} Operations with respect to $\xi$ in Eq.~\eqref{eq:LKFT_k12} are only present on the left-hand side, which resemble homogeneous first-order ordinary differential equations. In order to solve for $\mathcal{K}_j(s,s';\xi)$, consider the original definition of partial derivative:
	\begin{equation}
	\dfrac{\partial}{\partial\xi}\mathcal{K}(s,s';\xi)\equiv \lim\limits_{\Delta\rightarrow 0}\dfrac{\mathcal{K}(s,s';\xi+\Delta)-\mathcal{K}(s,s';\xi)}{\Delta}.
	\end{equation}
	Next, applying Eq.~\eqref{eq:LKFT_linearity_spectral_rep} many times gives
	\begin{align*}
	&\quad \rho(s,s';\xi+\Delta) =\int ds''\mathcal{K}(s,s'';\Delta)\rho(s'',s';\xi)\\
	& =\int ds''\int ds'''\mathcal{K}(s,s'';\Delta)\mathcal{K}(s'',s''';\xi)\rho(s'',s';0).
	\end{align*}
	
	Since the LKFT is independent of initial conditions,
	\begin{equation}
	\mathcal{K}(s,s';\xi+\Delta)=\int ds''\mathcal{K}(s,s'';\Delta)\mathcal{K}(s'',s';\xi).\label{eq:k_sum}
	\end{equation}
	Eq.~\eqref{eq:k_sum} should not come as a surprise given that the LKFT for the fermion propagator in a spectral representation is isomorphic to coordinate space LKFT. Eq.~\eqref{eq:LKFT_k12} then becomes
	\begin{align}
	&\quad \dfrac{\partial}{\partial\xi}\int ds\dfrac{\mathcal{K}(s,s';\xi)}{p^2-s+i\epsilon}\nonumber\\
	& =\lim\limits_{\Delta\rightarrow 0}\int ds\int ds''\dfrac{\mathcal{K}(s,s'';\Delta)-\delta(s-s'')}{(p^2-s+i\epsilon)\Delta}\mathcal{K}(s'',s';\xi)\nonumber\\[1.5mm]
	& =-\dfrac{\alpha}{4\pi}\int ds\dfrac{\Xi(p^2,s)}{p^2-s+i\epsilon}\mathcal{K}(s,s';\xi).\label{eq:dxi_Delta}
	\end{align}
	Taking the limit $\xi\rightarrow 0$ where $\mathcal{K}(s,s';\xi)$ becomes a  delta-function simplifies Eq.~\eqref{eq:dxi_Delta} into
	\begin{equation}
	\lim\limits_{\Delta\rightarrow 0}\dfrac{1}{\Delta}\int ds\dfrac{\mathcal{K}(s,s';\Delta)-\delta(s-s')}{p^2-s+i\epsilon}=-\dfrac{\alpha}{4\pi}\dfrac{\Xi(p^2,s')}{p^2-s'+i\epsilon}.\label{eq:dxi0_Delta}
	\end{equation}
	Eq.~\eqref{eq:dxi0_Delta} specifies how the distribution $\mathcal{K}(s,s';\xi)$ departs from its initial form (a delta-function) with infinitesimal $\xi$. Solving Eq.~\eqref{eq:dxi0_Delta} is sufficient to obtain $\mathcal{K}(s,s';\xi)$ with finite $\xi$, which, in principle, can be written as an infinite number of steps of distribution multiplication. Explicitly, this procedure is
	\begin{equation}
	\mathcal{K}(s,s';\xi)=\lim\limits_{N\rightarrow +\infty}\left[\prod_{n=0}^{N-1}\int ds_{n+1}~\mathcal{K}\left(s_n,s_{n+1};\dfrac{\xi}{N}\right) \right]\times\mathcal{K}\left(s_{N},s';0\right),\label{eq:k_Nsteps}
	\end{equation}
	with $s_0=s$.
	Formally Eq.~\eqref{eq:k_Nsteps} gives distributions $\mathcal{K}_j(s,s';\xi)$ with finite $\xi$, solving LKFT for $\rho_j(s;\xi)$. In practice one may prefer a closed form for the $\mathcal{K}_j$ with group multiplications in a minimal number of steps. Realizing Eq.~\eqref{eq:k_Nsteps} is the analogue of 
	\[\lim\limits_{N\rightarrow +\infty}\left(1+\dfrac{x}{N}\right)^N=e^x,\]
	and Eq.~\eqref{eq:LKFT_k12} is very  similar to $\frac{d}{dx}f(x)=af(x)$, we can assume the following form for $\mathcal{K}(s,s';\xi)$,
	\begin{equation}
	\mathcal{K}_j=\exp\left(-\dfrac{\alpha\xi}{4\pi}\Phi_j\right),\label{eq:k_exponential}
	\end{equation}
	where distributions $\Phi_j$ are independent of $\xi$. The exponential of a distribution is defined by
	\begin{equation}
	\exp\big\{\lambda\Phi\big\}=\sum_{n=0}^{+\infty}\dfrac{\lambda^n}{n!}\Phi^n=\delta(s-s')+\lambda\Phi+\dfrac{\lambda^2}{2!}\Phi^2+\dots,\label{eq:dist_exp}
	\end{equation}
	with distribution exponentiation given by
	\begin{equation}
	\Phi^n(s,s')=\int ds''\Phi(s,s'')\Phi^{n-1}(s'',s'),\label{eq:dist_exponentiation}
	\end{equation}
	for ${n\geq 1}$. And ${\Phi^0(s,s')=\delta(s-s')}$. One can check that $\mathcal{K}_j$ given by Eq.~\eqref{eq:k_exponential} satisfy Eq.~\eqref{eq:LKFT_k12} with initial conditions ${\mathcal{K}_j(s,s';0)=\delta(s-s')}$ given distributions $\Phi_j$ satisfy their own identities. To verify the exponential of distributions indeed solves Eq.~\eqref{eq:LKFT_k12} and find the identities $\Phi_j$ have to satisfy, let us start with
	\begin{equation*}
	\dfrac{\partial}{\partial\xi}\mathcal{K} =\dfrac{\partial}{\partial\xi}\exp\bigg\{-\dfrac{\alpha\xi}{4\pi}\Phi\bigg\}=-\dfrac{\alpha}{4\pi}\Phi\exp\bigg\{-\dfrac{\alpha\xi}{4\pi}\Phi\bigg\}=-\dfrac{\alpha}{4\pi}\Phi \mathcal{K},
	\end{equation*}
	then using Eq.~\eqref{eq:k_exponential}, the left hand side of Eq.~\eqref{eq:LKFT_k12} can be written as
	\begin{equation}
	\dfrac{\partial}{\partial\xi}\dfrac{1}{p^2-s+i\epsilon}\mathcal{K}=-\dfrac{\alpha}{4\pi}\dfrac{\Phi}{p^2-s+i\epsilon}\mathcal{K}.
	\end{equation}
	Comparing with its right hand side, one obtains after restoring the integration variables,
	\begin{equation}
	\int ds\,ds'\,\dfrac{\Phi(s,s')}{p^2-s+i\epsilon}\,\mathcal{K}(s',s'';\xi) =\int ds'\,\dfrac{\Xi(p^2,s')}{p^2-s'+i\epsilon}\,\mathcal{K}(s',s'';\xi),
	\end{equation}
	which indicates (or by multiplying $\mathcal{K}(s'',s''';-\xi)$ to the right)
	\begin{equation}
	\int ds\,\dfrac{\Phi_j(s,s')}{p^2-s+i\epsilon}\;=\;\dfrac{\Xi_j(p^2,s')}{p^2-s'+i\epsilon}.\label{eq:Phi_Xi}
	\end{equation}
	Therefore Eq.~\eqref{eq:LKFT_k12} is solved by Eq.~\eqref{eq:k_exponential} given $\Phi_j$ satisfy Eq.~\eqref{eq:Phi_Xi}. One can easily identify that distributions $\Phi_j$ are the generators for continuous groups defined by $\mathcal{K}_j$.
	\paragraph{Method 2: differential equations solved by multiplying inverse elements}
	From the group property of $\mathbf{K}=\{\mathcal{K}(s,s';\xi) \}$, the inverse element of $\mathcal{K}(s,s';\xi)$ is $\mathcal{K}(s,s';-\xi)$. This can be seen most easily from the isomorphic representation of LKFT in coordinate space. Or, in the language of distribution multiplication,
	\[\mathcal{K}^{-1}(s,s';\xi)=\mathcal{K}(s,s';-\xi). \]
	Multiplying this inverse element to the right of differential equation Eq.~\eqref{eq:LKFT_k12} gives
	\begin{align}
	&\quad\int ds\int ds'\dfrac{1}{p^2-s+i\epsilon}\left[\dfrac{\partial}{\partial\xi}\mathcal{K}(s,s';\xi)\right]\mathcal{K}(s',s'';-\xi)\nonumber\\
	&=-\dfrac{\alpha}{4\pi}\int ds\int ds'\dfrac{\Xi(p^2,s)}{p^2-s+i\epsilon}\mathcal{K}(s,s';\xi)\mathcal{K}(s',s'';-\xi),
	\end{align}
	or equivalently,
	\begin{equation}
	\int ds\dfrac{1}{p^2-s+i\epsilon}\dfrac{\partial}{\partial\xi}\ln~\mathcal{K}(s,s'';\xi) =-\dfrac{\alpha}{4\pi}\int ds\dfrac{\Xi(p^2,s'')}{p^2-s''+i\epsilon},\label{eq:log_k}
	\end{equation}
	where the logarithm of distribution $\mathcal{K}(s,s'';\xi)$ is taken in distributional sense, hence when spectral variables are omitted
	\begin{equation}
	\ln (\mathcal{K})=(\mathcal{K}-\delta)-\dfrac{1}{2}(\mathcal{K}-\delta)^2+\dfrac{1}{3}(\mathcal{K}-\delta)^3+\dots=\sum_{n=1}^{+\infty}\dfrac{(-1)^{n-1}}{n}(\mathcal{K}-\delta)^n,
	\end{equation}
	with distribution exponentiations defined by Eq.~\eqref{eq:dist_exponentiation}. To see that $(\partial_\xi~\mathcal{K})\mathcal{K}^{-1}$ is indeed $\partial_\xi\ln~\mathcal{K}$, denote $u=\ln \mathcal{K} \quad \mathcal{K}=e^u$. Then
	${\partial_\xi \mathcal{K}=(\partial_\xi u)e^u=(\partial_\xi u)\mathcal{K}}$. Therefore \[\partial_\xi \ln \mathcal{K}=\partial_\xi u=(\partial_\xi \mathcal{K})\mathcal{K}^{-1}.\]
	
	After clarifying the meaning of distribution logarithm, the null space of the spectral representation is supposed to be empty for the function space defined as the set of functions with analytic structures discussed in Chapter \ref{cp:spec_repr}. Therefore Eq.~\eqref{eq:log_k} indicates that when
	\begin{equation}
	\partial_\xi \ln~\mathcal{K}(s,s'';\xi)=-\dfrac{\alpha}{4\pi}\Phi(s,s''),
	\end{equation}
	with distribution $\Phi$ satisfying Eq.~\eqref{eq:Phi_Xi}, Eq.~\eqref{eq:Dxi_SF} is solved by Eq.~\eqref{eq:k_exponential}.
	
	Either through group operations or analogy with ordinary differential equations, we have formally found the Green's function specifying the $\xi$ dependence of the fermion propagator spectral functions. Because there are no dimension-odd operators in the LKFT for the fermion propagator, the Dirac vector and Dirac scalar components do not mix. The representation of linear operations by integrating distributions with spectral functions closely resembles matrices multiplying vectors as linear transforms.
	
	\section{LKFT in the spectral representation with arbitrary numbers of dimensions\label{ss:LKFT_spectral_rep}}
	Before applying these solutions to the LKFT in the form of Eq.~\eqref{eq:k_exponential} to calculate the $\xi$ dependence of the fermion propagator, we need to determine the  distributions $\Phi_j$ from Eq.~\eqref{eq:Phi_Xi}. To do so requires explicit expressions for the functions $\Xi_j(p^2,s)$. 
	
	We use standard perturbative techniques including the Feynman method for combining denominators and dimensional regularization. Then substituting the spectral representation of the fermion propagator Eq.~\eqref{eq:SF_Dirac_struct} into the LKFT for the momentum space fermion propagator, Eq.~\eqref{eq:Dxi_SF}, and comparing the resulting equation with the definition of $\Xi_j(p^2,s)$ in Eq.~\eqref{eq:LKFT_k12} gives, 
	\begin{align}
	&\quad  \Xi_1(p^2,s)=\int_{0}^{1}dx~2x\left[1-\epsilon+\dfrac{\epsilon}{1-xz}\right]\dfrac{\Gamma(\epsilon)(4\pi\mu^2/s)^\epsilon}{[(1-x)(1-xz)]^\epsilon}\label{eq:def_Xi_1}\\[2mm]
	&\quad \Xi_2(p^2,s)=\int_{0}^{1}dx~2x\left[1-\epsilon+\dfrac{\epsilon}{2}\dfrac{z+1}{1-xz}\right]\dfrac{\Gamma(\epsilon)(4\pi\mu^2/s)^\epsilon}{[(1-x)(1-xz)]^\epsilon},\label{eq:def_Xi_2}
	\end{align}
	where $z=p^2/s$ and the number of spacetime dimensions $d=4 -2 \epsilon$. Meanwhile, the dimension of $e^2/(4\pi)$ is carried by $\mu$ such that the coupling constant $\alpha$ remains dimensionless. 
	
	Results given in Eqs.~(\ref{eq:def_Xi_1},~\ref{eq:def_Xi_2}) characterize how the LKFT behaves in Minkowski space. Using the spectral representation there is no need to make a Wick rotation to perform the loop-type integral. This eliminates any ambiguity of which loop momentum should be integrated in Euclidean space. We use dimensional regularization (required when close to four dimensions) in one of two ways. We can follow Feynman and integrate the time component of the loop momentum to infinity first. We then have spherical symmetry in the $(d-1)$ spatial dimensions and use dimensional regularization only on the space components. Of course, we could instead Wick rotate, assuming this is valid and picks up no new singularities. One then has spherical symmetry in $d$ dimensions and regularize {\it \`a la} 't Hooft and Veltman \cite{tHooft:1972tcz}. The results are the same with or without Wick rotation, as discussed in Appendix \ref{ss:loop_Minkowski}. 

	Extensive use of definitions and properties of hypergeometric functions allows us to evaluate integrals over Feynman parameters in Eqs.~(\ref{eq:def_Xi_1},~\ref{eq:def_Xi_2}). Explicitly, we have 
	\begin{align}
	\dfrac{\Xi_1}{p^2-s}&=\dfrac{\Gamma(\epsilon)}{s}\left(\dfrac{4\pi\mu^2}{s}\right)^\epsilon\dfrac{-2}{(1-\epsilon)(2-\epsilon)}~_2F_1(\epsilon+1,3;3-\epsilon;z)\label{eq:Xi_1_reduced}\\
	\dfrac{\Xi_2}{p^2-s}&=\dfrac{\Gamma(\epsilon)}{s}\left(\dfrac{4\pi\mu^2}{s}\right)^\epsilon\dfrac{-1}{1-\epsilon}~_2F_1(\epsilon+1,2;2-\epsilon;z).\label{eq:Xi_2_reduced}
	\end{align}
	Since results given by Eqs.~(\ref{eq:Xi_2_reduced},~\ref{eq:Xi_2_reduced}) do not exist in the literature, their derivations are documented in Appendix~\ref{ss:Xi_12_epsilon}. Hypergeometric functions occurring in Eqs.~(\ref{eq:Xi_1_reduced},~\ref{eq:Xi_2_reduced}) are understood to be given by the integral definition Eq.~(15.3.1) in Abramowitz and Stegun \cite{abramowitz1964handbook}. For $\epsilon>0$, this integral definition is the analytic continuation of the series definition Eq.~\eqref{eq:2F1_Taylor} with a branch cut \cite{abramowitz1964handbook} on the real axis of $z$ from $1$ to $+\infty$, a property one would expect for corrections to the fermion propagator. The scenario where $\epsilon<0$ is beyond the scope of this article.
	
	Explicit calculation shows that in three dimensions
	\begin{align}
	& \lim\limits_{\epsilon\rightarrow 1/2}\Xi_1(p^2,s)=2\pi\sqrt{\dfrac{\mu^2}{s}}\bigg\{-\dfrac{z+1}{(z-1)z}+\dfrac{z-1}{z^{3/2}}\mathrm{arctanh}(\sqrt{z})\bigg\}\\
	& \lim\limits_{\epsilon\rightarrow 1/2}\Xi_2(p^2,s)=-\dfrac{4\pi}{z-1}\sqrt{\dfrac{\mu^2}{s}} \quad ,
	\end{align}
	while for small $\epsilon$, \textit{i.e.} approaching four dimensions:
	\begin{align}
	\Xi_1(p^2,s)&=\dfrac{1}{\epsilon}-\gamma_E+\ln\left(\dfrac{4\pi \mu^2}{s}\right)+1-\dfrac{1}{z}-\left(1+\dfrac{1}{z^2}\right)\ln(1-z)+\mathcal{O}(\epsilon^1)\\
	\Xi_2(p^2,s)&=\dfrac{1}{\epsilon}-\gamma_E+\ln\left(\dfrac{4\pi \mu^2}{s}\right)-\left(1+\dfrac{1}{z}\right)\ln(1-z)+\mathcal{O}(\epsilon).
	\end{align}
	The $\epsilon\rightarrow 1/2$ limits can be calculated using identities listed in Chapter 15 of Abramowitz and Stegun \cite{abramowitz1964handbook}. While the small $\epsilon$ expansions can be calculated according to Appendix \ref{ss:identities_2F1_epsilon}. Therefore $d=3$ and $4$ results have been recovered.
	
	With loop integrals $\Xi_j(p^2,s)$ calculated, the right-hand side of Eq.~\eqref{eq:Phi_Xi} is elegantly represented by Eqs.~(\ref{eq:Xi_1_reduced},~\ref{eq:Xi_2_reduced}). The remaining task is to find the corresponding distributions $\Phi_j$ that solve Eq.~\eqref{eq:Phi_Xi}. Since the distributions $\Phi_j$ are only allowed to be linear operators on the spectral variable $s$, solving Eq.~\eqref{eq:Phi_Xi} is equivalent to generating convoluted $p^2$ dependences embedded in hypergeometric functions from that of a free-particle propagator. For $\epsilon>0$, the behavior of functions $\Xi_j/(p^2-s+i\varepsilon)$ in the limit $p^2\rightarrow s$ is more singular than the free-particle propagator. In fact, this singularity behaves as 
	\begin{equation}
	\lim\limits_{p^2\rightarrow s}\dfrac{\Xi_j(p^2,s)}{p^2-s+i\varepsilon}=\Gamma(\epsilon)\left(\dfrac{4\pi\mu^2}{s}\right)^\epsilon \dfrac{4^\epsilon}{\sqrt{\pi}}\Gamma(1-\epsilon)\Gamma(1/2+\epsilon)\left(1-\dfrac{p^2}{s}-i\varepsilon\right)^{-1-2\epsilon},\label{eq:LKFT_sgl_Sigma}
	\end{equation}
	based on Eq.~(15.3.6) of Ref.~\cite{abramowitz1964handbook}. Therefore one can expect distributions $\Phi_j$ to be more singular than $\delta$-functions.
	\subsection{Exponent-preserving operations}
	Our task is to find out how to generate $p^2$ dependences in hypergeometric functions given by Eqs.~(\ref{eq:Xi_1_reduced},~\ref{eq:Xi_2_reduced}) from the free-particle propagator with only linear operations on the spectral variable $s$. It appears that the variable $z=p^2/s$ is more convenient than the spectral variable $s$ itself. In the process of finding the distributions $\Phi_j$, multiplication by $s$ can be regarded as a trivial linear operation. Therefore we are allowed to apply it as needed to make the remaining operations transparent. Meanwhile, having decided to work with the variable $z$ rather than the dispersive variable $s$, we are obligated to ensure that the net effect of operations on $z$ does not result in any operation on $p^2$. 
	
	Starting with the observation that the $p^2$ dependence of a free-particle propagator can be represented by 
	\[\dfrac{-s}{p^2-s}= \dfrac{1}{1-z}=~_2F_1(1,b;b;z) \; ,\]
	for any $b$.
	The factor $-s$ does not matter in this scenario because it is merely a multiplication factor. In addition, for any linear operation on the variable $z$, as long as the net effect does not act as multiplication by the variable $z$ or $z^\lambda$, such transforms can be written in terms of spectral variable $s$ independently of $p^2$. 
	
	To quantify this criterion, define the exponent $\lambda$ for linear transforms on the variable $z$. Starting with a simple multiplication factor $z^\lambda$, this has an exponent  $\lambda$, because it raises the index by $\lambda$ for every term in the series expansion of a function of $z$. 
	Thus the operation $z^md^{n}/dz^{n}$ has an exponent $\lambda=m-n$. 
	
	An identity involving transformations on the variable $z$ is called \textit{exponent-preserving} if the exponent of the transform on the left-hand side is identical to that on the right-hand side. For example, Eq.~(15.2.2) in Ref.~\cite{abramowitz1964handbook} is exponent-preserving for transformations on variable $z$ because exponents on both sides are identical. An identity that does not preserve exponents is called \textit{exponent-violating}. Exponent-violating transforms on variable $z$ cannot be translated into operations on the spectral variable $s$ only (not involving $p^2$).
	
	Next we need to determine the exponent-preserving linear transforms that generate any hypergeometric function $~_2F_1(a,b;c;z)$ from $~_2F_1(1,b;b;z)=1/(1-z)$. To accomplish this, one immediately thinks of Gauss' relations for contiguous functions. However, they only relate hypergeometric functions with integer differences of parameters $a,b$ and $c$. Meanwhile, not all of them are exponent-preserving. Another category of candidates is the differential relation for hypergeometric functions. These relations Eqs.~(15.2.3,~15.2.4) in Ref.~\cite{abramowitz1964handbook} are promising since they are exponent-preserving. However Eqs.~(15.2.3, 15.2.4) in Ref.~\cite{abramowitz1964handbook} can not be applied without generalization because they, similar to relations for contiguous functions, only raise or lower parameters $a,b$ or $c$ by integers.
	\subsection{Fractional calculus}
	To be able to solve LKFT in arbitrary dimensions, we need to overcome the limitation that Eqs.~(15.2.3, 15.2.4) in Ref.~\cite{abramowitz1964handbook} only work for integer differences in parameters for hypergeometric functions. Consequently we consider generalizing these to fractional orders of derivatives. We need to find out a version of fractional derivatives that applies to these differential relations for $~_2F_1(a,b;c;z)$. To do this, it is natural to consider the Riemann-Liouville definition of fractional calculus \cite{Riemann:fractional}:
	\begin{equation}
	I^\alpha f(z)=\dfrac{1}{\Gamma(\alpha)}\int_{\zeta}^{z}dz'(z-z')^{\alpha-1}f(z').\label{eq:def_Riemann_Liouville_I}
	\end{equation}
	For $\alpha>0$, the Riemann-Liouville fractional derivative is defined as
	\begin{equation}
	D^\alpha f(z)=\left(\dfrac{d}{dz}\right)^{\lceil \alpha \rceil}I^{\lceil\alpha\rceil-\alpha}f(z),
	\end{equation}
	where $\lceil \alpha \rceil$ is the smallest integer larger than $\alpha$, {\it i.e.} the ceiling function. Specifically for $\alpha \in (0,1)$, $\lceil \alpha \rceil=1$ and
	\begin{equation}
	D^\alpha f(z)=\dfrac{1}{\Gamma(1-\alpha)}\dfrac{d}{dz}\int_{\zeta}^{z}dz'(z-z')^{-\alpha}f(z').\label{eq:def_Riemann_Liouville_D}
	\end{equation}
	The lower limit $\zeta$ should be selected to reproduce Eqs.~(15.2.3, 15.2.4) in Ref.~\cite{abramowitz1964handbook} if the Riemann-Liouville formulation of fractional calculus is the expected version of fractional calculus that successfully generalizes them.
	
	To make an informed selection of $\zeta$, consider Eq.~\eqref{eq:2F1_Taylor}, the Taylor series expansion of hypergeometric functions. For $\alpha\in (0,1)$,
	\begin{equation}
	D^\alpha z^\beta=\dfrac{1}{\Gamma(1-\alpha)}\dfrac{d}{dz}\int_{\zeta}^{z}dz'(z-z')^{-\alpha}(z')^\beta.
	\end{equation}
	Since mixing among terms of the Taylor expansion after derivative operations is undesirable, we choose $\zeta=0$ and obtain
	\begin{equation}
	D^\alpha z^\beta=\dfrac{1}{\Gamma(1-\alpha)}\dfrac{d}{dz} \dfrac{\Gamma(1-\alpha)\Gamma(1+\beta)}{\Gamma(2-\alpha+\beta)}z^{1-\alpha+\beta}=\dfrac{\Gamma(1+\beta)}{\Gamma(1-\alpha+\beta)}z^{-\alpha+\beta},
	\end{equation}
	which applies when $\alpha<1,~\beta>-1$ and $z>0$. Since directly from the definition of Pochhammer symbol \[(1-\alpha+\beta)_\alpha=\Gamma(1+\beta)/\Gamma(1-\alpha+\beta),\]
	we have
	\begin{equation}
	D^\alpha z^\beta =(1-\alpha+\beta)_\alpha z^{-\alpha+\beta}.\label{eq:fractional_D_alpha}
	\end{equation}
	Notice that derivatives generalized this way to fractional orders also agree with integer order derivatives when $\alpha$ in Eq.~\eqref{eq:fractional_D_alpha} is an integer. This will be taken as the default definition of fractional calculus in this article. Eqs.~\eqref{eq:def_Riemann_Liouville_I} through \eqref{eq:fractional_D_alpha} are the standard formalism of the Riemann-Liouville fractional calculus. They form the bases of our solutions to Eq.~\eqref{eq:Phi_Xi} in the following half of this section.
	
	Showing Eqs.~(15.2.3, 15.2.4) in Ref.~\cite{abramowitz1964handbook} apply to fractional orders is straightforward starting with the application of Taylor series expansions of hypergeometric functions. Explicitly, 
	\begin{equation*}
	D^\alpha z^{a+\alpha-1}~_2F_1(a,b;c;z)=D^\alpha \sum_{n=0}^{+\infty}\dfrac{(a)_n(b)_n}{(c)_n n!}z^{n+a+\alpha-1}=\sum_{n=0}^{+\infty}\dfrac{(b)_n}{(c)_n n!}(a)_n(n+a)_\alpha z^{n+a-1}
	\end{equation*}
	Since
	\begin{equation*}
	(a)_n(n+a)_\alpha=\dfrac{\Gamma(a+n)\Gamma(a+n+a)}{\Gamma(a)\Gamma(n+a)}=\dfrac{\Gamma(a+\alpha)\Gamma(a+n+\alpha)}{\Gamma(a)\Gamma(a+\alpha)}=(a)_\alpha(a+\alpha)_n,
	\end{equation*}
	we have
	\begin{equation}
	D^\alpha z^{a+\alpha-1}~_2F_1(a,b;c;z)=\sum_{n=0}^{+\infty}(a)_\alpha z^{a-1}\dfrac{(a+\alpha)_n(b)_n}{(c)_n n!}z^n=(a)_\alpha z^{a-1}~_2F_1(a+\alpha,b;c;z).\label{eq:Da2F1}
	\end{equation}
	Therefore Eq.~(15.2.3) in Ref.~\cite{abramowitz1964handbook} has been generalized to accommodate fractional orders of derivatives. Meanwhile, Eq.~\eqref{eq:Da2F1} is exponent-preserving as one would expect. Therefore it is the generalization of Eq.~(15.2.3) we are seeking. Similar steps can be used to prove that Eq.~(15.2.4) in Ref.~\cite{abramowitz1964handbook} generalizes to fractional orders using our definition of fractional calculus as well;
	\begin{align}
	& \quad D^\alpha z^{c-1}~_2F_1(a,b;c;z)=D^\alpha\sum_{n=0}^{+\infty}\dfrac{(a)_n(b)_n}{(c)_n n!}z^{n+c-1}\nonumber\\
	&=\sum_{n=0}^{+\infty}\dfrac{(a)_n(b)_n}{(c)_nn!}(n+c-\alpha)_\alpha z^{n+c-1-\alpha} =\sum_{n=0}^{+\infty}\dfrac{(a)_n(b)_n}{n!}\dfrac{\Gamma(c)}{\Gamma(c+n)}\dfrac{\Gamma(n+c)}{\Gamma(n+c-\alpha)}z^{n+c-1-\alpha}\nonumber\\
	& =\sum_{n=0}^{+\infty}\dfrac{(a)_n(b)_n}{n!}\dfrac{\Gamma(c)}{\Gamma(c-\alpha)}\dfrac{\Gamma(c-\alpha)}{\Gamma(n+c-\alpha)}z^n z^{c-\alpha-1} =(c-\alpha)_\alpha z^{c-\alpha-1}~_2F_1(a,b;c-\alpha;z),\label{eq:Dc2Fa}
	\end{align}
	because the hypergeometric function $~_2F_1(a,b;c;z)$ is symmetric in parameters $a$ and $b$. Equipped with Eqs.~(\ref{eq:Da2F1},~\ref{eq:Dc2Fa}), any hypergeometric function $~_2F_1(a,b;c;z)$ can be linearly generated from the free-particle propagator with only a finite (up to two) steps of exponent-preserving linear operations. Explicitly, to generate the $z$ dependences of $~_2F_1(\epsilon+1,n;n-\epsilon;z)$ from a free particle propagator, consider the following linear operations
	\begin{equation}
	D^\epsilon z^\epsilon ~_2F_1(1,n;n;z)=(1)_\epsilon~_2F_1(1+\epsilon,n;n;z)\label{eq:Da_epsilon}
	\end{equation}
	and
	\begin{equation}
	D^\epsilon z^{n-1}~_2F_1(1+\epsilon,n;n;z)=(n-\epsilon)_\epsilon
	z^{n-\epsilon-1}~_2F_1(1+\epsilon,n;n-\epsilon;z)\label{eq:Dc_epsilon}.
	\end{equation}
	Therefore
	\begin{equation}
	~_2F_1(1+\epsilon,n;n-\epsilon;z)=\dfrac{\Gamma(n-\epsilon)}{\Gamma(n)\Gamma(1+\epsilon)}z^{\epsilon+1-n}D^\epsilon z^{n-1}D^\epsilon z^\epsilon~_2F_1(1,n;n;z),\label{eq:linear_trans_LKFT_epsilon}
	\end{equation}
	which after setting $n=2,3$ for $j=2,1$ respectively, recovers the exponent-preserving linear transforms required to generate the hypergeometric functions in ${\Xi_j(p^2,s)/(p^2-s+i\varepsilon)}$ from the free-particle propagator. To see explicitly how such a linear transform on $z$ can be written as that on $s$ not involving $p^2$, refer to Appendix \ref{ss:Example_epn_preserving} for an example. In addition, the exact order of component transforms given by Eqs.~(\ref{eq:Da_epsilon},~\ref{eq:Dc_epsilon}) should not matter because of the commutation relations for the hypergeometric function $~_2F_1(1,n;n;z)$
	\begin{equation}
	[z^{\epsilon+1-n}D^\epsilon z^{n-1},D^\epsilon z^\epsilon]~_2F_1(1,n;n;z)=0.
	\end{equation}
	This commutation relation is true for $~_2F_1(1,n;n;z)$ because Eqs.~(\ref{eq:Da_epsilon},~\ref{eq:Dc_epsilon}) acts on parameters $a$ and $c$ of the hypergeometric function $~_2F_1(a,b;c;z)$ independently.
	
	\subsection{Operator Exponentials\label{ss:Exp_k}}
	The combination of exponent-preserving requirement and fractional calculus allows us to solve for distributions $\Phi_j$ from Eq.~\eqref{eq:Phi_Xi}. The solution for the fermion propagator LKFT in spectral form is given by the exponential of distributions written formally as Eq.~\eqref{eq:k_exponential}. Using the definition of the distribution exponential in Eq.~\eqref{eq:dist_exp}, one can calculate $\mathcal{K}_j(s,s';\xi)$ to any order in $\xi$. However, such expansions only converge well for small $\alpha\xi$, and it is difficult to calculate at high orders. 
	
	The result for the distributions $\mathcal{K}_j(s,s';\xi)$ acting on an arbitrary function of spectral variables might be difficult to calculate. However, for massive fermion propagators, because their singularities do not occur before the mass threshold, Taylor expansions of such functions about $p^2=0$ always have finite radii of convergence. Therefore for the purpose of finding the gauge covariance condition for the fermion propagator, once we know how distributions $\mathcal{K}_j$ act on $z^\beta$, sufficiently with any $\beta\in \mathbf{Z}$, we know the distribution completely.
	
	To start, let us consider the following identity:
	\begin{equation}
	\dfrac{1}{p^2-s+i\varepsilon}=-\dfrac{z}{p^2}~_2F_1(1,b;b;z)\, ,\label{eq:free_particle_prop_2F1}
	\end{equation} 
	where recall $z \equiv p^2/s$.
	Because exponent-preserving operations on $z$ do not have any net effect on $p^2$, we are allowed to multiply by $p^2$ on both sides of Eq.~\eqref{eq:Phi_Xi}, which then becomes
	\begin{equation}
	\int ds'\,\Phi\dfrac{z}{z-1+i\varepsilon}=\dfrac{p^2\,\Xi(p^2,s)}{p^2-s+i\varepsilon}.
	\end{equation}
	We define the dimensionless operator $\phi$ such that at the operator level $\int ds'\Phi=\phi$. Then
	\begin{equation}
	\phi\dfrac{z}{z-1+i\varepsilon}=\dfrac{p^2\,\Xi}{p^2-s+\varepsilon}.\label{eq:phiXi}
	\end{equation}
	Next, substituting Eq.~\eqref{eq:free_particle_prop_2F1} into Eq.~\eqref{eq:phiXi} and combining the result with Eqs.~(\ref{eq:Xi_1_reduced},~\ref{eq:Xi_2_reduced}) and Eq.~\eqref{eq:linear_trans_LKFT_epsilon} gives
	\begin{equation}
	-\phi_n z~_2F_1(1,n;n;z)=\Gamma(\epsilon)\left(\dfrac{4\pi\mu^2}{p^2}\right)^\epsilon \dfrac{-\Gamma(2-\epsilon)}{(1-\epsilon)\Gamma(1+\epsilon)}z^{2\epsilon+2-n}D^\epsilon z^{n-1} D^\epsilon z^{\epsilon-1}z~_2F_1(1,n;n;z),
	\end{equation}
	from which we have
	\begin{equation}
	\phi_n=\Gamma(\epsilon)\left(\dfrac{4\pi\mu^2}{p^2}\right)^\epsilon \dfrac{\Gamma(1-\epsilon)}{\Gamma(1+\epsilon)}z^{2\epsilon+2-n}D^\epsilon z^{n-1}D^\epsilon z^{\epsilon-1}.\label{eq:def_phi}
	\end{equation}
	The distributions $\phi_n$ in Eq.~\eqref{eq:def_phi} correspond to $\Phi_j$ with $j=1,2$ when $n=3,2$ respectively.
	
	With the explicit form of $\Phi_j$ known as Eq.~\eqref{eq:def_phi}, we can proceed to calculate their exponentials. For convenience, define the operational part of $\phi_n$ as 
	\begin{equation}
	\overline{\phi}_n\equiv z^{2\epsilon+2-n}D^\epsilon z^{n-1}D^\epsilon z^{\epsilon-1}
	\end{equation}
	The action of $\overline{\phi}_n$ on $z^\beta$ can be calculated directly;
	\begin{align}
	\quad \overline{\phi}_nz^\beta &= z^{2\epsilon+2-n}D^\epsilon z^{n-1}D^\epsilon z^{\beta+\epsilon-1}=(\beta)_\epsilon z^{2\epsilon+2-n}D^\epsilon z^{n+\beta-2}\nonumber\\
	&=(\beta)_\epsilon (n+\beta-\epsilon-1)_\epsilon z^{\beta+\epsilon} =\dfrac{\Gamma(n+\beta-1)\Gamma(\beta+\epsilon)}{\Gamma(n+\beta-\epsilon-1)\Gamma(\beta)}z^\epsilon z^\beta.\label{eq:phibar_beta}
	\end{align}
	
	For the purpose of finding out how distributions $\mathcal{K}_j$ act on $z^\beta$, we need an explicit expression for $\overline{\phi}^m_nz^\beta$, which can be obtained by applying Eq.~\eqref{eq:phibar_beta} recursively,
	\begin{align}
	\overline{\phi}^m z^\beta &=\dfrac{\Gamma(n+\beta-1)\Gamma(\beta+\epsilon)}{\Gamma(n+\beta-\epsilon-1)\Gamma(\beta)}\phi^{m-1}z^{\beta+\epsilon}\nonumber\\
	& =\dfrac{\Gamma(n+\beta-1)\Gamma(\beta+\epsilon)}{\Gamma(n+\beta-\epsilon-1)\Gamma(\beta)}\dfrac{\Gamma(n+\beta+\epsilon-1)\Gamma(\beta+2\epsilon)}{\Gamma(n+\beta-1)\Gamma(\beta+\epsilon)}\dots\nonumber\\
	&\quad\quad\times \dfrac{\Gamma{(n+\beta+(m-1)\epsilon-1)}\Gamma(\beta+m\epsilon)}{\Gamma(n+\beta+(m-2)\epsilon-1)\Gamma(\beta+(m-1)\epsilon)}z^{\beta+m\epsilon}\nonumber\\
	& =z^{\beta+m\epsilon}\prod_{k=1}^{m}\dfrac{\Gamma(n+\beta+(k-1)\epsilon-1)\Gamma(\beta+k\epsilon)}{\Gamma(n+\beta+(k-2)\epsilon-1)\Gamma(\beta+(k-1)\epsilon)}\nonumber\\
	& =\dfrac{\Gamma(n+\beta+(m-1)\epsilon-1)\Gamma(\beta+m\epsilon)}{\Gamma(n+\beta-\epsilon-1)\Gamma(\beta)}z^{\beta+m\epsilon}.
	\end{align}	
	Alternatively, the calculation is more transparent by substituting $u=z^\epsilon, \quad \lambda=\beta/\epsilon$.
	\begin{align}
	\overline{\phi}^m_nu^{\lambda}
	& =u^{\lambda+m}\prod_{k=1}^{m}\dfrac{\Gamma(n+(\lambda+k-1)\epsilon)\Gamma((\lambda+k)\epsilon)}{\Gamma(n+(\lambda+k-2)\epsilon-1)\Gamma((\lambda+k-1)\epsilon)}\nonumber\\
	& =\dfrac{\Gamma(n+(\lambda+m-1)\epsilon-1)\Gamma((\lambda+m)\epsilon)}{\Gamma(n+(\lambda-1)\epsilon-1)\Gamma(\lambda\epsilon)}u^{\lambda+m}.
	\end{align}	
	After defining 
	\begin{equation}
	\overline{\alpha}\equiv \dfrac{\alpha\xi}{4\pi}\dfrac{\Gamma(\epsilon)\Gamma(1-\epsilon)}{\Gamma(1+\epsilon)}\left(\dfrac{4\pi\mu^2}{p^2}\right)^\epsilon,\label{eq:def_alpha}
	\end{equation}
	we obtain
	\begin{align}
	\mathcal{K}_j z^\beta&=\exp\left(-\dfrac{\alpha\xi}{4\pi}\phi_n\right)z^\beta
	=\exp\left(-\overline{\alpha}\overline{\phi}_n \right)z^\beta=\sum_{m=0}^{+\infty}\dfrac{(-\overline{\alpha})^m}{m!}\overline{\phi}^m_nz^\beta\nonumber\\
	& =\sum_{m=0}^{+\infty}\dfrac{(-\overline{\alpha})^m}{m!}\dfrac{\Gamma(n+\beta+(m-1)\epsilon-1)\Gamma(\beta+m\epsilon)}{\Gamma(n+\beta-\epsilon-1)\Gamma(\beta)}z^{\beta+m\epsilon},\label{eq:kn_zbeta}
	\end{align}	
	with $n=3,~2$ for $j=1,~2$ respectively. Eq.~\eqref{eq:kn_zbeta} specifies how $\mathcal{K}_n$ transforms one function of the spectral variable $s$ into another.  Notice $\overline{\alpha}$ always combines with $z^\epsilon$ to produce a factor of $(\mu^2/s)^\epsilon$, rendering $\mathcal{K}_j$ exponent-preserving. Therefore the action of $\mathcal{K}_j$ on any function can now be calculated as long as this function can be written as a linear combination of $s^{-\beta}$. This can be best understood through Mellin transforms. Effectively Eq.~\eqref{eq:kn_zbeta} tells us what the Mellin transform of $\mathcal{K}_j$ is. Since a Mellin transform disentangles multiplicative convolutions, the action of $\mathcal{K}_j$ on any function of spectral variable $s$ can be reconstructed though the inverse Mellin transform.
	
	Combining the spectral representation for the fermion propagator, Eq.~\eqref{eq:SF_Dirac_struct}, with the LKFT as a linear transform on spectral functions Eq.~\eqref{eq:LKFT_linearity_spectral_rep} produces
	\begin{equation}
	S_j(p^2;\xi)=\int ds\int ds'\,\dfrac{1}{p^2-s+i\varepsilon}\,\mathcal{K}_{j}(s,s';\xi)\,\rho_j(s';0).\label{eq:LKFT_linearity_momentum}
	\end{equation}
	Because the group multiplication of $\mathbf{K}$ is associative, it does not matter which spectral integral in Eq.~\eqref{eq:LKFT_linearity_momentum} is evaluated first. Looking only at the $s'$ integral, once the spectral functions of a fermion propagator at one covariant gauge is known, their counterparts at other covariant gauge can be calculated, which explains the meaning of Eq.~\eqref{eq:LKFT_linearity_spectral_rep}. 
	
	Alternatively when considering $\mathcal{K}_j$ acting on the free-particle propagator $1/(p^2-s+i\varepsilon)$, it transforms the free propagator into a function of both $p^2$ and $\xi$. Directly applying Eq.~\eqref{eq:kn_zbeta} gives
	\begin{equation}
	\mathcal{K}_j\dfrac{1}{p^2-s+i\varepsilon} =-\dfrac{1}{p^2}\sum_{\beta=1}^{+\infty}\sum_{m=0}^{+\infty}\dfrac{(-\overline{\alpha})^m}{m!}\dfrac{\Gamma(n+\beta+(m-1)\epsilon-1)}{\Gamma(n+\beta-\epsilon-1)}\dfrac{\Gamma(\beta+m\epsilon)}{\Gamma(\beta)}z^{\beta+m\epsilon}.\label{eq:kn_free_prop}
	\end{equation}
	Substituting Eq.~\eqref{eq:kn_free_prop} into Eq.~\eqref{eq:LKFT_linearity_momentum} then produces
	\begin{equation}
     S_j(p^2;\xi)=-\int ds \dfrac{1}{p^2}\sum_{\beta=1}^{+\infty}\sum_{m=0}^{+\infty}\dfrac{(-\overline{\alpha})^m}{m!}\dfrac{\Gamma(n+\beta+(m-1)\epsilon-1)}{\Gamma(n+\beta-\epsilon-1)}\dfrac{\Gamma(\beta+m\epsilon)}{\Gamma(\beta)} z^{\beta+m\epsilon}\rho_j(s;0),\label{eq:kn_SF}
	\end{equation}
	where as always $z=p^2/s$. Because for a given $\epsilon$, the imaginary part of Eq.~\eqref{eq:kn_SF} can be calculated, the result reveals to what linear operations $\mathcal{K}_j$ correspond.
	
	Eq.~\eqref{eq:kn_SF} is our solution to dependence on the covariant gauge-fixing parameter $\xi$ of the momentum space fermion propagator. For a specific number of dimensions, the function defined by Eq.~\eqref{eq:kn_free_prop} as a double series could potentially be simplified. Special cases of 3D and 4D will be discussed in the following two sections.
	\section{LKFT for the fermion propagator in 3D\label{ss:LKFT_fermion_3D}}
	When $d=3$, $\epsilon=1/2$, the effective one-loop integral in Eq.~\eqref{eq:Dxi_SF} is finite. Without the ambiguity caused by infinite renormalization, LKFT in 3D can be solved directly, serving as an example to test claims about the general properties of LKFT in Sections \ref{ss:LKFT_group} and \ref{ss:LKFT_spectral_rep}.
	
	Starting with Eq.~\eqref{eq:Dxi_SF}, after evaluating the effective loop-integral using the Feynman parameterization method, we obtain
	\begin{align}
	& \dfrac{\partial}{\partial\xi}\int ds\dfrac{\rho_1(s;\xi)}{p^2-s+i\epsilon} =\alpha\mu\int ds\bigg\{\dfrac{\sqrt{s}}{(p^2-s)^2} -\dfrac{\sqrt{s}}{2p^2(p^2-s)}-\dfrac{1}{2(p^2)^{3/2}}\mathrm{arctanh}(\sqrt{p^2/s}) \bigg\}\rho_1(s;\xi)\label{eq:LKFT_fermion_3D_1}\\[2mm]
	& \dfrac{\partial}{\partial\xi}\int ds\dfrac{\rho_2(s;\xi)}{p^2-s+i\epsilon}=\alpha\mu\int ds\dfrac{\sqrt{s}}{(p^2-s)^2}\rho_2(s;\xi).\label{eq:LKFT_fermion_3D_2}
	\end{align}
	Since Eq.~\eqref{eq:LKFT_fermion_3D_2} appears much simpler than Eq.~\eqref{eq:LKFT_fermion_3D_1}, let us consider its solution first.
	
	Utilizing Eq.~\eqref{eq:LKFT_linearity_spectral_rep}, the dependence of $\rho_2(s;\xi)$ on the covariant gauge parameter $\xi$ can be written as ${\rho_2(s;\xi)=\int ds' \mathcal{K}_2(s,s';\xi)\rho_2(s';0)}$. Since to generate $(p^2-s)^{-2}$ from $(p^2-s)^{-1}$ linearly involves a first order derivative, the distribution $\mathcal{K}_2$ should be given by 
	\begin{equation}
	\mathcal{K}_2(s,s';\xi)=\delta\left(s-\left(\sqrt{s'}+\alpha\mu\xi/2\right)^2\right),\label{eq:test_dist_2_3D}
	\end{equation}
	which corresponds to operations that shift and rescale the spectral function $\rho_2$. It is straightforward to show that the distribution $\mathcal{K}_2(s,s';\xi)$ satisfies its differential equation required by Eq.~\eqref{eq:LKFT_fermion_3D_2}. Meanwhile, it reduces to a simple $\delta$-function when $\xi=0$. Therefore Eq.~\eqref{eq:test_dist_2_3D} indeed specifies how $\rho_2$ changes from one covariant gauge to another. Additionally, $\mathcal{K}_2$ given by Eq.~\eqref{eq:test_dist_2_3D} satisfies group properties trivially. 
	
	While Eq.~\eqref{eq:LKFT_fermion_3D_1} is more complicated than Eq.~\eqref{eq:LKFT_fermion_3D_2}, Bashir and Raya~\cite{Bashir:2002sp} have solved the LKFT in coordinate space assuming that in the Landau gauge propagator is  free. They deduced using Fourier transforms, that under this assumption the Dirac vector part of fermion propagator in any covariant gauge is given by Eq.~(13) of Ref.~\cite{Bashir:2002sp} 
	\begin{align}
	&\quad S_1(p^2;\xi) =B(p_E;\xi)/p_E^2\nonumber\\
	& =\dfrac{-1}{p_E^2+(m+\alpha\mu\xi/2)^2}-\dfrac{\alpha\mu\xi}{2}\dfrac{m+\alpha\mu\xi/2}{p_E^2[p_E^2+(m+\alpha\mu\xi/2)^2]} +\dfrac{\alpha\mu\xi}{2p_E^3}\arctan\left(\dfrac{p_E}{m+\alpha\mu\xi/2} \right)\nonumber\\
	& =\dfrac{1}{(m+\alpha\mu\xi/2)^2}\dfrac{1}{x-1}-\dfrac{\alpha\mu\xi/2}{(m+\alpha\mu\xi/2)^3}\left(\dfrac{1}{x-1}-\dfrac{1}{x} \right) -\dfrac{\alpha\mu\xi/2}{(m+\alpha\mu\xi/2)^3}\dfrac{1}{x\sqrt{-x}}\arctan(\sqrt{-x})\nonumber\\
	& =\dfrac{m}{\left(m+\alpha\mu\xi/2\right)^3}\dfrac{1}{x-1}-\dfrac{\alpha\mu\xi/2}{\left(m+\alpha\mu\xi/2\right)^3}\dfrac{1}{x}\left[\dfrac{1}{\sqrt{x}}\mathrm{arctanh}(\sqrt{x})-1 \right],\label{eq:S1_3D_Bashir}
	\end{align}
	where ${p_E=\sqrt{-p^2}}$, ${x=p^2/(m+\alpha\mu\xi/2)^2}$, and ${\mathrm{arctanh}(u)=(1/2)\ln[(1+u)/(1-u)]}$. Meanwhile, since 
	\begin{align}
	& \dfrac{\mathrm{arctan}(\sqrt{-x})}{\sqrt{-x}}=\dfrac{\mathrm{arctanh}(\sqrt{x})}{\sqrt{x}},\\
	& -\dfrac{1}{\pi}\mathrm{Im}\Big\{\dfrac{1}{x-1+i\epsilon}\Big\}=\delta(x-1),
	\end{align}
	and
	\begin{equation} 
	-\dfrac{1}{\pi}\mathrm{Im}\Bigg\{\dfrac{1}{x+i\epsilon}\left[\dfrac{\mathrm{arctanh}(\sqrt{x+i\epsilon})}{\sqrt{x+i\epsilon}}-1 \right] \Bigg\}=-\dfrac{\theta(x-1)}{2x^{3/2}},
	\end{equation}
	we can take a shortcut of solving Eq.~\eqref{eq:LKFT_fermion_3D_1} by finding out the spectral function of Eq.~\eqref{eq:S1_3D_Bashir} as
	\begin{equation}
	-\dfrac{1}{\pi}\mathrm{Im}\{S_1 \} =\dfrac{m}{\left(m+\alpha\mu\xi/2\right)^3}\,\delta(x-1) +\dfrac{\alpha\mu\xi}{\left(m+\alpha\mu\xi/2 \right)^3}\,\dfrac{\theta(x-1)}{2x^{3/2}},\label{eq:rho1_3D_Bashir}
	\end{equation}
	with $x=s/(m+\alpha\mu\xi/2)^2$. The $\delta$-function term in Eq.~\eqref{eq:rho1_3D_Bashir} corresponds to the free-particle term in Eq.~\eqref{eq:S1_3D_Bashir}. While the $\theta$-function term in Eq.~\eqref{eq:rho1_3D_Bashir} comes from the inverse hyperbolic tangent function that generates a branch cut. 
	
	Since Eq.~\eqref{eq:rho1_3D_Bashir} reduces to a $\delta$-function when $\xi=0$, it also satisfies the differential equation Eq.~\eqref{eq:LKFT_fermion_3D_1}. Because its Fourier transform satisfies the coordinate equivalent of Eq.~\eqref{eq:LKFT_fermion_3D_1}. While Eq.~\eqref{eq:LKFT_fermion_3D_1} specifies exactly what conditions the distribution $\mathcal{K}_1(s,s';\xi)$ has to meet, $\mathcal{K}_1$ is given by Eq.~\eqref{eq:rho1_3D_Bashir} with the modification that $m\rightarrow \sqrt{s'}$. Explicitly,
	\begin{equation}
	\mathcal{K}_1(s,s';\xi)=\dfrac{\sqrt{s'}}{\sqrt{s'}+\dfrac{\alpha\mu\xi}{2}}\,\delta\left(s-\left(\sqrt{s'}+\dfrac{\alpha\mu\xi}{2}\right)^2\right) +\dfrac{\alpha\mu\xi}{4s^{3/2}}\,\theta\left(s-\left(\sqrt{s'}+\dfrac{\alpha\mu\xi}{2}\right)^2\right).\label{eq:test_dist_1_3D}
	\end{equation}
	We therefore obtain the LKFT for $\rho_1$ in 3D with ${\rho_1(s;\xi)=\int ds'\mathcal{K}_1(s,s';\xi)\rho_1(s';0)}$. The direct proof that Eq.~\eqref{eq:test_dist_1_3D} is the distribution we are seeking is lengthy. The detailed calculation is given in Appendix \ref{ss:k1_3D_diff}.
	
	Compared with $\mathcal{K}_2$ given by Eq.~\eqref{eq:test_dist_2_3D}, linear operations given by Eq.~\eqref{eq:test_dist_1_3D} are more convoluted. The $\delta$-function term in Eq.~\eqref{eq:test_dist_1_3D} corresponds to shift and rescale operations on $\rho_1$. The operation brought by the $\theta$-function term corresponds to a convolution with the spectral function $\rho_1$. It is trivial to show that $\mathcal{K}_2$ given by Eq.~\eqref{eq:test_dist_2_3D} meets the group properties listed in Section \ref{ss:LKFT_group}. While for $\mathcal{K}_1$ given by Eq.~\eqref{eq:test_dist_1_3D}, the associativity property is obvious. The identity element is found exactly when $\xi=0$. The closure property is proved in detail in Appendix \ref{ss:k1_3D_closure}. Once the closure property is satisfied, the inverse element of $\mathcal{K}_1(s,s';\xi)$ is just $\mathcal{K}_2(s,s';-\xi)$. Therefore we have shown that the sets of functions by $\mathcal{K}_j$ with multiplication defined for distributions  are indeed continuous groups with the gauge parameter $\xi$ working as the group parameter. Up until now, the LKFT for the fermion propagator in 3D has been obtained without using the general solution in the form of Eq.~\eqref{eq:k_exponential}. 
	
	Since in the special scenario when $n=2$ and $\epsilon=1/2$, Eq.~\eqref{eq:phibar_beta} simplifies to
	\begin{equation}
	\lim\limits_{\epsilon\rightarrow 1/2}\overline{\phi_2}z^\beta=\beta z^{\beta+1/2}=z^{3/2}\dfrac{d}{dz}z^\beta,
	\end{equation}
	where we have written the action of $\phi_2$ on $z^\beta$ as an operator independent of $\beta$. Then  $\phi_2$ given by Eq.~\eqref{eq:def_phi} reduces to
	\begin{equation}
	\lim\limits_{\epsilon\rightarrow 1/2}\phi_2=\dfrac{4\pi\mu}{\sqrt{p^2}}z^{3/2}\dfrac{d}{dz}=-2\pi\mu\dfrac{d}{ds^{1/2}},
	\end{equation}
	which, when combined with Eq.~\eqref{eq:k_exponential}, produces
	\begin{equation}
	\lim\limits_{\epsilon\rightarrow 1/2}\mathcal{K}_2=\exp\left(\dfrac{\alpha\xi\mu}{2}\dfrac{d}{ds^{1/2}}\right).\label{eq:k2_3D_exp}
	\end{equation}
	After identifying Eq.~\eqref{eq:k2_3D_exp} as the shifting operator for functions of $\sqrt{s}$ by $\alpha\xi\mu/2$, the result agrees with Eq.~\eqref{eq:test_dist_2_3D}. In principle, a similar calculation can be carried out for $\mathcal{K}_1$ in 3D as well. However, in practice, multiple operations are required to obtain the corresponding $\phi_1$, the calculation of whose exponential is nontrivial. 
	
	Alternatively, to verify that Eq.~\eqref{eq:k_exponential} with $\Phi_j$ given by Eq.~\eqref{eq:def_phi} solves LKFT in 3D, we only need to show that the imaginary part of Eq.~\eqref{eq:kn_SF} corresponds to distributions $\mathcal{K}_j$ in Eq.~\eqref{eq:test_dist_1_3D} and Eq.~\eqref{eq:test_dist_2_3D} in the limit $\epsilon\rightarrow 1/2$. 
	
	In the case $n=2$ and $\epsilon=1/2$, which leads to $\overline{\alpha}=\alpha\xi\mu/\sqrt{p^2}$, because of the duplication formula Eq.~(6.1.18) in Ref.~\cite{abramowitz1964handbook}, 
	\[\Gamma(2z)=(2\pi)^{-1/2}2^{2z-1/2}\Gamma(z)\Gamma(z+1/2),\] 
	Gamma functions in Eq.~\eqref{eq:def_phi} simplify to
	\begin{equation}
	\dfrac{\Gamma(n+\beta+(m-1)\epsilon-1)\Gamma(\beta+m\epsilon)}{m!\Gamma(n+\beta-\epsilon-1)\Gamma(\beta)}=\dfrac{\Gamma(2\beta+m)}{2^m\Gamma(2\beta)m!}.
	\end{equation}
	Next,
	\begin{equation}
	\sum_{m=0}^{+\infty}\dfrac{\Gamma(2\beta+m)}{2^m\Gamma(2\beta)m!}(-\overline{\alpha}\sqrt{z})^m=\left(1+\dfrac{\overline{\alpha}}{2}\sqrt{z} \right)^{-2\beta},
	\end{equation}
	and so we have
	\begin{equation}
	\mathcal{K}_2\dfrac{1}{p^2-s+i\varepsilon} =-\dfrac{1}{p^2}\sum_{\beta=1}^{+\infty}\left(1+\dfrac{\overline{\alpha}}{2}\sqrt{z} \right)^{-2\beta}z^\beta =\dfrac{-1}{p^2}\dfrac{z}{(1+\overline{\alpha}\sqrt{z})^2-z}=\dfrac{1}{p^2-(\sqrt{s}+\alpha\mu\xi/2)^2}.
	\end{equation}
	Since when operating on the free-particle propagator produces identical results, $\mathcal{K}_2$ given by Eq.~\eqref{eq:k_exponential} with $\Phi_2$ given by Eq.~\eqref{eq:def_phi} agrees with $\mathcal{K}_2$ given by Eq.~\eqref{eq:test_dist_2_3D}.
	
	Similarly for $\mathcal{K}_1$, when $n=3$ and $\epsilon=1/2$, the Gamma functions in Eq.~\eqref{eq:def_phi} simplify by firstly noting
	\begin{align}
	& \quad \dfrac{\Gamma(n+\beta+(m-1)\epsilon-1)\Gamma(\beta+m\epsilon)}{m!\Gamma(n+\beta-\epsilon-1)\Gamma(\beta)}\nonumber\\
	&=\dfrac{\Gamma(\beta+m/2+3/2)\Gamma(\beta+m/2)}{\Gamma(\beta+3/2)\Gamma(\beta)m!}\nonumber\\
	& =\dfrac{\beta+m/2+1/2}{\beta+1/2}\dfrac{\Gamma(\beta+m/2+1/2)\Gamma(\beta+m/2)}{\Gamma(\beta+1/2)\Gamma(\beta)m!}\nonumber\\
	&=\left(1+\dfrac{m}{2\beta+1} \right)\dfrac{\Gamma(2\beta+m)}{2^m\Gamma(2\beta)m!}.
	\end{align}
	In addition since
	\begin{equation} 
	\sum_{m=0}^{+\infty}\dfrac{m}{2\beta+1}\dfrac{\Gamma(2\beta+m)}{2^m\Gamma(2\beta)m!}(-\overline{\alpha}\sqrt{z})^m=\dfrac{-\beta}{1+2\beta}\overline{\alpha}\sqrt{z}\left(1+\dfrac{\overline{\alpha}}{2}\sqrt{z} \right)^{-1-2\beta}\nonumber
	\end{equation}
	and
	\begin{equation} \sum_{\beta=1}^{+\infty}\dfrac{-\beta}{1+2\beta}\overline{\alpha}\sqrt{z}\left(1+\dfrac{\overline{\alpha}}{2}\sqrt{z} \right)^{-1-2\beta}z^\beta=\dfrac{\overline{\alpha}}{2}\left[\dfrac{-\sqrt{z}\left( 1+\dfrac{\overline{\alpha}}{2}\sqrt{z}\right)}{\left(1+\dfrac{\overline{\alpha}}{2}\sqrt{z}\right)^2-z}+\mathrm{arctanh}\left(\dfrac{\sqrt{z}}{1+\overline{\alpha}\sqrt{z}/2} \right) \right],
	\end{equation}
	we have
	\begin{align}
	& \quad \mathcal{K}_1\dfrac{1}{p^2-s+i\varepsilon}=-\dfrac{1}{p^2}\sum_{\beta=1}^{+\infty}\Bigg\{\left(1+\dfrac{\overline{\alpha}}{2}\sqrt{z} \right)^{-2\beta}- \dfrac{\beta}{1+2\beta}\overline{\alpha}\sqrt{z}\left(1+\dfrac{\overline{\alpha}}{2}\sqrt{z} \right)^{-1-2\beta}\Bigg\}z^\beta\nonumber\\
	& =\dfrac{1}{p^2-(\sqrt{s}+\alpha\mu\xi/2)^2}-\dfrac{\alpha\xi\mu}{2p^2}\Bigg\{\dfrac{\sqrt{s}+\alpha\xi\mu/2}{p^2-(\sqrt{s}+\alpha\xi\mu/2)^2} +\dfrac{1}{\sqrt{p^2}}\mathrm{arctanh}\left(\dfrac{\sqrt{p^2}}{\sqrt{s}+\alpha\xi\mu/2} \right) \Bigg\},\label{eq:K_1_free_prop_4D}
	\end{align}
	which agrees with Eq.~\eqref{eq:S1_3D_Bashir}. Consequently it also agrees with Eq.~\eqref{eq:test_dist_1_3D}, as seen simply by taking the imaginary part of Eq.~\eqref{eq:S1_3D_Bashir}. 
	
	Through this analysis of the LKFT for the fermion propagator in 3D, we have established that solutions of $\mathcal{K}_j(s,s';\xi)$ directly from their differential equations satisfy group properties postulated in Section \ref{ss:LKFT_group} and agree with the general solution obtained in Section \ref{ss:LKFT_spectral_rep}. In Bashir and Raya \cite{Bashir:2002sp}, LKFT for the fermion propagator is solved through Fourier transforms to and from coordinate space assuming the propagator is the free-particle one in the Landau gauge. Therefore Eqs.~(16, 17) of Ref.~\cite{Bashir:2002sp} only apply under this assumption. However, through the spectral representation, any propagator function can be represented as a linear combination of free-particle propagators with different mass. Since the LKFT is also linear, results in Ref.~\cite{Bashir:2002sp} can be generalized to accommodate any initial conditions having spectral representations themselves. Consequently, Eq.~\eqref{eq:K_1_free_prop_4D} holds regardless of the assumed behavior of the propagator in the initial gauge. In our reduction of the exact solution to LKFT in any dimensions by Eq.~\eqref{eq:k_exponential} and Eq.~\eqref{eq:def_phi} to the special case of 3D, gauge covariance of the fermion propagator is solved directly in Minkowski momentum space through the language of spectral representation, and therefore is independent of the initial conditions specified in any one gauge.
	\section{LKFT for the fermion propagator in 4D\label{ss:LKFT_fermion_4D}}
	The group properties of LKFT for the fermion propagator spectral functions are maintained by Eq.~\eqref{eq:k_exponential}, for any positive $\epsilon$. However when $d\rightarrow 4$, only the leading expansions in $\epsilon$ are required. Therefore one expects distributions $\mathcal{K}_j$ to become simpler than Eq.~\eqref{eq:kn_free_prop} in this particular limit, as they did in 3D. However, to obtain the correct expansions, knowledge of the divergent part for the fermion propagator is required. By analyzing divergences alone, the LKFT specifies that the fermion propagator wavefunction renormalization \cite{Johnson:1959zz,Sonoda:2000kn} is $Z_2(\xi)=Z_2(0)\exp\left[-\alpha\xi/(4\pi\epsilon)\right]$. The same result can also be obtained by considering only the divergent parts in Eq.~\eqref{eq:Dxi_SF}. Additionally, based on Subsection \ref{ss:div_F} there is no ultraviolet divergence for the fermion self-energy in the Landau gauge, so we can take $Z_2(0)=1$. After dimensional regularization, any term at $\mathcal{O}(\epsilon^1)$ is regarded as higher order in the exponential. When Eq.~\eqref{eq:kn_free_prop} is convergent, the LKFT for the fermion propagator in 4D is found once the proper limit of $\epsilon\rightarrow 0$ is taken.
	
	To proceed to evaluating Eq.~\eqref{eq:kn_free_prop} with small $\epsilon$, consider the original definition of the Gamma function 
	\begin{equation}
	\Gamma(s)=\int_{0}^{+\infty}dx~x^{s-1}e^{-x}.\label{eq:def_Gamma}
	\end{equation}
	After reparameterizing Gamma functions in the numerator of the double series expansion of Eq.~\eqref{eq:kn_zbeta}, we obtain
	\begin{align}
	& \quad \sum_{m=0}^{+\infty}\dfrac{\Gamma(n+\beta+(m-1)\epsilon-1)\Gamma(\beta+m\epsilon)}{\Gamma(n+\beta-\epsilon-1)\Gamma(\beta)}\dfrac{(-\overline{\alpha}z^\epsilon)^m}{m!}\nonumber\\
	& =\sum_{m=0}^{+\infty}\int_{0}^{+\infty}dx\int_{0}^{+\infty}dy\dfrac{e^{-x-y}}{\Gamma(n+\beta-\epsilon-1)\Gamma(\beta)} x^{n+\beta-\epsilon-2} y^{\beta-1}\dfrac{[-(xyz)^\epsilon\overline{\alpha}]^m}{m!}\nonumber\\
	& =\int_{0}^{+\infty}dx\int_{0}^{+\infty}dy\dfrac{x^{n+\beta-\epsilon-2}y^{\beta-1}}{\Gamma(n+\beta-\epsilon-1)\Gamma(\beta)}\exp\left[-x-y-\overline{\alpha}\left(xyz\right)^\epsilon\right] \label{eq:kn_zbeta_para}
	\end{align}
	Eq.~\eqref{eq:kn_zbeta_para} is an alternative to Eq.~\eqref{eq:kn_zbeta}. 
	For any fermion propagator function $S_j(p^2)$ in 4D, having established that its divergent part is merely $Z_2=e^{-\alpha\xi/(4\pi\epsilon)}$, 
	it must also be the only divergence for $\mathcal{K}_nz^\beta$ for small $\epsilon$. With the renormalization factor in mind, $\mathcal{K}_nz^\beta$ is properly renormalized once its logarithm is truncated to $\mathcal{O}(\epsilon^0)$. Furthermore, integrals over parameters $x$ and $y$ do not modify the $1/\epsilon$ divergences of $\mathcal{K}_nz^\beta$ because the integral definition of the Gamma function by Eq.~\eqref{eq:def_Gamma} extends into the complex plane. Then for each integral element of Eq.~\eqref{eq:kn_zbeta_para},
	\begin{align}
	& \quad \ln\Bigg\{ \dfrac{x^{n+\beta-\epsilon-2}y^{\beta-1}}{\Gamma(n+\beta-\epsilon-1)\Gamma(\beta)}\exp[-x-y-\overline{\alpha}(xyz)^\epsilon]\Bigg\}\nonumber\\
	& =-x-y+(n+\beta-\epsilon-2)\ln x+(\beta-1)\ln y -\dfrac{\alpha\xi}{4\pi}\dfrac{\Gamma(\epsilon)\Gamma(1-\epsilon)}{\Gamma(1+\epsilon)}\left(\dfrac{4\pi\mu^2}{p^2}xyz \right)^\epsilon \nonumber\\
	& \quad-\ln\Gamma(n+\beta-\epsilon-1)-\ln\Gamma(\beta)\nonumber\\
	& =-x-y+(n+\beta-2)\ln x+(\beta-1)\ln y-\ln\Gamma(n+\beta-1) -\ln\Gamma(\beta)\nonumber\\
	& \quad -\dfrac{\alpha\xi}{4\pi}\left[\dfrac{1}{\epsilon}+\gamma_E+\ln\left(\dfrac{4\pi\mu^2}{p^2}xyz \right) \right]+\mathcal{O}(\epsilon^1).
	\end{align}
	After regularization,
	\begin{align}
	&\quad z^{-\beta}\mathcal{K}_j z^\beta \nonumber\\
	&=\exp\left[-\dfrac{\alpha\xi}{4\pi}\left(\dfrac{1}{\epsilon}+\gamma_E+\ln 4\pi+\mathcal{O}(\epsilon^1)\right) \right]\nonumber\\
	&\quad \times \dfrac{(\mu^2z/p^2)^{-\alpha\xi/(4\pi)}}{\Gamma(n+\beta-1)\Gamma(\beta)} \int_{0}^{+\infty}dx\int_{0}^{+\infty}dye^{-x-y}\times x^{n+\beta-2-\alpha\xi/(4\pi)}y^{\beta-1-\alpha\xi/(4\pi)}\nonumber\\
	& =\dfrac{\Gamma\left(n+\beta-1-\dfrac{\alpha\xi}{4\pi}\right)\Gamma\left(\beta-\dfrac{\alpha\xi}{4\pi}\right)}{\Gamma(n+\beta-1)\Gamma(\beta)}\left(\dfrac{\mu^2z}{p^2} \right)^{-\alpha\xi/(4\pi)}\nonumber\\
	&\quad\times \exp\left[-\dfrac{\alpha\xi}{4\pi}\left(\dfrac{1}{\epsilon}+\gamma_E+\ln 4\pi+\mathcal{O}(\epsilon^1) \right) \right].
	\end{align}
	While from the series definition of hypergeometric functions, we have
	\begin{align}
	& \quad \sum_{\beta=1}^{+\infty}\dfrac{\Gamma\left(n+\beta-1-\dfrac{\alpha\xi}{4\pi}\right)\Gamma\left(\beta-\dfrac{\alpha\xi}{4\pi}\right)}{\Gamma(n+\beta-1)\Gamma(\beta)}z^\beta\nonumber\\
	& =\dfrac{z}{\Gamma(n)}\Gamma\left(n -\dfrac{\alpha\xi}{4\pi}\right) \Gamma\left( 1-\dfrac{\alpha\xi}{4\pi}\right) \sum_{\beta=0}^{+\infty}\dfrac{\Gamma\left(n+\beta-\dfrac{\alpha\xi}{4\pi}\right)\Gamma\left(\beta+1 -\dfrac{\alpha\xi}{4\pi}\right)}{\Gamma\left(n -\dfrac{\alpha\xi}{4\pi}\right)\Gamma\left(1-\dfrac{\alpha\xi}{4\pi}\right)}\dfrac{\Gamma(n)}{\Gamma(n+\beta)}\dfrac{z^\beta}{\beta!}\nonumber\\
	& =\dfrac{z}{\Gamma(n)}\Gamma\left(n -\dfrac{\alpha\xi}{4\pi}\right) \Gamma\left( 1-\dfrac{\alpha\xi}{4\pi}\right)~_2F_1\left(1-\dfrac{\alpha\xi}{4\pi},n-\dfrac{\alpha\xi}{4\pi};n;z\right).
	\end{align}
	Therefore for $\mathcal{K}_j$ acting on the free-particle propagator, Eq.~\eqref{eq:kn_free_prop} becomes
	\begin{align}
	&\quad \mathcal{K}_j\dfrac{1}{p^2-s+i\varepsilon} \nonumber\\
	&=\dfrac{-1}{p^2}\left(\dfrac{\mu^2z}{p^2} \right)^{-\alpha\xi/(4\pi)}\exp\left[-\dfrac{\alpha\xi}{4\pi}\left(\dfrac{1}{\epsilon}+\gamma_E+\ln 4\pi+\mathcal{O}(\epsilon^1) \right)\right]\nonumber\\
	&\quad \hspace{5cm}\times  \sum_{\beta=1}^{+\infty}\dfrac{\Gamma\left(n+\beta-\dfrac{\alpha\xi}{4\pi}\right)\Gamma\left(\beta-\dfrac{\alpha\xi}{4\pi}\right)}{\Gamma(n+\beta-1)\Gamma(\beta)}z^\beta\nonumber\\
	& =\dfrac{-z}{p^2\Gamma(n)}\left(\dfrac{\mu^2z}{p^2} \right)^{-\alpha\xi/(4\pi)}\Gamma\left(n -\dfrac{\alpha\xi}{4\pi}\right) \Gamma\left( 1-\dfrac{\alpha\xi}{4\pi}\right)\nonumber\\
	&\quad\hspace{1cm} \times\exp\left[-\dfrac{\alpha\xi}{4\pi}\left(\dfrac{1}{\epsilon}+\gamma_E+\ln 4\pi+\mathcal{O}(\epsilon^1) \right)\right]~_2F_1\left(1-\dfrac{\alpha\xi}{4\pi},n-\dfrac{\alpha\xi}{4\pi};n;z\right).\label{eq:LKFT_rho_SF_4D}
	\end{align}
	This is our general result. In the special case when the propagator is assumed to be free in the Landau gauge, \textit{i.e.} $\rho_j$ are $\delta$-functions, Eq.~\eqref{eq:LKFT_rho_SF_4D} reduces to the results found by Bashir and Raya \cite{Bashir:2002sp} up to differences in renormalization schemes.
	
	To generate Eq.~\eqref{eq:LKFT_rho_SF_4D} from the free-particle propagator, consider a positive change in $\xi$. Naturally adopting the convention  $\nu=\alpha\xi/(4\pi)$ used in Ref.~\cite{Bashir:2002sp}, for instance, we have from Eq.~\eqref{eq:Dc2Fa}
	\begin{equation}
	I^{\nu}z^{-\nu}~_2F_1(1,n;n;z)=\Gamma(1-\nu)~_2F_1(1-\nu,n;n;z),
	\end{equation}
	and
	\begin{equation}  I^{\nu}z^{n-1-\nu}~_2F_1(1-\nu,n;b;z)=\dfrac{\Gamma(n-\nu)}{\Gamma(n)}z^{n-1}~_2F_1(1-\nu,n-\nu;n;z),
	\end{equation}
	since the hypergeometric $_2F_1(a,b;c;z)$ is symmetric in parameters $a$ and $b$, and	where the fractional differential operation in Eq.~\eqref{eq:Dc2Fa} becomes fractional integration for positive $\xi$. Combining these two identities gives
	\begin{equation}
	z^{n-1}I^{\nu}z^{n-1-\nu}I^{\nu}z^{-\nu}~_2F_1(1,n;n;z)=\dfrac{\Gamma(n-\nu)\Gamma(1-\nu)}{\Gamma(n)}~_2F_1(1-\nu,n-\nu;n;z).\label{eq:LKFT_SF_4D_generation}
	\end{equation}
	After representing the free-particle propagator as a hypergeometric function using Eq.~\eqref{eq:free_particle_prop_2F1}, comparing Eq.~\eqref{eq:LKFT_rho_SF_4D} with Eq.~\eqref{eq:LKFT_SF_4D_generation} yields
	\begin{equation}
	\mathcal{K}_j(\xi)=\left(\dfrac{\mu^2z}{p^2} \right)^{-\nu}\exp\bigg\{-\nu\left[\dfrac{1}{\epsilon}+\gamma_E+\ln 4\pi+\mathcal{O}(\epsilon^1) \right] \bigg\} z^{2-n}I^{\nu}z^{n-1-\nu}I^{\nu}z^{-\nu-1},\label{eq:kn_small_epsilon}
	\end{equation}
	again with $n=3,~2$ for $j=1,~2$. While for negative $\xi$, the fractional integration operators $I^\nu$ are replaced by derivative operators $D^{|\nu|}$. Eq.~\eqref{eq:kn_small_epsilon} then becomes 
	\begin{equation}
	\mathcal{K}_j(\xi) =\left(\dfrac{\mu^2z}{p^2} \right)^{|\nu|}\exp\bigg\{|\nu|\left[\dfrac{1}{\epsilon}+\gamma_E+\ln 4\pi+\mathcal{O}(\epsilon^1) \right] \bigg\}z^{2-n}D^{|\nu|}z^{n-1+|\nu|}D^{|\nu|}z^{|\nu|-1}.\label{eq:kn_small_epsilon_inverse}
	\end{equation}	
	One can verify that from Eqs.~(\ref{eq:kn_small_epsilon},~\ref{eq:kn_small_epsilon_inverse}), by acting on $z^\beta$, that ${\mathcal{K}_j(\xi_1)\mathcal{K}_j(\xi_2)=\mathcal{K}_j(\xi_1+\xi_2)}$ and ${\mathcal{K}_j^{-1}(\xi)=\mathcal{K}_j(-\xi)}$. Therefore the simplified form of LKFT for fermion propagator spectral functions in 4D also maintains group properties explicitly.
	\section{Summary of LKFT\label{ss:LKFT_conclusions}}
	Working in covariant gauges we have shown here that the Landau--Khalatnikov--Fradkin transformation (LKFT) defines a group of transformations parametrized by the gauge label $\xi$. These transformations define how a  propagator in one covariant gauge is related to that in any other. These transformations are readily studied if we assume the propagator satisfies a spectral reprsentation.  As an explicit example  we have investigated the fermion propagator in QED, which is expected to have the analytic properties required for such a representation. The LKFT then demands the spectral functions  obey exact  transformation properties to be gauge covariant. These hold in any dimension $d < 4$, naturally involving fractional calculus. In three dimensions when the calculus is of integer order, 
	we show how our results generalize those obtained earlier in a special case by Bashir and Raya \cite{Bashir:2002sp}. As we approach four dimensions, the general results can be expanded in powers of $\epsilon = 2-d/2$.
	The complexity of fractional calculus then becomes apparent. The solutions inevitably involve distributions with fractional orders of delta-functions and theta-functions. Nevertheless, considering arbitrary (non-integer) dimensions provides insights into how gauge covariance connects the properties of field theory Green's functions in different dimensions.	
	\chapter{The gauge covariance requirements on SDEs for the QED propagators\label{cp:GC_SDE_QED}}
\begin{figure}
	\centering
	\includegraphics[width=0.6\linewidth]{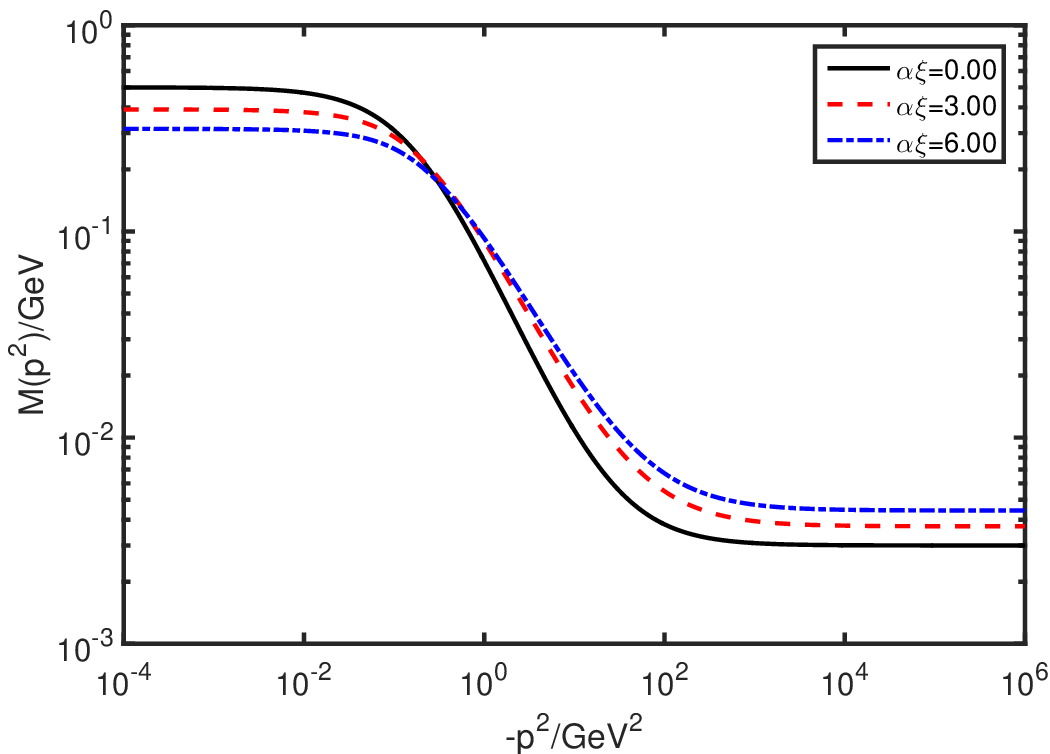}
	\caption{The dependence of the fermion propagator mass function $M(p^2)$ on $\xi$. The black solid line is the parametric form of $M(p^2)$ given by Eq.~(2.1) of Ref.~\cite{Pennington:2009vz} with $M_0=3~\mathrm{MeV}$, $c=1.239$ and $\Lambda_{\mathrm{QCD}}=401~\mathrm{MeV}$. The red dashed line and the blue dash-dot line correspond to what the mass function should be when $\alpha\xi=3$ and $\alpha\xi=6$ respectively. The red dashed line and the blue dash-dot line are obtained by the LKFT for the fermion propagator in 4D within the $\overline{\mathrm{MS}}$ renormalization scheme at the scale $\mu=\Lambda$, with $F(p^2)=1$ and $M(p^2)$ given by the black line as the initial conditions in the Landau gauge. Notice that gauge covariance produces a {\it node}-like feature, in this case at $p^2 \simeq -0.3$~GeV$^2$. Though this example is motivated by QCD in the choice of parameters, the calculations are from QED. }
	\label{fig:Mp2_xi_036}
\end{figure}

	Simple truncations schemes are known to violate gauge covariance. As an illustration let us consider a purely bare fermion-boson vertex in QED. This produces a mass function in the Landau gauge shown as the solid line in Fig.~\ref{fig:Mp2_xi_036}. The qualitative behavior of the QED mass function is very like that for a light quark in QCD in the Maris-Tandy model \cite{Maris:1999nt}. Solving the QED Schwinger--Dyson equation for the fermion in 4D with the same bare $\gamma^\mu$ vertex in the Feynman gauge (for instance) changes the mass function as in Fig.~11 of Ref.~\cite{Kizilersu:2013hea}. The corresponding 3D results are illustrated by Fig.~3.2 of Ref.~\cite{Williams:2007zzh}.

However, the gauge covariance of the fermion propagator is exactly specified by the Landau--Khalatnikov--Fradkin transformation discussed in Chapter \ref{cp:LKFT}. 
If one applies this to the 4D fermion mass functions shown in Fig.~2 of Ref.~\cite{Williams:2006vva} for the Landau gauge, one obtains the behavior in two other covariant gauges plotted here in Fig.~\ref{fig:Mp2_xi_036}. One sees that the mass function moves between gauges to produce what appears to be a {\it node}. This in fact ensures that any fermion condensate is gauge independent. The corresponding 3D results are given by Fig.~3.4 of Ref.~\cite{Williams:2007zzh}. However, solving the fermion Schwinger--Dyson equation in different gauges as shown in Fig.~11 of Ref.~\cite{Kizilersu:2013hea} gives mass functions that simply move up or down as the gauge changes. There is no hint of the nodal behavior required by the LKFT and seen in Fig.~\ref{fig:Mp2_xi_036}. This not surprisingly indicates that a vertex as simple as $\gamma^\mu$  cannot be appropriate in both the Landau and Feynman gauges in QED. Indeed, it may not hold in any covariant gauge.

In this chapter, the conditions that ensure the solutions of the Schwinger--Dyson equation (SDE) for the fermion propagator are gauge covariant will be presented \cite{Jia:2016udu}.

\section{SDE for fermion propagator spectral functions}
The SDE for the fermion propagator spectral functions can be most conveniently obtained from Eq.~\eqref{eq:SDE_fermion_momentum}. After decomposing this equation into its two Dirac components, the identity in Fig.~\ref{fig:DSE_fermion_rho} becomes
\begin{subequations}
	\label{eq:SDE_fermion_linear_rho}
	\begin{eqnarray}
	& p^2S_1(p^2)-mS_2(p^2)+\sigma_1(p^2)=1\\
	& S_2(p^2)-mS_1+\sigma_2(p^2)=0,
	\end{eqnarray}
\end{subequations}
where $\sigma_{j}(p^2)$ are the Dirac scalar and vector components of the loop integral. Equation \eqref{eq:SDE_fermion_linear_rho} is the bare version of Eqs.~(\ref{eq:SDE_R_Z2},~\ref{eq:SDE_R_Zm}).
For the second term on the right-hand side of Fig.~\ref{fig:DSE_fermion_rho}, recall the Ward identity states that for QED $Z_1=Z_2$~\cite{PhysRev.78.182}. Therefore the fermion propagator $S_F(p;\xi)$ shares the same renormalization constant with the fermion-photon vertex structure defined as $S_F(k)\Gamma^\mu(k,p)S_F(p)$, which indicates that the latter is also linear in $\rho_{j}(s;\xi)$. 

The gauge covariance of the solutions to the fermion and boson propagator Schwinger--Dyson equations will constrain the allowed forms of the fermion-boson vertex $\Gamma^\mu(k,p)$.
However the vertex in its full complexity with its 11 non-zero components is not required. Only the projections implied by the Schwinger--Dyson equation of
Figs.~\ref{fig:DSE_fermion_ori},~\ref{fig:DSE_fermion_rho}, and the corresponding equation for the inverse photon propagator ({\it i.e.} for the vacuum polarization) Fig.~\ref{fig:SDE_photon} are constrained. Thus it is this
{\it  effective} vertex that is restricted.  

One specific spectral construction of the vertex structure linear in $\rho_{j}(s;\xi)$ and satisfying the longitudinal Ward--Green--Takahashi identity is the Gauge Technique \cite{Delbourgo:1977jc}, which makes the Ansatz
\begin{equation}
S_F(p)\Gamma^\mu(k,p)S_F(p)=\int dW\dfrac{1}{\slashed{k}-W}\gamma^\mu\dfrac{1}{\slashed{p}-W}\,\rho(W),
\end{equation}
where $\rho(W)=\mathrm{sign}(W)[\rho_1(W^2)+W\rho_1(W^2)]$, resulting in Eq.~\eqref{eq:def_sigma_j_SDE}. 
Transverse supplements to the Gauge Technique are required to meet various principles of QED, including renormalizablility \cite{Curtis:1990zs,Kizilersu:2009kg}, gauge covariance \cite{Bashir:2004mu} and transverse Ward--Green--Takahashi identities \cite{He:1999hb,He:2000we,He:2001cu,He:2006my}. However, from the equality $Z_1=Z_2$ \cite{PhysRev.78.182} we can further assume that such modifications are also linear in $\rho_{j}(s;\xi)$, and once known, allow us to calculate the loop integral in Fig.~\ref{fig:DSE_fermion_rho}, resulting in a function of $p$ as a linear functional of $\rho_j(s;\xi)$. Since this one-loop integral reduces to corrections to the fermion propagator in perturbative calculations, such $p^2$ dependences must be linearly generated from the free-particle propagator, as discussed in Subsection \ref{ss:sr_SDE_SF}. Therefore after taking the imaginary part of Fig.~\ref{fig:DSE_fermion_rho}, or equivalently that of Eq.~\eqref{eq:SDE_fermion_linear_rho}, we obtain
\begin{subequations}
	\label{eq:SDE_fermion_rho_s}
	\begin{align}
	s\rho_1(s;\xi)-m_B\rho_2(s;\xi)-\dfrac{1}{\pi}\mathrm{Im}\big\{ \sigma_1(s+i\varepsilon;\xi)\big\}=0,\label{eq:SDE_fermion_rho_1_s}\\
	\rho_2(s;\xi)-m_B\rho_1(s;\xi)-\dfrac{1}{\pi}\mathrm{Im}\big\{\sigma_2(s+i\varepsilon;\xi)\big\}=0.\label{eq:SDE_fermion_rho_2_s}
	\end{align}
\end{subequations}
The real constant term on the left-hand side disappears. After dividing Eq.~\eqref{eq:SDE_fermion_rho_1_s} by $s$, Eq.~\eqref{eq:SDE_fermion_rho_s} can be rewritten as
\begin{equation}
\int ds'
\begin{pmatrix}
\Omega_{11}(s,s';\xi) & \Omega_{12}(s,s';\xi) \\ 
\Omega_{21}(s,s';\xi) & \Omega_{22}(s,s';\xi)
\end{pmatrix} 
\begin{pmatrix}
\rho_1(s';\xi) \\ 
\rho_2(s';\xi)
\end{pmatrix}
+
\begin{pmatrix}
\rho_1(s;\xi) \\ 
\rho_2(s;\xi)
\end{pmatrix}
=
\begin{pmatrix}
0 \\ 
0
\end{pmatrix},
\label{eq:SDE_fermion_rho_itg}
\end{equation}
where the $\Omega_{ij}(s,s';\xi)$ encode all required linear operations on the spectral functions $\rho_{j}(s;\xi)$, which are obtained by functional derivatives similar to
\begin{equation}
\Omega(s,s')=-\dfrac{\delta}{\delta \rho(s')}\dfrac{1}{\pi}\mathrm{Im}\big\{\sigma(s+i\varepsilon)\big\}.\label{eq:delta_delta_rho}
\end{equation}
The bare mass coupling in Eq.~\eqref{eq:SDE_fermion_rho_s} is explicitly included in the off-diagonal terms of $\Omega_{ij}(s,s';\xi)$. When the fermion-photon vertex is given by the Gauge Technique the resulting $\sigma_{j}$ is given by Eq.~\eqref{eq:def_sigma_j_SDE}. Then $m_B$ is the only coupling between equations for $\rho_1$ and $\rho_2$. However, when dimension-odd operators are allowed to enter the expression for $S_F(k)\Gamma^\mu(k,p)S_F(p)$, they will contribute additionally to off-diagonal elements of $\Omega_{ij}$.

For a given Ansatz for the fermion-photon vertex that ensures $S_F(k)\Gamma^\mu(k,p)S_F(p)$ being linear in $\rho_{j}(s;\xi)$, there is a corresponding $\Omega$. It is the matrix $\Omega$ that is constrained by gauge covariance. Regardless of the photon being quenched or not, the SDE for fermion propagator spectral functions takes the form of Eq.~\eqref{eq:SDE_fermion_rho_itg}. Solutions to Eq.~\eqref{eq:SDE_fermion_rho_itg} found in different covariant gauges are, of course, different because the fermion propagator is not a physical observable. However any Ansatz for the fermion-photon vertex that respects Eq.~\eqref{eq:SDE_fermion_rho_itg} is expected to be gauge covariant. Satisfying the Ward--Green--Takahashi identity, a consequence of gauge invariance, however, is not sufficient to ensure the gauge covariance of solutions to Eq.~\eqref{eq:SDE_fermion_rho_itg}, as we will see explicitly later on. 
In order to explore the conditions on the  $\Omega_{ij}(s,s';\xi)$ that ensure gauge covariance of solutions to Eq.~\eqref{eq:SDE_fermion_rho_itg}, results in Chapter \ref{cp:LKFT} on the LKFT for the fermion propagator spectral function will be applied. 
\section{Gauge covariance requirements for the propagator SDEs\label{ss:GC_requirements}}
\subsection{Gauge covariance requirement for the fermion propagator SDE}
For notational convenience, when two distributions are multiplied together, the integration over the spectral variable is implied. This convention agrees with the group multiplication defined in Subsection \ref{ss:rep_LKFT}. After adopting this notation, only dependences on $\xi$ are required to be written explicitly. Therefore Eq.~\eqref{eq:SDE_fermion_rho_itg} becomes
\begin{equation}
\begin{pmatrix}
\rho_1(\xi) \\ 
\rho_2(\xi)
\end{pmatrix} +
\begin{pmatrix}
\Omega_{11}(\xi) & \Omega_{12}(\xi) \\ 
\Omega_{21}(\xi) & \Omega_{22}(\xi)
\end{pmatrix} 
\begin{pmatrix}
\rho_1(\xi) \\ 
\rho_2(\xi)
\end{pmatrix}=
\begin{pmatrix}
0 \\
0
\end{pmatrix}
.\label{eq:DSE_xi_matrix}
\end{equation}
Since LKFT does not couple $\rho_1$ with $\rho_2$, we have the following abbreviated versions of Eq.~\eqref{eq:LKFT_linearity_spectral_rep},
\begin{equation}
\rho_j(\xi)=\mathcal{K}_j(\xi)\rho_j(0).\label{eq:LKFT_linearity_spectral_rep_abb}
\end{equation}
Substituting Eq.~\eqref{eq:LKFT_linearity_spectral_rep_abb} into Eq.~\eqref{eq:DSE_xi_matrix} gives
\begin{equation}
\begin{pmatrix}
\mathcal{K}_{1}(\xi) & \\ 
& \mathcal{K}_{2}(\xi)
\end{pmatrix} 
\begin{pmatrix}
\rho_1(0) \\ 
\rho_2(0)
\end{pmatrix}+
\begin{pmatrix}
\Omega_{11}(\xi) & \Omega_{12}(\xi) \\ 
\Omega_{21}(\xi) & \Omega_{22}(\xi)
\end{pmatrix} 
\begin{pmatrix}
\mathcal{K}_1(\xi) &  \\ 
& \mathcal{K}_2(\xi)
\end{pmatrix} 
\begin{pmatrix}
\rho_1(0) \\ 
\rho_2(0)
\end{pmatrix}=
\begin{pmatrix}
0\\ 
0
\end{pmatrix}
.\label{eq:DSE_xi_matrix_rho0}
\end{equation}
Since obviously \[(\mathrm{diag}\{\mathcal{K}_1(\xi),~\mathcal{K}_2(\xi) \})^{-1}=\mathrm{diag}\{\mathcal{K}_1(-\xi),~\mathcal{K}_2(-\xi) \}\] with matrix inversion defined by regular matrix multiplication and distribution inversion defined by distribution multiplication that gives a $\delta$-function. Combining this result with Eq.~\eqref{eq:DSE_xi_matrix} in the Landau gauge,
\begin{equation}
\begin{pmatrix}
\rho_1(0) \\ 
\rho_2(0)
\end{pmatrix} +
\begin{pmatrix}
\Omega_{11}(0) & \Omega_{12}(0) \\ 
\Omega_{21}(0) & \Omega_{22}(0)
\end{pmatrix} 
\begin{pmatrix}
\rho_1(0) \\ 
\rho_2(0)
\end{pmatrix}=
\begin{pmatrix}
0 \\
0
\end{pmatrix}
,\label{eq:DSE_0_matrix}
\end{equation}
yields
\begin{equation}
\begin{pmatrix}
\Omega_{11}(0) & \Omega_{12}(0) \\ 
\Omega_{21}(0) & \Omega_{22}(0)
\end{pmatrix} =
\begin{pmatrix}
\mathcal{K}_1(-\xi) & \\ 
& \mathcal{K}_2(-\xi)
\end{pmatrix} 
\begin{pmatrix}
\Omega_{11}(\xi) & \Omega_{12}(\xi) \\ 
\Omega_{21}(\xi) & \Omega_{22}(\xi)
\end{pmatrix} 
\begin{pmatrix}
\mathcal{K}_1(\xi) & \\ 
& \mathcal{K}_2(\xi)
\end{pmatrix}.\label{eq:consistency_fermion_prop_SDE_LKFT}
\end{equation}
Since for different Ansatz the Landau gauge solutions $\rho(s;0)$ are allowed to be different, Eq.~\eqref{eq:consistency_fermion_prop_SDE_LKFT} is the necessary condition for solutions to the SDE for the fermion propagator to be consistent with its LKFT. 

Meanwhile, when $\Omega(0)$ is given by Eq.~\eqref{eq:consistency_fermion_prop_SDE_LKFT}, Eq.~\eqref{eq:DSE_0_matrix} becomes Eq.~\eqref{eq:DSE_xi_matrix_rho0}, which, when viewed as equations for $\mathcal{K}_1(\xi)\rho_1(0)$ and $\mathcal{K}_2(\xi)\rho_2(0)$, is identical to Eq.~\eqref{eq:DSE_xi_matrix}. Therefore Eq.~\eqref{eq:consistency_fermion_prop_SDE_LKFT} is also the sufficient condition for solutions to the fermion propagator SDE to be consistent with LKFT. Therefore solutions of the SDE for fermion propagator are consistent with LKFT if and only if Eq.~\eqref{eq:consistency_fermion_prop_SDE_LKFT} is satisfied.
\subsection{Gauge covariance requirement for the photon propagator SDE\label{ss:GC_requirement_photon}}
After gauge fixing, the photon propagator $D^{\mu\nu}(q)$ takes the form of Eq.~\eqref{eq:def_photon_propagator}.
While the Landau gauge photon propagator $\Delta^{\mu\nu}(q)$ is given by Eq.~\eqref{eq:def_Delta_munu}. The dressing function $G(q^2)$ is determined by the SDE for the photon propagator. As illustrated in Fig.~\ref{fig:SDE_photon}, the same vertex structure $S_F(k)\Gamma^\mu(k,p)S_F(p)$ appears in the SDE for the photon propagator. This allows us to derive the gauge covariance requirement on the photon propagator SDE. Meanwhile, the spectral representation ensures the transversality of the vacuum polarization tensor through the translational invariance of the loop momentum. 
To start with, the dependence of the photon propagator $D^{\mu\nu}(q)$ on the covariant gauge parameter $\xi$ is completely specified by the $\xi q^\mu q^\nu/q^4$ term, as a direct consequence of which, $G(q^2)$ of Eq.~\eqref{eq:def_photon_propagator} and the transverse vacuum polarization tensor $\Pi^{\mu\nu}(q^2)=(g^{\mu\nu}q^2-q^\mu q^\nu)\Pi(q^2)$ are required to be independent of $\xi$.

Based on the analytic structures of $\Pi(q^2)$, the spectral representation of the photon propagator is covered in Section \ref{sc:sr_photon}. The SDE for the photon propagator spectral function is discussed in Section \ref{sc:SDE_rho_gamma}. Although the analytic structures of the photon propagator differ from those for the fermion propagator, 
we can still proceed by keeping the external momentum dependence explicit without introducing a spectral function for the photons. Therefore the consistency requirement for the photon propagator SDE is simply given by $\partial_\xi \Pi(q^2)=0$. Since the vacuum polarization function $\Pi(q^2)$ is linear in the fermion propagator spectral functions. One can write
\begin{equation}
\Pi(q^2) =\int dW\,\Omega^\gamma(q^2,W;\xi)\,\rho(W;\xi) =\int ds\, (\Omega^\gamma_1(q^2,s;\xi),~\Omega^\gamma_2(q^2,s;\xi))\,
\begin{pmatrix}
\rho_1(s;\xi) \\ 
\rho_2(s;\xi)
\end{pmatrix}. 
\end{equation}
With the $\xi$ dependence of $\rho_j(s;\xi)$ given by Eq.~\eqref{eq:k_exponential}, the $\xi$ independence of $\Pi(q^2)$ specifies
\begin{equation}
\Omega_{j}^\gamma(q^2,s;\xi)=\int ds'\,\Omega_{j}^\gamma(q^2,s';0)\,\exp\left[\dfrac{\alpha\xi}{4\pi}\Phi_{j}(s',s) \right],\label{eq:consistency_gamma}
\end{equation}
or at the operator level $\Omega_\xi^\gamma=\Omega_0^\gamma e^{\nu\Phi}$. This is the consistency requirement between the photon SDE and the LKFT.
\section{The decomposition of $\Omega$\label{ss:decomposition_Omega}}
The operator $\Omega$ can be decomposed into components from the fermion mass, and the longitudinal and transverse parts of the photon propagator. Some of these contributions can be calculated exactly. In the quenched approximation, $G(q^2)=1$ and the photon propagator is known exactly. When photons are unquenched, the vacuum polarization produces a nontrivial $G(q^2)$ in Eq.~\eqref{eq:def_photon_propagator}. 
Meanwhile, since the longitudinal part of the fermion-photon vertex is fixed by the Ward--Green--Takahashi identity, contributions from the $\xi q^\mu q^\nu/q^4$ term to $\Omega$ are known exactly regardless of either the dressing of the photon propagator or the transverse part of the fermion-photon vertex. 

While the bare mass $m_B$ contributes to off-diagonal terms of $\Omega_{ij}$ containing terms at most linear in $m_B$, allowing the following decomposition of $\Omega$,
\begin{equation}
\Omega=\Omega^m+\Omega^\xi+\Omega^\Delta,\label{eq:Omega_decomposition}
\end{equation}
where
\begin{equation}
\Omega^m(s,s')=
\begin{pmatrix}
& -\dfrac{m_B}{s}\delta(s-s') \\ 
-m_B\delta(s-s') & 
\end{pmatrix} 
\end{equation}
stands for the operation linear in $m_B$ that is also independent of $\xi$. Furthermore, denoting by $\Omega^\xi$ the contribution from the longitudinal component of the photon propagator $\xi q^\mu q^\nu/q^4$, this can be readily computed exactly. While $\Omega^\Delta$ is calculated with the $\Delta^{\mu\nu}(q)$ term of the photon propagator in Eq.~\eqref{eq:def_photon_propagator}, which remains unknown without either the photon dressing function or the transverse part of the fermion-photon vertex. 

$\Omega^\xi$, being linear in $\xi$, vanishes in the Landau gauge. While $\Omega^\Delta$ depends on the gauge because of the transverse aspects of $S_F(k)\Gamma^\mu(k,p)S_F(p)$. These need not be zero in the Landau gauge, despite this being commonly assumed.
\subsection{Exact expressions for $\Omega^\xi$}
In order to calculate $\Omega^\xi$ in any dimensions, based on Eq.~\eqref{eq:delta_delta_rho} we need to calculate the contribution to $\sigma_j(p^2;\xi)$ as functionals of $\rho_j(s;\xi)$ with explicit dependence on the number of spacetime dimensions $d=4-2\epsilon$. We denote by $\sigma_j^\xi$ the contribution to $\sigma_j$ from the longitudinal component of the photon propagator. After replacing $D_{\mu\nu}(q)$ by $\xi q_\mu q_\nu/q^4$, we have
	\begin{align}
	& \quad \sigma_1^{\xi}(p^2)+\slashed{p}\sigma_2^{\xi}(p^2)=ie^2\xi\int dW\int d\underline{k}\,\slashed{q}\,\dfrac{1}{\slashed{k}-W}\,\slashed{q}\,\dfrac{1}{q^4}\,\dfrac{\rho(W)}{\slashed{p}-W}\nonumber\\
	& =\dfrac{-\alpha\xi}{4\pi}\int dW\int_{0}^{1}dy\,\Gamma(\epsilon)\left(\dfrac{4\pi\mu^2}{s} \right)^\epsilon\Bigg\{\dfrac{-\epsilon(1-y)y}{(1-yz)\mathcal{D}^\epsilon}\slashed{p}+\dfrac{y[3\epsilon-4+(3-2\epsilon)y]}{\mathcal{D}^\epsilon}\slashed{p}+\dfrac{W}{\mathcal{D}^\epsilon}\Bigg\}\dfrac{\rho(W)}{\slashed{p}-W},\label{eq:def_sigma_overline_Delta}
	\end{align}
	with $y$ being the Feynman parameter, $z=p^2/s$ and the combined denominator given by\\ ${\mathcal{D}=(1-y)(1-y z)}$. After applying the integral definition of hypergometric functions \cite{abramowitz1964handbook}, we have 
	\begin{equation}
	\int_{0}^{1}dy\,\dfrac{y^p(1-y)^q}{(1-yz)^a}=\dfrac{\Gamma(p+1)\Gamma(q+1)}{\Gamma(p+q+2)}~_2F_1(a,p+1;p+q+2;z).
	\end{equation}
	Then the loop-integral factor of Eq.~\eqref{eq:def_sigma_overline_Delta} becomes
	\begin{align}
	&\quad ie^2\int d\underline{k}\,\slashed{q}\,\dfrac{1}{\slashed{k}-W}\,\dfrac{\slashed{q}}{q^4}
	\nonumber\\
	&= -\dfrac{\alpha}{4\pi}\Gamma(\epsilon)\left(\dfrac{4\pi\mu^2}{s} \right)^\epsilon\Bigg\{\dfrac{-\epsilon\slashed{p}}{(3-\epsilon)(2-\epsilon)}~_2F_1(1+\epsilon,2;4-\epsilon;z) +\dfrac{(3\epsilon-4)\slashed{p}}{(2-\epsilon)(1-\epsilon)}~_2F_1(\epsilon,2;3-\epsilon;z)\nonumber\\
	& \quad + \dfrac{2(3-2\epsilon)\slashed{p}}{(3-\epsilon)(2-\epsilon)(1-\epsilon)}~_2F_1(\epsilon,3;4-\epsilon;z)+\dfrac{W}{1-\epsilon}~_2F_1(\epsilon,1;2;z)  \Bigg\}.\label{eq:sigma_overline_delta_loop}
	\end{align}
Since $\sigma_{j}^{\xi}$ are properly formulated Feynman diagrams corresponding to loop-corrections to the fermion propagator where the $\rho_j(s;\xi)$ are given by $\delta$-functions, one expects that linear combinations of hypergeometric functions in Eq.~\eqref{eq:sigma_overline_delta_loop} are finite (at least in 4D) when $z\rightarrow 1$ such that there are contributions to fermion propagator functions no more singular than those of a free particle.

After numerous applications of contiguous relations for hypergeometric functions $~_2F_1(a,b;c;z)$, Eq.~\eqref{eq:def_sigma_overline_Delta} becomes
\begin{subequations}\label{eq:sigma_xi}
	\begin{align}
	& \sigma_1^{\xi}(p^2)=\dfrac{\alpha\xi}{4\pi}\int ds\left(\dfrac{4\pi\mu^2}{s} \right)^\epsilon \dfrac{\Gamma(\epsilon)}{1-\epsilon}~_2F_1(\epsilon,2;2-\epsilon;z)\,\rho_1(s),\label{eq:sigma_1_xi}\\
	& \sigma_2^{\xi}(p^2)=\dfrac{\alpha\xi}{4\pi}\int ds\left(\dfrac{4\pi\mu^2}{s} \right)^\epsilon \dfrac{\epsilon\Gamma(\epsilon)}{(2-\epsilon)(1-\epsilon)}~_2F_1(\epsilon+1,2;3-\epsilon;z)\dfrac{1}{s}\,\rho_2(s).\label{eq:sigma_2_xi}
	\end{align}
\end{subequations}
Details of the intermediate steps can be found in Appendix \ref{ss:simplify_sigma_xi}.

Next, since $\Omega^\xi$ is only linear in $\xi$, we define $\Theta$ as 
\begin{equation}
\Theta=-\nu^{-1}\Omega^\xi \; ,\label{eq:def_Theta}
\end{equation}
(recalling $\nu\equiv \alpha\xi/4\pi$) such that the distribution $\Theta$ is independent of $\xi$. Apparently only diagonal elements of $\Theta_{ij}$ survive, therefore
\begin{subequations}\label{eq:def_Theta_dd}
	\begin{align}
	&  -\nu\Theta_{11}(s,s';\xi)=-\dfrac{1}{s}\dfrac{\delta}{\delta\rho_1(s';\xi)}\dfrac{1}{\pi}\mathrm{Im}\big\{\sigma_1^\xi(s+i\varepsilon;\xi)\big\}.\label{eq:def_Theta_11}\\
	& -\nu\Theta_{22}(s,s';\xi)=-\dfrac{\delta}{\delta\rho_2(s';\xi)}\dfrac{1}{\pi}\mathrm{Im}\big\{\sigma_2^\xi(s+i\varepsilon;\xi)\big\}.\label{eq:def_Theta_22}
	\end{align}
\end{subequations}
Let us define at the operator level $\int ds\Theta=\mathrm{diag}\big\{\theta_1,\theta_2\big\}$. Since
\[\dfrac{1}{p^2-s+i\varepsilon}=-\dfrac{1}{p^{2}}\sum_{\beta=1}^{+\infty}z^\beta,\] Equations~(\ref{eq:sigma_xi},~\ref{eq:def_Theta_dd}) imply
\begin{subequations}
	\begin{align}
	-\nu\theta_1\dfrac{1}{p^2-s+i\varepsilon}& =\nu\Gamma(\epsilon)\left(\dfrac{4\pi\mu^2}{p^2} \right)^\epsilon\dfrac{1}{p^2}\dfrac{1}{1-\epsilon}\sum_{\beta=1}^{+\infty}\dfrac{(\epsilon)_{\beta-1}(2)_{\beta-1}}{(2-\epsilon)_{\beta-1}(\beta-1)!}z^{\beta+\epsilon}\\
	-\nu\theta_2\dfrac{1}{p^2-s+i\varepsilon}& =\nu\Gamma(\epsilon)\left(\dfrac{4\pi\mu^2}{p^2} \right)^\epsilon\dfrac{1}{p^2}\dfrac{\epsilon}{(2-\epsilon)(1-\epsilon)} \sum_{\beta=1}^{+\infty}\dfrac{(\epsilon+1)_{\beta-1}(2)_{\beta-1}}{(3-\epsilon)_{\beta-1}(\beta-1)!}z^{\beta+\epsilon}.
	\end{align}
\end{subequations}
Therefore we have the following identities for $\theta_j$,
\begin{subequations}
	\begin{align}
	& \theta_1 z^\beta=\left(\dfrac{4\pi\mu^2}{p^2} \right)^\epsilon\dfrac{\Gamma(1-\epsilon)\Gamma(\beta+\epsilon-1)\beta}{\Gamma(1+\beta-\epsilon)}z^{\beta+\epsilon},\\
	& \theta_2 z^\beta=\left(\dfrac{4\pi\mu^2}{p^2}\right)^\epsilon\dfrac{\Gamma(1-\epsilon)\Gamma(\beta+\epsilon)\beta}{\Gamma(2+\beta-\epsilon)}z^{\beta+\epsilon},
	\end{align}
\end{subequations}
which completely specify $\Theta$, and consequently $\Omega^\xi$.
\subsection{Consistency requirement as recurrence relations}
Based on previous analysis, for a given Ansatz for the fermion-photon vertex that ensures the vertex structure $S_F(k)\Gamma^\mu(k,p)S_F(p)$ being linear in $\rho(W)$, the corresponding distributions $\Omega_{ij}$ can be calculated. Such an Ansatz is consistent with LKFT if and only if Eq.~\eqref{eq:consistency_fermion_prop_SDE_LKFT} is satisfied. Independent of any Ansatz, two terms $\Omega^m$ and $\Omega^\xi$ are now known exactly. 

In this subsection we explore how Eq.~\eqref{eq:consistency_fermion_prop_SDE_LKFT} is satisfied incorporating $\Omega^\Delta$, \textit{i.e.} with $\Omega^m$ and $\Omega^{\xi}$ explicitly included. Straightforwardly, one could substitute Eq.~\eqref{eq:Omega_decomposition} with known components into the consistency requirement Eq.~\eqref{eq:consistency_fermion_prop_SDE_LKFT}, and obtain 
\begin{equation}
\Omega^\Delta=e^{-\nu\Phi}(\Omega^m+\Omega^\Delta_0)e^{\nu\Phi}-\Omega^m+\nu\Phi,
\end{equation}
as the consistency requirement on $\Omega^\Delta$. Alternatively, with LKFT for fermion propagator spectral functions given by Eq.~\eqref{eq:k_exponential}, we have
\begin{equation}
\Omega_\xi=e^{-\nu\Phi}\Omega_0 e^{\nu\Phi},\label{eq:Omega_xi_0}
\end{equation}
where the subscript of $\Omega_\xi$ highlights the $\xi$ dependent $\Omega$ in Eq.~\eqref{eq:Omega_decomposition}, therefore $\Omega_0=\lim\limits_{\xi\rightarrow 0}\Omega_\xi$. To see how infinitesimal changes in $\xi$ affect $\Omega^\Delta$, consider taking the derivative with respect to $\nu$ (effectively $\xi$) of Eq.~\eqref{eq:Omega_xi_0},
\[\partial_\nu\Omega_\xi=-\Phi e^{-\nu\Phi}\Omega_0 e^{\nu\Phi}+  e^{-\nu\Phi}\Omega_0 e^{\nu\Phi}\Phi=[\Omega_\xi,\Phi].\]
Substituting in Eq.~\eqref{eq:Omega_decomposition} and Eq.~\eqref{eq:def_Theta} produces
\begin{equation}
\partial_\nu\Omega^\Delta+[\Phi,\Omega^\Delta]=-[\Phi,\Omega^m]+\Theta+\nu[\Phi,\Theta].\label{eq:D_xi_Omega_Delta}
\end{equation}
In order to recover the corresponding terms using the spectral representation for the fermion propagator, one calculates
\begin{equation}
\int ds\,ds'\dfrac{1}{p^2-s+i\varepsilon}\,\Omega(s,s')\, \rho(s').\label{eq:convention_S_rho}
\end{equation}
Since the $z^\beta$ expansion is in fact the $p^2/s$ expansion of the free-particle propagator, commutators of operations on $z^\beta$ should be calculated with $z^\beta$ to the left. There exists an alternative convention to Eq.~\eqref{eq:convention_S_rho} that locates the free-particle propagator to the right of the operation, which subsequently modifies Eq.~\eqref{eq:consistency_fermion_prop_SDE_LKFT}. The net effect of adopting the alternative convention to solutions of Eq.~\eqref{eq:D_xi_Omega_Delta} is, however, zero compared with the convention given by Eq.~\eqref{eq:convention_S_rho} because deriving Eq.~\eqref{eq:consistency_fermion_prop_SDE_LKFT} using the alternative convention for the location of free-particle propagator leads to exchanging $\mathcal{K}_j^\xi$ with $\mathcal{K}_j^{-\xi}$.

Within this convention of locations, the right-hand side of Eq.~\eqref{eq:D_xi_Omega_Delta}, operating on $z^\beta$ can be calculated according to Eqs.~(\ref{eq:zbeta_phi_theta},~ \ref{eq:zbeta_phi}).

Since physical $\Omega^\Delta$ are generated by loop-corrections to the fermion propagator, the following criteria apply:
\begin{itemize}
	\item while the dependence of $\Omega^\Delta$ on $\nu=\alpha\xi/(4\pi)$ is allowed to be any order, $\Omega^\Delta$ cannot depend on the bare coupling alone because of the renormalizability of fermion propagator SDE, the bare and renormalized forms of $\alpha\xi$ being identical.
	\item for diagonal elements of $\Omega^\Delta$, a trivial solution exists with $\Omega^\Delta=\nu \Theta=-\Omega^\xi$. However, in this case there is no correction to the free-particle propagator.
\end{itemize}
More generally, we define 
\begin{equation}
\Omega^\Delta=
\begin{pmatrix}
\Omega^\Delta_{11} & -\dfrac{m_B}{p^2}\Omega^\Delta_{12} \\ 
-m_B\Omega^\Delta_{21} & \Omega^\Delta_{22}
\end{pmatrix} \label{eq:Omega_Delta_matrix}
\end{equation}
such that $\Omega^\Delta_{ij}$ correspond to dimensionless transforms. In addition, from Eq.~\eqref{eq:k_exponential}, one can easily verify ${\partial_\nu \mathcal{K}_\xi+\Phi\mathcal{K}_\xi=0}$. The similarity of Eq.~\eqref{eq:D_xi_Omega_Delta} to this differential equation for $\mathcal{K}_\xi$ indicates the following expansions for $\Omega^\Delta_{ij}$,
\begin{subequations}
	\label{eq:Omegaij_series}
	\begin{align}
	& \Omega_{ij}^\Delta \,z^\beta=\sum_{m=0}^{+\infty}\dfrac{(-\nu)^m}{m!}\left(\dfrac{4\pi\mu^2}{p^2} \right)^{m\epsilon}\omega_{ij}(\beta,m)\,z^{\beta+m\epsilon}\quad\quad \mathrm{for}~(ij)\neq (12),\label{eq:Omega12_series}\\
	& \Omega_{12}^\Delta\, z^\beta=\sum_{m=0}^{+\infty}\dfrac{(-\nu)^m}{m!}\left(\dfrac{4\pi\mu^2}{p^2} \right)^{m\epsilon}\omega_{12}(\beta,m)\,z^{\beta+m\epsilon+1}.\label{eq:Omegaij/12_series}
	\end{align}
\end{subequations}
where the expansion coefficients $\omega_{ij}(\beta,m)$ are allowed to implicitly depend on $\epsilon$. The \lq 12' component of $\Omega^\Delta$ is expanded differently from other components to ensure that $\Omega^\Delta$ given by Eq.~\eqref{eq:Omega_Delta_matrix} translates into operations solely on the spectral variables.

With Eq.~\eqref{eq:Omegaij_series}, the left-hand side of Eq.~\eqref{eq:D_xi_Omega_Delta} can be calculated according to Eq.~\eqref{eq:zbeta_Omega_Delta}. Then recurrence relations for $\omega_{ij}(\beta,m)$ are obtained by the comparison of $\mathcal{O}(\nu^m)$ terms in Eq.~\eqref{eq:D_xi_Omega_Delta}. As a result, we have
	\begin{subequations}
		\label{eq:rec_omega_ij}
		\begin{align}
		& \quad -\omega_{11}(\beta,m+1)+\Gamma(\epsilon)\dfrac{\Gamma(1-\epsilon)}{\Gamma(1+\epsilon)}\Bigg\{\dfrac{\Gamma(2+\beta)\Gamma(\beta+\epsilon)}{\Gamma(2+\beta-\epsilon)\Gamma(\beta)}\omega_{11}(\beta+\epsilon,m) \nonumber\\
		&\hspace{5cm}-\omega_{11}(\beta,m)\dfrac{\Gamma(2+\beta+m\epsilon)\Gamma(\beta+(m+1)\epsilon)}{\Gamma(2+\beta+(m-1)\epsilon)\Gamma(\beta+m\epsilon)} \Bigg\}\nonumber\\[1.0mm]
		& =\begin{cases}
		\Gamma(1-\epsilon)\dfrac{\Gamma(\beta+\epsilon-1)\Gamma(\beta+1)}{\Gamma(1+\beta-\epsilon)\Gamma(\beta)} & \mathrm{for}~m=0,\\[2.5mm]
		-\dfrac{\Gamma(\epsilon)[\Gamma(1-\epsilon)]^2}{\Gamma(1+\epsilon)}\left(\dfrac{\beta+1}{\beta+2\epsilon}-\dfrac{1+\beta+\epsilon}{\beta+1} \right)\dfrac{\Gamma(\beta+2\epsilon)\Gamma(\beta+\epsilon+1)}{\Gamma(\beta-\epsilon+2)\Gamma(\beta)} & \mathrm{for}~m=1,\\[2.5mm]
		0 & \mathrm{for}~m\geq 2.
		\end{cases}\label{eq:rec_omega_11}
		\end{align}
		\begin{align}
		&\quad  -\omega_{12}(\beta,m+1)+\Gamma(\epsilon)\dfrac{\Gamma(1-\epsilon)}{\Gamma(1+\epsilon)}\Bigg\{\dfrac{\Gamma(2+\beta)\Gamma(\beta+\epsilon)}{\Gamma(2+\beta-\epsilon)\Gamma(\beta)}\omega_{12}(\beta+\epsilon,m)\nonumber\\
		&\hspace{5cm}-\omega_{12}(\beta,m)\dfrac{\Gamma(2+\beta+m\epsilon)\Gamma(\beta+1+(m+1)\epsilon)}{\Gamma(2+\beta+(m-1)\epsilon)\Gamma(\beta+1+m\epsilon)}\Bigg\}\nonumber\\[1.0mm]
		& =\begin{cases}
		\Gamma(1-\epsilon)\dfrac{\Gamma(2+\beta)\Gamma(\beta+\epsilon)}{\Gamma(2+\beta-\epsilon)\Gamma(\beta+1)} & \mathrm{for}~m=0,\\[2.5mm]
		0 & \mathrm{for}~m\geq 1.
		\end{cases}\label{eq:rec_omega_12}
		\end{align}
		\begin{align}
		&\quad  -\omega_{21}(\beta,m+1)+\Gamma(\epsilon)\dfrac{\Gamma(1-\epsilon)}{\Gamma(1+\epsilon)}\Bigg\{\dfrac{\Gamma(1+\beta)\Gamma(\beta+\epsilon)}{\Gamma(1+\beta-\epsilon)\Gamma(\beta)}\omega_{21}(\beta+\epsilon,m)\nonumber\\
		&\hspace{5cm}-\omega_{21}(\beta,m)\dfrac{\Gamma(2+\beta+m\epsilon)\Gamma(\beta+(m+1)\epsilon)}{\Gamma(2+\beta+(m-1)\epsilon)\Gamma(\beta+m\epsilon)}\Bigg\}\nonumber\\[1.0mm]
		&=\begin{cases}
		\Gamma(1-\epsilon)\dfrac{\Gamma(1+\beta)\Gamma(\beta+\epsilon)}{\Gamma(2+\beta-\epsilon)\Gamma(\beta)} & \mathrm{for}~m=0,\\[2.5mm]
		0 & \mathrm{for}~m\geq 1.
		\end{cases}\label{eq:rec_omega_21}
		\end{align}
		\begin{align}
		& \quad -\omega_{22}(\beta,m+1)+\Gamma(\epsilon)\dfrac{\Gamma(1-\epsilon)}{\Gamma(1+\epsilon)}\Bigg\{\dfrac{\Gamma(1+\beta)\Gamma(\beta+\epsilon)}{\Gamma(1+\beta-\epsilon)\Gamma(\beta)}\omega_{22}(\beta+\epsilon,m) \nonumber\\
		&\hspace{5cm}-\omega_{22}(\beta,m)\dfrac{\Gamma(1+\beta+m\epsilon)\Gamma(\beta+(m+1)\epsilon)}{\Gamma(1+\beta+(m-1)\epsilon)\Gamma(\beta+m\epsilon)}\Bigg\}\nonumber\\[1.0mm]
		& =\begin{cases}
		\Gamma(1-\epsilon)\dfrac{\Gamma(\beta+\epsilon)\Gamma(\beta+1)}{\Gamma(2+\beta-\epsilon)\Gamma(\beta)} & \mathrm{for}~m=0,\\[2.5mm]
		-\dfrac{\Gamma(\epsilon)[\Gamma(1-\epsilon)]^2}{\Gamma(1+\epsilon)}\left(\dfrac{1}{\beta+1}-\dfrac{1}{1+\beta-\epsilon} \right)\dfrac{\Gamma(\beta+2\epsilon)\Gamma(\beta+\epsilon+1)}{\Gamma(\beta-\epsilon+1)\Gamma(\beta)} & \mathrm{for}~m=1,\\[2.5mm]
		0 & \mathrm{for}~m\geq 2.
		\end{cases}\label{eq:rec_omega_22}
		\end{align}
	\end{subequations}
These recurrence relations specify how gauge covariance is satisfied when distributions $\Omega^\Delta_{ij}$ are expanded as Taylor series in $\nu=\alpha\xi/4\pi$ written in Eq.~\eqref{eq:Omegaij_series}. On one hand, when the $\Omega^\Delta_{ij}$ are only known in the Landau gauge, Eq.~\eqref{eq:rec_omega_ij} can be used to calculate  $\Omega^\Delta_{ij}$  in any other covariant gauge. On the other hand, when an Ansatz for $S_F(k)\Gamma^\mu(k,p)S_F(p)$ is known, the operations of $\Omega^\Delta_{ij}$ on $z^\beta$ can be calculated. Eq.~\eqref{eq:rec_omega_ij} then works to verify if this Ansatz ensures that solutions to fermion propagator SDE are consistent with LKFT.
\subsection{Example: The Gauge Technique in the quenched approximation in 4D}
In the quenched approximation with the Gauge Technique Ansatz for $S_F(k)\Gamma^\mu(k,p)S_F(p)$ \cite{Delbourgo:1977jc}, based on Eqs.~(\ref{eq:Gauge_Technique_sigma_1},~\ref{eq:Gauge_Technique_sigma_2}) we deduce the $\Omega_{ij}$ to be
\begin{align}
\Omega_{11}(s,s';\xi)& =-\dfrac{3\alpha}{4\pi}\bigg\{\left(\dfrac{1}{\epsilon}-\gamma_E+\ln 4\pi+\dfrac{4}{3}+\ln\dfrac{\mu^2}{s}\right)\delta(s-s')\nonumber\\
& \quad-\dfrac{s'}{s^2}\theta(s-s') \bigg\}-\dfrac{\alpha\xi}{4\pi}\dfrac{1}{s}\theta(s-s'),\nonumber\\[1mm]
\Omega_{12}(s,s';\xi)& =-\dfrac{m_B}{s}\delta(s-s'),\nonumber\\[1mm]
\Omega_{21}(s,s';\xi)& =-m_B \delta(s-s'),\nonumber\\[1mm]
\Omega_{22}(s,s';\xi)&=-\dfrac{3\alpha}{4\pi}\bigg\{\left(\dfrac{1}{\epsilon}-\gamma_E+\ln 4\pi+\dfrac{4}{3}+\ln\dfrac{\mu^2}{s}\right)\times\delta(s-s')\nonumber\\
& \quad-\dfrac{1}{s}\theta(s-s') \bigg\}-\dfrac{\alpha\xi}{4\pi}\dfrac{s'}{s^2}\theta(s-s'),\label{eq:Omega_GT}
\end{align}
where $m_B$ is the bare mass and $d=4-2\epsilon$. Equivalently written as operators on $z$, $\Omega_{ij}$ become
\begin{align}
\Omega_{11}(\xi)& =-\dfrac{3\alpha}{4\pi}\left[\tilde{C}+\ln(z) -z^{-1}I\right]-\dfrac{\alpha\xi}{4\pi}Iz^{-1},\nonumber\\[1mm]
\Omega_{12}& =-\dfrac{m_B}{p^2}z,\nonumber\\
\Omega_{21}& =-m_B,\nonumber\\[1mm]
\Omega_{22}(\xi)& =-\dfrac{3\alpha}{4\pi}\left[\tilde{C}+\ln(z) -Iz^{-1}\right]-\dfrac{\alpha\xi}{4\pi}z^{-1}I,\label{eq:Omega_GT_z}
\end{align}
where $\tilde{C}=1/\epsilon-\gamma_E+\ln(4\pi\mu^2/p^2)+4/3$.

Meanwhile, since in 4D the LKFT for the fermion propagator reduces to Eq.~\eqref{eq:kn_small_epsilon}, we have, 
\begin{subequations}
	\begin{align}
	z^\beta \mathcal{K}_1(\xi)& =\left(\dfrac{\mu^2}{p^2}\right)^{-\nu}\exp\bigg\{-\nu\left[ \dfrac{1}{\epsilon}+\gamma_E+\ln(4\pi)\right]\bigg\}\dfrac{\Gamma(\beta-\nu)\Gamma(2+\beta-\nu)}{\Gamma(\beta)\Gamma(2+\beta)}z^{\beta-\nu},\\[1.5mm]
	z^\beta \mathcal{K}_2(\xi)& =\left(\dfrac{\mu^2}{p^2}\right)^{-\nu}\exp\bigg\{-\nu\left[ \dfrac{1}{\epsilon}+\gamma_E+\ln(4\pi)\right]\bigg\}\dfrac{\Gamma(\beta-\nu)\Gamma(1+\beta-\nu)}{\Gamma(\beta)\Gamma(1+\beta)}z^{\beta-\nu}.
	\end{align}
\end{subequations}
For the consistency requirement, it is more convenient to write Eq.~\eqref{eq:consistency_fermion_prop_SDE_LKFT} as 
\begin{equation}
\begin{pmatrix}
\Omega_{11}(\xi) & \Omega_{12}(\xi) \\ 
\Omega_{21}(\xi) & \Omega_{22}(\xi)
\end{pmatrix}=
\begin{pmatrix}
\mathcal{K}_1 (\xi)\Omega_{11}(0)\mathcal{K}_1(-\xi) & \mathcal{K}_1 (\xi)\Omega_{12}(0)\mathcal{K}_2(-\xi) \\ 
\mathcal{K}_2 (\xi)\Omega_{21}(0)\mathcal{K}_1(-\xi) & \mathcal{K}_2(\xi)\Omega_{22}(0)\mathcal{K}_2(-\xi)
\end{pmatrix} .\label{eq:consistency_fermion_prop_SDE_LKFT_inverse}
\end{equation}
With the assistance of the following four identities for fractional calculus, 
\begin{subequations}
	\begin{align}
	I^\alpha z^\beta &=\dfrac{\Gamma(\beta+1)}{\Gamma(\alpha+\beta+1)}z^{\alpha+\beta},\\
	D^\alpha z^\beta &=\dfrac{\Gamma(\beta+1)}{\Gamma(-\alpha+\beta+1)}z^{-\alpha+\beta},\\
	I^\alpha z^\beta \ln(z)& =\dfrac{\Gamma(\beta+1)}{\Gamma(\alpha+\beta+1)}\big\{\psi(\beta+1)-\psi(\alpha+\beta+1) +\ln(z) \big\}z^{\alpha+\beta},\\
	D^\alpha z^\beta \ln(z)& =\dfrac{\Gamma(\beta+1)}{\Gamma(-\alpha+\beta+1)}\big\{\psi(\beta+1)-\psi(-\alpha+\beta+1) +\ln(z)\big\}z^{-\alpha+\beta},
	\end{align}
\end{subequations}
where $\psi(\beta)$ is the digamma function, one then obtains
\begin{subequations}
	\label{eq:K_Omega0_Kinv_zbeta}
	\begin{align}
	& \quad z^\beta \mathcal{K}_1(\xi)\Omega_{11}(0)\mathcal{K}_1(-\xi)\nonumber\\
	& =-\dfrac{3\alpha}{4\pi}\bigg\{\tilde{C}-\dfrac{1}{\beta-\nu+1}+\psi(\beta)-\psi(\beta-\nu)+\psi(\beta+2) -\psi(\beta+2-\nu)+\ln z \bigg\}z^\beta,
	\end{align}
	\begin{align}
	& z^\beta \mathcal{K}_1(\xi)\Omega_{12}(0)\mathcal{K}_2(-\xi)=-\dfrac{m_B}{p^2}\dfrac{\beta}{\beta-\nu}z^{\beta+1},\\[3mm]
	& z^\beta \mathcal{K}_2(\xi)\Omega_{21}(0)\mathcal{K}_1(-\xi)=-m_B\dfrac{\beta+1}{\beta+1-\nu}z^\beta,
	\end{align}
	\begin{align}
	& z^\beta \mathcal{K}_2(\xi)\Omega_{22}(0)\mathcal{K}_2(-\xi)\nonumber\\
	& =-\dfrac{3\alpha}{4\pi}\bigg\{\tilde{C}-\dfrac{1}{\beta-\nu}+\psi(\beta)-\psi(\beta-\nu)+\psi(\beta+1)-\psi(\beta+1-\nu)+\ln z \bigg\}z^\beta.
	\end{align}
\end{subequations}
While from Eq.~\eqref{eq:Omega_GT_z}, we have
\begin{subequations}
	\label{eq:Omega_xi_zbeta}
	\begin{align}
	& z^\beta \Omega_{11}(\xi)=\bigg\{-\dfrac{3\alpha}{4\pi}\left[\tilde{C}-\dfrac{1}{\beta+1}+\ln z\right]-\dfrac{\nu}{\beta}\bigg\}z^{\beta},\\
	& z^\beta\Omega_{12}(\xi)=-\dfrac{m_B}{p^2}z^{\beta+1},\\
	& z^\beta\Omega_{21}(\xi)=-m_B z^\beta,\\
	& z^\beta\Omega_{22}(\xi)=\bigg\{-\dfrac{3\alpha}{4\pi}\left[\tilde{C}-\dfrac{1}{\beta}+\ln z\right]-\dfrac{\nu}{\beta+1}\bigg\}z^{\beta}
	\end{align}
\end{subequations}
Observe that the digamma functions only occur in  Eq.~\eqref{eq:K_Omega0_Kinv_zbeta},  not in Eq.~\eqref{eq:Omega_xi_zbeta}. Additionally, the dependence on $\nu$ is only linear in Eq.~\eqref{eq:Omega_xi_zbeta}, but not in Eq.~\eqref{eq:K_Omega0_Kinv_zbeta}. Therefore the consistency requirement given by Eq.~\eqref{eq:consistency_fermion_prop_SDE_LKFT_inverse} is not satisfied by the Gauge Technique in 4D. The same conclusion has been realized by Delbourgo, Keck and Parker \cite{Delbourgo:1980vc} in a completely different approach.
\section{Summary of gauge covariance requirements\label{ss:GC_SDE_summary}}
In this chapter we have formulated the fermion propagator SDE in terms of propagator spectral functions. With the fermion-photon vertex structure $S_F(k)\Gamma^\mu(k,p)S_F(p)$ being linear in the $\rho_j(s;\xi)$ as implied by the equality of renormalization factors $Z_1=Z_2$, we have derived the necessary and sufficient condition for the solutions of the fermion propagator SDE to be consistent with LKFT in covariant gauges. With known contributions to the fermion propagator SDE calculated, this reduces the consistency requirement to that for the contribution to $\Omega$ in Eq.~\eqref{eq:SDE_fermion_rho_itg} from the Landau gauge photon propagator. 
Next, an expansion of the operator $\Omega^\Delta$ (defined in Eq.\eqref{eq:Omega_decomposition}), similar to that
of $\mathcal{K}_j$ in Eq.~\eqref{eq:kn_zbeta}, has been postulated in Eq.~\eqref{eq:Omegaij_series}. The consistency
requirements can then be converted into the form of recurrence relations of this expansion,
shown in Eq.~\eqref{eq:rec_omega_ij}.
The requirement on $S_F(k)\Gamma^\mu(k,p)S_F(p)$ to ensure the gauge invariance of $\Pi(q^2)$ was also derived. 

We observe that the Gauge Technique \cite{PhysRev.130.1287,PhysRev.135.B1398,PhysRev.135.B1428,Delbourgo:1977jc} does not ensure gauge covariance for the fermion propagator in QED. In fact, when fermions are massive, dimension-odd operators are required in $S_F(k)\Gamma^\mu(k,p)S_F(p)$ to ensure gauge covariance. Our formalism for the SDEs using a spectral representation allows propagators to be solved in Minkowski space, one attemp of which has been made in Chapter \ref{ch:SDE_oGT}. Furthermore, our consistency requirements can be used as criteria for truncating the SDEs for QED propagators. 

Importantly, our calculations have been performed in arbitary dimensions.  Keeping $\epsilon = 2- d/2$ explicit to the end turns out to give concise and meaningful results in the case of the $\sigma_{j}^{\xi}$ in Eq.~\eqref{eq:def_sigma_overline_Delta}, 
the fermion Schwinger--Dyson equation, as well as the LKFT for the fermion propagator. Results are concise in the sense that one hypergeometric function describes the $p^2$ dependence for each Dirac component of every loop integral. Meaningful in the sense that the results apply to any number of spacetime dimensions as long as hypergeometric functions converge. Based on these two merits, one might suspect that dimensional regularization evaluated by keeping $\epsilon$ explicit to the last step is intrinsic to QED itself.

This work marks a path towards ensuring consistent truncations of the Schwinger--Dyson equations for the fermion and boson propagators yield gauge covariant fermion mass functions like that in Fig.~\ref{fig:Mp2_xi_036}: an essential requirement for validating any truncation scheme used.
	\chapter{Conclusions and outlook\label{cp:conclusions}}
	We started with the path integral formulation and deduced the SDEs for the generating functionals. We have seen that SDEs for Green's functions are recurrence relations for the Taylor series expansions of the generating functional. Unlike recurrence relations for the expansions of a function, solving such recurrence relations for a functional requires a truncation scheme. Insights into the proper trunctions of SDEs for the propagators are expected from the WGTIs. However other than the longitudinal one, these identities do not form a closed system because they relate the QED vertices not only to the fermion propagator, but also to other unknowns. Alternatively we have seen that the analytic structures of the propagators promote the spectral representations. Combined with the longitudinal WGTI, this representation results in the Gauge Technique. Although the Gauge Technique Ansatz violates renormalizability, it maintains the analytic structures of the propagators. Furthermore the spectral representation allows us to solve the LKFT and the SDEs in Minkowski space. Based on these solutions, the requirements to maintain gauge covariance for any truncation scheme have been derived. The next step would be to construct an Ansatz that respects these gauge covariance requirements. At the same time, this anzatz also needs to preserve the analytic structures of the propagators.

The divergences of a renormalizable theory like QED observe their own patterns. With dimensional regularization and a mass-independent scheme, these patterns are given explicitly as recurrence relations. Multiplicative renormalizability of QED is preserved once these recurrence relations are satisfied by the truncation of SDEs. This renormalizability constraint and the gauge covariance requirements supplement each other. Hopefully the truncation scheme that satisfies both conditions represents closely to the true vertex projected onto the SDEs for the propagators. The equivalence of SDEs for the generating functional to all-order perturbation theory may provide further insight into the proper truncation of SDEs for the Green's functions.

The analytic structures of QCD propagators form another interesting topic. Because of confinement, quark and gluon propagators are not allowed to contain free-particle poles in the timelike region. Therefore their spectral functions, if they exist, are not allowed to have delta-function components. It is likely that the analytic structures of the QCD propagators are limited to branch-cuts in the timelike region only. However the exact mechanism responsible for such structures requires better understandings of the quark-gluon interaction, the gauge fixing, and the self-coupling of gluons. Exploring this topic is also likely to involve the renormalization of QCD.

After solving the SDEs for QCD propagators, the phenomenology of QCD bound states is then given by the Bethe--Salpeter equations and the Faddeev equations. Formulating SDEs in the Minkowski space may suggest that Bethe--Salpeter and Faddeev equations are also better understood in Minkowski space. However the multivariable nature of these two types of equations complicates the spectral representations of their solutions~\cite{Jia:2023imt,Jia:2024dbp}. One promising candidate is the Nakanishi representation for the Bethe--Salpeter amplitude \cite{nakanishi1971graph}. With the Nakanishi representation, the formulations and solutions of the Bethe--Salpeter equations in Minkowski space can also be found in Refs.~\cite{Kusaka:1997xd,Kusaka:1995za,Kusaka:1995wb,Kusaka:1995nv}.
	\appendix
	\chapter{Mathematical Methods}
	\section{The spectral representation of complex functions\label{ss:sr_fz}}
	\subsection{Uniqueness of the spectral representation}
When solving for $D(p^2)$ from its SDE using the spectral representation of Green's functions, we are motivated to understand the mathematical properties of K\"{a}ll\'{e}n--Lehmann spectral representation given by Eq.~\eqref{eq:KLSR_scalar}. Specifically,
\begin{itemize}
	\item how does spectral function $\rho(s)$ uniquely determine its propagator function in the Minkowski space of $p^2$, or $p^2\in \{z|z=x\pm i\epsilon,~x\in \mathbf{R},~\epsilon\geq 0\}$;
	\item how to extract spectral function $\rho(s)$ when $D(p^2)$ is known.
\end{itemize}

Before moving to calculations, there are some assumptions to be made on propagator function $D(p^2)$. First, consider the dimensionless function $f(z)=\mu^D D(p^2/\mu^2)$, with an arbitrary scale $\mu$ and $D$ being the dimension of propagator function $D$. Since the word ``analytic" has been abused in the literature, a more rigorous description on properties of $f(z)$ is required. Here are several assumptions on $f(z)$:
\begin{itemize}
	\item The function $f(z)$ is defined on the complex plane with at most a branch cut and at a finite number of poles.
	\item The function $f(z)$ has a branch cut on the non-negative real axis.
	\item The function $f(z)$ is real on the real axis except for the branch cut and at most a finite number of poles.
	\item The conjugation of the argument of $f(z)$ is equivalent of the complex conjugation of $f(z)$ itself: $f(z^*)=\overline{f}(z)$.
	\item $f(z)$ is holomorphic within the domain of its definition.
\end{itemize}
The branch cuts and poles correspond to the production of real particles within quantum loop corrections of the propagator. The conjugation property and the holomoerphic requirement allow us to determine $f(z)$, up to a real constant, once the imaginary part of $f(z)$ along the branch cut is known.

Recall the decomposition of a complex function
\[f(x+iy)=u(x,y)+iv(x,y). \]
When $f(z)$ is holomorphic on some domain $z\in \mathrm{D}$, the following Cauchy--Riemann equation applies
\begin{equation}
\Bigg\{ 
\begin{array}{l}
\dfrac{\partial u}{\partial x}=\dfrac{\partial v}{\partial y} \\ 
\dfrac{\partial u}{\partial y}=-\dfrac{\partial v}{\partial x}
\end{array} ,
\end{equation}
or written in polar coordinates $\rho e^{i\phi}=x+iy$, $\rho\in [0,+\infty),~\phi \in [0,2\pi]$,
\begin{equation}
\Bigg\{ 
\begin{array}{l}
\rho\dfrac{\partial}{\partial \rho}u=\dfrac{\partial}{\partial \phi}v \\ 
\dfrac{\partial}{\partial \phi}u=-\rho\dfrac{\partial}{\partial\rho}v
\end{array} .
\end{equation}
Because of Cauchy--Riemann equation, both $u$ and $v$ satisfy two-dimensional Laplace equations
\begin{equation}
\nabla^2u=0,\quad \nabla v^2=0.
\end{equation}
\subsection{Example: eigenfunction expansion of the spectral representation}
If the only known of $f(z)$ is the imaginary part of itself close to the branch cut, how should we proceed to solve for $f(z)$ in its holomorphic domain $\mathrm{D}$? The answer is given by the boundary value problems of Laplace equations. Since $f(z^*)=\overline{f}(z)$, we know 
\begin{equation}
\lim\limits_{\phi\rightarrow 0}v(\rho,\phi)=-\lim\limits_{\phi\rightarrow 2\pi}v(\rho,\phi).\label{eq:boundary_branch_cut}
\end{equation}
Meanwhile, we need the knowledge of the asymptotic behavior of $v$ when $\rho\rightarrow +\infty$. For simplicity, consider $\lim\limits_{\rho\rightarrow +\infty}v(\rho,\phi)=0$.

Notice boundary conditions at the branch cut are inhomogeneous, in order to solve for $v(\rho,\phi)$ that satisfies its Laplace equation
\begin{equation}
\nabla^2v=\dfrac{1}{\rho}\partial_\rho \rho\partial_\rho v(\rho,\phi)+\dfrac{1}{\rho^2}\partial^2_\phi v(\rho,\phi)=0,
\end{equation}
we need to find out a specific solution $V(\rho,\phi)$ that satisfies the Laplace equation with the boundary condition given by Eq.~\eqref{eq:boundary_branch_cut} and combine it with a solution to the Laplace equation with homogeneous boundary conditions. The inclusion of the homogeneous solution is to provide extra free parameters to match at boundaries when the domain $\mathbf{D}$ of $f(z)$ is naturally separated into subsets.

Using the standard technique of variable separation, the homogeneous solution of the Laplace equation is 
\begin{equation}
v_{\{a,b\}}(\rho,\phi)=\sum_{n=1}^{+\infty}\left(a_n\rho^n+\dfrac{b_n}{\rho^n}\right)\sin(n\phi).
\end{equation}
We cannot proceed without the specific knowledge of $v(\rho,\phi)$ in the vicinity of the branch cut. Therefore, consider a simple example where $v(\rho,0)=\pi\theta(\rho-1)$ with $\theta(x)$ being the Heaviside step function. We know this corresponds to a polylogarithm function $f(z)=-\ln(1-z)$. In this example, domain $\mathbf{D}$ is naturally separated into $\rho<1$ and $\rho\geq 1$ subsets.

Consider $\rho\geq 1$, the specific solution $V(\rho,\phi)$ can be easily found because dependences on $\rho$ and $\phi$ variables naturally separate;
\begin{equation}
V(\rho,\phi)=R(\rho)\Phi(\phi).
\end{equation}
When such separation does not happen, the expansion of $R(\rho)$ in terms of eigenfunctions of the differential equation 
\[ \rho\dfrac{d}{d\rho}\rho\dfrac{d}{d\rho}R(\rho)+\lambda R(\rho)=0 \]
is required.

Assuming $R(\rho)=\pi$, from the Laplace equation we obtain
\begin{equation}
\dfrac{1}{R(\rho)}\rho\dfrac{d}{d\rho}\rho\dfrac{d}{d\rho}R(\rho)+\dfrac{1}{\Phi(\phi)}\dfrac{d^2}{d\phi^2}\Phi(\phi)=0,
\end{equation}
which indicates
\begin{equation}
\Phi(\phi)=a\phi+b.
\end{equation}
Parameters $a$ and $b$ are determined by boundary conditions for $\Phi$: $\Phi(0)=1,~\Phi(2\pi)=-1$. Therefore when $\rho\geq 1$,
\begin{equation}
V(\rho,\phi)=1-\phi/\pi.
\end{equation}

Next, homogeneous solutions are added on;
\begin{equation}
v(\rho,\phi)=\Bigg\{
\begin{array}{ll}
\pi-\phi+\sum_{n=1}^{+\infty}\dfrac{b_n}{\rho^n}\sin(n\phi) & (\rho\geq 1) \\ 
\sum_{n=1}^{+\infty}a_n\rho^n\sin(n\phi) & (0 \leq \rho<0)
\end{array} .
\end{equation}
To match boundary conditions at $\rho=1$, we need the following Fourier expansion
\begin{equation}
\pi-\phi=\sum_{n=1}^{+\infty}\dfrac{2}{n}\sin(n\phi) \quad (\phi\in[0,2\pi]).
\end{equation}
Therefore
\begin{equation}
v(\rho,\phi)=\Bigg\{
\begin{array}{ll}
\sum_{n=1}^{+\infty}\left(\dfrac{2}{n}+\dfrac{b_n}{\rho^n}\right)\sin(n\phi) & (\rho\geq 1) \\ 
\sum_{n=1}^{+\infty}a_n\rho^n\sin(n\phi) & (0 \leq \rho<0)
\end{array} .
\end{equation}
From
\begin{align}
& \lim\limits_{\rho\rightarrow 1^+}v=\lim\limits_{\rho\rightarrow 1^-}v\\
& \lim\limits_{\rho\rightarrow 1^+}\rho\partial_\rho v=\lim\limits_{\rho\rightarrow 1^-}\rho\partial_\rho v,
\end{align}
we get
\begin{equation}
2/n+b_n=a_n\quad a_n+b_n=0,
\end{equation}
or $a_n=-b_n=1/n$.

Then, the corresponding solution of $u(\rho,\phi)$ is needed. First, consider the corresponding $u_n(\rho,\phi)$ of
\begin{equation}
v_n(\rho,\phi)=\Bigg\{
\begin{array}{ll}
\dfrac{b_n}{\rho^n}\sin(n\phi) & (\rho\geq 1) \\ 
a_n\rho^n\sin(n\phi) & (0\leq \rho <1)
\end{array} .
\end{equation}
From one of the Cauchy--Riemann equations $\rho\partial_\rho u_n=\partial_\phi v_n$, we have
\begin{equation}
u_n(\rho,\phi)=\Bigg\{
\begin{array}{ll}
-\dfrac{b_n}{\rho^n}\cos(n\phi)+ g(\phi) & (\rho\geq 1)\\ 
a_n\rho^n\cos(n\phi) +h(\phi) & (0\leq \rho<1)
\end{array} .
\end{equation}
And from the other Cauchy--Riemann equation $\partial_\phi u=-\rho\partial_\rho v$ we obtain
\begin{equation}
\dfrac{d}{d\phi}g(\phi)=0,\quad \dfrac{d}{d\phi}h(\phi)=0,
\end{equation}
or $g(\phi)=\tilde{g},~h(\phi)=\tilde{h}$, where $\tilde{g}$ and $\tilde{h}$ are real numbers.

Similarly using Cauchy--Riemann equations, the corresponding real part function to the inhomogeneous solution $V(\rho,\phi)$ is
\begin{equation}
U(\rho,\phi)=-\ln\rho+\tilde{k},
\end{equation}
with $\tilde{k}\in \mathrm{R}$.
Finally, using 
\[ \sum_{n=1}^{+\infty}\dfrac{x^n}{n}=-\ln(1-x)\]
to sum up homogeneous terms, we have 
\begin{equation}
f(\rho,\phi)=u(\rho,\phi)+iv(\rho,\phi)=\Bigg\{
\begin{array}{ll}
-\ln\rho +i(\pi-\phi)-\ln\left(1-\dfrac{1}{\rho e^{i\phi}}\right)+\tilde{k}+\tilde{g} & (\rho\geq 1)\\ 
-\ln(1-\rho e^{i\phi})+\tilde{h} & (0\leq \rho<1)
\end{array} .
\end{equation}
After applying the default branch cut for logarithm and taking the continues requirement at $\rho=1$, we obtain 
\begin{equation}
f(z)=-\ln(1-z)+\tilde{h}.
\end{equation}
In summary, we have shown that the imaginary part of function $f(z)$ at $\phi=0^+$ given by $\pi\theta(\rho-1)$ specifies a function $f(z)=-\ln(1-z)+\tilde{h}$ based on assumptions of $f(z)$ made in Chapter \ref{cp:spec_repr}.
	\section{The Mellin transform and its relation to the analytic continuation\label{ss:Mellin}}
	Consider the propagator function of a scalar field, its spectral representation is given by Eq.~\eqref{eq:KLSR_scalar}. For massive particles, their spectral functions contain $\delta$-function terms as the on-shell components and $\theta$-function terms corresponding to real-particle productions through quantum loop corrections. 

For the propagator function with simple poles and branch cuts in the timelike region as its singularities, there exists a unique $\rho(s)$ given by Eq.~\eqref{eq:rho_D}.
In this scenario, the spectral representation is bijective.

\begin{figure}
\centering
\includegraphics[width=1\linewidth]{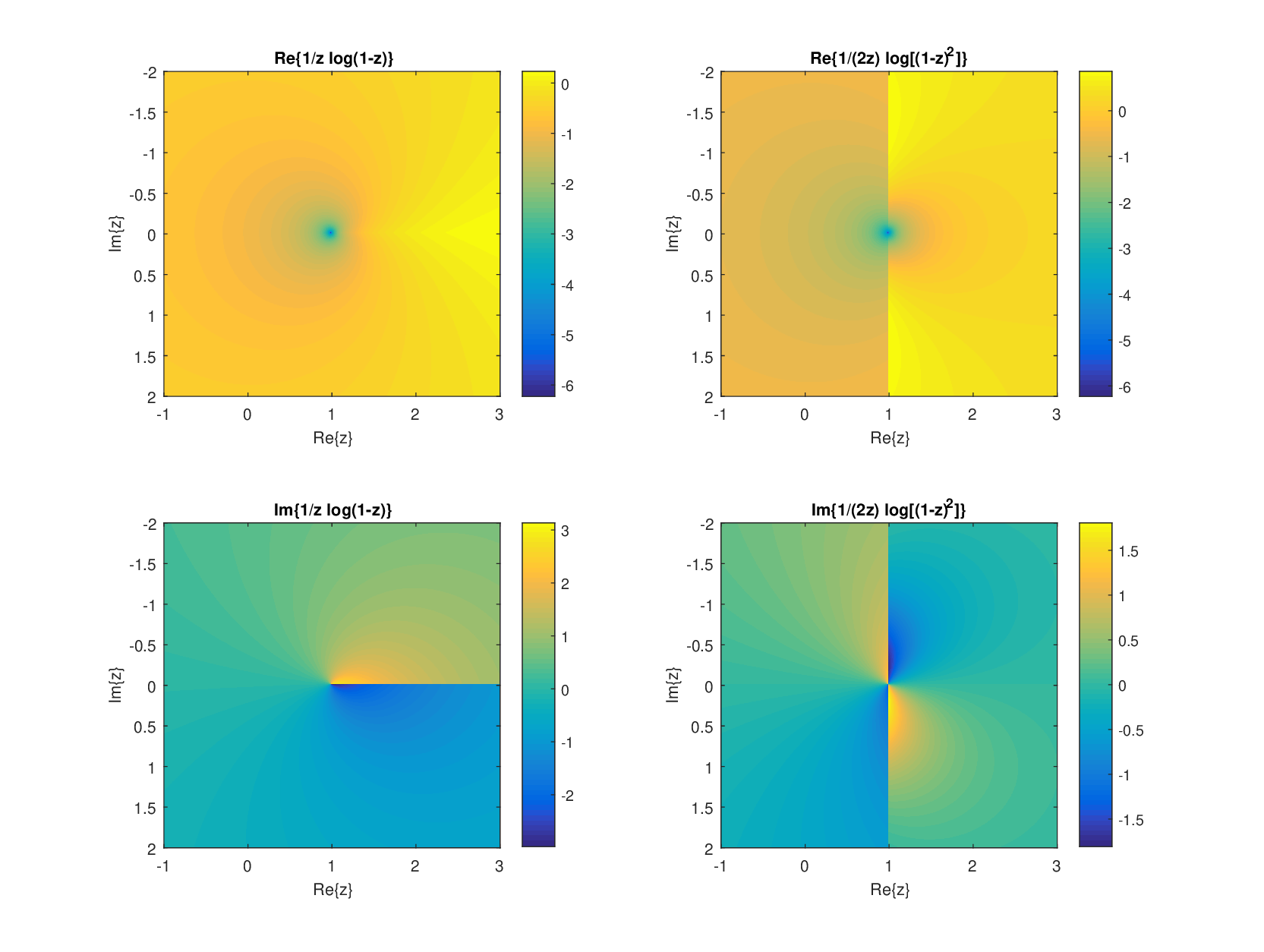}
\caption{The comparison of two functions defined in Eq.~\eqref{eq:f1f2_branches}. They are identical for $z<1$ ($z\in \mathbf{R}$), but different else where on the complex plane.}
\label{fig:f1f2branchcutsarticle}
\end{figure}
In general, knowing the propagator $D(p^2)$ only in the spacelike region ($p^2<0$) is insufficient to uniquely determine $\rho(s)$ because there are infinite ways to perform analytic continuations. One simple example is given by 
\begin{equation}
f_1(z)=\dfrac{1}{z}\ln(1-z),\quad f_2(z)=\dfrac{1}{2z}\ln\left[\left(1-z\right)^2 \right],\label{eq:f1f2_branches}
\end{equation}
illustrated in Fig.~\ref{fig:f1f2branchcutsarticle}. Functions $f_1(z)$ and $f_2(z)$ are identical in the Euclidean space ($z<0$ and real), but different else where on the complex plane. However, when singularities of $D(p^2)$ are only allowed to exist in the timelike region and nowhere else, spectral functions $\rho(s)$ can be constructed with the assistance of Mellin transform.

First, consider the Mellin transform of the free-particle propagator. After defining $\zeta=-p^2=p_E^2$, we then have 
\begin{equation}
\int_{0}^{+\infty}d\zeta\dfrac{\zeta^{\beta-1}}{-\zeta-s+i\varepsilon}=-\pi s^{\beta-1}\csc(\pi\beta), \quad \mathrm{Re}\{\beta\}\in (0,1).\label{eq:Mellin_free_prop}
\end{equation}
The inverse transform is given by 
\begin{equation}
D(-\zeta)=\dfrac{-1}{2\pi is}\int_{c-i\infty}^{c+i\infty}d\beta\, \zeta^{-\beta}s^{\beta}\pi\csc(\pi\beta),\quad c\in (0,1).
\end{equation}
When $|\zeta|<1$, $\zeta^{-\beta}=\exp(-\beta\ln\zeta)\rightarrow 0$ when $\mathrm{Re}\{\beta\}\rightarrow-\infty$. Therefore, $D(-\zeta)$ is obtained with contributions from residuals of $-\beta\in\mathbf{Z}$ poles. In this case we have 
\begin{equation}
D(-\zeta)=\dfrac{-1}{s}\sum_{n=0}^{+\infty}(-\zeta/s)^{n}=\dfrac{1}{-\zeta-s}.
\end{equation}
When $|\zeta|>1$, $\zeta^{-\beta}=\exp(-\beta\ln\zeta)\rightarrow 0$ when $\mathrm{Re}\{\beta\}\rightarrow+\infty$. Notice in this case the contour direction is clockwise, resulting in an extra minus sign. Therefore we have 
\begin{equation}
D(-\zeta)=\dfrac{1}{s}\sum_{n=1}^{+\infty}(-s/\zeta)^n=\dfrac{1}{-\zeta-s}.
\end{equation}
We have seen that the Mellin transform successfully reconstructed the free-particle propagator by only sampling the spacelike region. This is done correctly by using the variable $\zeta=-p^2$ such that during the inverse transform, singularities are allowed to occur in the timelike region only. The importance of ensuring the correct positioning of singularities is more apparent when the propagator is more complicated than the free-particle propagator. As we will see in the next example.

In some scenarios the propagator functions can be written as hypergeometric functions:
\begin{equation}
D(p^2)=\dfrac{1}{\mu^2}~_2F_1\left(a,b;c;\dfrac{p^2}{\mu^2}\right),\label{eq:Dp2_2F1}
\end{equation}
where $\mu$ is a mass scale. Parameters $a,~b$ and $c$ cannot be arbitrary because only when $\mathrm{Re}\{c\}>\mathrm{Re}\{c\}>0$ does the integral representation given by Eq.~\eqref{eq:2F1_int_z1} converge.
This integral representation ensures the branch cut of $~_2F_1$ lays along $z>1$. Therefore it is the desired analytic continuation of the hypergeometric series given by Eq.~\eqref{eq:2F1_Taylor},
which is only convergent for $|z|<1$.

The Mellin transform for the kernel function of this integral representation is given by
\begin{equation}
\int_{0}^{+\infty}d\zeta\dfrac{\zeta^{\beta-1}}{(1+t\zeta/\mu^2)^a}=t^{-\beta}\mu^{2\beta}\dfrac{\Gamma(a-\beta)\Gamma(\beta)}{\Gamma(a)},\quad \mathrm{Re}\{\beta\}\in (0,\mathrm{Re}\{a\}).
\end{equation}
While
\begin{equation}
\int_{0}^{1}dt\,t^{b-\beta-1}(1-t)^{c-b-1}=\dfrac{\Gamma(b-\beta)\Gamma(c-b)}{\Gamma(c-\beta)},\quad \mathrm{Re}\{\beta\}<\mathrm{Re}\{b\},
\end{equation}
we have the following representation of Eq.~\eqref{eq:Dp2_2F1},
\begin{equation}
\dfrac{1}{\mu^2}~_2F_1\left(a,b;c;\dfrac{p^2}{\mu^2}\right)=\dfrac{1}{2\pi i\mu^2}\dfrac{\Gamma(c)}{\Gamma(a)\Gamma(b)}\int_{\tilde{c}-i\infty}^{\tilde{c}+i\infty}d\beta\,\left(\dfrac{-p^2}{\mu^2}\right)^{-\beta}\dfrac{\Gamma(a-\beta)\Gamma(b-\beta)}{\Gamma(c-\beta)}\Gamma(\beta),\label{eq:Dp2_inv_Mellin}
\end{equation}
with $\tilde{c}\in\left(0,\min\big\{\mathrm{Re}\{a\},~\mathrm{Re}\{b\} \big\}\right)$, and $\mathrm{Re}\{c\}>\mathrm{Re}\{b\}$. One can easily verify that when $|p^2/\mu^2|<1$,  Eq.~\eqref{eq:Dp2_inv_Mellin} reduces to Eq.~\eqref{eq:2F1_Taylor}. Meanwhile, since Eq.~\eqref{eq:Dp2_inv_Mellin} ensures the branch cut to be located on the positive real axis, it also agrees with Eq.~\eqref{eq:2F1_int_z1}.

When the spectral representation is applied to solve QFT equations for propagators, the question on how to reduce $p^2$ dependences of these equations into that of the free-particle propagator naturally arises. To answer this question, consider the following identity 
\begin{equation}
\int ds\dfrac{\Phi(s,s')}{p^2-s+i\varepsilon}=\dfrac{1}{s'}K(p^2,s'),\label{eq:Phi_K}
\end{equation}
where $K(p^2,s)$ is a dimensionless known function and $\Phi(s,s')$ is the distribution one tries to solve. 
After applying Eq.~\eqref{eq:Mellin_free_prop}, we have the Mellin transform of Eq.~\eqref{eq:Phi_K} as 
\begin{equation}
\dfrac{1}{2\pi i }\int_{c_0-i\infty}^{c_0+i\infty}d\beta\,\zeta^{-\beta}\int ds\,\Phi(s,s')s^{\beta-1}(-\pi)\csc(\pi\beta)=\dfrac{1}{2\pi i}\int_{c_1-i\infty}^{c_1+i\infty}d\beta\,\zeta^{-\beta}(s')^{\beta-1}\kappa(\beta),
\end{equation}
where 
\begin{equation}
(s')^{\beta}\kappa(\beta)=\int_{0}^{+\infty}d\zeta\,\zeta^{\beta-1}K(-\zeta,s').
\end{equation}
Therefore $\kappa(\beta)$ is recognized as the Mellin transform of $K(-\zeta,s')$. Eq.~\eqref{eq:Phi_K} is solved by
\begin{equation}
-\pi\csc(\pi\beta)\int ds\,\Phi(s,s')s^{\beta-1}=(s')^{\beta-1}\kappa(\beta),\label{eq:Mellin_Phi_s_sp}
\end{equation}
as long as the holomorphic region of $\kappa(\beta)$ overlaps with $\mathrm{Re}\{\beta\}\in(0,1)$. When this condition is satisfied, the Mellin transform of $\Phi(s,s')$ is obtained readily from Eq.~\eqref{eq:Mellin_Phi_s_sp}. Then through the inverse transform, Eq.~\eqref{eq:Mellin_Phi_s_sp} determines $\Phi(s,s')$.
	\section{Frobenius method for homogeneous equations\label{SS:Froenius}}
In this section of the appendix, define $x=(s-m^2)/m^2$. Eq \eqref{eq:ODE_f1} then becomes
\begin{equation}
\left[(x+1)\dfrac{d^2}{dx^2}+\dfrac{d}{dx}-\dfrac{a(a+1)}{x^2}\right]f_1=0.\label{eq:ODE_f1_Frobenius}
\end{equation}
Since Eq.~\eqref{eq:ODE_f1_Frobenius} is regularly singular when $x\rightarrow 0$, the following series expansions of solutions exist:
\begin{equation}
f_1(x)=\sum_{n=0}^{+\infty}a_nx^{n+r}.
\end{equation}
Straightforwardly, we have 
\begin{align}
& \dfrac{d}{dx}f_1(x)=\sum_{n=0}^{+\infty}(n+r)a_nx^{n+r-1},\\
& \dfrac{d^2}{dx^2}f_1(x)=\sum_{n=0}^{+\infty}(n+r)(n+r-1)a_nx^{n+r-2}.
\end{align}
Substituting these expansions into Eq.~\eqref{eq:ODE_f1_Frobenius} give
\begin{align}
& \quad \sum_{n=-1}^{+\infty}(n+r+1)(n+r)a_{n+1}x^{n+r}+\sum_{n=-2}^{+\infty}(n+r+1)(n+r+2)a_{n+2}x^{n+r}\nonumber\\
& +\sum_{n=-1}^{+\infty}(n+r+1)a_{n+1}x^{n+r}-a(a+1)\sum_{n=-2}^{+\infty}a_{n+2}x^{n+r}=0.\label{eq:ODE_f1_Frobenius_Series}
\end{align}
At the order of $x^r$, $(r-1)ra_0-a(a+1)a_0=0$.
The indicial equation,
\begin{equation}
(r+1)(r-a-1)=0,
\end{equation}
has two solutions $r=-a$ and $r=a+1$. 

Assuming $s\,r_1(s)$ is finite at $s=m^2$, given $g_2(s)=[m^2/(s-m^2)]^{a+1}$, the root is chosen to be $r=a+1$. In this case, the recurrence relation for $a_n$ is
\begin{equation}
a_n=\dfrac{(n+a)^2}{a(a+1)-(n+a)(n+a+1)}a_{n-1},\quad n\geq 1,
\end{equation}
which gives 
\begin{equation}
\dfrac{a_n}{a_0}=\dfrac{(-1)^n(a+1)((a+2)_{n-1})^2}{2(2)_{n-1}(2a+3)_{n-1}},\quad n\geq 0.
\end{equation}
Then we obtain the solution
\begin{equation}
f_1(x)=x^{a+1}\sum_{n=0}^{+\infty}a_nx^n=a_0x^{a+1}~_2F_1(a+1,a+1;2a+2;-x).\label{eq:f_1_homogeneous}
\end{equation}
	\chapter{Identities for loop integrals, hypergeometric functions, and Landau--Khalatnikov--Fradkin transformation}
		\section{Evaluating Loop Integrals in Minkowski Space \label{ss:loop_Minkowski}}	
	For a given loop integral in quantum field theory, after Feynman parameterization, one possible form of the integral is,
	\begin{equation}
	L_{0n}(\Delta,\epsilon)=\int d\underline{l}\dfrac{1}{(l^2-\Delta+i\varepsilon)^n},
	\end{equation}
	with $d\underline{l}\equiv d^d l/(2\pi)^d$. Here $\Delta$ is the mass function for the combined denominator, and $\varepsilon$ denotes the Feynman prescription for timelike integrals.
	\begin{figure}
	\centering
	\includegraphics[width=0.85\linewidth]{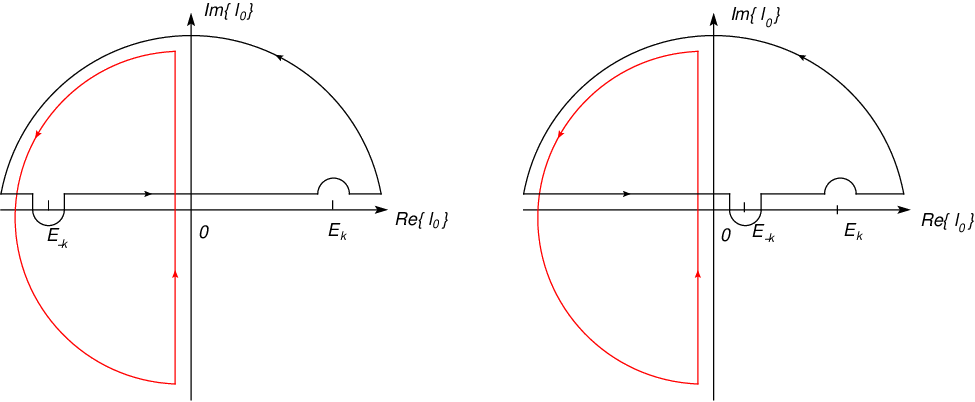}
	\caption{Illustrations of different scenarios for the loop integration in the time component.}
	\label{fig:wickrotation}
	\end{figure}
	The textbook version of evaluating $L_{0n}$ is to apply Wick rotation directly as $l_0=il_4$, then evaluate $L_{0n}$ using dimensional regularization (or other regularization schemes). We want to explore the possibility of evaluating loop integrals directly in Minkowski space without Wick rotation, while still employing dimensional regularization. 

	Since $l^2=l_0^2-\overrightarrow{l}^2=l_0^2-\overrightarrow{l}\cdot \overrightarrow{l}$, where $l_0$ is the time component of loop momentum while $\overrightarrow{l}$ represents all spatial components. The number of components described by $\overrightarrow{l}$ is related to the number of spacetime dimensions. We take the convention that dimensional regularization is only allowed to change spatial dimensions, leaving the time component alone. As illustrated on the left figure of Fig. \ref{fig:wickrotation}, when evaluating the contour for the time component of loop integral, the Feynman prescription tells us that when the contour is closed above, only the $l_0=-E_{\overrightarrow{l}}=-(\overrightarrow{l}^2+\Delta)^{1/2}$ pole is included, the residue of which is the result of the time integral. As expected, an identical result is obtained if instead the contour is closed from below, encircling the pole at $l_0=+E_{\overrightarrow{l}}$. Alternatively, one is allowed to close the contour from below and incorporate the $l_0=+E_{\overrightarrow{l}}$ pole, the result of which agrees with that obtained by closing the contour from above. 
	With Wick rotation $l_0=il_4$, one can easily verify that the contour for time integration is rotated $90^{\circ}$ counterclockwise around the origin, rendering the same pole encompassed in the contour as required by Feynman prescription for Minkowski space time integrals, therefore producing identical results. However, when the two poles locate on the same side of the imaginary $l_0$ axis, Wick rotation incorporates different poles than evaluating the time integral directly. In this scenario, the results with and without Wick rotation will be different. When the loop integral is spherical symmetric, these two poles always locate symmetrically about the imaginary $l_0$ axis.
	
	To see how to evaluate $L_{0n}$ in Minkowski space directly, consider its time integration first. Because contributions from the infinite radius arc vanish for large enough $n$, we have
	\begin{equation}
	\int dl_0\dfrac{1}{(l_0^2-E_{\overrightarrow{l}}^2)^n}=2\pi i~\mathrm{Res}_{l_0\rightarrow -E_{\overrightarrow{l}}}\dfrac{1}{(l_0+E_{\overrightarrow{l}})^n(l_0-E_{\overrightarrow{l}})^n},
	\end{equation}
	where $E_{\overrightarrow{l}}=(\overrightarrow{l}^2+\Delta)^{1/2}$. Next, since the order of the pole at $-E_{\overrightarrow{l}}$ is $n$,
	\begin{align}
	& \quad \mathrm{Res}_{l_0\rightarrow -E_{\overrightarrow{l}}}\dfrac{1}{(l_0+E_{\overrightarrow{l}})^n(l_0-E_{\overrightarrow{l}})^n}\nonumber\\
	& =\lim\limits_{l_0\rightarrow -E_{\overrightarrow{l}}}\dfrac{1}{(n-1)!}\left(\dfrac{d}{dl_0} \right)^{n-1}(l_0+E_{\overrightarrow{l}})^n\dfrac{1}{(l_0^2-E_{\overrightarrow{l}}^2)^n}\nonumber\\
	& =\lim\limits_{l_0\rightarrow -E_{\overrightarrow{l}}}\dfrac{1}{(n-1)!}\left(\dfrac{d}{dl_0} \right)^{n-1}(l_0-E_{\overrightarrow{l}})^{-n}\nonumber\\
	& =\lim\limits_{l_0\rightarrow -E_{\overrightarrow{l}}}\dfrac{(-1)^{n-1}}{(n-1)!}\dfrac{\Gamma(2n-1)}{\Gamma(n)}(l_0-E_{\overrightarrow{l}})^{-2n+1}\nonumber\\
	& =(-1)^{n}2^{-2n+1}\dfrac{\Gamma(2n-1)}{[\Gamma(n)]^2E_{\overrightarrow{l}}^{2n-1}}.
	\end{align}
	While for the spatial integration, dimensional regularization is applied such that 
	\[\int d\overrightarrow{l}=\int d\Omega_{d-1}\int_{0}^{+\infty}d|\overrightarrow{l}|~|\overrightarrow{l}|^{d-2},\]
	where $|\overrightarrow{l}|=\sqrt{\overrightarrow{l}^2}$ and for spherical symmetric kernels ${\int d\Omega_{d-1}=2\pi^{(d-1)/2}/\Gamma((d-1)/2)}$.
	Therefore
	\begin{align}
	L_{0n}(\Delta,\epsilon)& =\dfrac{1}{(2\pi)^d}\int d\overrightarrow{l}\int dl_0\dfrac{1}{(l_0^2-E_{\overrightarrow{l}}^2)^n} =\dfrac{2\pi i}{(2\pi)^d}\int d\overrightarrow{l}\dfrac{(-1)^n 2^{-2n+1}\Gamma(2n-1)}{[\Gamma(n)]^2\left(\overrightarrow{l}^2+\Delta\right)^{n-1/2}}\nonumber\\
	& =\dfrac{i(-1)^n2^{-2n+1}\Gamma(2n-1)}{(2\pi)^{d-1}[\Gamma(n)]^2}\dfrac{\pi^{(d-1)/2}}{\Gamma\left(\dfrac{d-1}{2} \right)}\int_{0}^{+\infty}d\overrightarrow{l}^2\dfrac{\left(\overrightarrow{l}^2\right)^{(d-3)/2}}{\left(\overrightarrow{l}^2+\Delta\right)^{n-1/2}}.
	\end{align} 
	Substituting $x=(\overrightarrow{l}^2/\Delta+1 )^{-1}$ for the integration variable, we have
	\begin{equation}
	L_{0n}(\Delta,\epsilon) =\dfrac{i(-1)^n2^{-2n+2-d}\Gamma(2n-1)}{\pi^{(d-1)/2}[\Gamma(n)]^2\Delta^{n-d/2}}\dfrac{1}{\Gamma\left(\dfrac{d-1}{2} \right)}\int_{0}^{1}dx~x^{n-d/2-1}(1-x)^{(d-1)/2-1}.
	\end{equation}
	This integral over $x$ is then just the Euler Beta function ${B(n-d/2,(d-1)/2)}$.
	Noting that
	\[\dfrac{\Gamma(2n-1)}{\Gamma(n)\Gamma(n-1/2)}=\dfrac{2^{2n-2}}{\sqrt{\pi}}, \]
	we arrive at
	\begin{equation}
	L_{0n}(\Delta,\epsilon)=\dfrac{i(-1)^n}{(4\pi)^{d/2}}\dfrac{\Gamma(n-d/2)}{\Gamma(n)\Delta^{n-d/2}},
	\end{equation}
	which agrees with (A.44) of Ref.~\cite{peskin1995introduction} for the Wick-rotated result.
	
	The more general integrals 
	\[{L_{mn}(\Delta,\epsilon)=\int d\underline{l}~l^{2m}/(l^2-\Delta+i\varepsilon)^n}\] 
	are determined by combinations of $L_{0r}(\Delta,\epsilon)$. Consequently, the result for $L_{mn}(\Delta,\epsilon)$ is 
	\begin{equation}
	L_{mn}(\Delta,\epsilon) =\dfrac{i(-1)^{n-m}}{(4\pi)^{d/2}}\dfrac{\Gamma(n-d/2-m)}{\Gamma(n)\Delta^{n-d/2-m}}\prod_{m'=1}^{m}\left(\dfrac{d}{2}+m'-1\right),\label{eq:L_mn_Delta_epsilon}
	\end{equation}
	with $m\geq 1$ and $n\geq m+1$ to ensure the convergence of the $l_0$ integral. This also agrees with Ref.~\cite{peskin1995introduction}. 
	
	While in the special case of $\Delta=0$, singularities of $l_0$ integrals are modified from the case of $\Delta\neq 0$. Therefore integrations for the massless case require a separate discussion, which is not needed in this article.
	
	To see how Eq.~\eqref{eq:L_mn_Delta_epsilon} is obtained, consider $m=1$ first. For the time integral, we can calculate the corresponding residue with a similar approach to that for $L_{0n}$. However, the $l^2$ term on the numerator makes the process of finding the residue cumbersome. To circumvent this difficulty, consider separating the numerator such that
	\begin{equation}
	L_{1,n}(\Delta,\epsilon)=\int d\underline{l}\dfrac{l^2}{(l^2-\Delta+i\varepsilon)^n}=\int d\underline{l}\dfrac{l^2-\Delta+\Delta}{(l^2-\Delta+i\varepsilon)^n}=L_{0,n-1}(\Delta,\epsilon)+\Delta L_{0,n}(\Delta,\epsilon).
	\end{equation}
	We have reduced our integral into those solved before. Next, applying properties of Gamma functions, we obtain
	\begin{align}
	L_{1,n}(\Delta,\epsilon)& =\dfrac{i(-1)^{n-1}}{(4\pi)^{d/2}}\dfrac{\Gamma(n-d/2-1)}{\Gamma(n-1)\Delta^{n-d/2-1}}+\dfrac{i(-1)^{n}}{(4\pi)^{d/2}}\dfrac{\Gamma(n-d/2)}{\Gamma(n)\Delta^{n-d/2-1}}\nonumber\\
	& =\dfrac{i(-1)^{n-1}}{(4\pi)^{d/2}}\dfrac{\Gamma(n-d/2-1)}{\Gamma(n)\Delta^{n-d/2-1}}\dfrac{d}{2},
	\end{align}
	which agrees with (A.45) of Ref.~\cite{peskin1995introduction}. In this case, the $l_0$ integral is convergent only if the integer $n\geq~2$. Otherwise contributions from the infinite radius arcs do not vanish.
	Applying the procedure of separating the numerator once more, one obtain
	\begin{equation}
	L_{2,n}(\Delta,\epsilon)=L_{1,n-1}(\Delta,\epsilon)+\Delta L_{1,n}(\Delta,\epsilon)=\dfrac{i(-1)^{n-2}}{(4\pi)^{d/2}}\dfrac{\Gamma(n-d/2-2)}{\Gamma(n)\Delta^{n-d/2-2}}\dfrac{d}{2}\left(\dfrac{d}{2}+1\right),
	\end{equation}
	which agrees with (A.47) of Ref.~\cite{peskin1995introduction}.
	\section{$\Xi_j(p^2,s)$ with $d=4-2\epsilon$ as hypergeometric functions\label{ss:Xi_12_epsilon}}	
	From the Euler type integral definition of hypergeometric functions \cite{abramowitz1964handbook}, 
	\begin{equation}
	\int_{0}^{1}dx~x^{b-1}(1-x)^{c-b-1}(1-zx)^{-a}=\dfrac{\Gamma(b)\Gamma(c-b)}{\Gamma(c)}~_2F_1(a,b;c;z),
	\end{equation}
	we express the following two integrals as hypergeometric functions,
	\begin{align}
	I_0(z,\epsilon)& \equiv \int_{0}^{1}dx~\dfrac{2x}{(1-x)^\epsilon (1-xz)^\epsilon}=\dfrac{2~_2F_1(\epsilon,2;3-\epsilon;z)}{(1-\epsilon)(2-\epsilon)},\\
	I_1(z,\epsilon)& \equiv \int_{0}^{1}dx~\dfrac{2x}{(1-x)^\epsilon (1-xz)^{1+\epsilon}}=\dfrac{2~_2F_1(\epsilon+1,2;3-\epsilon;z)}{(2-\epsilon)(1-\epsilon)}\nonumber\\
	& =\dfrac{-2}{(1-\epsilon)(z-1)}+ \dfrac{2[1-\epsilon(z+1)]}{(2-\epsilon)(1-\epsilon)(z-1)} ~_2F_1(1,1+\epsilon;3-\epsilon;z).
	\end{align}
	Applying this result to Eqs.~(\ref{eq:def_Xi_1},~\ref{eq:def_Xi_2}) gives
	\begin{align}
	\Xi_1(p^2,s)& =\Gamma(\epsilon)\left(\dfrac{4\pi\mu^2}{s} \right)^\epsilon [(1-\epsilon)I_0(z,\epsilon)+\epsilon I_1(z,\epsilon)]\nonumber\\
	\Xi_2(p^2,s)& =\Gamma(\epsilon)\left(\dfrac{4\pi\mu^2}{s} \right)^\epsilon\bigg\{(1-\epsilon)I_0(z,\epsilon)+\dfrac{\epsilon(z+1)}{2}I_1(z,\epsilon)\bigg\}.
	\end{align}
	Using Eq.~(15.2.10) in Ref.~\cite{abramowitz1964handbook}, with ${a=\epsilon+1}$, ${b=2,~c=3-\epsilon}$, we obtain,
	\begin{align*}
	& \quad (1-\epsilon)~_2F_1(\epsilon,2;3-\epsilon;z)\nonumber\\
	&=-\dfrac{\epsilon+1}{2}(z-1)~_2F_1(\epsilon+2,2;3-\epsilon;z)- \dfrac{1}{2}[3\epsilon-1+(1-\epsilon)z]~_2F_1(1+\epsilon,2;3-\epsilon;z).
	\end{align*}
	While applying Eqs.~(15.2.14,~15.2.17) with ${a=\epsilon+1}$, $b=2$ and $c=3-\epsilon$ respectively, we have, 
	\begin{equation*}
	2~_2F_1(\epsilon+1,3;3-\epsilon;z)=(1-\epsilon)~_2F_1(\epsilon+1,2;3-\epsilon;z) +(\epsilon+1)~_2F_1(\epsilon+2,2;3-\epsilon;z)
	\end{equation*}
	and
	\begin{equation*}
	(2-\epsilon)~_2F_1(\epsilon+1,2;2-\epsilon;z)= (1-2\epsilon)~_2F_1(\epsilon+1,2;3-\epsilon;z) +(\epsilon+1)~_2F_1(\epsilon+2,2;3-\epsilon;z).
	\end{equation*}
	Therefore
	\begin{align*}
	&\quad (1-\epsilon)~_2F_1(\epsilon,2;3-\epsilon;z)+\epsilon~_2F_1(\epsilon+1,2;\epsilon-2;z)\nonumber\\
	& =-\dfrac{\epsilon+1}{2}(z-1)~_2F_1(\epsilon+2,2;3-\epsilon;z) -\dfrac{(1-\epsilon)}{2}(z-1)~_2F_1(\epsilon+1,2,3-\epsilon,z),\\
	& =(1-z)~_2F_1(\epsilon+1,3;3-\epsilon;z),
	\end{align*}
	and
	\begin{align*}
	& \quad (1-\epsilon)~_2F_1(\epsilon,2;3-\epsilon;z)+\dfrac{\epsilon}{2}(z+1)~_2F_1(\epsilon+1,2;3-\epsilon;z)\nonumber\\
	& =-\dfrac{\epsilon+1}{2}(z-1)~_2F_1(\epsilon+2,2;3-\epsilon;z) -\dfrac{1-2\epsilon}{2}(z-1)~_2F_1(\epsilon+1,2;3-\epsilon;z)\\
	& =(1-z)\dfrac{2-\epsilon}{2}~_2F_1(\epsilon+1,2;2-\epsilon;z).
	\end{align*}
	Then the $z$ dependences of $\Xi_j/(p^2-s)$ combine as
	\begin{align}
	\dfrac{\Xi_1}{p^2-s}& =\dfrac{\Gamma(\epsilon)}{s(z-1)}\left(\dfrac{4\pi\mu^2}{s}\right)^\epsilon\dfrac{2}{(1-\epsilon)(2-\epsilon)}\bigg\{(1-\epsilon)~_2F_1(\epsilon,2;3-\epsilon;z)+\epsilon ~_2F_1(\epsilon+1,2;3-\epsilon;z)\bigg\}\nonumber\\
	& =\dfrac{\Gamma(\epsilon)}{s}\left(\dfrac{4\pi\mu^2}{s}\right)^\epsilon\dfrac{-2}{(1-\epsilon)(2-\epsilon)}~_2F_1(\epsilon+1,3;3-\epsilon;z)\\
	\dfrac{\Xi_2}{p^2-s}&=\dfrac{\Gamma(\epsilon)}{s(z-1)}\left(\dfrac{4\pi\mu^2}{s}\right)^\epsilon\dfrac{2}{(1-\epsilon)(2-\epsilon)} \bigg\{(1-\epsilon) ~_2F_1(\epsilon,2;3-\epsilon;z)\nonumber\\
	& \quad\hspace{7cm} +\dfrac{\epsilon(z+1)}{2} ~_2F_1(\epsilon+1,2;3-\epsilon;z)\bigg\}\nonumber\\
	& =\dfrac{\Gamma(\epsilon)}{s}\left(\dfrac{4\pi\mu^2}{s}\right)^\epsilon\dfrac{-1}{1-\epsilon}~_2F_1(\epsilon+1,2;2-\epsilon;z).
	\end{align}
	Using results in Abramowitz and Stegun \cite{abramowitz1964handbook} and Appendix \ref{ss:identities_2F1_epsilon}, one can verify that Eq.~\eqref{eq:Xi_1_reduced} and Eq.~\eqref{eq:Xi_2_reduced} reduce to results by the direct calculation of integrations over Feynman parameters after taking the ${\epsilon=1/2}$ limit and the $\epsilon\rightarrow 0$ expansion, respectively. 
	\section{Useful identities for hypergeometric functions $~_2F_1$\label{ss:identities_2F1}}
	\subsection{Definitions}
	We collect identities we have used from Abramowitz and Stegun \cite{abramowitz1964handbook}. The series definition of the hypergeometric function $~_2F_1$ is 
	\begin{equation}
	~_2F_1(a,b;c;z)=~_2F_1(a,b,c,z)=F(a,b;c;z)=\sum_{n=0}^{+\infty}\dfrac{(a)_n (b)_b}{(c)_n n!}z^n,\label{eq:2F1_Taylor}
	\end{equation}
	where $(a)_n$ is the Pochharmer symbol given by
	\begin{equation}
	(a)_n=a(a+1)(a+2)\dots (a+n-1)=\dfrac{\Gamma(a+n)}{\Gamma(a)}.
	\end{equation} 
	Additionally, the Gamma function definition of Pochharmer symbol applies even when $n$ is not an integer.
	\subsection{Identities for $_2F_1(a,b;c;z)$\label{ss:identities_2F1_AS}}
	Identities listed in this subsection are selected equations from Abramowitz and Stegun \cite{abramowitz1964handbook}. Equations numbered from the left are labeled by their the original numbers in Ref.~\cite{abramowitz1964handbook}.
	\paragraph{Special Elementary Cases of Gauss Series}
	\begin{align*}
	& (15.1.4)\quad F\left(\dfrac{1}{2},1;\dfrac{3}{2};z^2\right)=\dfrac{1}{2z}\ln\left(\dfrac{1+z}{1-z}\right)=\dfrac{\mathrm{arctanh}(z)}{z}\\
	& (15.1.5)\quad F\left(\dfrac{1}{2},1;\dfrac{3}{2};z^2\right)=\dfrac{\mathrm{arctan}(z)}{z}\\
	& (15.1.8)\quad F(a,b;b;z)=(1-z)^{-a}\\
	\end{align*}
	\paragraph{Differentiation Formulas}
	\begin{align*}
	& (15.2.3)\quad \dfrac{d^n}{dz^n}[z^{a+n-1}F(a,b;c;z)]=(a)_n z^{a-1}F(a+n,b;c;z)\\
	& (15.2.4)\quad \dfrac{d^n}{dz^n}[z^{c-1}F(a,b;c;z)]=(c-n)_n z^{c-n-1}F(a,b;c-n;z)\\
	\end{align*}
	\paragraph{Gauss' relations for contiguous functions}
	\begin{align*}
	& (15.2.10)\quad (c-a)F(a-1,b;c;z)+(2a-c-az+bz)F(a,b;c;z)+a(z-1)F(a+1,b;c;z)=0\\
	& (15.2.14)\quad (b-a)F(a,b;c;z)+aF(a+1,b;c;z)-bF(a,b+1;c;z)=0\\
	& (15.2.17)\quad (c-a-1)F(a,b;c;z)+aF(a+1,b;c;z)-(c-1)F(a,b;c-1;z)=0
	\end{align*}
	\paragraph{Integral Representations and Transformation Formulas}
	\begin{equation}
	(15.3.1)\quad F(a,b;c;z)=\dfrac{\Gamma(c)}{\Gamma(b)\Gamma(c-b)}\int_{0}^{1}dt~t^{b-1}(1-t)^{c-b-1}(1-tz)^{-a}\label{eq:2F1_int_z1}
	\end{equation}
	with $\mathrm{Re}\{c\}>\mathrm{Re}\{b\}>0$.
	\begin{align}
	(15.3.5)\quad F(a,b;c;z)& =(1-z)^{-b}F(b,c-a;c;z/(z-1))\nonumber\\
	(15.3.6)\quad F(a,b;c;z)& =\dfrac{\Gamma(c)\Gamma(c-a-b)}{\Gamma(c-a)\Gamma(c-b)}F(a,b;a+b-c+1;1-z)\label{eq:2F1_singular_z1}\\
	& \quad +(1-z)^{c-a-b}\dfrac{\Gamma(c)\Gamma(a+b+c)}{\Gamma(a)\Gamma(b)}F(c-a,c-b;c-a-b+1;1-z),\nonumber
	\end{align}
	with $c-a-b\notin \mathbf{N}$. When $b-c=m\in \mathbf{N}^*$, we have
	\begin{align*}
	(15.3.14)&\quad F(a,a+m;c;z) =F(a+m,a;c;z)\\
	& =\dfrac{\Gamma(c)(-z)^{-a-m}}{\Gamma(a+m)\Gamma(c-a)}\sum_{n=0}^{+\infty}\dfrac{(a)_{n+m}(1-c+a)_{n+m}}{n!(n+m)!}z^{-n}\big\{\ln(-z)\\
	& \quad +\psi(1+m+n)+\psi(1+n)-\psi(a+m+n)-\psi(c-a-m-n)\big\}\\
	& \quad+ (-z)^{-a}\dfrac{\Gamma(c)}{\Gamma(a+m)}\sum_{n=0}^{m-1}\dfrac{\Gamma(m-n)(a)_n}{n!\Gamma(c-a-n)}z^{-n}\\
	& \quad (\mathrm{for}~|\mathrm{arg}(-z)|<\pi,~|z|>1,~(c-a)\neq \mathbf{Z}).
	\end{align*}
	\section{Leading expansions on small parameters\label{ss:identities_2F1_epsilon}}
	The definition of derivatives on parameters is given by 
	\begin{equation}
	~_2F_1^{(l,m,n,0)}(\alpha,\beta;\gamma;z)\equiv\lim\limits_{(a,b,c)\rightarrow (\alpha,\beta,\gamma)}\dfrac{\partial^{l+m+n}}{\partial a^l\partial b^m\partial c^n}~_2F_1(a,b;c;z).
	\end{equation}
	For the purpose of calculating $\epsilon\rightarrow 0$ limits, only first order derivatives are required. One simple example that is relevant to the $\epsilon\rightarrow 0$ limit of the LKFT is
	\begin{equation}
	~_2F_1^{(1,0,0,0)}(1,n;n;z)=\lim\limits_{a\rightarrow 1}\dfrac{\partial}{\partial a}(1-z)^{-a}=-\dfrac{\ln(1-z)}{1-z}.
	\end{equation}
	
	A straightforward way to calculate these leading derivatives is to use the following series definition in Eq.~\eqref{eq:2F1_Taylor}.
	First, consider the derivative of the Pochhammer symbol
	\begin{equation}
	\dfrac{\partial}{\partial a}(a)_n=\dfrac{\partial}{\partial a}\dfrac{\Gamma(a+n)}{\Gamma(a)} =\dfrac{\Gamma(a+n)}{\Gamma(a)}\left[\dfrac{\partial}{\partial a}\ln\Gamma(a+n)-\dfrac{\partial}{\partial a}\ln\Gamma(a) \right]=(a)_n\left[\psi(a+n)-\psi(a)\right],
	\end{equation}
	where $\psi(z)=d\ln\Gamma(z)/dz$ is the digamma function, and
	\begin{equation}
	\psi(z+1)=\psi(z)+1/z.
	\end{equation}
	For integer $n$, $\psi(n)=H_{n-1}-\gamma_E$, where the harmonic number is defined by ${H_{n-1}=\sum_{m=1}^{n-1}\frac{1}{m}}$. Then
	\begin{align}
	\psi(a+n)-\psi(a)& =\dfrac{1}{a+n-1}+\dfrac{1}{a+n-2}+\dots +\dfrac{1}{a}\nonumber\\
	& =\sum_{m=0}^{n-1}\dfrac{1}{a+m},\quad \mathrm{for}~n\in \mathbf{N^*}.
	\end{align}
	
	In order to calculate ${~_2F_1^{(0,0,1,0)}(1,3;3;z)}$ and ${~_2F_1^{(0,0,1,0)}(1,2;2;z)}$, consider the following series expansion:
	\begin{equation}
	\lim\limits_{c\rightarrow b}\dfrac{\partial}{\partial c}~_2F_1(1,b;c;z) =\lim\limits_{c\rightarrow b}\dfrac{\partial}{\partial c}\sum_{n=0}^{+\infty}\dfrac{(b)_n}{(c)_n}z^n=\sum_{n=1}^{+\infty}[\psi(b)-\psi(b+n)]z^n.
	\end{equation}
	Then we have
	\begin{align}
	& ~_2F_1^{(0,0,1,0)}(1,3;3;z)=-\dfrac{z+z^2/2+\ln(1-z)}{z^2(z-1)}\\
	& ~_2F_1^{(0,0,1,0)}(1,2;2;z)=-\dfrac{z+\ln(1-z)}{z(z-1)},
	\end{align}
	from which we finally obtain
	\begin{align}
	&~_2F_1(1-\epsilon,3;3-\epsilon;z) =\dfrac{-1}{z-1}+\epsilon\left[\dfrac{\ln(1-z)}{z-1}+\dfrac{z+z^2/2+\ln(1-z)}{z^2(z-1)} \right]+\mathcal{O}(\epsilon^1)\\
	& ~_2F_1(1-\epsilon,2;2-\epsilon;z)=\dfrac{-1}{z-1}+\epsilon\left[\dfrac{\ln(1-z)}{z-1}+\dfrac{z+\ln(1-z)}{z(z-1)} \right]+\mathcal{O}(\epsilon^1).
	\end{align}
	\section{Example: the exponent-preserving effect of Eq.~\eqref{eq:linear_trans_LKFT_epsilon} \label{ss:Example_epn_preserving}}
	Operations constructed to generate $p^2$ dependences from the free-particle propagator using exponent-preserving linear transforms are free from operations on momentum variable $p^2$, an essential criterion for the application of spectral representation of propagators to solve the LKFT. If all operations are exponent-preserving on the variable $z=p^2/s$, after the integral variable transform $dz=-p^2s^{-2}ds$ there is no residual $p^2$ multiplication factors. This can be verified by the following example corresponding to the linear transform in Eq.~\eqref{eq:linear_trans_LKFT_epsilon}. 
	Explicitly, consider the operation 
	\begin{align}
	& \quad z^{\epsilon+1-n}D^\epsilon z^{n-1}D^\epsilon z^\epsilon\nonumber\\
	& =\dfrac{z^{\epsilon+1-n}}{\Gamma(1-\epsilon)}\dfrac{d}{dz}\int_{0}^{z}dz'(z-z')^{-\epsilon}\dfrac{(z')^{n-1}}{\Gamma(1-\epsilon)}\dfrac{d}{dz'}\int_{0}^{z'}dz''(z'-z'')^{-\epsilon}(z'')^\epsilon\nonumber\\
	& =\left(\dfrac{p^2}{s}\right)^{\epsilon+1-n}\dfrac{s^2}{\Gamma(1-\epsilon)p^2}\dfrac{d}{ds}\int_{s}^{+\infty}ds'\dfrac{p^2}{(s')^2}\left(\dfrac{p^2}{s}-\dfrac{p^2}{s'} \right)^{-\epsilon}\left(\dfrac{p^2}{s'} \right)^{n-1}\dfrac{(s')^2}{\Gamma(1-\epsilon)p^2}\nonumber\\
	&\quad\hspace{5cm} \quad  \times\dfrac{d}{ds'}\int_{s'}^{+\infty}ds''\dfrac{p^2}{(s'')^2}\left(\dfrac{p^2}{s'}-\dfrac{p^2}{s''}\right)^{-\epsilon}\left(\dfrac{p^2}{s''}\right)^\epsilon\nonumber\\
	&= \dfrac{s^{1+n-\epsilon}}{(\Gamma(1-\epsilon))^2}\dfrac{d}{ds}\int_{s}^{+\infty}ds'(s')^{1-n+\epsilon}\left(\dfrac{s'}{s}-1 \right)^{-\epsilon} \dfrac{d}{ds'}\int_{s'}^{+\infty}ds''(s'')^{-2}\left(\dfrac{s''}{s'}-1\right)^{-\epsilon},
	\end{align}
	which being exponent-preserving is independent of $p^2$.
	\section{Properties of the distribution $\mathcal{K}_1$ in 3D}
	\subsection{As the solution to its differential equation\label{ss:k1_3D_diff}}
	Apparently Eq.~\eqref{eq:test_dist_1_3D} reduces to a simple delta-function when $\xi=0$. To see Eq.~\eqref{eq:test_dist_1_3D} also satisfies its differential equation, namely Eq.~\eqref{eq:LKFT_k12} for $\mathcal{K}_j$ with $j=1$ and $\epsilon=1/2$, which is explicitly written as
	\begin{align}
	& \quad \dfrac{\partial}{\partial\xi}\int ds\dfrac{\mathcal{K}_1(s,s';\xi)}{p^2-s+i\epsilon}\nonumber\\
	& =\alpha\mu\int ds\bigg\{\dfrac{\sqrt{s}}{(p^2-s)^2}-\dfrac{\sqrt{s}}{2p^2(p^2-s)}-\dfrac{1}{2(p^2)^{3/2}}\mathrm{arctanh}(\sqrt{p^2/s}) \bigg\}\mathcal{K}_1(s,s';\xi),\label{eq:LKFT_fermion_3D_k1}
	\end{align}
	we start with the following helpful relations, 
	\begin{equation}
	\int_{s_{th}}^{+\infty}ds\dfrac{1}{(p^2-s)s^{3/2}}=\dfrac{2}{(p^2)^{3/2}}\left[\sqrt{p^2/s_{th}}-\mathrm{arctanh}\sqrt{p^2/s_{th}} \right],
	\end{equation}
	and
	\begin{equation} \int_{s_{th}}^{+\infty}ds\left[\dfrac{1}{\sqrt{s}}-\dfrac{1}{\sqrt{p^2}}\mathrm{arctanh}\sqrt{\dfrac{p^2}{s}} \right]\dfrac{1}{s^{3/2}}=\dfrac{1}{s_{th}}-\dfrac{2}{\sqrt{s_{th}p^2}}\mathrm{arctanh}\sqrt{\dfrac{p^2}{s_{th}}}-\dfrac{1}{p^2}\ln\left(1-\dfrac{p^2}{s_{th}}\right),
	\end{equation}
	where $s_{th}=(\sqrt{s'}+\alpha\mu\xi/2)^2$. Next, applying Eq.~\eqref{eq:test_dist_1_3D} and writing $1+\frac{\alpha\mu\xi}{2\sqrt{s'}}=\frac{\sqrt{s_{th}}}{\sqrt{s'}}$ produce
	\begin{align}
	& \quad \int ds\left[\dfrac{\sqrt{s}}{(p^2-s)^2}-\dfrac{\sqrt{s}}{2p^2(p^2-s)}-\dfrac{1}{2(p^2)^{3/2}}\mathrm{arctanh}(\sqrt{p^2/s}) \right]\mathcal{K}_1(s,s';\xi)\nonumber\\
	& =\left(1+\dfrac{\alpha\mu\xi}{2\sqrt{s'}}\right)^{-1}\left[\dfrac{\sqrt{s_{th}}}{(p^2-s_{th})^2}-\dfrac{1}{2\sqrt{s_{th}}(p^2-s_{th})}+\dfrac{1}{2p^2}\left(\dfrac{1}{\sqrt{s_{th}}}-\dfrac{1}{\sqrt{p^2}}\mathrm{arctanh}\sqrt{\dfrac{p^2}{s_{th}}} \right) \right]\nonumber\\
	&\quad +\dfrac{\alpha\mu\xi}{4}\Bigg\{\dfrac{-1}{s_{th}(p^2-s_{th})}+\dfrac{1}{s_{th}p^2}+\dfrac{1}{p^4}\ln\left(1-\dfrac{p^2}{s_{th}}\right)-\dfrac{1}{2}\left[\dfrac{1}{s_{th}p^2}+\dfrac{1}{p^4}\ln\left(1-\dfrac{p^2}{s_{th}}\right) \right]\nonumber\\
	&\quad  +\dfrac{1}{2p^2}\left[\dfrac{1}{s_{th}}-\dfrac{1}{p^2}\ln\left(1-\dfrac{p^2}{s_{th}} \right)-\dfrac{2}{\sqrt{s_{th}p^2}}\mathrm{arctanh}\sqrt{\dfrac{p^2}{s_{th}}} \right] \Bigg\}\nonumber\\
	& =\dfrac{\sqrt{s'}}{(p^2-s_{th})^2}-\dfrac{1}{2\sqrt{s_{th}}(p^2-s_{th})}+\dfrac{1}{2\sqrt{s_{th}}p^2}-\dfrac{1}{2(p^2)^{3/2}}\mathrm{arctanh}\sqrt{\dfrac{p^2}{s_{th}}}.\label{eq:rhs_LKFT_fermion_3D_k1}
	\end{align}
	Meanwhile, since $\frac{\partial}{\partial\xi}\sqrt{s_{th}}=\frac{\alpha\mu}{2}$,
	\begin{align}
	&\quad \dfrac{\partial}{\partial\xi}\int ds\dfrac{1}{p^2-s}\mathcal{K}_1(s,s';\xi)=\dfrac{\partial}{\partial\xi}\left[\dfrac{\sqrt{s'}}{\sqrt{s_{th}}(p^2-s_{th})}+\dfrac{\alpha\mu\xi}{2p^2}\left(\dfrac{1}{\sqrt{s_{th}}}-\dfrac{1}{p^2}\mathrm{arctanh}\sqrt{\dfrac{p^2}{s_{th}}} \right) \right]\nonumber\\
	& =-\dfrac{\alpha\mu\sqrt{s'}}{2s_{th}(p^2-s_{th})}+\dfrac{\sqrt{s'}}{\sqrt{s_{th}}}\dfrac{\alpha\mu\sqrt{s_{th}}}{(p^2-s_{th})^2} +\dfrac{\alpha\mu}{2p^2}\left(\dfrac{1}{\sqrt{s_{th}}}-\dfrac{1}{\sqrt{p^2}}\mathrm{arctanh}\sqrt{\dfrac{p^2}{s_{th}}} \right)\nonumber\\
	& \hspace{7cm} -\dfrac{\alpha\mu\xi}{2p^2}\left(\dfrac{\alpha\mu}{2s_{th}}-\dfrac{1}{\sqrt{p^2}}\dfrac{\sqrt{p^2}\dfrac{\alpha\mu}{2s_{th}}}{1-\dfrac{p^2}{s_{th}}} \right)\nonumber\\
	& =\alpha\mu\left[\dfrac{\sqrt{s'}}{(p^2-s_{th})^2}-\dfrac{1}{2\sqrt{s_{th}}(p^2-s_{th})}+\dfrac{1}{2p^2\sqrt{s_{th}}}-\dfrac{1}{2(p^2)^{3/2}}\mathrm{arctanh}\sqrt{\dfrac{p^2}{s_{th}}} \right].\label{eq:lhs_LKFT_fermion_3D_k1}
	\end{align}
	The combination of Eq.~\eqref{eq:lhs_LKFT_fermion_3D_k1} with Eq.~\eqref{eq:rhs_LKFT_fermion_3D_k1} explicitly shows that $\mathcal{K}_1(s,s';\xi)$ given by Eq.~\eqref{eq:test_dist_1_3D} indeed satisfies Eq.~\eqref{eq:LKFT_fermion_3D_k1}. 
	\subsection{The closure property\label{ss:k1_3D_closure}}
	While for the group $\mathbf{K}$ defined by Eq.~\eqref{eq:test_dist_1_3D},
	\begin{align}
	& \quad \int ds'\mathcal{K}_1(s,s';\xi)\mathcal{K}_1(s',s'';\xi')\nonumber\\
	& =\int ds'\Bigg\{ \left(1+\dfrac{\alpha\mu\xi}{2\sqrt{s'}}\right)^{-1}\delta\left(s-\left(\sqrt{s'}+\dfrac{\alpha\mu\xi}{2}\right)^2 \right)\left(1+\dfrac{\alpha\mu\xi'}{2\sqrt{s''}}\right)^{-1}\delta\left(s'-\left(\sqrt{s''}+\dfrac{\alpha\mu\xi'}{2} \right)^2 \right) \nonumber\\
	& \quad +\left(1+\dfrac{\alpha\mu\xi}{2\sqrt{s'}}\right)^{-1}\delta\left(s-\left(\sqrt{s'}+\dfrac{\alpha\mu\xi}{2}\right)^2 \right)\dfrac{\alpha\mu\xi'}{4(s')^{3/2}}\theta\left(s'-\left(\sqrt{s''}+\dfrac{\alpha\mu\xi'}{2} \right)^2 \right)\nonumber\\
	& \quad +\dfrac{\alpha\mu\xi}{4s^{3/2}}\theta\left(s-\left(\sqrt{s'}+\dfrac{\alpha\mu\xi}{2}\right)^2 \right)\left(1+\dfrac{\alpha\mu\xi'}{2\sqrt{s''}}\right)^{-1}\delta\left(s'-\left(\sqrt{s''}+\dfrac{\alpha\mu\xi'}{2} \right)^2 \right)\nonumber\\
	& \quad +\dfrac{\alpha\mu\xi}{4s^{3/2}}\theta\left(s-\left(\sqrt{s'}+\dfrac{\alpha\mu\xi}{2}\right)^2 \right)\dfrac{\alpha\mu\xi'}{4(s')^{3/2}}\theta\left(s'-\left(\sqrt{s''}+\dfrac{\alpha\mu\xi'}{2} \right)^2 \right) \Bigg\}.\label{eq:Group_k1_3D}
	\end{align}
	Integrals for the first and third terms on the right-hand side of Eq.~\eqref{eq:Group_k1_3D} are obvious. While for the second term, since $\sqrt{s'}>0$ and $\sqrt{s}>\alpha\mu\xi/2$
	\begin{equation}
	\delta\left(s-\left(\sqrt{s'}+\dfrac{\alpha\mu\xi}{2}\right)^2 \right)=\left(1+\dfrac{\alpha\mu\xi}{2\sqrt{s'}}\right)^{-1}\delta\left(s'-\left(\sqrt{s}-\dfrac{\alpha\mu\xi}{2}\right)^2 \right),
	\end{equation}
	and the theta-function is not zero only when $\sqrt{s'}\geq \sqrt{s''}+\alpha\mu\xi'/2$. Therefore 
	\begin{align}
	& \quad \int ds'\left(1+\dfrac{\alpha\mu\xi}{2\sqrt{s'}}\right)^{-1}\delta\left(s-\left(\sqrt{s'}+\dfrac{\alpha\mu\xi}{2}\right)^2 \right)\dfrac{\alpha\mu\xi'}{4(s')^{3/2}}\,\theta\left(s'-\left(\sqrt{s''}+\dfrac{\alpha\mu\xi'}{2} \right)^2 \right)\nonumber\\
	& =\dfrac{\alpha\mu\xi'}{4s^{3/2}\left(1-\dfrac{\alpha\mu\xi}{2\sqrt{s}} \right)}\,\theta\left(s-\left[\sqrt{s''}+\dfrac{\alpha\mu}{2}(\xi+\xi') \right]^2 \right).
	\end{align}
	For the fourth term, two theta-functions overlap only if $s\geq [\sqrt{s''}+\alpha\mu(\xi+\xi')/2]^2$. Then
	\begin{align}
	& \quad \dfrac{\alpha\mu\xi}{4s^{3/2}}\,\theta\left(s-\left(\sqrt{s'}+\dfrac{\alpha\mu\xi}{2}\right)^2 \right)\dfrac{\alpha\mu\xi'}{4(s')^{3/2}}\,\theta\left(s'-\left(\sqrt{s''}+\dfrac{\alpha\mu\xi'}{2} \right)^2 \right)\nonumber\\
	& =\theta\left(s-\left[\sqrt{s''}+\dfrac{\alpha\mu}{2}(\xi+\xi') \right]^2 \right)\int_{(\sqrt{s''}+\alpha\mu\xi'/2)^2}^{(\sqrt{s}-\alpha\mu\xi/2)^2}ds'\dfrac{(\alpha\mu)^2\xi\xi'}{16(ss')^{3/2}}\nonumber\\
	& =-\dfrac{\xi\xi'(\alpha\mu)^2}{8s^{3/2}}\left[\left(\sqrt{s}-\dfrac{\alpha\mu\xi}{2}\right)^{-1} -\left(\sqrt{s''}+\dfrac{\alpha\mu\xi'}{2}\right)^{-1}\right].
	\end{align}
	Therefore in the end, we obtain
	\begin{align}
	& \quad \int ds'\mathcal{K}_1(s,s';\xi)\mathcal{K}_1(s',s'';\xi')\nonumber\\
	& =\left[1+\dfrac{\alpha\mu}{2\sqrt{s''}}\left(\xi+\xi'\right) \right]^{-1}\delta\left(s-\left[\sqrt{s''}+\dfrac{\alpha\mu}{2}(\xi+\xi') \right]^2 \right)\nonumber\\
	& \quad +\theta\left(s-\left[\sqrt{s''}+\dfrac{\alpha\mu}{2}(\xi+\xi') \right]^2 \right) \Bigg\{\dfrac{\alpha\mu\xi'}{4s^{3/2}}\left(1-\dfrac{\alpha\mu\xi}{2\sqrt{s}} \right)^{-1}+\dfrac{\alpha\mu\xi}{4s^{3/2}}\left(1+\dfrac{\alpha\mu\xi'}{2\sqrt{s''}} \right)^{-1}\nonumber\\
	& \quad{\hspace{5cm}}  -\dfrac{\xi\xi'(\alpha\mu)^2}{8s^{3/2}}\left[\left(\sqrt{s}-\dfrac{\alpha\mu\xi}{2}\right)^{-1} -\left(\sqrt{s''}+\dfrac{\alpha\mu\xi'}{2}\right)^{-1}\right] \Bigg\}\nonumber\\
	& =\left[1+\dfrac{\alpha\mu}{2\sqrt{s''}}\left(\xi+\xi'\right) \right]^{-1}\delta\left(s-\left[\sqrt{s''}+\dfrac{\alpha\mu}{2}(\xi+\xi') \right]^2 \right) \nonumber\\
	&\quad{\hspace{5cm}}+\dfrac{\alpha\mu(\xi+\xi')}{4s^{3/2}}\,\theta\left(s-\left[\sqrt{s''}+\dfrac{\alpha\mu}{2}(\xi+\xi') \right]^2 \right)\nonumber\\
	& =\mathcal{K}_1(s,s'';\xi+\xi').
	\end{align}
	So $\mathbf{K}$ defined by $\mathcal{K}_1(s,s';\xi)$ given by Eq.~\eqref{eq:test_dist_1_3D} satisfies the closure property of a group.
	\chapter{Known contributions to the fermion propagator SDE}
	\section{Simplification of $\sigma_j^\xi(p^2;\xi)$\label{ss:simplify_sigma_xi}}
To simplify Eq.~\eqref{eq:sigma_overline_delta_loop}, we will need contiguous relations for hypergeometric functions from Ref.~\cite{abramowitz1964handbook} and the following identity
\begin{equation}
(A\slashed{p}+BW)(\slashed{p}+W)=s(Az+B)+(A+B)\slashed{p}W
=s(z-1)\left[\left(A+\dfrac{A+B}{z-1} \right)+\dfrac{A+B}{s(z-1)}\slashed{p}W\right].
\end{equation}
Equations referred to by Eq.~(15.2.XX) are identities in Ref.~\cite{abramowitz1964handbook}. With ${ a=3,~b=1+\epsilon,~c=4-\epsilon}$, Eq.~(15.2.19) becomes
\begin{align}
\dfrac{2(3-2\epsilon)}{(3-\epsilon)(2-\epsilon)(1-\epsilon)}~_2F_1(\epsilon,3;4-\epsilon;z)\ & =\dfrac{2}{(3-\epsilon)(2-\epsilon)}~_2F_1(2,1+\epsilon;4-\epsilon;z)\nonumber\\
& \quad+\dfrac{2(1-z)}{(3-\epsilon)(1-\epsilon)}~_2F_1(3,1+\epsilon;4-\epsilon;z).\label{eq:15.2.19.1}
\end{align}
Explicitly then,
\begin{align}
A & =\dfrac{-\epsilon}{(3-\epsilon)(2-\epsilon)}~_2F_1(1+\epsilon,2;4-\epsilon;z)+  \dfrac{3\epsilon-4}{(2-\epsilon)(1-\epsilon)}~_2F_1(\epsilon,2;3-\epsilon;z) \nonumber\\
& \quad +\dfrac{2(3-2\epsilon)}{(3-\epsilon)(2-\epsilon)(1-\epsilon)}~_2F_1(\epsilon,3;4-\epsilon;z) \nonumber\\
& =\dfrac{2(1-z)}{(3-\epsilon)(1-\epsilon)}~_2F_1(1+\epsilon,3;4-\epsilon;z)+\dfrac{1}{3-\epsilon}~_2F_1(1+\epsilon,2;4-\epsilon;z) \label{eq:15.2.19.1.a}\\
& \quad +\dfrac{3\epsilon-4}{(2-\epsilon)(1-\epsilon)}~_2F_1(\epsilon,2;3-\epsilon;z),\nonumber
\end{align}
where Eq.~\eqref{eq:15.2.19.1} is used to derive Eq.~\eqref{eq:15.2.19.1.a}.
From Eq.~(15.2.17) with $a=1,~b=\epsilon,~c=3-\epsilon$ we have
\begin{equation}
B =\dfrac{1}{1-\epsilon}~_2F_1(\epsilon,1;2-\epsilon;z)=\dfrac{1}{2-\epsilon}~_2F_1(\epsilon,1;3-\epsilon;z)+ \dfrac{1}{(2-\epsilon)(1-\epsilon)}~_2F_1(\epsilon,2;3-\epsilon;z).
\end{equation}
Next, with ${a=\epsilon,~b=2,~c=3-\epsilon}$, Eq.~(15.2.15) becomes
\begin{equation}
\dfrac{1}{2-\epsilon}~_2F_1(\epsilon,1;3-\epsilon;z) =\dfrac{1-2\epsilon}{(2-\epsilon)(1-\epsilon)}~_2F_1(2,\epsilon;3-\epsilon;z)+ \dfrac{\epsilon(1-z)}{(2-\epsilon)(1-\epsilon)}~_2F_1(2,\epsilon+1;3-\epsilon;z).\label{eq:15.2.15.1}
\end{equation}
With ${a=\epsilon,~b=3,~c=4-\epsilon}$, Eq.~(15.2.17) becomes
\begin{equation}
\dfrac{-1}{1-\epsilon}~_2F_1(\epsilon,2;3-\epsilon;z) =\dfrac{-(3-2\epsilon)}{(3-\epsilon)(1-\epsilon)}~_2F_1(\epsilon,2;4-\epsilon;z)+ \dfrac{-\epsilon}{(3-\epsilon)(1-\epsilon)}~_2F_1(\epsilon+1,2;4-\epsilon;z). \label{eq:15.2.17.1}
\end{equation}
With ${a=2,~b=1+\epsilon,~c=4-\epsilon}$, Eq.~(15.2.15) becomes
\begin{equation}
-(3-2\epsilon)~_2F_1(\epsilon,2;4-\epsilon;z)+ (1-2\epsilon)~_2F_1(\epsilon+1,2;4-\epsilon;z)=-2(1-z)~_2F_1(\epsilon+1,3;4-\epsilon;z).\label{eq:15.2.15.2}
\end{equation}
Therefore
\begin{subequations}
\begin{align}
& \quad A+B \nonumber\\
& =\dfrac{2(1-z)}{(3-\epsilon)(1-\epsilon)}~_2F_1(1+\epsilon,3;4-\epsilon;z)+\dfrac{\epsilon(1-z)}{(2-\epsilon)(1-\epsilon)}~_2F_1(\epsilon+1,2;3-\epsilon;z) \label{eq:15.2.15.1.a}\\
& \quad + \dfrac{-1}{1-\epsilon}~_2F_1(\epsilon,2;3-\epsilon;z)+ \dfrac{1}{3-\epsilon}~_2F_1(1+\epsilon,2;4-\epsilon;z) \nonumber\\
& =\dfrac{2(1-\epsilon)}{(3-\epsilon)(1-\epsilon)}~_2F_1(1+\epsilon,3;4-\epsilon;z)+  \dfrac{\epsilon(1-z)}{(2-\epsilon)(1-\epsilon)}~_2F_1(\epsilon+1,2;3-\epsilon;z) \label{eq:15.2.17.1.a}\\
& \quad +\dfrac{-(3-2\epsilon)}{(3-\epsilon)(1-\epsilon)}~_2F_1(\epsilon,2;4-\epsilon;z)+ \dfrac{1-2\epsilon}{(3-\epsilon)(1-\epsilon)}~_2F_1(\epsilon+1,2;4-\epsilon;z) \nonumber\\
& =\dfrac{\epsilon(1-z)}{(2-\epsilon)(1-\epsilon)}~_2F_1(\epsilon+1,2;3-\epsilon;z), \label{eq:15.2.15.2.a}
\end{align}
\end{subequations}
where Eqs.~(\ref{eq:15.2.15.1},~\ref{eq:15.2.17.1},~\ref{eq:15.2.15.2}) are used to derive Eqs.~(\ref{eq:15.2.15.1.a},~\ref{eq:15.2.17.1.a},~\ref{eq:15.2.15.2.a}), respectively. In addition, with ${a=\epsilon,~b=2,~c=4-\epsilon}$, Eq.~(15.2.14) becomes
\begin{equation}
\epsilon ~_2F_1(\epsilon+1,2;4-\epsilon;z)=2~_2F_1(\epsilon,3;4-\epsilon;z)- (2-\epsilon)~_2F_1(\epsilon,2;4-\epsilon;z).\label{eq:15.2.14.1}
\end{equation}
With ${a=\epsilon,~b=2,~c=3-\epsilon}$, Eq.~(15.2.14) becomes
\begin{equation}
\epsilon ~_2F_1(\epsilon+1,2;3-\epsilon;z) =2~_2F_1(\epsilon,3;3-\epsilon;z)- (2-\epsilon)~_2F_1(\epsilon,2;4-\epsilon;z).\label{eq:15.2.14.2}
\end{equation}
With ${a=\epsilon,~b=2,~c=4-\epsilon}$, Eq.~(15.2.24) becomes
\begin{equation}
(1-\epsilon)~_2F_1(\epsilon,2;4-\epsilon;z)+ 2~_2F_1(\epsilon,3;4-\epsilon;z)=(3-\epsilon)~_2F_1(\epsilon,2;3-\epsilon;z).\label{eq:15.2.24.1}
\end{equation}
With ${a=\epsilon,~b=2,~c=3-\epsilon}$, Eq.~(15.2.24) becomes
\begin{equation}
-\epsilon ~_2F_1(\epsilon,2;3-\epsilon;z)+2~_2F_1(\epsilon,2;3-\epsilon;z)=(2-\epsilon)~_2F_1(\epsilon,2;2-\epsilon;z).\label{eq:15.2.24.2}
\end{equation}
Then
\begin{subequations}
\begin{align}
A + \frac{A+B}{(z-1)}& =\dfrac{-\epsilon}{(3-\epsilon)(2-\epsilon)}~_2F_1(1+\epsilon,2;4-\epsilon;z)+\dfrac{3\epsilon-4}{(2-\epsilon)(1-\epsilon)}~_2F_1(\epsilon,2;3-\epsilon;z)+ \nonumber\\
& \quad \dfrac{2(3-2\epsilon)}{(3-\epsilon)(2-\epsilon)(1-\epsilon)}~_2F_1(\epsilon,3;4-\epsilon;z)+  \dfrac{-\epsilon}{(2-\epsilon)(1-\epsilon)}~_2F_1(\epsilon+1,2;3-\epsilon;z) \nonumber\\
& =\dfrac{1}{(3-\epsilon)(2-\epsilon)}\bigg\{\dfrac{2(2-\epsilon)}{1-\epsilon}~_2F_1(\epsilon,3;4-\epsilon;z)+(2-\epsilon)~_2F_1(\epsilon,2;4-\epsilon;z)\bigg\} \label{eq:15.2.14.1a}\\
& \quad +\dfrac{1}{(2-\epsilon)(1-\epsilon)}\big\{-2 ~_2F_1(\epsilon,3;3-\epsilon;z)+ 2(\epsilon-1)~_2F_1(\epsilon,2;3-\epsilon;z)\big\}\nonumber\\
& = \dfrac{2}{(3-\epsilon)(1-\epsilon)}~_2F_1(\epsilon,3;4-\epsilon;z)+ \dfrac{1}{3-\epsilon}~_2F_1(\epsilon,2;4-\epsilon;z)\label{eq:15.2.14.2a}\\
& \quad+\dfrac{-2}{(2-\epsilon)(1-\epsilon)}~_2F_1(\epsilon,3;3-\epsilon;z)+ \dfrac{-2}{2-\epsilon}~_2F_1(\epsilon,2;3-\epsilon;z) \nonumber\\
& =\dfrac{\epsilon}{(1-\epsilon)(2-\epsilon)}~_2F_1(\epsilon,2;3-\epsilon;z)+ \dfrac{-2}{(2-\epsilon)(1-\epsilon)}~_2F_1(\epsilon,3;3-\epsilon;z) \label{eq:15.2.24.1a}\\
& =\dfrac{-1}{1-\epsilon}~_2F_1(\epsilon,2;2-\epsilon;z), \label{eq:15.2.24.2a}
\end{align}
\end{subequations}
where Eqs.~(\ref{eq:15.2.14.1},~\ref{eq:15.2.14.2},~\ref{eq:15.2.24.1},~\ref{eq:15.2.24.2}) have been utilized to derive Eqs.~(\ref{eq:15.2.14.1a},~\ref{eq:15.2.14.2a},~\ref{eq:15.2.24.1a},~\ref{eq:15.2.24.2a}), respectively.
Finally we obtain,
\begin{align}
ie^2\xi\int d\underline{k}\slashed{q}\dfrac{1}{\slashed{k}-W}\dfrac{\slashed{q}}{q^4}\dfrac{1}{\slashed{p}-W}&=\dfrac{-\alpha\xi}{4\pi}\Gamma(\epsilon)\left(\dfrac{4\pi\mu^2}{s} \right)^\epsilon \Bigg\{\dfrac{-1}{1-\epsilon}~_2F_1(\epsilon,2;2-\epsilon;z)\nonumber\\
& \quad\hspace{2cm} +\dfrac{-\epsilon}{(2-\epsilon)(1-\epsilon)}~_2F_1(\epsilon+1,2;3-\epsilon;z)\dfrac{\slashed{p}}{W} \Bigg\}.
\end{align}
\section{Operations on $z^\beta$ from terms in Eq.~\eqref{eq:D_xi_Omega_Delta}\label{ss:recurrence_z_beta}}
For commutators on the right-hand side of Eq.~\eqref{eq:D_xi_Omega_Delta}, explicit calculation shows that
\begin{subequations}\label{eq:zbeta_phi_theta}
\begin{align}
& z^\beta \phi_3\theta_1=\dfrac{\Gamma(\epsilon)[\Gamma(1-\epsilon)]^2}{\Gamma(1+\epsilon)}\left(\dfrac{4\pi\mu^2}{p^2} \right)^{2\epsilon} \dfrac{\beta+1}{\beta+2\epsilon} \dfrac{\Gamma(\beta+2\epsilon)\Gamma(\beta+\epsilon+1)}{\Gamma(\beta-\epsilon+2)\Gamma(\beta)}z^{\beta+2\epsilon},\\[1mm]
& z^\beta \theta_1\phi_3=\dfrac{\Gamma(\epsilon)[\Gamma(1-\epsilon)]^2}{\Gamma(1+\epsilon)}\left(\dfrac{4\pi\mu^2}{p^2} \right)^{2\epsilon} \dfrac{1+\beta+\epsilon}{\beta+1} \dfrac{\Gamma(\beta+2\epsilon)\Gamma(\beta+\epsilon+1)}{\Gamma(\beta-\epsilon+2)\Gamma(\beta)}z^{\beta+2\epsilon},\\[1mm]
& z^\beta \phi_2\theta_2=\dfrac{\Gamma(\epsilon)[\Gamma(1-\epsilon)]^2}{\Gamma(1+\epsilon)}\left(\dfrac{4\pi\mu^2}{p^2} \right)^{2\epsilon} \dfrac{1}{\beta+1} \dfrac{\Gamma(\beta+2\epsilon)\Gamma(\beta+\epsilon+1)}{\Gamma(\beta-\epsilon+1)\Gamma(\beta)}z^{\beta+2\epsilon},\\[1mm]
& z^\beta \theta_2\phi_2=\dfrac{\Gamma(\epsilon)[\Gamma(1-\epsilon)]^2}{\Gamma(1+\epsilon)}\left(\dfrac{4\pi\mu^2}{p^2} \right)^{2\epsilon} \dfrac{1}{1+\beta-\epsilon} \dfrac{\Gamma(\beta+2\epsilon)\Gamma(\beta+\epsilon+1)}{\Gamma(\beta-\epsilon+1)\Gamma(\beta)}z^{\beta+2\epsilon},
\end{align}
\end{subequations}
and
\begin{subequations}\label{eq:zbeta_phi}
\begin{align}
z^\beta(\phi_3z-z\phi_3)& =\Gamma(\epsilon)\left(\dfrac{4\pi\mu^2}{p^2} \right)^\epsilon \dfrac{\Gamma(1-\epsilon)}{\Gamma(1+\epsilon)}\left[\dfrac{\Gamma(2+\beta)\Gamma(\beta+\epsilon)}{\Gamma(2+\beta-\epsilon)\Gamma(\beta)}-\dfrac{\Gamma(2+\beta)\Gamma(\beta+1+\epsilon)}{\Gamma(2+\beta-\epsilon)\Gamma(\beta+1)} \right]z^{\beta+\epsilon+1}\nonumber\\[1mm]
& =-\Gamma(1-\epsilon)\left(\dfrac{4\pi\mu^2}{p^2} \right)^\epsilon\dfrac{\Gamma(2+\beta)\Gamma(\beta+\epsilon)}{\Gamma(2+\beta-\epsilon)\Gamma(\beta+1)}z^{\beta+\epsilon+1}\\[1mm]
z^\beta(\phi_2-\phi_3)& =\Gamma(\epsilon)\left(\dfrac{4\pi\mu^2}{p^2} \right)^\epsilon \dfrac{\Gamma(1-\epsilon)}{\Gamma(1+\epsilon)}\left[\dfrac{\Gamma(1+\beta)\Gamma(\beta+\epsilon)}{\Gamma(1+\beta-\epsilon)\Gamma(\beta)}-\dfrac{\Gamma(2+\beta)\Gamma(\beta+\epsilon)}{\Gamma(2+\beta-\epsilon)\Gamma(\beta)} \right]z^{\beta+\epsilon}\nonumber\\[1mm]
& =-\Gamma(1-\epsilon)\left(\dfrac{4\pi\mu^2}{p^2} \right)^\epsilon\dfrac{\Gamma(1+\beta)\Gamma(\beta+\epsilon)}{\Gamma(2+\beta-\epsilon)\Gamma(\beta)}z^{\beta+\epsilon}.
\end{align}
\end{subequations}
Up until now all terms on the right-hand side of Eq.~\eqref{eq:D_xi_Omega_Delta} are explicit. For the left-hand side, we have
\begin{subequations}\label{eq:zbeta_Omega_Delta}
\begin{align}
z^\beta\partial_\nu \Omega_{ij}^\Delta=\sum_{m=0}^{+\infty}&\dfrac{(-\nu)^m}{m!}\left(\dfrac{4\pi\mu^2}{p^2}\right)^{(m+1)\epsilon}[-\omega_{ij}(\beta,m+1)]z^{\beta+(m+1)\epsilon}\quad (i,j)\neq (1,2),\\[1mm]
z^\beta\partial_\nu \Omega_{12}^\Delta&=\sum_{m=0}^{+\infty}\dfrac{(-\nu)^m}{m!}\left(\dfrac{4\pi\mu^2}{p^2} \right)^{(m+1)\epsilon}\left[-\omega_{12}(\beta,m+1)\right]z^{1+\beta+(m+1)\epsilon}\\[1mm]
z^\beta(\phi_3\Omega_{11}^\Delta-\Omega_{11}^\Delta\phi_3)& =\sum_{m=0}^{+\infty}\dfrac{(-\nu)^m}{m!}\left(\dfrac{4\pi\mu^2}{p^2} \right)^{(m+1)\epsilon}\Gamma(\epsilon)\dfrac{\Gamma(1-\epsilon)}{\Gamma(1+\epsilon)}\Bigg\{\dfrac{\Gamma(2+\beta)\Gamma(\beta+\epsilon)}{\Gamma(2+\beta-\epsilon)\Gamma(\beta)}\omega_{11}(\beta+\epsilon,m)\nonumber\\[1mm]
& \quad -\omega_{11}(\beta,m)\dfrac{\Gamma(2+\beta+m\epsilon)\Gamma(\beta+(m+1)\epsilon)}{\Gamma(2+\beta+(m-1)\epsilon)\Gamma(\beta+m\epsilon)} \Bigg\}z^{\beta+(m+1)\epsilon}\\[1mm]
z^\beta(\phi_3\Omega_{12}^\Delta-\Omega_{12}^\Delta\phi_2)& =\sum_{m=0}^{+\infty}\dfrac{(-\nu)^m}{m!}\left(\dfrac{4\pi\mu^2}{p^2} \right)^{(m+1)\epsilon}\Gamma(\epsilon)\dfrac{\Gamma(1-\epsilon)}{\Gamma(1+\epsilon)}\Bigg\{\dfrac{\Gamma(2+\beta)\Gamma(\beta+\epsilon)}{\Gamma(2+\beta-\epsilon)\Gamma(\beta)}\omega_{12}(\beta+\epsilon,m)\nonumber\\[1mm]
& \quad -\omega_{12}(\beta,m)\dfrac{\Gamma(2+\beta+m\epsilon)\Gamma(\beta+1+(m+1)\epsilon)}{\Gamma(2+\beta+(m-1)\epsilon)\Gamma(\beta+1+m\epsilon)}\Bigg\}z^{1+\beta+(m+1)\epsilon}\\[1mm]
z^\beta(\phi_2\Omega_{21}^\Delta-\Omega_{21}^\Delta\phi_3) &=\sum_{m=0}^{+\infty}\dfrac{(-\nu)^m}{m!}\left(\dfrac{4\pi\mu^2}{p^2} \right)^{(m+1)\epsilon}\Gamma(\epsilon)\dfrac{\Gamma(1-\epsilon)}{\Gamma(1+\epsilon)}\Bigg\{\dfrac{\Gamma(1+\beta)\Gamma(\beta+\epsilon)}{\Gamma(1+\beta-\epsilon)\Gamma(\beta)}\omega_{21}(\beta+\epsilon,m)\nonumber\\[1mm]
& \quad -\omega_{21}(\beta,m)\dfrac{\Gamma(2+\beta+m\epsilon)\Gamma(\beta+(m+1)\epsilon)}{\Gamma(2+\beta+(m-1)\epsilon)\Gamma(\beta+m\epsilon)}\Bigg\}z^{\beta+(m+1)\epsilon}\\[1mm]
z^\beta(\phi_2\Omega_{22}^\Delta-\Omega_{22}^\Delta\phi_2)& =\sum_{m=0}^{+\infty}\dfrac{(-\nu)^m}{m!}\left(\dfrac{4\pi\mu^2}{p^2} \right)^{(m+1)\epsilon}\Gamma(\epsilon)\dfrac{\Gamma(1-\epsilon)}{\Gamma(1+\epsilon)}\Bigg\{\dfrac{\Gamma(1+\beta)\Gamma(\beta+\epsilon)}{\Gamma(1+\beta-\epsilon)\Gamma(\beta)}\omega_{22}(\beta+\epsilon,m)\nonumber\\[1mm]
& \quad -\omega_{22}(\beta,m)\dfrac{\Gamma(1+\beta+m\epsilon)\Gamma(\beta+(m+1)\epsilon)}{\Gamma(1+\beta+(m-1)\epsilon)\Gamma(\beta+m\epsilon)}\Bigg\}z^{\beta+(m+1)\epsilon}.
\end{align}
\end{subequations}
	\addcontentsline{toc}{chapter}{Bibliography}
	\bibliographystyle{unsrt}
	\bibliography{SDE_QEDsp_bib}
\end{document}